\RequirePackage[l2tabu, orthodox]{nag} 

\documentclass[british,a4paper,11pt,twoside]{StyleThese}

\newcommand{\uzl}{University Hospital Leuven}
\newcommand{\uzlshort}{UHL}

\newcommand{\vienna}{Medical University of Vienna}
\newcommand{\viennashort}{MUV}

\newcommand{\kcl}{King's College London}
\newcommand{\kclshort}{KCL}

\newcommand{\uclh}{University College London Hospital}
\newcommand{\uclhshort}{UCLH}

\newcommand{\manchester}{Manchester}
\newcommand{\manchestershort}{MCT}

\newcommand{\belfast}{Belfast}
\newcommand{\belfastshort}{BFT}

\newcommand{\cork}{Cork}
\newcommand{\corkshort}{CRK}

\newcommand{\newcastle}{Newcastle}
\newcommand{\newcastleshort}{NCS}

\newcommand{\liverpool}{Liverpool}
\newcommand{\liverpoolshort}{LVP}

\newcommand{\numraters}{$8$}  % include people supervising scoring
\newcommand{\numscoring}{$4$}
\def\TWAI{\textrm{TWAI}}
\def\AI{\textrm{AI}}
\def\fallback{\textrm{fallback}}
\def\failsafe{\textrm{fail-safe}}
\def\anatomy{\textrm{anatomy}}
\def\intensity{\textrm{intensity}}

\newcommand{\codetwai}{\href{https://github.com/LucasFidon/trustworthy-ai-fetal-brain-segmentation}{trustworthy-ai-fetal-brain-segmentation}}
\usepackage{amsmath,amsfonts,bm}
\usepackage{amsthm}
\usepackage{amssymb}
\usepackage{dsfont}
\usepackage{mathtools}
\usepackage{physics}
\usepackage{algorithm}
\usepackage[noend]{algpseudocode}
\usepackage{accents}
\usepackage{upgreek}

\newcommand{\vardbtilde}[1]{\tilde{\raisebox{0pt}[0.85\height]{$\tilde{#1}$}}}
\def\pset{{P\left(\mathbf{L}\right)^N}}
\def\gset{{\left(2^{\mathbf{L}}\right)^N}}
\newcommand{\de}{{\beta}}
\newtheorem{theorem}{Theorem}[section]
\newtheorem{lemma}[theorem]{Lemma}

\newtheorem{assumption}{Assumption}[section]
\newtheorem{definition}{Definition}[section]
\newtheorem{example}{Example}[section]

\DeclareMathOperator*{\gd}{\mathfrak{d}}
\DeclareMathOperator*{\cL}{\mathcal{L}}

\def\Figref#1{Figure~\ref{#1}}

\def\secref#1{section~\ref{#1}}
\def\chapref#1{chapter~\ref{#1}}
\def\Chapref#1{Chapter~\ref{#1}}
\def\Algref#1{Algorithm~\ref{#1}}

\def\1{\bm{1}}

\def\rx{{\textnormal{x}}}
\def\ry{{\textnormal{y}}}

\def\vtheta{{\bm{\theta}}}
\def\va{{\bm{a}}}

\def\vg{{\bm{g}}}
\def\vh{{\bm{h}}}

\def\vL{{\bm{L}}}

\def\vp{{\bm{p}}}
\def\vq{{\bm{q}}}

\def\vv{{\bm{v}}}

\def\vx{{\bm{x}}}
\def\vy{{\bm{y}}}
\def\vz{{\bm{z}}}

\DeclareMathAlphabet{\mathsfit}{\encodingdefault}{\sfdefault}{m}{sl}
\SetMathAlphabet{\mathsfit}{bold}{\encodingdefault}{\sfdefault}{bx}{n}

\def\sR{{\mathbb{R}}}

\newcommand{\E}{\mathbb{E}}
\newcommand{\V}{\mathbb{V}}

\newcommand{\softmax}{\mathrm{softmax}}
\newcommand{\relu}{\mathrm{ReLU}}

\newcommand{\KL}{D_{\mathrm{KL}}}

\DeclareMathOperator*{\argmax}{arg\,max}
\DeclareMathOperator*{\argmin}{arg\,min}

\usepackage{graphicx}
\usepackage[dvipsnames]{xcolor}
\usepackage{tabularx,booktabs,subcaption}

\newcolumntype{Y}{>{\centering\arraybackslash}X}
\newcolumntype{C}[1]{>{\centering\let\newline\\\arraybackslash\hspace{0pt}}m{#1}}

\makeatletter
\title{Trustworthy Deep Learning for Medical Image Segmentation}\let\thetitle\@title
\author{Lucas Fidon}\let\theauthor\@author
\makeatother

\usepackage{ifpdf}
\usepackage{bibentry}
\nobibliography*
\usepackage{amsmath} % load before txfonts
\usepackage{amssymb}

\usepackage[utf8]{inputenc}

\usepackage[T1]{fontenc}

\usepackage{textcomp}

\usepackage{csquotes}  % For quotes with \enquote{}
\usepackage{fontawesome}

\usepackage{placeins} % for \FloatBarrier
\usepackage{graphicx}

\usepackage{multirow}

\usepackage{diagbox}

\usepackage{xspace}
\makeatletter
\DeclareRobustCommand\onedot{\futurelet\@let@token\@onedot}
\def\@onedot{\ifx\@let@token.\else.\null\fi\xspace}

\makeatother

\providecommand{\Figref}[1]{Figure~\ref{#1}}
\providecommand{\secref}[1]{Section~\ref{#1}}

\providecommand{\chapref}[1]{Chapter~\ref{#1}}
\providecommand{\Chapref}[1]{Chapter~\ref{#1}}

\providecommand{\Algref}[1]{Algorithm~\ref{#1}}

\usepackage[left=1.5in,right=1.3in,top=1.1in,bottom=1.1in,includefoot,includehead,headheight=13.6pt]{geometry}
\usepackage{setspace}
\onehalfspace

\providecommand{\draftnote}{}

\usepackage{fancyhdr}
\makeatletter

\def\chaptermark#1{%
 \markboth {\MakeUppercase{%
 \ifnum \c@secnumdepth >\m@ne
  \if@mainmatter
   \@chapapp\ \thechapter. \ %
  \fi
 \fi
#1}}
{\ifnum \c@secnumdepth >\m@ne
  \if@mainmatter
   \@chapapp\ \thechapter. \ %
  \fi
 \fi
#1}}%

\if@twoside
 \fancypagestyle{plain}{
   \fancyhf{} % clear all header and footer fields
   \fancyfoot[C]{\bfseries\thepage}
   \fancyfoot[LE,RO]{\draftnote}
   
 }

 \fancypagestyle{empty}{
   \fancyhf{} % clear all header and footer fields
   \fancyfoot[LE,RO]{\draftnote}
   
 }
\else
 \fancypagestyle{plain}{
   \fancyhf{} % clear all header and footer fields
   \fancyfoot[C]{\bfseries\thepage}
   \fancyfoot[R]{\draftnote}
   
 }
 \fancypagestyle{empty}{
   \fancyhf{} % clear all header and footer fields
   \fancyfoot[R]{\draftnote}
   
 }
\fi
\makeatother

\makeatletter
\def\cleardoublepage{\clearpage\if@twoside \ifodd\c@page\else%
  \hbox{}%
  \thispagestyle{empty}%              % Empty header styles
  \newpage%
  \if@twocolumn\hbox{}\newpage\fi\fi\fi}
\makeatother

\newcommand{\copyrightpage}[2]{
\thispagestyle{empty}%              % Empty header styles
~\\
\vfill
\begin{center}
 {\copyright}#1\\
 #2\\
 All Rights Reserved
\end{center}
\clearpage
}

\newcommand{\declarationpage}[1]{
\thispagestyle{plain}%
I, #1, confirm that the work presented in this thesis is my own. Where
information has been derived from other sources, I confirm that this
has been indicated in the thesis.
\cleardoublepage
}

\usepackage{fancybox}
\usepackage{wallpaper}

\usepackage[nohints]{minitoc}

\mtcsettitle{minitoc}{\contentsname}

\usepackage{lmodern}
\usepackage{color}
\definecolor{linkcol}{rgb}{0,0,0}
\definecolor{citecol}{rgb}{0,0,0}

\ifpdf
 \usepackage[backref=page,pdfusetitle,pdfpagelabels=true]
{hyperref}
\else
 \usepackage[backref=page,dvips,pdfusetitle,pdfpagelabels=true]
{hyperref}
\fi

\hypersetup
{
bookmarksdepth=3,
bookmarksopenlevel=0,
bookmarksnumbered=true,
draft=false, % force to use hyperref even in draft mode
plainpages=false,
pdfhighlight=/O, %effet d'un clic sur un lien hypertexte
colorlinks=true, %couleurs sur les liens hypertextes
pdfdisplaydoctitle=true,
pdflang=en-US, % Does not seem to work but does not harm
pdfpagemode=UseOutlines,
pdfstartview=FitH,
linkcolor=linkcol, %couleur des liens hypertextes internes
citecolor=citecol, %couleur des liens pour les citations
urlcolor=linkcol, %couleur des liens pour les url
pdfkeywords={keyword 1, keyword 2},
pdfsubject={PhD Thesis}
}

\begin{document}

\frontmatter

\begin{titlepage}
\pdfbookmark[0]{Tile page}{titlepage} % Sets a PDF bookmark

\pagestyle{empty}

\ThisURCornerWallPaper{0.25}{figs/kcllogo}

\vspace*{1.5cm}

\begin{center}
	{\LARGE{\theauthor}\par}
\end{center}
\vspace{0.6cm}
\begin{center}
        {\huge\baselineskip=0.95em plus 1pt \expandafter{
        \textbf{\thetitle}
        \par}}
\end{center}

\vspace{3cm}

\begin{center}
	\LARGE{\expandafter{\textrm{King's College London}}}\par
	\expandafter{\Large{School of Biomedical Engineering \& Imaging Sciences}\par}
\end{center}

\vspace{1.5cm}

\begin{center}
	A dissertation submitted in partial fulfilment 
	\\
	of the requirements for the degree of
	\\ 
	\textbf{Doctor of Philosophy}	
\end{center}

\vspace{0.2cm}
\begin{center}
	\today
\end{center}

\vspace{1.0cm}

\begin{center}
\begin{tabular}{l p{3.3cm} l l}
	
Technical Supervisor & & Technical Co-Supervisor \\
\textbf{Pr. Tom Vercauteren} & & \textbf{Pr. S\'ebastien Ourselin}  \\
 & & \\
Clinical Supervisor & & \\
\textbf{Pr. Jan Deprest} & & 
	
\end{tabular}
\end{center}

\end{titlepage} 

\dominitoc

\copyrightpage{\the\year}{\theauthor}
\declarationpage{\theauthor}

\cleardoublepage
\chapter{Abstract}

Despite the recent success of deep learning methods at achieving new state-of-the-art accuracy for medical image segmentation, 
some major limitations are still restricting their deployment into clinics.
One major limitation
of deep learning-based segmentation methods
is their lack of robustness to variability in the image acquisition protocol and in the imaged anatomy that were not represented or were underrepresented in the training dataset.
This suggests 
% a solution consisting in 
adding new manually segmented images to the training dataset to better cover the image variability.
However, in most cases, the manual segmentation of medical images requires highly skilled raters and is time-consuming, making this solution prohibitively expensive.
Even when manually segmented images from different sources are available, they are rarely annotated for exactly the same regions of interest. 
This poses an additional challenge for current state-of-the-art deep learning segmentation methods that rely on supervised learning and therefore require all the regions of interest to be segmented for all the images to be used for training.

This thesis introduces new mathematical and optimization methods to mitigate those limitations.
Our contributions are threefold.

We introduce the mathematical framework of \emph{label-set loss functions}. This is an infinitely large family of loss functions that can be used to train a deep neural network using partially annotated images, that is images for which some but not necessarily all the regions of interest were segmented or images for which the segmentation of some of the regions of interest were grouped.
We propose a method for converting any existing loss function into a label-set loss function and in addition we propose a new label-set loss function that we found to lead to improve segmentation accuracy.

We introduce an optimization algorithm for training deep neural networks using distributionally robust optimization (DRO). We propose to use DRO to improve the generalization of deep neural networks after training on underrepresented populations in the training dataset.
We formally prove the convergence of the algorithm when applied to deep neural networks.
We also give mathematical insights into the connection between DRO and the maximization of the worst-case performance.

We propose an approach for trustworthy deep learning for medical image segmentation based on Dempster-Shaffer theory.
Trustworthy AI is a concept that has been increasingly discussed in sociology and in guidelines for the safe deployment of emerging technologies such as the deep learning methods used for medical image segmentation.
We present the first concrete mathematical framework and implementation of trustworthy AI for medical image segmentation.
Our method aims at making a backbone deep learning segmentation algorithm trustworthy with respect to criteria, called \emph{contracts of trust}, that are derived from expert knowledge and correspond to expectations of radiologists expert in the segmentation task at hand.

We illustrate and evaluate the proposed methods on several segmentation tasks with a focus on fetal brain 3D T2w MRI segmentation.
The segmentation of fetal brain MRI is essential for the study of normal and abnormal fetal brain development. Reliable analysis and evaluation of fetal brain structures could also support diagnosis of central nervous system pathology, patient selection for fetal surgery, evaluation and prediction of outcome, hence also parental counselling. 
Fetal brain 3D T2w MRI segmentation presents multiple challenges that align with the limitations of current deep learning algorithms that this thesis aims to mitigate.
The manually segmented fetal brain 3D T2w MRIs available are scarce and their level of annotations varies.
There are variations in T2w MRI protocols used across clinical centers.
Last but not least, there is a spectacular variation of the fetal brain anatomy across gestational ages and across normal and abnormal fetal brain anatomy.

% \cleardoublepage
% \chapter{Impact statement}
% An impact statement is often required.

\cleardoublepage
\chapter{Acknowledgments}
First of all, I would like to thank my supervisors Tom Vercauteren, S\'ebastien Ourselin, and Jan Deprest for their unconditioned and consistent support since the first day we met.
I wish to thank Tom Vercauteren for our weekly discussions about science that were particularly fruitful,
and Michael Aertsen for teaching me so much about neuroanatomy and the development of the fetal brain.

I would like to thank all my colleagues and friends from London, Leuven, Vienna, Munich, and all the TRABIT family.
I also would like to acknowledge the European Union for providing the funding for my PhD project under the Marie Sk{\l}odowska-Curie grant agreement TRABIT No 765148.

I am thankful to my parents, Alain and Marianne, and to my siblings, Alexia, Camille, Julia, and Pierre, for letting me follow my passion for science.
I used to organize scientific conferences about dinosaurs for them when I was $5$ years old.
Little did we know that $23$ years later I would defend a PhD about artificial intelligence applied to medical image segmentation.

\cleardoublepage
\chapter{List of Publications}
\section*{First-author Articles in International\\ Peer-reviewed Journals} % (fold)
\label{sec:first_journal_articles}
\begin{enumerate}
    \item \bibentry{fidon2021atlas}.
    \item \bibentry{fidon2020distributionally}.
    \item \bibentry{fidon2022trustworthy}.
\end{enumerate}

\section*{First-author Articles in International\\ Peer-reviewed Conferences with Published Proceedings}
\label{sec:first_peer_reviewed_conference_papers}
\begin{enumerate}
    \item \bibentry{fidon2021label}.
    \item \bibentry{fidon2019incompressible}.
\end{enumerate}

\section*{First-author Articles in International\\ Peer-reviewed Workshops with Published Proceedings}
\label{sec:first_peer_reviewed_workshop_papers}
\begin{enumerate}
    \item \bibentry{fidon2020generalized}. \textbf{4th Place of the BraTS 2020 challenge}.
    \item \bibentry{fidon2021generalized}.
    \item \bibentry{fidon2021distributionally}.
    \item \bibentry{fidon2021partial}.
\end{enumerate}

\section*{Co-author Articles in International Peer-reviewed Journals} % (fold)
\label{sec:journal_articles}
\begin{enumerate}
    \item \bibentry{deprest2022application}.
    \item \bibentry{emam2021longitudinal}.
    \item \bibentry{mufti2021cortical}.
    \item \bibentry{mufti2023}.
    \item \bibentry{garcia2021image}.
    \item \bibentry{kofler2021robust}.
    \item \bibentry{payette2022fetal}.
    \item \bibentry{eisenmann2022biomedical}.
\end{enumerate}

\section*{Co-author Articles in International\\ Peer-reviewed Conferences with Published Proceedings}
\label{sec:peer_reviewed_conference_papers}
\begin{enumerate}    
    \item \bibentry{asad2022econet}.
    \item \bibentry{kofler2022blob}.
\end{enumerate}

\section*{Co-author Articles in International\\ Peer-reviewed Workshops with Published Proceedings}
\label{sec:peer_reviewed_workshop_papers}
\begin{enumerate}
    \item \bibentry{irzan2020min}.
    \item \bibentry{kofler2022deep}.
\end{enumerate}

\cleardoublepage
\tableofcontents

% KCL and UCL want a list of tables and figures
\clearpage
\listoftables
\clearpage
\listoffigures

\mainmatter

\chapter{Introduction}
\label{chap:intro}
\minitoc

\begin{center}
\begin{minipage}[b]{0.9\linewidth}
\small
\textbf{Foreword\,}
In this introductory chapter, the background on deep learning for medical image segmentation has used elements from \cite{fidon2020generalized}. The review of related work in medical image segmentation using deep learning has been extended and additional references have been added.
The background on trustworthy deep learning has used elements from \cite{fidon2021distributionally,fidon2022trustworthy,fidon2020distributionally}.
\end{minipage}
\end{center}

\section{Background on Deep Learning for Medical Image Segmentation}

Medical image segmentation is the extraction of regions of interest (ROIs) from images, e.g. Magnetic Resonance Imaging (MRI).
Segmentation is required for the computation of volume and surface-based biomarkers and therefore aims at the quantification of the anatomy.
It has many applications in clinical research and clinical practice.
Examples of applications of segmentation in fetal brain MRI can be found in \Chapref{chap:fetaldataset}.
Manual medical image segmentation is time-consuming, requires highly skilled raters, and is subject to inter-rater and intra-rater variability.
Automatic methods for medical image segmentation methods are therefore required.

Formally, automatic medical image segmentation consists in finding a function $h$, computable by a computer
\begin{equation}
\begin{aligned}
    h:\,\, \mathds{R}^{N\times M}& \xrightarrow{} \mathbf{L}^N\\
        \vx &\mapsto \vy
\end{aligned}
\end{equation}
where the variable $\vx$ corresponds to an image or volume, with $M$ imaging modalities, to be segmented, $N$ is the number of voxels in each image, the variable $\vy$ is the automatic segmentation computed by $h$, and $\mathbf{L}$ is a finite set of labels associated with the regions of interest to be segmented.

Deep learning for medical image segmentation is the field of research that studies automatic medical image segmentation in the case $h$ is a deep neural network.
This approach is the current state of the art in automatic medical image segmentation~\cite{isensee2021nnu}.
% In medical image segmentation using deep learning, $h$ is a deep neural network with a set of parameters $\vtheta$ that typically consists of millions of parameters.
A deep neural network $h$ is parameterized by a set of parameters $\vtheta$ that typically consists of millions of parameters.
So as to produce relevant and accurate automatic segmentations, the parameters $\vtheta$ of the deep neural network needs to be tuned automatically.
This is performed during the \textit{training} process of the deep neural network.

The ideal optimization problem that the training process aims to solve is
\begin{equation}
    \label{eq:intro-training_ideal}
    \vtheta^* =
    \argmin_{\vtheta \in \Theta} \E_{(\vx,\vy) \sim \mathcal{X}\times \mathcal{Y}}
    \left[\cL \left(h(\vx;\vtheta), \vy\right)\right]
\end{equation}
where $\Theta$ is the space of parameters of the deep neural network $h$, $\mathcal{X}\times \mathcal{Y}$ is the joint probability distribution of all possible pairs $(\vx,\vy)$ of input image or volume $\vx$ and its associated segmentation $\vy$, and $\cL$ is a smooth per-volume loss function, i.e. a function that evaluates how good is the fit between any $h(\vx;\vtheta)$ and $\vy$ for $(\vx,\vy) \sim \mathcal{X}\times \mathcal{Y}$ with lower value of the per-volume loss function indicating a better fit.

However, in practice it is impossible compute the expectation in \eqref{eq:intro-training_ideal}, or to draw samples exactly from the joint distribution $\mathcal{X}\times \mathcal{Y}$, or even to have access to all the possible pairs $(\vx,\vy)$ with a non-zeros probability under the joint probability $\mathcal{X}\times \mathcal{Y}$.
Instead the pairs $(\vx,\vy)$ available for training are limited to a predefined finite set of pairs $\left\{(\vx_i, \vy_i)\right\}_{i=1}^n$ where the $\vx_i$ are images or volumes that were acquired at one or a few acquisition centers and the segmentations $\vy_i$ have been manually segmented by experts to approximate the true segmentations.
The set $\left\{(\vx_i, \vy_i)\right\}_{i=1}^n$ is called the \textit{training dataset}.
In current state-of-the-art deep learning pipelines for medical image segmentation, the ideal training optimization problem \eqref{eq:intro-training_ideal} is therefore approximated by the empirical risk minimization (ERM) problem to which regularization terms can be added
\begin{equation}
    \label{eq:intro-erm}
    \vtheta^*_{ERM}
    = \argmin_{\vtheta \in \Theta} \frac{1}{n} \sum_{i=1}^n \cL \left(h(\vx_i;\vtheta), \vy_i\right)
\end{equation}
% where $h$ is a deep neural network with parameters $\vtheta$, $\cL$ is a smooth per-volume loss function, and $\left\{(\vx_i, \vy_i)\right\}_{i=1}^n$ is the training dataset. 
% $\vx_i$ are the input images or volumes, and $\vy_i$ are the ground-truth manual segmentations.

\begin{figure}[t]
    \centering
    \includegraphics[width=\linewidth]{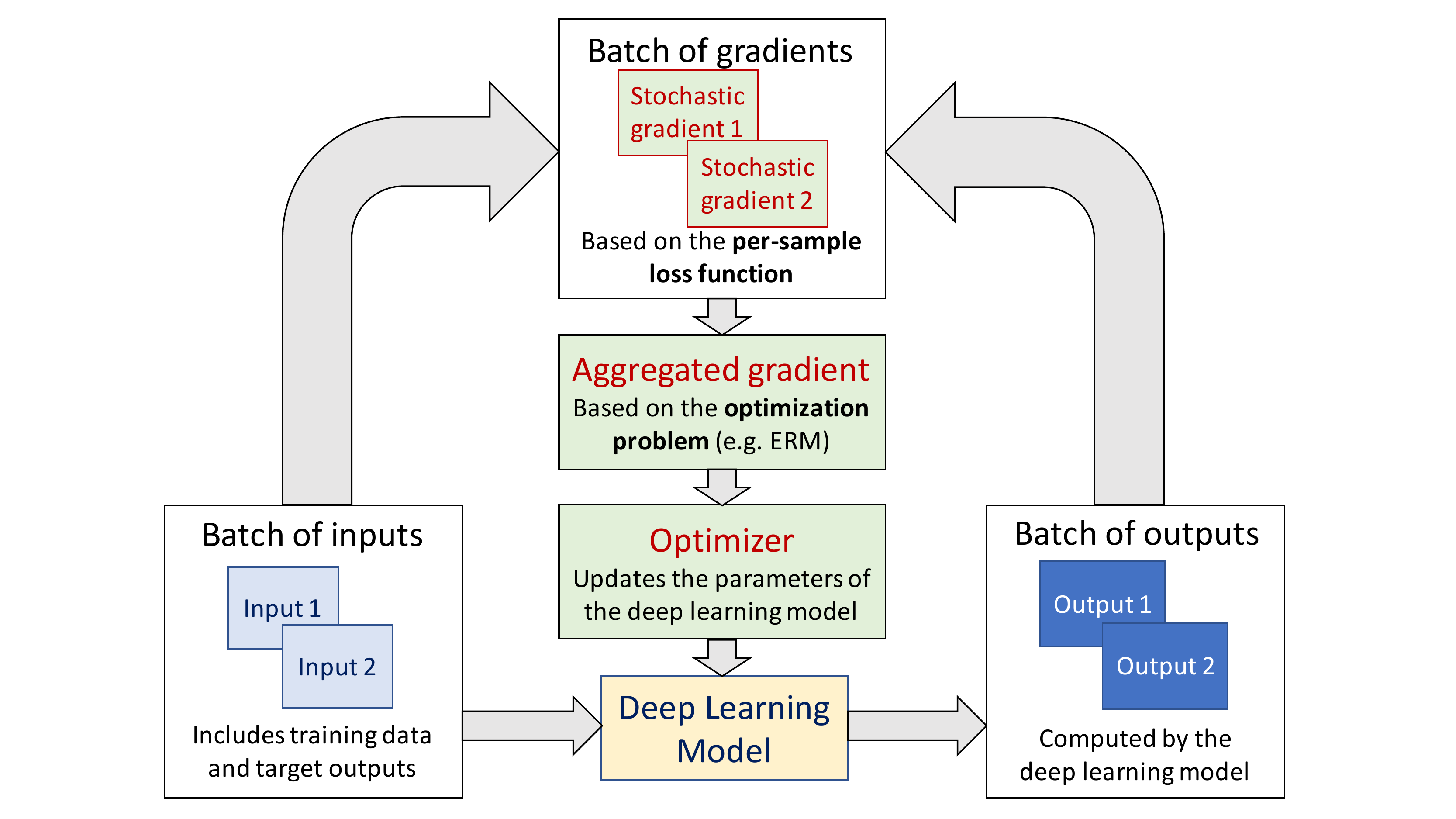}
    \caption{
    \label{fig:diagram_deep_pipeline}
    \textbf{Diagram of a deep learning optimization pipeline.}
    Deep learning optimization methods are made of four main components:
    1) the design of the deep neural network architecture,
    2) the \textbf{per-sample loss function} (e.g. the Dice loss) that determines the stochastic gradient,
    3) the \textbf{optimization problem} (e.g. the empirical risk) that determines how to merge the stochastic gradients into one aggregated gradient,
    4) the \textbf{optimizer} (e.g. SGD, Adam) that determines how the aggregated gradient is used to update the parameters of the deep neural network at each training iteration.
    In this work, we explore variants for the per-sample loss function, the population loss function and the optimizer for application to automatic brain tumor segmentation.
    }
\end{figure}

The main ingredients of the optimization problem \eqref{eq:intro-erm} are:
1) the deep neural network architecture for $h$,
2) the per-sample loss function $\cL$,
3) the optimization problem, here \textit{empirical risk minimization} (ERM) corresponding to the minimization of the mean of the per-sample loss functions,
and 4) the optimizer which is the algorithm that allows finding an approximation of $\vtheta^*_{ERM}$.
A graphical representation is given in Fig.~\ref{fig:diagram_deep_pipeline}.

The optimization problem can include data augmentation, as formalized in vicinal risk minimization~\cite{chapelle2000vicinal}.
In a work that I have coauthored, we have formally shown how the optimization problem can also be extended to include synthetic data~\cite{garcia2021image}.
The optimizer includes important hyper-parameters such as the batch size, the learning rate schedule, and the sampling strategy.

% CNN ARCHITECTURES
In recent years, important research efforts have been put into the development and optimization of deep neural network architectures for medical image segmentation.
Research in deep neural network architectures in the context of medical image segmentation has been reviewed in~\cite{lei2020medical,liu2021review,lundervold2019overview,siddique2021u}.
In the following I give a very brief overview of the literature of deep neural network architecture for medical image segmentation and highlight my contribution to this field.

Innovations on convolutional neural network (CNN) architectures, have led to significant improvement in segmentation accuracy~\cite{bakas2018identifying,cciccek20163d,kamnitsas2016deepmedic,Ronneberger2015,wang2017automatic}, including papers I have authored~\cite{fidon2017scalable,fidon2017generalised} and papers I have coauthored~\cite{li2017compactness,niftynet,chandra2018context,garcia2017toolnet,wang2022system}.
The development of nnU-Net ('no-new-Net')~\cite{isensee2021nnu} has shown that a well-tuned 2D U-Net~\cite{Ronneberger2015} or 3D U-Net~\cite{cciccek20163d} can achieve state-of-the-art results for a large set of medical image segmentation problems and datasets.
% TODO TRANSFORMERS
The most recent developments, there has been several attempts to include transformer layers~\cite{vaswani2017attention} to deep neural network architectures for medical image segmentation~\cite{chen2021transunet,hatamizadeh2022unetr,wang2021transbts,xie2021cotr}.
The nnFormer ('not-another transFormer')~\cite{zhou2021nnformer} has been found to outperform previous transformer-based architectures on three segmentation tasks.
It is also the first transformer-based architecture to compete with nnU-Net in terms of segmentation accuracy.
However, nnFormer does not significantly outperform nnU-Net~\cite{zhou2021nnformer}.
In my participation to the BraTS challenge 2021~\cite{fidon2021generalized}, we have experimented with a transformer-based architecture, TransUNet~\cite{chen2021transunet}, and found that it leads to lower Dice scores and higher Hausdorff distances than the 3D U-Net~\cite{cciccek20163d} used in nnU-Net~\cite{isensee2021nnu} on the BraTS challenge 2021 dataset.
This is consistent with the observation that transformer-based architectures have not yet achieved a new state-of-the-art in medical image segmentation.
In computer vision, recent results also suggest that pure convolutional neural networks can also outperform the most recent transformer-based architectures in this field~\cite{liu2022convnet}.
The 2D U-Net and 3D U-Net used in the current state-of-the-art pipeline nnU-Net were among the first convolutional neural network architectures proposed for medical image segmentation.
This suggests that the improvement that the design of the deep neural network can still bring to medical image segmentation is more limited than what was previously thought.
% 
% ADD LITERATURE REVIEW ABOUT LOSS FUNCTION???

In contrast to the deep neural networks architecture, relatively little attention has been paid to the other aspects of the deep learning optimization problems for medical image segmentation \eqref{eq:intro-erm}, namely the per-sample loss function, the population loss function, and the optimizer.
The recent advances in per-sample loss functions have been reviewed in~\cite{ma2021loss}.
In my research work prior to my PhD, I have contributed the generalized Wasserstein Dice loss~\cite{fidon2017generalised} which is a per-sample loss function that allows taking advantage of the hierarchical structure of the regions of interest during training.

Another relevant research topic in medical image segmentation is the study of the gap between the ideal training problem \eqref{eq:intro-training_ideal} and the surrogate empirical risk minimization problem \eqref{eq:intro-erm}.
We introduce this topic in the next section.

% I hypothesize that innovation regarding those three aspects can help improving the robustness and segmentation accuracy of current state-of-the-art deep neural network architectures.
% % 
% 
% % 
% In this thesis, I propose a new family of per-sample loss function, the \textit{label-set loss functions} in \Chapref{chap:partialsup} that are robust to missing regions in the manual segmentations of the training dataset.
% % 
% In \Chapref{chap:dro}, 
% % 
% The 3D U-Net~\cite{cciccek20163d}, as implemented in nnU-Net~\cite{isensee2021nnu}, will be used in all the deep learning-based pipelines for medical image segmentation in this thesis.

\section[Trustworthy Deep Learning]{Deep Learning for Medical Image Segmentation is not Trustworthy Yet}

Deep learning for medical image segmentation can reach super-human accuracy on average~\cite{isensee2021nnu} and yet most radiologists do not trust them~\cite{allen20212020,cabitza2019biases}.
This is partly because, for some cases, deep learning algorithms fail spectacularly with errors that violate expert knowledge about the segmentation task when the deep learning was applied across imaging protocol and anatomical pathologies~\cite{allen20212020,fidon2021distributionally,gonzalez2021detecting}.
This sense of distrust is exacerbated by the current lack of clear fit-for-purpose regulatory requirements for deep learning-based medical image software~\cite{van2021artificial}.
The legal framework for the deployment in clinics of deep learning tools for medical segmentation is likely to soon become more stringent once the European Union has proposed its Artificial Intelligence Act to regulate AI and in particular deep learning~\cite{regulations2021}.
A key requirement for AI algorithms for medical image segmentation prior to their deployment on the EU market will be to trustworthy~\cite{ethics,regulations2021}.

One of the possible causes of the lack of robustness of the current state-of-the-art deep learning methods for medical image segmentation is the gap between the ideal training problem \eqref{eq:intro-training_ideal} and the surrogate empirical risk minimization (ERM) problem \eqref{eq:intro-erm} used for training deep learning models in practice.
Datasets used to train deep neural networks typically contain some underrepresented subsets of cases.
These cases are not specifically dealt with by the training algorithms currently used for deep neural networks.
This problem has been referred to as hidden stratification~\cite{oakden2020hidden}.
Hidden stratification has been shown to lead to deep learning models with good average performance but poor performance on underrepresented but clinically relevant subsets of the population~\cite{larrazabal2020gender,oakden2020hidden,puyol2021fairness}.
In standard deep learning pipelines, this hidden stratification is ignored and the model is trained to minimize the mean per-example loss, which corresponds to the standard ERM problem as in \eqref{eq:intro-erm}.
As a result, models trained with ERM are more likely to underperform on those examples from the underrepresented subdomains, seen as hard examples.
This may lead to \textit{unfair} AI systems~\cite{larrazabal2020gender,puyol2021fairness}.
For example, state-of-the-art deep learning models for brain tumor segmentation (currently trained using ERM) underperform for cases with confounding effects, such as low grade gliomas, despite achieving good average and median performance~\cite{bakas2018identifying}.
For safety-critical systems, such as those used in healthcare, this greatly limits their usage as ethics guidelines of regulators such as~\cite{ethics} require AI systems to be technically robust and fair prior to their deployment in hospitals.

% It is important to be able to exploit all available data.

% But more data might not be enough.

\section{Thesis Organization and Contributions}

This thesis is organized in eight chapters.

The \textbf{current chapter} (\nameref{chap:intro}) provides a brief introduction on deep learning for medical image segmentation and trustworthiness of AI in this context.
I started from the optimization problem that the training of deep neural networks attempts to solve.
I identified four main components of this optimization: the deep neural network architecture, the per-sample loss function, the optimization problem, and the optimizer.
The first component is the one that has been the most extensively studied in the literature. I gave a brief overview of the methods that have been proposed to tailor deep neural architectures for medical image segmentation.
Last but not least, an introduction to the problem of trustworthy AI in deep learning and its importance for medical image segmentation was proposed.

In \textbf{\Chapref{chap:relatedwork}} (\nameref{chap:relatedwork}), we give the definition of trustworthy AI used in this thesis and discuss previous work on trustworthy AI and on fetal brain MRI segmentation.

In \textbf{\Chapref{chap:fetaldataset}} (\nameref{chap:fetaldataset}), we introduce the largest fetal brain 3D MRI segmentation dataset to date.
The dataset contains $540$ fetal brain 3D MRIs from $13$ hospitals and manually segmented into seven tissue types.
I discuss why trustworthiness is particularly important and challenging for this segmentation task.
Fetal brain 3D MRI segmentation is the main application studied in this thesis.
I also review the literature of automatic methods for fetal brain MRI segmentation.
Furthermore, I give a detailed description of our fetal brain MRI segmentation dataset and how we have created it.
The inter-rater variability for manual fetal brain 3D T2 weighted MRI is also evaluated.

As discussed in the current chapter, in contrast to the deep neural network architecture, the other three main components defining the training process in deep learning has been less studied in medical image segmentation.
I hypothesize that innovation regarding those three aspects can help improving the robustness and segmentation accuracy of current state-of-the-art deep neural network architectures.
In this thesis, a state-of-the-art 3D U-Net~\cite{cciccek20163d}, as implemented in nnU-Net~\cite{isensee2021nnu}, is the backbone deep neural network architecture used in all the deep learning-based pipelines for medical image segmentation.
In \Chapref{chap:partialsup} and \Chapref{chap:dro} we propose new methods regarding the per-sample loss, the optimization problem and the optimizer that can be used in \eqref{eq:intro-erm}.

% FULLY-SUP
In equation \eqref{eq:intro-erm} it is assumed assume that a full manual segmentations $\vy_i$ is available for each training sample $\vx_i$.
This assumption characterizes \textbf{fully-supervised learning}.
In \textbf{\Chapref{chap:partialsup}}  (\nameref{chap:partialsup}), we will relax this assumption and allow the training segmentations $\vy_i$ to be partial, i.e. some but not all the regions of interest may be manually segmented in $\vy_i$.
This setting corresponds to \textbf{partially supervised learning}.
Partial manual segmentations are often present in medical image segmentation where larger training datasets are obtained by pooling data from different sources where images will have been manually segmented for different regions of interest.
We found partially supervised learning to be a very useful tool while building our segmentation dataset for fetal brain 3D MRI.
In \Chapref{chap:partialsup}, we show that training with partially segmented images is possible by changing the family of possible per-sample loss functions used in \eqref{eq:intro-erm}.
We call this new family of per-sample loss functions the \textit{label-set loss functions}.

In \textbf{\Chapref{chap:dro}} (\nameref{chap:dro}), we propose an optimizer for training any deep neural network using \textbf{distributionally robust optimization} (DRO), a generalization of \textbf{empirical risk minimization} (ERM) used in \eqref{eq:intro-erm}.
ERM weights every training sample equally.
When subpopulations are underrepresented in the training dataset, as is typically the case in medical image segmentation, deep neural networks trained using ERM might underperform for new cases from those subpopulations.
DRO aims to mitigate this problem by replacing the minimization problem \eqref{eq:intro-erm} by a min-max optimization problem.
Our method consists of a new sampling strategy that we call \textit{hardness weighted sampling} and that can be used to convert an existing optimizer for ERM, e.g. Adam or SGD, into an optimizer for DRO.
We show that the hardness weighted sampling is both computationally efficient and theoretically sound.
We compared DRO to ERM on the segmentation tasks of fetal brain MRI segmentation and brain tumor segmentation using MRI.
We found that DRO leads to improvement of some of the worst-case segmentation results.

As discussed in the previous section, the lack of trustworthiness of deep neural networks trained with the optimization problem \eqref{eq:intro-erm} may be related to the gap between this optimization problem and the optimization problem \eqref{eq:intro-training_ideal}.
The methods proposed in \Chapref{chap:partialsup} and \Chapref{chap:dro} aim at improving the use of all available annotated data during the training of deep neural networks.
However, it is currently unknown what amount of data would required to train trustworthy deep neural networks and if collecting this amount of data would be feasible.
It is therefore interesting to look for other kind of approaches than increasing the amount of annotated medical images to improve the trustworthiness of deep neural networks.
In particular, expert knowledge about properties of the anatomy and properties of the image formation appears as a promising complement to annotated data.
% 
% Expert knowledge had a central role in the best performing segmentation methods developed before deep learning became the state of the art.

% In \textbf{\Chapref{chap:incompressible}} (\nameref{chap:incompressible}), we study incompressible registration to illustrate how optimization methods can provide theoretical guarantees regarding known anatomical properties of the myocardium.
% 
% We propose a constrained optimisation framework for incompressible diffeomorphic registration.
% 
% As an efficient means to parameterize discrete stationary velocity field (SVF) for this problem, we introduce multivariate divergence-conforming B-spline.
% 
% We demonstrate that the properties of divergence-conforming B-splines can be exploited to impose bounds on the divergence of the SVF over the entire continuous space using sparse linear constraints on its finite parameters.
% 
% We validated the proposed method for multi-modal incompressible registration of synthetically deformed brains and on the STACOM’11 myocardial tracking challenge dataset.
% Our incompressible registration method achieves similar registration accuracy as a state-of-the-art method while better retaining the incompressibility.

In \textbf{\Chapref{chap:atlas}} (\nameref{chap:atlas}), we present our method to compute the first spatio-temporal brain T2 weighted MRI segmentation atlas for the developing fetal brain with spina bifida aperta.
Previous fetal brain atlases were limited to population of fetuses with a normal development.
Brain MRI atlases are the state-of-the-art representation for anatomical prior in a population of normal or pathological human brains.
Spina bifida aperta (SBA), as described in \Chapref{chap:fetaldataset}, is the most common fetal brain disorder and the population of fetuses with abnormal brain development that we study the most in depth in this thesis.
We propose a protocol for the annotation of anatomical landmarks in brain 3D MRI of fetuses with SBA with the aim of making spatial alignment of abnormal fetal brain MRIs more robust. 
In addition, we propose a weighted generalized Procrustes method based on the anatomical landmarks for the initialization of the atlas. The proposed weighted generalized Procrustes can handle temporal regularization and missing annotations. After initialization, the atlas is refined iteratively using non-linear image registration based on the image intensity and the anatomical landmarks.
A semi-automatic method is used to obtain the segmentations of our fetal brain atlas.
Our evaluation shows that using the proposed fetal brain atlas for SBA for an atlas-based segmentation method leads to improve segmentation accuracy on brain MRIs of fetuses with SBA as compared to using a normal fetal brain atlas.

In \textbf{\Chapref{chap:twai}} (\nameref{chap:twai}), we propose the first general method to make deep learning algorithms for medical image segmentation trustworthy.
We propose a principled method for trustworthy AI that lies at the intersection of the mathematical theory of Dempster-Shafer and the psychological theory of human-AI trust.
Our method, in which we propose the first principled and actionable implementation of a fallback and a fail-safe mechanism for deep learning algorithms, is a direct response for medical image segmentation to the EU guidelines for trustworthy AI published in 2019 and that upheld the AI act.
Fetal brain atlases, such as the one developed in \Chapref{chap:atlas}, play a central role in the proposed implementation of our trustworthy AI framework for fetal brain MRI segmentation.
We report significant improvement on the fetal brain MRI segmentation using the largest fetal brain MRI segmentation dataset to date described in \Chapref{chap:fetaldataset}.

Finally, \textbf{\Chapref{chap:conclusion}} (\nameref{chap:conclusion}) concludes this thesis and provides avenues for future work.

\chapter[Related Work]{Related Work on Fetal Brain Segmentation \& Trustworthy AI}
\label{chap:relatedwork}
\minitoc
\begin{center}
	\begin{minipage}[b]{0.9\linewidth}
		\small
		\textbf{Foreword\,}
		This chapter presents previous work on trustworthy AI and fetal brain segmentation.
	\end{minipage}
\end{center}

\section{Trustworthy AI}
In this section, we start by giving the definition of trustworthy AI used in this thesis.
We then discuss the most related work.

\subsection{Background on Human-AI trust}\label{sec:humanAItrust}
In this section, we describe the definition of human-AI trust and trustworthiness that we use in this thesis.

\textbf{Artificial intelligence} is defined as any automation perceived by the individual using it as having an intent~\cite{jacovi2021formalizing}.

Human-AI trust is multi-dimensional.
For example, the user can trust an AI medical image segmentation algorithm for a given tissue type, for images coming from a given type of scanner, or for a given population and not another.
This observation that trust has several facets and is context-dependent~\cite{hoffman2017taxonomy} has motivated the introduction of \emph{contractual trust}~\cite{jacovi2021formalizing}.
A \textbf{contract of trust} is an attribute of the AI algorithm which, if not fulfilled, causes a risk in using the AI algorithm.
A contract of trust is not necessarily related to the accuracy of the AI algorithm for the task at hand.

For the automatic segmentation of the heart on chest CT images, one contract of trust could be:
"The heart labels are always on the left side of the body."
This contract is not directly related to the accuracy of the AI segmentation algorithm.
The AI algorithm can fulfil this contract and yet compute an inaccurate segmentation of the heart.
This contract can also be restricted to CT images of sufficient quality.
This allows to model that contract may hold in the context of CT images of sufficiently high quality but not otherwise.
Context can be added to the contract and several contracts can be derived from the contract above for different contexts.
In addition, this contract does not apply to individuals with dextrocardia (their heart is located on the right).

An AI algorithm is defined as \textbf{trustworthy} with respect to a contract of trust
if it provides guarantees that it will abide by the obligations of said contract~\cite{jacovi2021formalizing}.

The requirements proposed in the EU guidelines for trustworthy AI~\cite{ethics} are examples of contract or set of contracts of trust.
One important requirement of trustworthy AI is the technical robustness and safety~\cite{ethics} which is the focus of our work.
The EU guidelines propose to achieve trustworthiness in practice using a \emph{fallback plan}.
However, no technical means of implementing such plan have been provided or published. 
Problems with the backbone AI algorithm should be detected using a \emph{fail-safe} algorithm and a \emph{fallback} algorithm should be available.

In \Chapref{chap:twai}, we will present a theoretical framework for the implementation of a trustworthy AI system leveraging Dempster-Shafer theory to implement a failsafe fallback plan.
And we will show how our framework can be used to maintain several concrete contracts of trust for fetal brain MRI segmentation.

\subsection{Domain generalization}
Domain generalization aims at improving out-of-distribution robustness without access to a labelled or unlabeled test distribution at training time~\cite{bahmani2022semantic,wang2022generalizing}.

Trustworthy AI, as defined above, can fit into this definition and is therefore closely related to domain generalization.
However, trustworthy AI aims at improving robustness not only for out-of-distribution data but also for in-distribution data.
It depends on the contracts of trust that are considered.

Recent domain generalization methods for image segmentation are based on data augmentation during training~\cite{zhang2020generalizing}, test-time augmentation~\cite{bahmani2022semantic,choi2021robustnet}, adaptation of the batch normalization layers~\cite{bahmani2022semantic}, denoising auto-encoder~\cite{karani2021test}, specific loss functions~\cite{choi2021robustnet}, and shape prior~\cite{liu2022single}.
For a recent review on domain generalization we refer the reader to \cite{wang2022generalizing}.

The method the most related to our trustworthy AI method, described in \Chapref{chap:twai}, is the method based on shape prior of \cite{liu2022single}.
In the test-time adaptation from shape dictionary method~\cite{liu2022single},
they first train a dictionary of explicit shape priors from the source domain.
second, a regression branch is added to the segmentation network during training. The output of the regression branch are the coefficients that parameterize the shape prior. The shape prior is then concatenated with the deep features of the segmentation network before the last convolutional block.
Finally, they use consistency and data augmentation to adapt the model parameters at test time. Only the batch normalization layers are updated via one step of gradient descent.
However, The method is only evaluated for 2D binary segmentation with region of interests consisting of only one connected component per image and simple shapes.
It is unclear how such method would generalize to multiple classes and to fetal brain segmentation without replacing the shape dictionary with a fetal brain atlas as proposed in \Chapref{chap:atlas} and \Chapref{chap:twai}.

\subsection{Information fusion for trustworthy medical image segmentation}

Information fusion methods based on probability theory have been proposed to combine different segmentations~\cite{warfield2004simultaneous,welinder2010multidimensional}.
The Simulataneous Truth And Performance Level Estimation (STAPLE) algorithm weight each segmentation by estimating the sensitivity and specificity of each segmentations~\cite{warfield2004simultaneous}.
In particular, these methods define only image-wise weights to combine the segmentations.
Fusion methods with weights varying spatially have been proposed for the special case of atlas-based algorithms~\cite{cardoso2015geodesic}, but not in general as in our method.
In the context of deep learning-based segmentation methods, simple averaging is used in state-of-the-art pipelines~\cite{isensee2021nnu}.
Perhaps more importantly, fusion methods based on probability theory only cannot model imprecise or partial prior expert-knowledge~\cite{warfield2004simultaneous}.
In contrast, the use of Dempster-Shafer theory in our method, described in \Chapref{chap:twai}, allows us a larger diversity of prior knowledge that is typically robust but imprecise.
We show that our approach based on Dempster-Shafer can model prior given by either atlases or voxel intensity prior distributions in the case of fetal brain segmentation and more priors could be modelled as well in other segmentation tasks.

\section{Fetal Brain MRI Segmentation}

To tackle the inaccuracy and limited reproducibility of manual measurements in fetal brain MRI~\cite{aertsen2019reliability}, automatic methods for the segmentation of the fetal brain into a set of tissue types are necessary.

Fetal MRI data are specifically challenging to segment automatically due to several factors:
\begin{itemize}
    \item the evolution of the shape and appearance of the fetal brain across gestational ages. 
    The shape of the fetal brain surface changes during the gyrification process. The size and intensity contrast change across gestational ages and individuals during the myelination process~\cite{dubois2014early}.
    \item the strong neuro-morphology variability in fetuses with spina bifida compared to controls (Chiari II malformation, ventriculomegaly, etc...)
    \item MRI sequences used for fetuses vary a lot from one hospital to the other, and even in the same hospital it changes continuously over time
    \item the presence of motion artifacts, especially before 3D super-resolution and reconstruction (SRR)
    \item the lack of standard SRR method, and all SRR used hyper-parameters that sometimes need to be adjusted manually to get an optimal outcome. In addition there is no consensus on the metric to use to evaluate SRR methods, which means that SRR hyper-parameters cannot be optimized automatically.
\end{itemize}
As a result, methods previously validated for adult brain MRIs or even neonatal brain MRIs can fail when applied to fetal brain MRIs.

Methods for automatic fetal brain segmentation using MRI have been reviewed in ~\cite{benkarim2017toward,makropoulos2018review,rousseau2016vivo,torrents2019segmentation}.
We provide here an overview of the methods that have been proposed for the segmentation of the fetal brain into several clinically relevant tissue types.
Unlike some of those reviews, we focus on fetuses and do not include segmentation methods developed and evaluated for neonates or preterms.
There are two main groups of methods that have been proposed for fetal brain MRI segmentation: atlas-based and deep learning-based methods.

\subsection{Atlas-based fetal brain MRI segmentation methods.}
%The first methods proposed to address the problem of fetal brain MRI segmentation were atlas-based methods\cite{habas2010spatiotemporal,serag2012multi,dittrich2014spatio,gholipour2017normative}.
%
An atlas is a pair of two volumes: an image (e.g. a T2-weighted MRI) representing an average anatomy for a population and a label map (or probability map for a probabilistic atlas) associated with this image that indicates the delineation of structures in the image.
%
% A fetal brain atlas aims at approximating the average anatomy of a population.
%
In the context of MRI segmentation, atlases can be used as reference to transfer atlases' label maps to new MRI.

In \cite{habas2010atlas} the feasibility of atlas-based segmentation for fetal brain 3D MRI segmentation using a 3D atlas has been demonstrated for fetuses with a normal brain and aged between $20.5$ weeks and $22.5$ weeks of gestation at the time of MRI.
They propose to use an atlas-based segmentation method based on Expectation-Maximization (EM) based on methods previously developed for adult brain MRI segmentation~\cite{van1999automated,wells1996adaptive}.
Their method segment the cortical gray matter, the white matter, the germinal matrix and the lateral ventricles using 3D reconstructed MRI of the fetal brain.
Their evaluation on a set of $14$ normal fetal brain MRI suggests that using the segmentation atlas leads to more anatomically plausible and accurate segmentations as compared to using only the intensity and voxel neighborhood prior.

However, due to the change of the fetal brain anatomy across gestational ages, using a normative 3D atlas is not satisfactory for segmenting all the fetal brain MRI that are typically acquired from $20$ weeks to $40$ weeks of gestation.
Instead, \cite{habas2010spatiotemporal} proposed to use a spatio-temporal atlas for fetal brain segmentation.
In this case, the atlas is 4D instead of 3D and corresponds to the average growth of the fetal brain with one age-specific averaged volume per week of gestation.
Spatio-temporal atlases allow to improve the propagation of the segmentation to fetal brain MRIs by registering it to the closest age-specific volume of the 4D atlas.
They compute a spatio-temporal fetal brain MRI atlas for gestational ages $21$ weeks to $24.5$ weeks using $20$ MRI examinations of fetuses with a normal brain.
Using the same atlas-based registration algorithm as in their previous work, Habas et al found that using a spatio-temporal atlas rather than a 3D atlas as in~\cite{habas2010atlas} leads to improve segmentation accuracy~\cite{habas2010atlas}.
Spatio-temporal atlases are used by all the atlas-based fetal brain segmentation methods that followed~\cite{habas2010spatiotemporal}.

In \cite{serag2012multi}, a spatio-temporal atlas for the normal fetal brain T2 weighted MRI from $23$ weeks to $37$ weeks is proposed.
Brain hemispheres, ventricles, cortex and cerebrospinal fluid (CSF) are segmented in the atlas.
Their method to compute a fetal brain spatio-temporal atlas is based on non-rigid registration parameterized using cubic B-splines~\cite{rueckert1999nonrigid} and temporal kernel regression~\cite{davis2010population} with an adaptive width~\cite{serag2012construction}.
%
% First each image is registered to all other images; then the spatial transformation are average with weighted sum based on temporal kernels; this averaged transformation gives an estimation of the mapping of the current image to the atlas space (see LLE: Locally Linear Embedding); at last the atlas is obtained by temporal kernel regression of the images in the atlas space.
%
The dataset used to compute the atlas contains $80$ T2 MRI studies of fetuses with a normal brain development at gestational ages from $21.7$ weeks to $38.7$ weeks.
They reported qualitative and quantitative differences between age-matched spatio-temporal atlases computed using fetal and neonatal 3D T2 MRI.
However, they do not evaluate the segmentation accuracy of atlas-based fetal brain segmentation using their fetal brain atlas.
Their atlas is publicly available at \url{https://brain-development.org/brain-atlases/fetal-brain-atlases/}.

The work of \cite{gholipour2012multi} is the first to study the segmentation of abnormal fetal brain using MRI.
Specifically, they study the automatic segmentation of the lateral ventricles in fetal brain MRI of fetuses either with a normal brain or with dilated ventricles, a condition called ventriculomegaly.
They propose a new multi-atlas multi-shape method with soft constraints to avoid intersection among structures.
In this case, the atlas consists of a dataset of fetal brain MRIs with manual segmentation.
Segmentations from multiple atlas volumes are registered, propagated, and finally fused using the simultaneous truth and performance level estimation (STAPLE) algorithm~\cite{warfield2004simultaneous}.
In their experiments they use a set of $25$ fetal brain MRIs of fetuses with normal brain or ventriculomegaly and gestational age between $19$ weeks and $38$ weeks.
They find that their multi-atlas method leads to more robust segmentation of the ventricles as compared to a single atlas method.

The problem of computing a spatio-temporal atlas using a few segmented 3D MRI and a majority of non-annotated 3D MRI is studied in \cite{dittrich2014spatio}.
They propose a method for the semi-supervised learning of a spatio-temporal latent atlas of fetal brain development with segmentation for the ventricles and the cortex.
Their dataset consists of $44$ MRI, including $32$ fetuses with a normal brain and $13$ fetuses with lissencephaly, a malformation of the cortical development.
They applied their method to the computation of a spatio-temporal fetal brain atlas for gestational age from $20$ weeks to $30$ weeks.

% TODO : ADD FUSION METHOD

In later work, new spatio-temporal fetal brain T2w MRI atlases have been proposed with improved sharpness, more tissue types segmented, and more specific fetal populations~\cite{gholipour2017normative,wu2021age,fidon2021atlas}.
In \cite{gholipour2017normative}, a spatio-temporal MRI atlas of the normal fetal brain from $21$ weeks to $37$ weeks of gestation is proposed.
The hippocampi, amygdala, fornix, cerebellum, brainstem, caudate nuclei, thalami, subthalamic nuclei, lentiform nuclei, corpus callosum, lateral ventricles, developing white matter, cortical plate, and cerebrospinal fluid (CSF) are segmented in the proposed atlas.
The same segmentation protocol as in the ALBERT neonatal brain atlas~\cite{gousias2012magnetic,gousias2013magnetic} has been used as a baseline and adapted when necessary for the lowest gestational ages using~\cite{bayer2005human}.
The main components of their pipeline are symmetric diffeomorphic free-form deformation and kernel-based temporal regression like in~\cite{serag2012multi}.
% 
% They used an alternated optimization between estimation of the 4D atlas by weighted average and registration between all the volumes and the atlas.
%
Their spatio-temporal atlas has been computed using $81$ MRIs of fetuses with a normal brain and gestational age between $19$ weeks and $39$ weeks.
They show qualitatively that their atlas is sharper and better preserves anatomical details as compared to a previous fetal brain atlas.
The proposed atlas is publicly available at \url{http://crl.med.harvard.edu/research/fetal_brain_atlas/}.
Recently, a spatio-temporal fetal brain MRI atlas for the normal brain in the chinese population for gestational ages from $22$ weeks to $35$ weeks has been proposed~\cite{wu2021age}.
They used $115$ fetal brain 3D MRIs of Chinese fetuses with a normal brain development with gestational ages from $22$ weeks to $35$ weeks to compute the atlas.
Their method to compute the atlas is similar to the one used in~\cite{gholipour2017normative}.
They report quantitative anatomical differences between the proposed fetal brain atlas for the Chinese population and the fetal brain atlas~\cite{gholipour2017normative}.
In parallel, we have proposed the first spatio-temporal fetal brain MRI atlas for spina bifida aperta for gestational ages from $21$ weeks to $34$ weeks~\cite{fidon2021atlas}.
The atlas has been computed using $90$ fetal brain 3D MRIs of operated and non-operated fetuses with spina bifida aperta and gestational ages from $21$ weeks to $35$ weeks.
This is the first fetal brain atlas for a population with abnormal brains.
Our experiments show that using the proposed atlas rather than the normal brain atlas~\cite{gholipour2017normative} in an atlas-based segmentation method results in a higher segmentation accuracy for fetal brain MRI of fetuses with spina bifida aperta.
The description of our method and experiments can be found in \Chapref{chap:atlas}.

\subsection{Deep learning-based fetal brain MRI segmentation methods.}
% DEEP LEARNING
We now give a brief review of the methods for fetal brain MRI segmentation based on deep learning.
An introduction to deep learning for medical image segmentation can be found in \Chapref{chap:intro}.

Recently, the first methods using Convolutional Neural Networks (CNNs) for fetal brain MRI segmentation have been proposed~\cite{khalili2019automatic,payette2019longitudinal}.
In \cite{khalili2019automatic}, the fetal brain is segmented into seven tissue types: the cerebelum, the basal ganglia, the basal thalami, the ventricular system, the white matter, the brainstem, the cortical gray matter, and the extracerebelar cerebrospina fluid.
The motivation of~\cite{khalili2019automatic} for using a CNN for fetal brain MRI segmentation is that compared to atlas-based methods requiring intensity inhomogeneity correction and complex preprocessing, CNNs can learn to be invariant to these inhomogeneities.
Their CNN-based method segment the fetal brain directly in the low resolution 2D MRI slices and does not require to use a super-resolution and reconstruction algorithm prior to the automatic segmentation.
The main technical contribution of this work is a data augmentation method based on simulated intensity inhomogeneity artifacts for training the CNNs.
Similarly to~\cite{wang2017automatic}, they use a cascade of two 2D Convolutional Neural Networks (CNNs).
The first CNN segments the intra-cranial volume in the original 2D slices.
The second CNN takes as input a smaller image cropped around the intra-cranial volume segmented by the first CNN and segment the brain into seven tissue types.
The two CNNs are based on the 2D U-net architecture~\cite{Ronneberger2015}.
They train and evaluate their method using fetal brain MRI of $12$ fetuses, all acquired in the same center.
In addition, they used slices from $9$ neonatal 3D MRIs to increase the size of their dataset.
In \cite{payette2019longitudinal}, a 2D U-net~\cite{Ronneberger2015} is also used to segment the ventricular system in fetal brain MRIs.
In addition, the model is trained and evaluated on fetuses with both normal brain and with spina bifida aperta, referred to as open spina bifida in \cite{payette2019longitudinal}.
They train their CNN using $46$ fetal brain 3D MRIs, all acquired at the same center.
However, the accuracy of the segmentation is evaluated on only $3$ 3D MRIs.
In contrast to~\cite{khalili2019automatic}, the segmentation is performed after super-resolution and reconstruction to obtain high-resolution fetal brain 3D MRIs.
The use of 3D MRIs allows the longitudinal analysis of the ventricular system using the automatic segmentations.
All the work following \cite{payette2019longitudinal} performed automatic fetal brain MRI segmentation using 3D reconstructed MRI.
In \cite{payette2020efficient}, the problem of the robustness of deep learning methods for fetal brain MRI to the choice of the 3D super-resolution and reconstruction (SRR) algorithm is studied.
The use different SRR algorithms for transfer learning is proposed to make the deep neural network more robust to the SRR method used.
In \cite{hong2020fetal}, a 2.5D deep learning approach with multi-view aggregation is proposed for the segmentation of the cortical gray matter.
In contrast to previous work, they also propose to use a hybrid loss function that is the sum of the logarithmic Dice loss~\cite{wong20183d} and a novel boundary logarithmic Dice loss instead of the Dice loss~\cite{milletari2016v}.
In \cite{fidon2020distributionally}, we have studied the problem of robustness of deep learning to abnormal anatomy underrepresented in the training dataset.
We proposed to mitigate this problem by replacing the training optimization problem commonly used in deep learning, namely empirical risk minimization, by distributionally robust optimization.
Our method and experiments are detailed in \Chapref{chap:dro}.

Recently, the first publicly available fetal brain 3D MRI dataset with manual segmentations has been released~\cite{payette2021automatic}.
The dataset, called Fetal Tissue Annotation and Segmentation Dataset (FeTA),
contains $50$ fetal brain 3D MRIs in the first data release and $40$ 3D MRIs in the second data release. All 3D MRIs were acquired at the University Children’s Hospital Zurich.
More details about the FeTA dataset can be found in the next section.
The FeTA dataset has allowed the organization of the first international challenge for automatic fetal brain MRI segmentation (\url{https://www.synapse.org/#!Synapse:syn25649159/wiki/610007}).
All the top-performing methods in the FeTA challenge 2021 were based on deep learning and the 3D U-Net architecture~\cite{cciccek20163d}.

Several work studied the problem of efficient annotation of fetal brain 3D MRIs for training a deep learning-based segmentation algorithm~\cite{fetit2020deep,fidon2021label,fidon2021partial}.
In \cite{fetit2020deep}, a human-in-the-loop approach is proposed.
They alternate between automatic segmentation and manual refinement, and training a deep neural network based on the deep medic architecture~\cite{kamnitsas2016deepmedic}. % 
An atlas-based method for neonatal fetal brain MRI~\cite{makropoulos2014automatic} has been used to initialize the first automatic segmentations. 
They used a total of $249$ fetal brain 3D MRIs for training but their final deep learning-based segmentation model is not evaluated quantitatively.
In \cite{fidon2021label}, we proposed a family of loss functions, \textit{the label-set loss functions}, is proposed to allow to train any deep neural network using partially segmented fetal brain 3D MRI.
Partially segmented MRIs are MRIs in which some but not all the tissue types have been segmented manually.
When present, the manual segmentation of a tissue type is supposed to be complete.
A total of $146$ partially segmented fetal brain 3D MRIs were used to train a 3D U-Net~\cite{cciccek20163d} using different loss functions and an independent testing set with $100$ fetal brain 3D MRI, including MRI acquired from multiple centers, have been used to compare the different loss functions quantitatively.
Details can be found in \Chapref{chap:partialsup} of this thesis.

\subsection{Comparison of atlas-based and deep learning-based methods}
The experiments of \cite{khalili2019automatic} suggest that their deep learning pipeline achieves similar segmentation accuracy to the atlas-based methods~\cite{habas2010spatiotemporal,serag2012multi}.
However, the different methods were not evaluated on the same data which limits this comparison.
In \cite{zhao2022automated}, a 3D U-Net~\cite{cciccek20163d} is used to segment the cerebrospinal fuild, the cortical gray matter, the white matter, the deep gray matter, the cerebellum, and the brainstem in fetal brain 3D MRIs.
The 3D U-Net is shown to outperform an atlas-based segmentation method~\cite{makropoulos2014automatic} for MRIs of fetuses with a normal brain and acquired at the same center as the MRIs used to train the 3D U-Net.
However, the selection of an atlas-based method~\cite{makropoulos2014automatic} was originally developed for neonatale MRI biases this comparison.
Despite the findings of \cite{zhao2022automated}, 
when an atlas-based method tailored to fetal brain MRI segmentation is used,
empirical results in other work suggest that atlas-based method can outperform state-of-the-art deep learning methods when evaluated on MRIs
with a low quality~\cite{payette2020efficient} 
and on MRIs acquired at a different center than during training~\cite{fidon2022trustworthy}.

\subsection{Combining atlas-based and deep learning-based methods.}
In the early work of \cite{sanroma2018learning}, two ensembling strategies to combine a multi-atlas segmentation method and a machine learning approach, like deep learning, have been studied.
In \cite{li2021cas}, three deep neural network are trained jointly.
The first network learns to perform fetal brain 3D MRI segmentation,
The second compute a spatio-temporal fetal brain MRI atlas,
and the third network learns to register the atlas to the input and result in a second automatic segmentation of the output.
In addition, a convolutional layer that combines the two automatic segmentation is trained end-to-end with the three previous networks.
In \cite{fidon2022trustworthy} and \Chapref{chap:twai},
a method based on Dempster-Shafer theory is proposed to combine atlas-based and deep learning-based methods for fetal brain 3D MRI segmentation.

\chapter[Fetal Brain 3D MRI Segmentation Dataset]{Fetal Brain 3D MRI Segmentation Dataset}
\label{chap:fetaldataset}
\minitoc
\begin{center}
	\begin{minipage}[b]{0.9\linewidth}
		\small
		\textbf{Foreword\,}
		This chapter presents the fetal brain MRI segmentation dataset that was created and used in this thesis.

		\textbf{Contributions:}
		\begin{itemize}
		    \item We have created the largest fetal brain 3D T2w MRI segmentation dataset to date.
		    \item The construction of the fetal brain T2w MRI dataset has been lead by Michael Aertsen and myself under the supervision of Tom Vercauteren and Jan Deprest.
		    \item I estimated the inter-rater variability in fetal brain MRI segmentation for: 
		    the white matter, the intra-axial cerebrospinal fluid (CSF), the cerebellum, the extra-axial CSF, the cortical gray matter, the deep gray matter, and the brainstem. 
		    \item The 3D reconstructed MRIs, except for the data from FeTA~\cite{payette2021automatic}, were computed using NiftyMIC~\cite{ebner2020automated} by:
            Michael Aertsen, Fr\'ed\'eric Guffens, and myself.
            \item The class hierarchy for fetal brain segmentation has been contributed by:
            Michael Aertsen, Ernst Schwartz, and myself.
            \item The manual segmentations of the fetal brain 3D MRIs have been contributed by:
            Michael Aertsen, Philippe Demaerel, Thomas Deprest, Doaa Emam, Fr\'ed\'eric Guffens, Nada Mufti, Esther van Elslander, the consortium of the FeTA dataset, and myself.
		    \item Several persons have contributed fetal brain MRIs:
            Jan Deprest and Michael Aertsen for data from University Hospital Leuven,
            Anna David, Andrew Melbourne, and Nada Mufti for the data from University College London Hospital and various centers in the United Kingdom,
            Mary Rutherford for the data from King's College London,
            Gregor Kasprian and Daniela Prayer for the data from Medical University of Vienna,
            and Kelly Payette and Andras Jakab for the data from University Children's Hospital Zurich that they made publicly available in the FeTA dataset~\cite{payette2021automatic}.
		\end{itemize}
	\end{minipage}
\end{center}

\section{Introduction}

\begin{figure}[tb!]
    \centering
    \includegraphics[width=\textwidth]{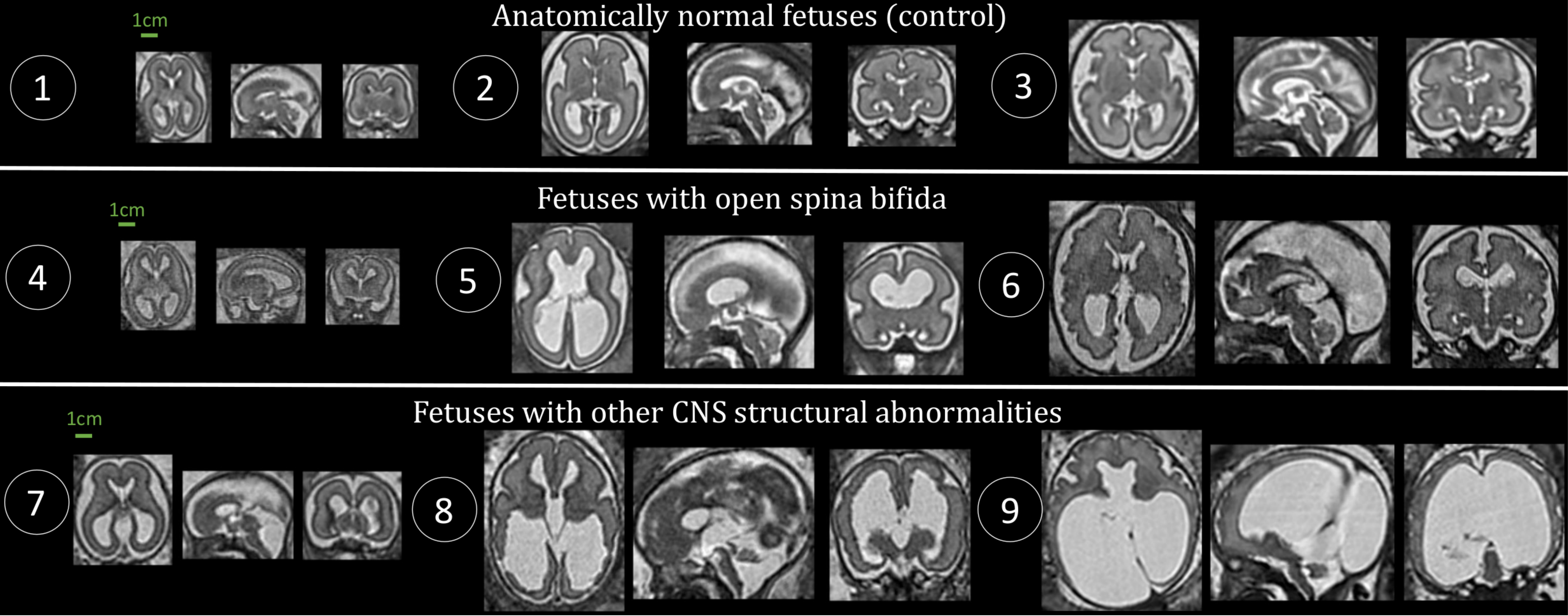}
    \caption{Illustration of the anatomical variability in fetal brain across gestational ages and diagnostics.
    1: Control (22 weeks);  % AUS00215
    2: Control (26 weeks);  % AUS00008
    3: Control (29 weeks);  % AUS00113
    4: Spina bifida (19 weeks);
    5: Spina bifida (26 weeks);
    6: Spina bifida (32 weeks);
    7: Dandy-walker malformation with corpus callosum abnormality (23 weeks);
    8: Dandy-walker malformation with ventriculomegaly and periventricular nodular heterotopia (27 weeks);
    9: Aqueductal stenosis (34 weeks).
    }
    \label{fig:anatomy_variability2}
\end{figure}

% Accurate assessment of the fetus before and after surgery is of upmost importance~\cite{adzick2011randomized,aertsen2019reliability}.
% 
The two main imaging modalities used in paediatric radiology are Ultrasound (US) and Magnetic Resonance Imaging (MRI).
US is the standard imaging modality because it is widely available in clinic, real-time, low-cost, non-invasive and high-resolution.
However, it has a small field-of-view, limited soft tissue contrast and poor image quality.
As a result, MRI is used to complement ultrasound imaging and detect any additional potential abnormality that may be harder to visualize by the former.
It has been shown that MRI allows to improve the management of fetuses with visible abnormalities in US~\cite{simon2000fast,kubik2000ultrafast,levine2003fast}.
In the study of~\cite{levine2003fast}, for $31.7\%$ of the fetuses with abnormal US findings, MRI findings lead to change of prognosis.

MRI is the current clinical standard to aid decision making, such as for prenatal surgery for spina bifida fetuses~\cite{aertsen2019reliability}.
However, the motion of the fetus is a challenge for MRI as usual acquisition processes are slow in comparison to the movement of the fetus\footnote{See this \href{https://www.youtube.com/watch?v=djJnsC_CddI\&list=PLerImOteTaCZXME50snRZrj2hcG8l2LJK\&index=2\&t=0s}{\textcolor{blue}{video}} for an illustration of the motion of a fetus.}.
Motion artefacts are addressed by using specific MRI acquisition protocols such as Single-Shot Fast Spin Echo (SSFSE)~\cite{yamashita1997mr}.
SSFSE is a fast imaging method to acquire thick, low resolution stacks of 2D MRI slices to reduce motion artifacts.
    Each stack of 2D MRI consists of a volume with high resolution and low amount of motion along two dimensions, and low resolution and high level of motion along the third dimension.
    Several stacks are acquired along orthogonal directions with the aim that every part of the fetal brain is visible in high resolution in at least one stack.
% Several stacks are acquired along different axis of the fetal anatomy. Ideally, the 3D volume is oversampled to make sure that nothing is missing.
One second or less per slice is required.
Super-resolution and reconstruction (SRR) algorithms can be used to compute a high-resolution and motion-free reconstructed 3D MRI from the stacks of 2D MRI slices~\cite{ebner2020automated}.
    More details about SRR algorithms is given in section~\ref{s:srr}.
Manual measurements can then be obtained using the 2D MRI slices or a reconstructed 3D MRI selected by an expert~\cite{aertsen2019reliability,khawam2021fetal}.
Many anatomical biomarkers of spina bifida have been proposed based on the knowledge of the clinicians, but the choice of standard anatomical biomarkers for the diagnosis is still unclear~\cite{aertsen2019reliability}. Manual anatomical biomarkers measurements also suffer from large inter-rater and intra-rater variability, and is time-consuming.

In particular, studying shape and volume of different brain tissues is essential for 
the study of fetal brain development~\cite{benkarim2017toward}.
Reliable analysis and evaluation of fetal brain structures
could also support diagnosis of central nervous system pathology, patient selection for fetal surgery, evaluation and prediction of outcome, hence also parental counselling~\cite{aertsen2019reliability,danzer2020fetal,moise2016current,sacco2019fetal,zarutskie2019prenatal}.
Therefore, automatic methods for fetal brain segmentation in 3D is needed.

In this chapter, 
we give an introduction to 3D super-resolution and reconstruction algorithms that are required to obtain coherent and high-resolution isotropic MRI of the fetal brain for shape and volume analysis.
We also review the literature of automatic fetal brain MRI segmentation.
Our dataset for fetal brain 3D MRI segmentation is described.
In particular, we evaluate the inter-rater variability in manual fetal brain MRI segmentation for the first time.

\subsection{Spina Bifida Aperta}
In this section, we give a short introduction of spina bifida aperta which is the brain defect studied the most in depth in this thesis.

Spina bifida aperta (SBA), also referred to as open spina bifida,  open spinal dysraphism, or simply spina bifida in a context where there is no ambiguity with other sorts of spina bifida,
is the most prevalent fetal brain defect.
The prevelance of spina bifida is approximately 4.9 per 10,000 live births in Europe, and 3.17 in USA~\cite{aertsen2019reliability,khoshnood2015long}.
Spina bifida aperta occurs when the neural tube fails to close in the first four weeks after conception~\cite{adzick2011randomized,aertsen2019reliability}. 
SBA is divided into two types: myelomeningocele and myeloschisis. 
Myelomeningocele (MMC) is the most frequent form of spina bifida aperta, where the spinal defect causes an extrusion of the spinal cord into a fluid filled sac.
In both cases, SBA leads to an abnormal development of the central nervous system caused by leak of amniotic fluid.
Most cases of SBA are accompanied by severe anatomical brain abnormalities~\cite{pollenus2020impact}
with enlargement of the ventricles and a type II Chiari malformation being most prevalent.
The Chiari malformation type II is characterized by a small posterior fossa and hindbrain herniation in which the medulla, cerebellum, and fourth ventricule are displaced caudally into the direction of the spinal canal~\cite{naidich1980computed}.
This downwards displacement causes accumulation of fluid in the brain termed \emph{hydrocephalus}.
This is due to the  obstruction of the normal flow of cerebrospinal fluid (CSF) around the body.
This hydrocephalus often requires treatment after birth via means of a ventriculoperitoneal shunt in $80\%$ of infants~\cite{kahn2014fetal}.
However, shunts can get infected, and can fail requiring multiple surgical revisions which can lead to developmental delay~\cite{geerdink2012essential,sutton1999improvement}.
Motivated by the Management of Mylelomeningocele Study (MOMS) randomized control trial published in 2011~\cite{adzick2011randomized}, patients are now evaluated and treated by surgery before birth.
The corpus callosum of fetuses with SBA is also abnormal~\cite{kunpalin2021incidence,pollenus2020impact} and has been found to be significantly smaller for fetuses with SBA than for normal fetuses~\cite{crawley2014structure,dennis2016white,kunpalin2021incidence}.
SBA fetuses have also smaller hippocampus~\cite{treble2015prospective}, 
abnormal cortical thickness and gyrification~\cite{mufti2021cortical,treble2013functional}, and smaller deep gray matter volume and total brain volume~\cite{hasan2008quantitative,mandell2015volumetric}.
The main neuroanatomy vocabulary specific to SBA can be found in Appendix~\ref{sec:neuroanatomy}.
For all those reasons the anatomy of the brain of fetuses with SBA differs from the normal fetal brain anatomy.
In addition, the mechanisms underlying those anatomical brain abnormalities remain incompletely understood~\cite{danzer2020fetal}.

The benefits of in-utero surgery have to be balanced with the higher risk of preterm birth and possible long term consequences for the mother~\cite{moldenhauer2017fetal}.
However, an optimum patient selection criteria for predicting which patients would benefit more from the surgery are still being optimized~\cite{moldenhauer2017fetal}.
The segmentation of fetal brain MRI is essential for 
the quantitative study of fetal brain development~\cite{benkarim2017toward}.
Reliable analysis and evaluation of fetal brain structures using segmentation and registration algorithms
could also support diagnosis of central nervous system pathology, patient selection for fetal surgery, evaluation and prediction of outcome, hence also parental counselling~\cite{aertsen2019reliability,danzer2020fetal,moise2016current,sacco2019fetal,zarutskie2019prenatal}.
Literature on computer-assisted fetal brain MRI processing is briefly reviewed in the next section of this chapter.

\subsection{Super-Resolution and Reconstruction: From 2D to 3D}\label{s:srr}
One of the main challenges for qualitative and quantitative analysis using fetal brain MRI is the motion of the fetus.
This motivated the creation of specific MRI sequences for fetal MRI that are designed to produce multiple stacks of 2D slices rather than a 3D image.
However, those 2D MRI slices typically have lower through-plane resolution, suffers from motion between neighboring slices, motion artefact, and suboptimal cross-section.
The inaccuracy of manual measurements in 2D MRI slices is also due to the manual selection of a few slices~\cite{aertsen2019reliability}.

To tackle this issue, 3D super resolution and reconstruction (SRR) algorithms have been proposed to improve the resolution, and remove motion between neighboring slices and motion artefacts present in the original 2D MRI slices~\cite{ebner2020automated,tourbier2017automated,gholipour2010robust,rousseau2006registration,kainz2015fast,kim2009intersection,kuklisova2012reconstruction,alansary2017pvr,hou20183}.
The output of the 3D super resolution and reconstruction algorithm is a reconstructed 3D MRI of the fetal brain with an isotropic image resolution.
A detailed review of the literature of SRR algorithms for fetal brain MRI can be found in \cite[see Chapter 2]{ebner2019volumetric}.

Briefly, SRR algorithms aim at combining a set of low-resolution stacks of 2D MRI slices that are acquired for a patient into a unique high-resolution 3D MRI volumes.
Current state-of-the-art SRR algorithms all have the same structure:
\begin{enumerate}
    \item localization and segmentation of the fetal brain in the 2D MRI slices
    \item iteratively alternate between motion correction and volumetric reconstruction
    % 3D interpolation between the 2D slices available and the spatial repositioning of the 2D slices in the current 3D reconstruction
    % \item use of heuristics for automatic outlier slices rejection
\end{enumerate}
The localization and segmentation step is sometimes manual in some methods.

In this thesis, I have used the fully-automatic SRR pipeline proposed in~\cite{ebner2020automated} which is one of the current state of the art SRR methods.

It is hypothesized that the reconstructed 3D MRIs facilitates the qualitative and quantitative anatomical analysis of fetal brain MRI as compared to the original stacks of 2D MRI slices.
It has been shown that it leads to better visualization and could become an essential tool in the spina bifida clinical pipeline \cite{zarutskie2019prenatal}.
A comparison of manual biometric measurements performed on both the original 2D MRIs and the reconstructed 3D MRIs for the same studies has shown a good agreement~\cite{khawam2021fetal}.
This suggests that manual measurements can be safely performed on reconstructed 3D MRIs.
Reconstructed 3D MRIs are also likly to facilitate the manual delineation of the fetal brain structures compared to the original stacks of 2D MRI slices.

\section[Fetal Brain MRI Dataset]{A Large and Multi-centric Fetal Brain 3D MRI Segmentation Dataset}

In this section, I describe the fetal brain MRI dataset that we have built during my PhD and that supports the work that is described in the following chapters.

% \subsection{Ethics Statement}
% The MRI data were automatically pseudonymized  using the \texttt{GIFT-Cloud} data sharing platform~\cite{doel2017gift} prior to using them for research.
% % 
% \\At University Hospitals Leuven, ethical approval to use the data for research was given by the Ethics Committee University Hospitals Leuven (ethical approval S63598).
% % 
% A retrospective study does not fall under the Belgian law of May 7, 2004 regarding experiments on the human person.
% % 
% However, given the use of potentially identifying MRIs in the study, the requirements set forth in the EU Regulation 2016/679 (General Data Protection Regulation, GDPR) must be met. The sponsor of this study is University Hospitals Leuven, and University Hospitals Leuven maintains "public interest" as the legal basis for data processing. 
% % 
% Article 14 of the GDPR mentions the information obligation of the data controller (= sponsor of the study) to the data subject whose personal data are collected. An information obligation is therefore sufficient according to GDPR, and informed consent is not legally required for the use of the MRIs for illustrative purposes. All snapshots of fetal MRIs used in our figures are based on MRI acquired at Leuven.
% % 
% \\At University College London Hospital (UCLH) the study was approved by the Caldicott guardian at UCLH and patient consent was not required as these images were acquired for clinical purposes and the data used retrospectively.

\subsection{Underlying MRI data}

\begin{figure}
    \centering
    \includegraphics[width=\linewidth]{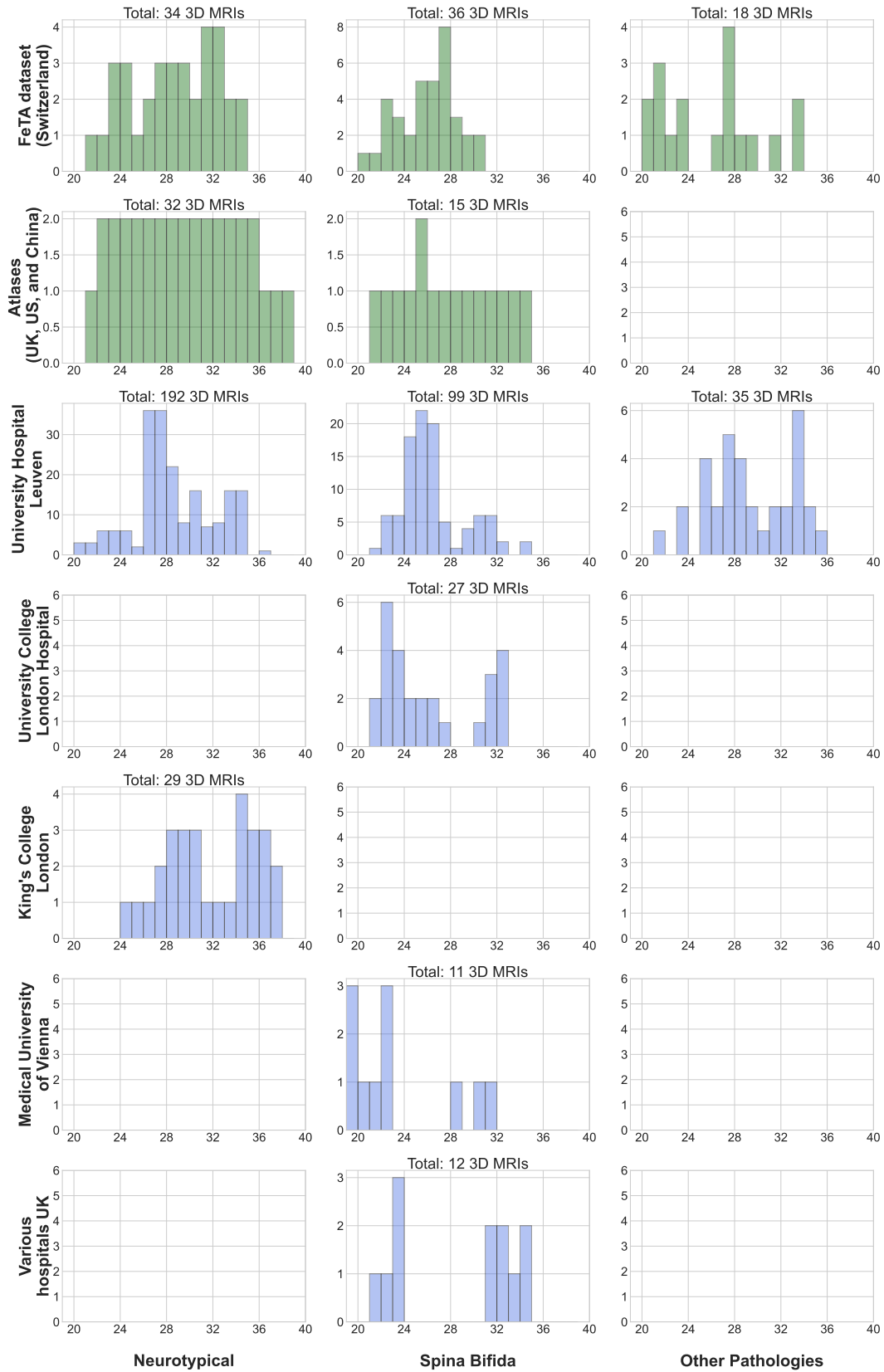}
    \caption{Histogram of the gestational ages for each group of centers and for neurotypical, spina bifida, and other pathologies for the full dataset.}
    \label{fig:full_dataset_stats}
\end{figure}

We have collected a dataset with a total of $540$ fetal brain 3D MRIs with neurotypical or abnormal brain development and from $13$ sources of data across $6$ countries.
A graphical summary of the dataset can be found in Fig.~\ref{fig:full_dataset_stats}.

The dataset consists of $326$ 3D MRIs acquired at \uzl{} (\uzlshort{}),
$88$ 3D MRIs from the FeTA dataset~\cite{payette2021automatic} (data release 1 and 2),
$11$ 3D MRIs acquired at \vienna{} (\viennashort{}),
$29$ 3D MRIs acquired at \kcl{} (\kclshort{}),
$27$ 3D MRIs acquired at \uclh{} (\uclhshort{}),
$4$ 3D MRIs acquired at \manchester{} (\manchestershort{}),
$4$ 3D MRIs acquired at \belfast{} (\belfastshort{}),
$2$ 3D MRIs acquired at \cork{} (\corkshort{}),
$1$ 3D MRIs acquired at \newcastle{} (\newcastleshort{}),
$1$ 3D MRIs acquired at \liverpool{} (\liverpoolshort{}),
and $47$ 3D MRIs from three fetal brain brain atlases.
The three open-access fetal brain spatio-temporal atlases consist of
$18$ population-averaged 3D MRIs computed from fetal neurotypical brain MRIs acquired at Boston Children's Hospital, USA~\cite{gholipour2017normative},
$14$ population-averaged 3D MRIs computed from fetal neurotypical brain MRIs acquired in China~\cite{wu2021age},
and $15$ population-averaged 3D MRIs computed from fetal spina bifida brain MRIs acquired at \uzlshort{} and \uclhshort{}~\cite{fidon2021atlas}.

Data from \uzlshort{} includes $192$ 3D MRIs of neurotypical fetuses, $99$ 3D MRIs of fetuses with spina bifida aperta, and $35$ 3D MRIs of fetuses with an abnormal brain anatomy due to a condition other than spina bifida.
The majority of the neurotypical fetuses was scanned for a suspected abnormality somewhere else than in the brain, while a minority was scanned for screening of brain abnormality but was proven neurotypical after MRI.
The $35$ 3D MRIs of fetuses with other abnormalities consisted of:
$3$ examinations of a case with an enlarged subarachnoid space,
$3$ cases of intraventricular hemmorhage,
$1$ cases of intracranial hemmorhage,
$1$ case with a partial rhombencefalosynapsis,
$1$ case with a closed lip Schizencephaly,
$4$ cases with dandy-walk malformation,
$1$ case with an unilateral ventriculomegaly due to a hemmorhage,
$1$ case with choroid plexus pappiloma,
$1$ case with high flow Dura isinus malformation,
$7$ cases with corpus callosum agnesis,
$1$ case with corpus callosum agenesis with interhemispheric cyst, temporal cysts and delayed gyration,
$2$ cases with tuberosis sclerosis,
$1$ case with a Blake's pouch cyst,
$2$ cases with aqueductal stenosis,
$1$ case with an idiopathic dilatation of the lateral ventricles,
$2$ cases with cytomegalovirus encephalitis,
and $1$ case with parenchyma loss due to a vasculo-ischemic insult.

Data from the open-access FeTA dataset~\cite{payette2021automatic} includes $34$ 3D MRIs of fetuses with a normal brain development, $36$ 3D MRIs of fetuses with spina bifida aperta, and $18$ 3D MRIs of fetuses with an abnormal brain anatomy due to conditions other than spina bifida.
Those $18$ 3D MRIs of fetuses with other abnormalities consisted of: 
$3$ cases with heterotopia, $8$ cases with ventriculomegaly withour spina bifida, $2$ cases with acqueductal stenosis, $2$ cases with interhemispheric cyst, $1$ case with cerebellar hemmorhage, $1$ case with a Dural sinus malformation, and $1$ case with Bilateral subependymal cysts and temporal cysts.

Data from \kclshort{} consists exclusively of brain 3D MRIs of fetuses with a normal brain development.
Data from \viennashort{}, \uclhshort{}, \manchestershort{}, \belfastshort{}, \corkshort{}, \newcastleshort{}, and \liverpoolshort{}
consist only of 3D MRIs of fetuses diagnosed with spina bifida aperta.

\subsection{Fetal MRI Acquisition Details}
All the MRI examinations at \uzlshort{}, \uclhshort{}, \manchestershort{}, \corkshort{}, \newcastleshort{}, \belfastshort{}, \liverpoolshort{}, \kclshort{}, and \viennashort{} were performed as part of clinical routine following abnormal findings during ultrasound examination.
For each study, at least three orthogonal T2-weighted HASTE series of the fetal brain were collected on a $1.5$T scanner, except for the MRI acquired at \viennashort{} for which T2-weighted TSE series were used.
The MRIs from \uzlshort{} and \uclhshort{} were acquired with an echo time of $133$ms, a repetition time of $1000$ms, with no slice overlap nor gap, pixel size $0.39$mm to $1.48$mm, and slice thickness $2.50$mm to $4.40$mm.
The MRIs from \manchestershort{}, \corkshort{}, \newcastleshort{}, \belfastshort{}, and \liverpoolshort{} were acquired with an echo time of $145$ms, a repetition time of $9953$ms, with no slice overlap nor gap, pixel size $0.56$mm to $1.25$mm, and slice thickness $3$mm to $4$mm.
The MRI from \kclshort{} were acquired with an echo time of $160$ms, a repetition time of $15,000$ms, a slice overlap of $1.5$mm, and a slice thickness of $2.5$mm.
For patients imaged at \viennashort{}, a 1.5T scanner was used, applying an echo time of $140$ms, a repetition time of $53,189$ms, slice gap of $0.4$mm, pixel size $0.9$mm to $1.37$mm, and slice thickness $3$mm to $4$mm.
In all acquisition centers, a radiologist attended all the acquisitions for quality control.
The dataset contains longitudinal MRI examinations with up to $5$ examinations per fetus.
Fetuses with spina bifida were diagnosed with spina bifida aperta at fetal ultrasound examinations.
In addition, MRI examinations for spina bifida fetuses were performed either for unoperated fetuses or after open fetal surgery performed before $26$ weeks of gestation to close the spina bifida aperta defect.
In particular, the MRI studies from \viennashort{} all correspond to unoperated spina bifida fetuses and longitudinal data.
% The distribution of gestational ages for MRI examinations and whether they were done before or after surgery can be found in Figure~\ref{fig:ga}.

\subsection{Automatic Brain Extraction}
Automatic segmentation of the fetal brain in the raw 2D MRIs are obtained using a deep learning-based method~\cite{martaisbi2021}.
Those brain masks are an input required by the 3D super resolution and reconstruction algorithm described below.
A public implementation of the deep learning pipeline \href{https://github.com/gift-surg/MONAIfbs}{\texttt{MONAIfbs}}~\cite{martaisbi2021} is used in this study to obtain the brain masks.
The main git branch, commit $bcab52a$, was used.

\subsection{Super-Resolution and Reconstruction}\label{fetaldataset:srr}
The reconstructed fetal brain 3D MRIs were obtained using \href{https://github.com/gift-surg/NiftyMIC}{\texttt{NiftyMIC}}~\cite{ebner2020automated}
a state-of-the-art super resolution and reconstruction algorithm. The volumes were all reconstructed to a resolution of $0.8$ mm isotropic.
\href{https://github.com/gift-surg/NiftyMIC}{\texttt{NiftyMIC}} version $0.8$ with Python $3.8$ has been used.
The 2D MRIs were also corrected for image intensity bias field as implemented in \href{https://simpleitk.org}{SimpleITK} version $1.2.4$.
Our pre-processing improves the resolution, and removes motion between neighboring slices and motion artefacts present in the original 2D slices~\cite{ebner2020automated}.

The 3D super resolution and reconstruction algorithm~\cite{ebner2020automated} also combines the 2D brain masks. This results in a 3D brain mask for the 3D reconstructed MRI that is computed fully-automatically.

All the reconstructed 3D MRIs and 3D brain masks were rigidly aligned to a spatio-temporal fetal brain atlas~\cite{gholipour2017normative} as implemented in \href{https://github.com/gift-surg/NiftyMIC}{\texttt{NiftyMIC}}~\cite{ebner2020automated} version $0.8$.
All the reconstructed 3D MRIs are therefore aligned to a standard clinical view in which the axes are aligned with the axial, sagittal, and coronal planes of the fetal brain. This facilitates the manual delineation and annotation of the fetal brain structures.
The target time-point in the neurotypical 4D atlas is chosen based on the brain volume computed using the automatic 3D brain mask.

\subsection{Semi-automatic Fetal Brain 3D MRI Segmentations}
In this section, we describe the method that we have used to perform the manual segmentation of the fetal brain 3D MRIs.
In this work, we call \textbf{manual segmentation} every segmentation that has been checked by an expert for every slice and corrected wherever necessary. 
In particular, initialization of the segmentations was obtained using an automatic segmentation methods and/or from another, potentially less experienced, raters.

\subsubsection{Manual Segmentation}
\begin{figure}
    \centering
    \includegraphics[width=\linewidth]{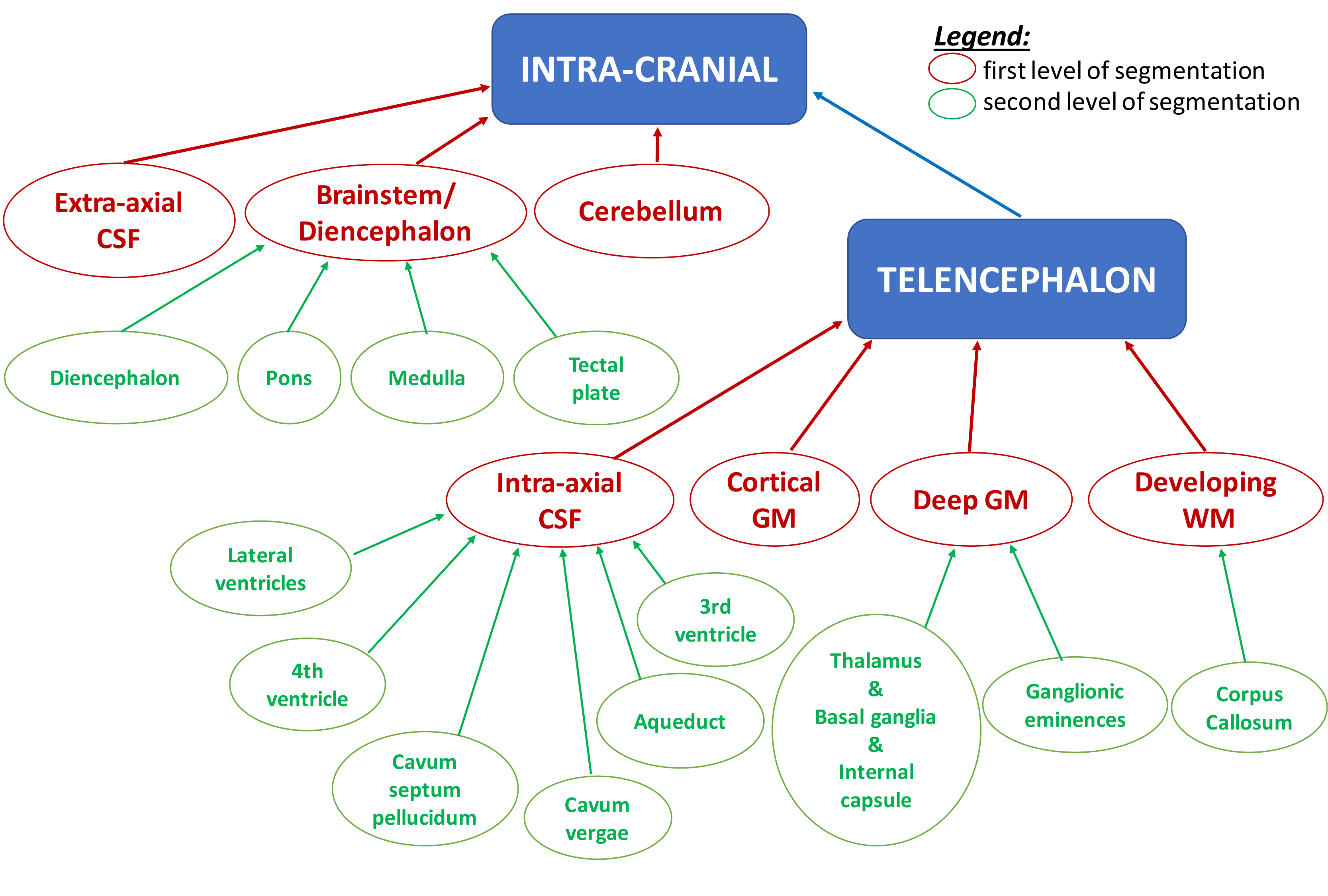}
    \caption{Proposed brain classes hierarchy for the developing fetal brain.}
    \label{fig:class_hierarchy}
\end{figure}

The 3D reconstructed fetal brain MRIs were segmented for eight tissue types:
\begin{enumerate}
    \item the white matter (excluding the corpus callosum; see below),
    \item the intra-axial cerebrospinal fluid (ventricular
system with the cavum septi pellucidi and cavum verga),
    \item the cerebellum,
    \item the extra-axial cerebrospinal fluid,
    \item the cortical gray matter,
    \item the deep gray matter,
    \item the brainstem,
    \item the corpus callosum
\end{enumerate}
Those tissue types follow the hierarchy that we propose in Fig.~\ref{fig:class_hierarchy}.

The labelling protocol used for white matter, intra-axial CSF, cerebellum, extra-axial CSF, cortical gray matter, deep gray matter, and brainstem is based on the segmentation guidelines that were used for the FeTA dataset~\cite{payette2021automatic}.
We have made the following amendments to the FeTA segmentation guidelines.
The corpus callosum, that is not present in FeTA, was segmented.
As a result, the corpus callosum was excluded from the white matter manual segmentations. 
We use the term \textit{intra-axial CSF} rather than \textit{ventricular system} as in FeTA because, in addition to the lateral ventricles, third ventricle, and forth ventricle, this region also contains the cavum septum pellucidum and the cavum vergae that are not part of the ventricular system~\cite{tubbs2011cavum}.
We also add that the cortical gray matter separates the white matter from the extra-axial CSF.
In particular, the white matter should never be directly in contact with the extra-axial CSF.
The two tissue types are always separated by a fine layer of cortical gray matter.

The eight tissue types were segmented for our dataset by 
Michael Aertsen, a paediatric radiologist at \uzl{} specialized in fetal brain anatomy,
and by
Thomas Deprest,
Doee Emam,
Esther Van Elslander,
Fr\'ed\'eris Guffens,
Lucas Fidon,
and Nada Mufti
under the supervision of Michael Aertsen who trained them, and quality controlled and corrected their manual segmentations.
% 
% Philippe Demaerel, professor in radiology at \uzl{}, gave advice on the deep gray matter segmentation protocol.

Manual segmentations were all performed using the software ITK-SNAP~\cite{yushkevich2016itk}.
The segmentation was performed iteratively in the three views (axial, sagittal, and coronal).
The rigid registration of the 3D MRI to a normalized atlas space during the super-resolution and reconstruction step (see section~\ref{fetaldataset:srr}) made this process easier for the raters.
The manual segmentation of the different tissue types was performed jointly with the different tissue types represented as different layers in the segmentation nifti file.
We found in preliminary work, that this sped up the manual segmentation process and avoided inconsistencies between layers as compared to working with separate binary segmentation files with one file per tissue type.
It was observed that raters tended to under-segment the tissue types when they performed manual segmentations using separated files for the different tissue types.
We attribute this to a form of imposter syndrome.
In case experts are not entirely sure about the assignment of a voxel to a given tissue types, experts will prefer to not segment the voxel.
However, this implicitly attributes those missed voxels to the background segmentation layer and lead to under-segmentation.
In contrast, when performing the segmentation for all the tissue types jointly, the experts are forced to chose the label the most likely for every voxel inside the brain.

\subsubsection{AI-assisted Manual Segmentation}

To reduce the time required for the manual segmentation, the segmentation were all initialized using automatic segmentation methods.
As the number of manual segmentations was increasing, the automatic segmentation methods was progressively refined.
The more manual segmentations were available, the better were the automatic initial segmentations, and the faster it was to correct the automatic segmentations to obtain more manual segmentations.

The first automatic initial segmentations have been computed using an atlas-based segmentation method originally developed for the brain MRI segmentation of infants born preterm~\cite{orasanu2016cortical}.
Once a dozen of manual segmentations were available we trained a deep learning-based segmentation algorithm based on 3D U-Net~\cite{cciccek20163d} for fetal brain 3D MRI segmentation.
The deep learning method has been progressively re-trained and improved, and has progressively replaced the atlas-based method.

The registration-based approach sometimes lead to unsatisfactory inital segmentations.
This was in particular the case for the youngest fetuses at the time of MRI acquisition and for the fetuses with an abnormal brain anatomy.
We hypothesize that the use of a neonatal brain atlas~\cite{kuklisova2011dynamic} in the atlas-based method~\cite{orasanu2016cortical} may account for this.
This was one of the motivations for developing our fetal brain spatio-temporal atlas for the developing fetal brain with spina bifida aperta~\cite{fidon2021atlas} (see \chapref{chap:atlas}).

It is worth noting that the deep learning pipeline has also been found to compute segmentations that were suboptimal, yet superior to the atlas-based method on average, for the youngest fetuses and the fetuses with an abnormal brain anatomy.
We hypothesized that it was partly due to the underrepresentation of abnormal brain anatomy in the training dataset.
To mitigate this problem, we have added 3D MRIs of fetal brains with various and severe brain abnormal anatomy early in the construction of the dataset.
In addition, we have developed a new training method for deep neural network based on distributionally robust optimization~\cite{fidon2020distributionally,fidon2021distributionally} (see \chapref{chap:dro}).

The manual segmentation was also stratified in terms of the tissue types that were segmented.
The first $200$ fetal brain 3D MRIs were, initially, segmented only for the white matter, the intra-axial CSF, and the cerebellum.
Later, some of those fetal brain MRI have been also segmented for the extra-axial CSF.
Eventually, the remaining tissue types were added and new fetal brain MRIs were systematically segmented for all the tissue types, including control and correction by an expert.

The manual segmentations publicly available as part of the FeTA~\cite{payette2021automatic} dataset were also corrected~\cite{fidon2021partial,fidon2021distributionally} to reduce the variability against the published segmentation guidelines that was released with the FeTA dataset.
The corpus callosum was added to the manual segmentations of the FeTA dataset 3D MRIs.
%
%Two volumes of spina bifida aperta cases were excluded because the poor quality of the 3D reconstruction (\texttt{sub-feta007} and \texttt{sub-feta009}) did not allow to segment them reliably for the seven tissue types.

\subsubsection{Inter-rater Variability}
We have estimated the inter-rater variability using all the 3D MRIs from the public fetal brain dataset FeTA~\cite{payette2021automatic} by comparing our refined manual segmentations against the original manual segmentations available with the FeTA dataset.

In~\cite{payette2021automatic}, the inter-rater variability has been previously estimated for only $9$ fetal brain 3D MRIs.

\begin{figure}[t]
	\centering
	\includegraphics[width=\linewidth]{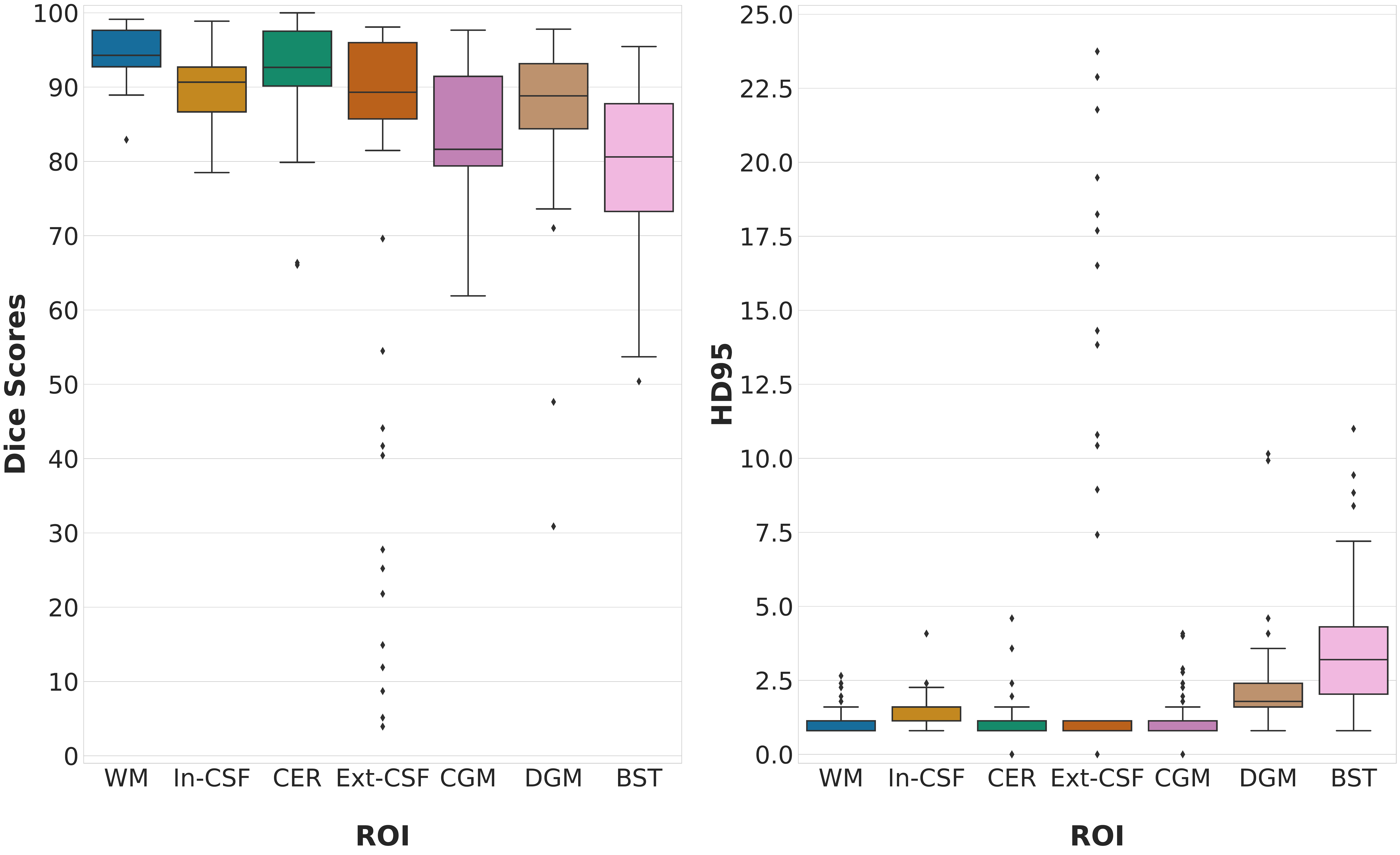}
	\caption{
		\textbf{Manual Segmentation Inter-rater Variability}.
		The evaluation is performed using two manual segmentations for $78$ 3D MRIs from the FeTA dataset~\cite{payette2021automatic}.
		(left) the Dice score (in percentage) and (right) Hausdorff distance at $95\%$ percentiles (HD95; in mm).
		WM: white matter,
		in-CSF: intra-axial cerebrospinal fluid,
		CER: cerebellum,
		ext-CSF: extra-axial cerebrospinal fluid,
		CGM: cortical gray matter,
		DGM: deep gray matter,
		BST: brainstem.
	}
	\label{fig:inter-rater-var}
\end{figure}

The inter-rater variability is evaluated for all the tissue types except for the corpus callosum that was not present in the original FeTA manual segmentations using the Dice scores and the Hausdorff distance at $95\%$ percentiles.
The results of the inter-rater variability can be found in Fig.~\ref{fig:inter-rater-var}.

We obtained a mean$\pm$standard deviation of Dice scores (in \%) of
$94.7\pm3.1$ for the white matter,
$89.9\pm4.5$ for the intra-axial CSF,
$92.5\pm6.2$ for the cerebellum,
$81.0\pm25.4$ for the extra-axial CSF,
$83.7\pm7.9$ for the cortical gray matter,
$87.2\pm10.0$ for the deep gray matter,
and $79.6\pm10.7$ for the brainstem.

We also obtained a mean$\pm$standard deviation of Hausdorff distance at $95\%$ percentiles (in mm) of
$1.1\pm0.4$ for the white matter,
$1.3\pm0.5$ for the intra-axial CSF,
$1.0\pm0.7$ for the cerebellum,
$3.3\pm6.0$ for the extra-axial CSF,
$1.1\pm0.7$ for the cortical gray matter,
$2.2\pm1.6$ for the deep gray matter,
and $3.6\pm2.1$ for the brainstem.

Those average metric performances are similar to the one achieved by a state-of-the-art deep learning pipeline (nnU-Net~\cite{isensee2021nnu}) when trained and evaluated on MRI data acquired in the same center (see \chapref{chap:twai} and Fig.~\ref{fig:twai-dice_roi}~\ref{fig:twai-hausdorff_roi}).
This suggests that current state-of-the-art deep learning methods achieve human-level performance when evaluated on MRI data acquired at the same center as the data used for training.
It is worth noting that here we compare segmentation performances measured on different 3D MRIs.

The inter-rater variability for the extra-axial CSF, cortical gray matter, deep gray matter, and brainstem appears higher than for the white matter, intra-axial CSF, and cerebellum.
In particular, outliers with high variations are observed for the extra-axial CSF, deep gray matter, and brainstem.
Those results are consistent with the large inter-rater variability previously measured for the extra-axial CSF in~\cite{payette2021automatic}.

For the original FeTA manual segmentations, the raters annotated some of the tissue types in an indirect manner. 
% They segmented manually superclasses that were intersected and combined to obtain the final tissue types segmentations.
% 
For example, the extra-axial CSF was not segmented directly but instead was obtained by subtracting all the other tissue types to the manually segmented intracranial space.
For neurotypical fetal brains and most abnormal fetal brains the extra-axial CSF delimits the intra-cranial space, which justifies this strategy.
However, for the 3D MRIs of spina bifida with gestational ages of $27$ weeks or less we observed that this is often not the case.
This may account for the outliers observed for the extra-axial CSF in Fig.~\ref{fig:inter-rater-var}.
% 
% Those outliers correspond exclusively to spina bifida cases.
% with gestational ages between $x$ weeks and $x$ weeks.

It is worth noting that our refined manual segmentation are based on the original manual segmentations.
This might bias the refined manual segmentation towards the original manual segmentations.
In addition, different annotation tools were used for the original FeTA segmentations~\cite{payette2021automatic} and our refined segmentations.
The original FeTA segmentations were performed using the draw tool in 3D slicer, in only one plane for each tissue type, and the annotations were performed manually only on every second or third slice.
In addition, no spatial alignment to a standard clinical view using an atlas was performed prior to manual segmentation.
In contrast, our refined manual segmentations have been performed after rigid registration to a fetal brain atlas~\cite{gholipour2017normative} and annotation were performed densely using the three planes and the paint tool in ITK-SNAP.
Therefore, the variability measured in Fig.\ref{fig:inter-rater-var} is not only due to the different raters.
We also hypothesize that the refined manual segmentation are more accurate than the original FeTA manual segmentations.

\subsection{Data availability}
% Readers and reviewers can email the corresponding author (lucas.fidon@kcl.ac.uk) to request access to the private fetal brain MRI data.
% 
Access to the data from \uzlshort{} will require approval by the ethics committee at \uzlshort{}.
Access to the data acquired at \uclhshort{}, \manchestershort{}, \belfastshort{}, \corkshort{}, \newcastleshort{}, and \liverpoolshort{} will require approval by the Caldicott guardian at \uclhshort{}. 
Access to the data from \viennashort{} can be requested using the contact details available at \url{https://www.cir.meduniwien.ac.at/research/fetal/}.
Access to the data from \kclshort{} will require approval by the ethics committee at \kclshort{}.

The FeTA dataset~\cite{payette2021automatic} is publicly available on Synapse:  \url{https://doi.org/10.7303/syn23747212}. Access requires registration to Synapse and agreement to the terms of use.
The manual segmentations for the fetal brain MRI of FeTA dataset, that we have contributed in this work and our previous work~\cite{fidon2021distributionally,fidon2021partial}, are publicly available on Zenodo: \url{https://doi.org/10.5281/zenodo.5148612} under the term of the \href{https://creativecommons.org/licenses/by-nc-nd/3.0/}{Creative Commons Attribution-NonCommercial-NoDerivs 3.0 Unported} license (CC BY-NC-ND 3.0). Access to the data is restricted. Readers and reviewers can apply for access to the data by filling in a form. The only requirement is to acknowledge that the applicant will not use those data for commercial purposes.

The first spatio-temporal atlas of the normal developing fetal brain~\cite{gholipour2017normative} that we have used is publicly available at \url{http://crl.med.harvard.edu/research/fetal_brain_atlas/}. Access requires readers to fill in an access form. Alternatively, one can download the fetal brain atlas directly from the \href{https://github.com/gift-surg/NiftyMIC/tree/master/data/templates}{\texttt{NiftyMIC}}~\cite{ebner2020automated} GitHub repository.
The second spatio-temporal atlas of the normal developing fetal brain~\cite{wu2021age}, that was based on a Chinese population, is publicly available at \url{https://github.com/DeepBMI/FBA-Chinese}.
The spina bifida spatio-temporal atlas~\cite{fidon2021atlas} is publicly available on Synapse under the terms of the \href{https://creativecommons.org/publicdomain/zero/1.0/}{Creative Commons Zero "No rights reserved" data waiver} (CC0 1.0 Public domain dedication): \url{https://www.synapse.org/#!Synapse:syn25887675/wiki/611424}.
Access requires registration to Synapse and agreement to the terms of use.

The fetal brain 3D T2w MRI segmentation protocol that was used for the manual segmentations can be found at \url{http://neuroimaging.ch/feta}.

\newpage
\section{Conclusion}

In this chapter, we have reviewed the literature on fetal brain MRI segmentation.
We identified two main groups of methods: the atlas-based methods and the deep learning-based methods.
We have outlined how the work presented in this thesis has contributed to atlas-based methods (\Chapref{chap:atlas}), to deep learning-based methods (\Chapref{chap:partialsup} and \Chapref{chap:dro}), and to the comparison and combination of atlas-based and deep learning-based methods (\Chapref{chap:twai}).

Our fetal brain 3D MRI dataset and the pre-processing applied to all the the 3D MRIs have also been described.
We will therefore whenever possible refer to this chapter in the rest of this thesis for the fetal brain data used in our experiments.
With $540$ fetal brain 3D MRIs from a total of $13$ hospitals, our dataset is the largest fetal brain MRI segmentation dataset to date.
Only subsets of the dataset are used in some chapters because we had not completed the dataset at the time of publication or submission.

\paragraph{Future work.}
The dataset described in this chapter is for a large part private.
Releasing publicly all or part of the private fetal brain MRI data is part of future work.
For now, we were able to make public our refined manual segmentations for the FeTA 2021 dataset.

The high inter-rater variability measured in this chapter for some tissue types also suggest that a precise labeling protocol for fetal brain MRI is still needed.
% where to end the fourth ventricle mask?
% where to end the brainstem mask?
% where to stop the corpus callosum?
% How to define the limit of the deep gray matter using visual landmarks?

\chapter[Label-set Loss Functions for Partial Supervision]{Label-set Loss Functions for Partial Supervision}
\label{chap:partialsup}
\minitoc
\begin{center}
	\begin{minipage}[b]{0.9\linewidth}
		\small
		\textbf{Foreword\,}
		This chapter is to a large extent an \emph{in extenso} reproduction of \cite{fidon2021label}.
		However, the experiments on BraTS in this chapter are new and have not been previously published.
	\end{minipage}
\end{center}

% Content
\section{Introduction to partially supervised learning}
As detailed in \Chapref{chap:fetaldataset}, 
the parcellation of fetal brain MRI is essential for 
the study of fetal brain development~\cite{benkarim2017toward}.
Reliable analysis and evaluation of fetal brain structures
could also support diagnosis of central nervous system pathology, patient selection for fetal surgery, evaluation and prediction of outcome, hence also parental counselling~\cite{aertsen2019reliability,danzer2020fetal,moise2016current,sacco2019fetal,zarutskie2019prenatal}.
Deep learning sets the state of the art for the automatic parcellation of fetal brain MRI~\cite{fetit2020deep,khalili2019automatic,payette2020efficient,payette2019longitudinal}.
Training a deep learning model requires a large amount of accurately annotated data.
However, manual parcellation of fetal brain 3D MRI requires highly skilled raters and is time-consuming.

Training a deep neural network for segmentation with partially segmented images is known as partially supervised learning~\cite{zhou2019prior}.
Recent studies have proposed to use partially supervised learning for body segmentation in CT~\cite{dmitriev2019learning,fang2020multi,shi2021marginal,zhou2019prior} and for the joined segmentation of brain tissues and lesions in MRI~\cite{dorent2021learning,roulet2019joint}.
One of the main challenges in partially supervised learning is to define loss functions that can handle partially segmented images.
Several previous studies have proposed to adapt existing loss functions for fully supervised learning using a somewhat ad hoc marginalization method~\cite{fang2020multi,roulet2019joint,shi2021marginal}. Theoretical motivations for 
%this method are still unknown
such marginalisation were missing.
%and it 
%is unclear if it
It also
remains unclear whether it
is the only way to build loss functions for partially supervised learning.

\begin{figure}[bt]
    \centering
    \includegraphics[width=\textwidth, trim={0.1cm 5.12cm 0.1cm 5.8cm},clip]{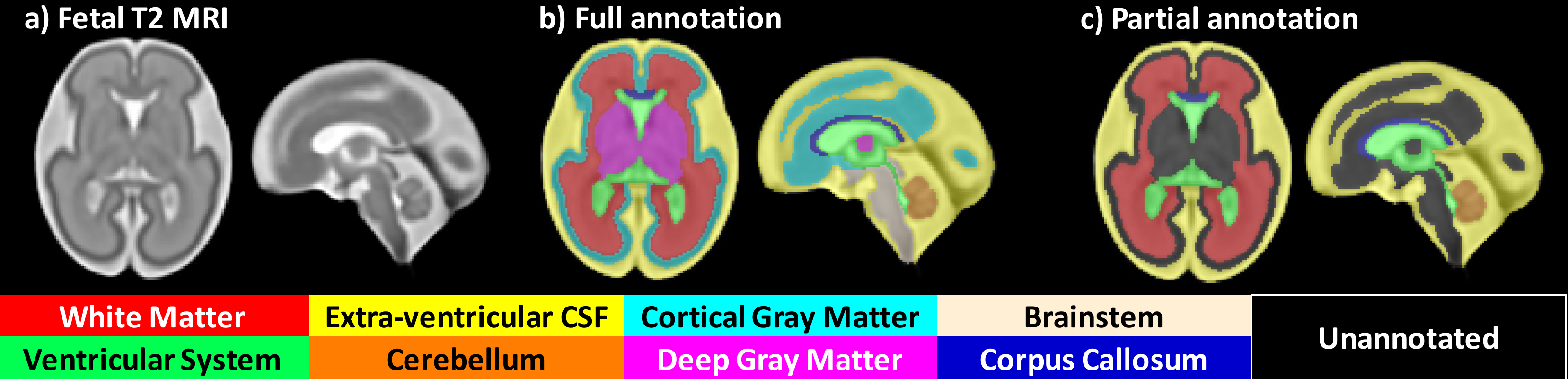}
    \caption{
    Illustration of partial annotations on a control fetal brain MRI~\cite{gholipour2017normative}.
    b) all the leaf-labels are annotated.
    c) partial annotations where cortical gray matter (CGM), deep gray matter (DGM), and brainstem (B) are not annotated.
    In this cases, unannotated voxels have the label-set annotation \{CGM, DGM, B\}.
    }
    \label{fig:partial_annotations}
\end{figure}

In this chapter, we give the first theoretical framework for loss functions in partially supervised learning.
We call those losses \emph{label-set loss functions}.
While in a fully supervised scenario, each voxel is assigned a single label, which we refer to as a \emph{leaf-label} hereafter to avoid ambiguity;
with partial supervision, each voxel is assigned a combined label, which we refer to as a \emph{label-set}.
As illustrated in Fig.~\ref{fig:partial_annotations}, a typical example of partial supervision arises when there are missing leaf-label annotations.
In which case the regions that were not segmented manually are grouped under one label-set (\emph{unannotated}).
Our theoretical contributions are threefold:
% CONTRIBUTIONS 1
1) We introduce an axiom that label-set loss functions must satisfy to guarantee 
compatibility across label-sets and leaf-labels;
% CONTRIBUTIONS 2
2) we propose a generalization of the Dice loss, leaf-Dice, that satisfies our axiom for the common case of missing leaf-labels; and
% CONTRIBUTIONS 3
3) we demonstrate that there is one and only one way to convert a classical segmentation loss for fully supervised learning into a loss function for partially supervised learning that complies with our axiom.
This theoretically justifies the marginalization method used in previous work~\cite{fang2020multi,roulet2019joint,shi2021marginal}.

In our experiments, we propose the first application of
%partially supervised learning
partial supervision
to fetal brain 3D MRI segmentation.
We use
%a total of
$244$ fetal brain volumes from 3 clinical centers, to evaluate the automatic segmentation of 8 tissue types for both normal fetuses and fetuses with open spina bifida.
We compare the proposed leaf-Dice to another label-set loss~\cite{shi2021marginal} and to two other baselines.
Our results support the superiority of label-set losses that comply with the proposed axiom and show that the leaf-Dice loss significantly outperforms the three other methods.

\section{Theory of Label-set Loss Functions}

In fully-supervised learning, we learn from
ground-truth
segmentations $\textbf{g} \in \mathbf{L}^N$
where $N$ is the number of voxels and $\mathbf{L}$ is the set of 
%leaf-labels
final labels
(e.g. white matter, 
ventricular system).
We denote elements of $\mathbf{L}$ as \emph{leaf-labels}.
In contrast,
%in partially-supervised learning,
with partial supervision,
$\textbf{g} \in \left(2^{\mathbf{L}}\right)^N$, where $2^{\mathbf{L}}$ is the set of subsets of $\mathbf{L}$.
In other words, 
%rather than a single leaf-label,
each voxel 
annotation is
%is annotated with 
a \emph{set} of leaf-labels 
(a \emph{label-set}).
%called \emph{label-set}.
The label-set of an annotated white matter voxel would be simply \{white matter\}, and the label-set of 
a
%an unannotated
voxel that could be either white matter or deep gray matter would be 
%annotated as
\{white matter, deep gray matter\}.
In both fully-supervised and partially-supervised learning,
%for a volume,
the network is trained to
perform full segmentation predictions: 
%for a volume is
$\textbf{p}\in\pset$, where $P(\mathbf{L})$ is the space of probability vectors for leaf-labels.

\subsection{Label-set Loss Functions}\label{sec:axiom}

A loss function 
$\mathcal{L}_{partial}(\cdot,\cdot)$
for partially supervised learning can be any differentiable function
that compares a proposed probabilistic network output for leaf-labels, $\textbf{p} \in \pset$, and a partial label-set ground-truth annotation, $\textbf{g} \in \gset$,
\begin{equation}
\begin{split}
    \mathcal{L}_{partial}:\,\,& \pset \times \gset \xrightarrow{} \mathds{R}\\
\end{split}
\end{equation}
However, such $\mathcal{L}_{partial}$, in general, may consider all possible label-sets as independent and ignore the relationship between a label-set and its constituent leaf-labels.
We claim that segmentation loss functions for partially supervised learning must be compatible with label-set inclusion.
The terms \emph{set} and \emph{inclusion} refer to the definitions of \emph{set} and \emph{inclusion} in set theory~\cite{bourbaki2004theory}.
For example, in the case of three leaf-labels $\textbf{L}=\{l_1, l_2, l_3\}$, let voxel $i$ be labeled with the label-set $g_i = \{l_1, l_2\}$. 
We know that the true leaf-label is either $l_1$ or $l_2$.
Therefore the exemplar leaf-label probability vectors $\textbf{p}_i = (0.4, 0.4, 0.2)$, $\textbf{q}_i = (0.8, 0, 0.2)$, and $\textbf{h}_i = (0, 0.8, 0.2)$ need to be equivalent conditionally to the ground-truth partial annotation $g_i$.
That is, the value of the loss function should be the same whether the predicted leaf-label probability vector for voxel $i$ is $\textbf{p}_i$, $\textbf{q}_i$, or $\textbf{h}_i$.

Formally, let us define the marginalization function $\Phi$ as 
\begin{equation*}
    \begin{split}
        \Phi: \,\,& P\left(\mathbf{L}\right)^N \times \left(2^{\mathbf{L}}\right)^N 
        \xrightarrow{} P\left(\mathbf{L}\right)^N\\
         & (\textbf{p}, \textbf{g}) \mapsto
         (\tilde{p}_{i,c})_{i=1\ldots N,\,c\in \mathbf{L}}
    \end{split}
     \,\,\textup{s.t.}\quad \forall i,c,\,
         \left\{
            \begin{array}{cc}
              \tilde{p}_{i,c} = \frac{1}{|g_i|} \sum_{c' \in g_i} p_{i,c'}& \textup{if}\,\, c \in g_i\\
              \tilde{p}_{i,c} = p_{i,c} &  \textup{if}\,\, c \not \in g_i
            \end{array}
         \right.
\end{equation*}
where $|.|$ is the number of elements in a set.
In the previous example, 
$
\Phi(\textbf{p};\textbf{g})_i
= \Phi(\textbf{q};\textbf{g})_i
= \Phi(\textbf{h};\textbf{g})_i
= (0.4, 0.4, 0.2).
$

We define \textbf{label-set loss functions} as the functions $\mathcal{L}_{partial}$ that satisfy the axiom
\begin{equation}
    \forall \textbf{g},
    % \in \gset,\,
    \forall \textbf{p},\textbf{q},\quad
    % \in \pset,\quad
    \Phi(\textbf{p}; \textbf{g}) = \Phi(\textbf{q}; \textbf{g})
    \implies 
    \mathcal{L}_{partial}(\textbf{p}, \textbf{g})
    = \mathcal{L}_{partial}(\textbf{q}, \textbf{g})
    \label{eq:axiom}
\end{equation}
We demonstrate that a loss $\mathcal{L}$ is a label-set loss function if and only if
\begin{equation}
    \forall (\textbf{p},\textbf{g}), \quad \mathcal{L}(\textbf{p}, \textbf{g}) = \mathcal{L}\left(\Phi(\textbf{p}; \textbf{g}), \textbf{g}\right)
    \label{eq:lemma_general}
\end{equation}
See the supplementary material for a proof of this equivalence.

\subsection{Leaf-Dice: A Label-set Generalization of the Dice Loss}\label{sec:leafdice}

In this section, as per previous work~\cite{dmitriev2019learning,dorent2021learning,fang2020multi,roulet2019joint,shi2021marginal,zhou2019prior}, we consider the particular case in which, per training example,
there is only one label-set 
that is not a singleton and contains all the leaf-labels that were not manually segmented in this example.
An illustration for fetal brain segmentation can be found in Fig.\ref{fig:partial_annotations}.

We propose a generalization of the mean class Dice Loss~\cite{fidon2017generalised,milletari2016v} for this particular case
and prove that it satisfies our axiom~\eqref{eq:axiom}.

Let $\textbf{g} \in \gset$ be a partial label-set segmentation such that there exists a label-set $\mathbf{L}_{\textbf{g}}' \subsetneq \mathbf{L}$ that contains all the leaf-labels that were not manually segmented for 
this subject.
Therefore, $\textbf{g}$ takes its values in 
$\{\mathbf{L}_{\textbf{g}}'\} \cup \left\{\{c\}\,|\,c \in \mathbf{L}\setminus\mathbf{L}_{\textbf{g}}'\right\}$.
We demonstrate that the leaf-Dice loss defined below is a label-set loss function 
\begin{equation}
    \forall \textbf{p},\quad
    \mathcal{L}_{Leaf-Dice}(\textbf{p}, \textbf{g}) =
    1 - 
    \frac{1}{|\mathbf{L}|} 
    \sum_{c \in \mathbf{L}} 
    \frac{2 \sum_i \mathds{1}(g_i =\{c\})\,p_{i,c}}{
    \sum_i \mathds{1}(g_i =\{c\})^{\alpha}
    + \sum_{i} p_{i,c}^{\alpha}
    +\epsilon
    }
    \label{eq:ls_dice}
\end{equation}
where $\alpha \in \{1,2\}$
(in line with the variants of soft Dice encountered in practice), 
and
$\epsilon > 0$ is a small constant.
A proof that $\mathcal{L}_{Leaf-Dice}$ satisfies~\eqref{eq:lemma_general} can be found in the supplementary material.
It is worth noting that using
$\mathcal{L}_{Leaf-Dice}$ is not equivalent to just masking out the unannotated voxels,
i.e. the voxels
$i$ such that $g_i=\mathbf{L'}_{\textbf{g}}$.
Indeed,
for all the $c \in \mathbf{L}\setminus \mathbf{L}_{\textbf{g}}'$,
the term $\sum_i p_{i,c}^{\alpha}$ in the denominator pushes $p_{i,c}$ towards $0$ for all the voxels indices $i$ including the indices $i$ for which $g_i=\mathbf{L}_{\textbf{g}}'$.
As a result, when $g_i=\mathbf{L}_{\textbf{g}}'$, only the $p_{i,c}^{\alpha}$ for $c \not \in \mathbf{L}_{\textbf{g}}'$ are pushed toward $0$, which in return pushes the $p_{i,c}^{\alpha}$ for $c \in \mathbf{L}_{\textbf{g}}'$ towards $1$ since $\sum_{c \in \mathbf{L}} p_{i,c}=1$.

\subsection{Converting Classical Loss Functions to Label-set Loss Functions}\label{sec:conversion}
In this section, we demonstrate that there is one and only one canonical method to convert a segmentation loss function for fully supervised learning into a label-set loss function for partially supervised learning satisfying~\eqref{eq:lemma_general}.

Formally, a label-set segmentation $\textbf{g}$ can be seen as a soft segmentation using an injective function
$
\Psi: \left(2^{\mathbf{L}}\right)^N \hookrightarrow{} P\left(\mathbf{L}\right)^N
$
that satisfies the relation
\begin{equation}
    \forall \textbf{g} \in \gset,\,\forall i,c,\quad 
    % \left(
        [\Psi(\textbf{g})]_{i,c} > 0 \implies c \in g_i
    % \right)
\end{equation}
Note however, that the function $\Psi$ is not unique. 
Following the maximum entropy principle leads to choose
\begin{equation}
    \Psi_0: (g_i) \mapsto (\tilde{p}_{i,c})
    \quad \textup{s.t.}\quad \forall i,c\,
         \left\{
            \begin{array}{cc}
              \tilde{p}_{i,c} = \frac{1}{|g_i|} & \textup{if}\,\, c \in g_i\\
              \tilde{p}_{i,c} = 0 &  \textup{otherwise}
            \end{array}
         \right.
    \label{eq:psi0}
\end{equation}
We are interested in converting a loss function for fully-supervised learning
$
\mathcal{L}_{fully}:\,\, P\left(\mathbf{L}\right)^N \times P\left(\mathbf{L}\right)^N \xrightarrow{} \mathds{R}
$
into a label-set loss function for partial supervision defined as
$\mathcal{L}_{partial}(\textbf{p}, \textbf{g}) = \mathcal{L}_{fully}(\textbf{p}, \Psi\left(\textbf{g}\right))$.

Assuming that $\mathcal{L}_{fully}$ is minimal if and only if the predicted segmentation and the soft ground-truth segmentation are equal,
we demonstrate in the supplementary material that $ \mathcal{L}_{partial}$ is a label-set loss function if and only if
\begin{equation}
\forall (\textbf{p}, \textbf{g})\in \pset \times \gset,\quad
\left\{
\begin{aligned}
    \mathcal{L}_{partial}(\textbf{p}, \textbf{g})
    &= \mathcal{L}_{fully}\left(\Phi(\textbf{p}; \textbf{g}), \Psi(\textbf{g})\right)\\
    \Psi(\textbf{g}) &= \Psi_0(\textbf{g})
\end{aligned}
\right.
\label{eq:partial_conversion_unique}
\end{equation}
% 
% A proof can be found in the supplementary material.
% 
Therefore, 
% when dealing with partial annotations,
the only way to convert a fully-supervised loss to a loss for partially-supervised learning that complies with our axiom \eqref{eq:axiom} %fully supervised learning into a label-set loss function 
is to use the marginalization function $\Phi$ on the predicted segmentation and $\Psi_0$ on the ground-truth partial segmentation.
For clarity, we emphasise that the Leaf-Dice is a generalisation rather than a conversion of the Dice loss.

% \subsubsection{Related work.}
\subsubsection{Related Work.}
When $\mathcal{L}_{fully}$ is the mean class Dice loss~\cite{fidon2017generalised,milletari2016v} and the values of $\textbf{g}$ are a partition of $\mathbf{L}$,
like in section~\ref{sec:leafdice},
one can prove that $\mathcal{L}_{partial}$ is the marginal Dice loss~\cite{shi2021marginal} (proof in supplementary material).
Similarly, when $\mathcal{L}_{fully}$ is the cross entropy loss, $\mathcal{L}_{partial}$ is the marginal cross entropy loss~\cite{fang2020multi,roulet2019joint,shi2021marginal}.
Note that in~\cite{fang2020multi,roulet2019joint,shi2021marginal}, the marginalization approach is proposed as a possible method to convert the loss function.
We prove that this is the only method.
% was proposed ad hoc while we prove that it is a sufficient and necessary condition.

\section{MRI Datasets}\label{sec:dataset}
\subsection{Fetal Brain MRI Data with Partial Segmentations}

In this section, we describe the fetal brain 3D MRI datasets that were used.
This is a subset of the dataset described in \Chapref{chap:fetaldataset}.
Details about the MRI data acquisition and pre-processing are described in \Chapref{chap:fetaldataset}.

\begin{table}[bt]
    \caption{
	\textbf{
	Number of 3D MRI and number of manual segmentations available per tissue types.}
	\textcolor{red}{\textbf{WM}}: white matter,
	\textcolor{ForestGreen}{\textbf{Vent}}: ventricular system,
	\textcolor{orange}{\textbf{Cer}}: cerebellum,
	\textcolor{Goldenrod}{\textbf{ECSF}}: extra-ventricular CSF,
	\textcolor{Cyan}{\textbf{CGM}}: cortical gray matter,
	\textcolor{Purple}{\textbf{DGM}}: deep gray matter,
	\textcolor{Gray}{\textbf{BS}}: brainstem,
	\textcolor{blue}{\textbf{CC}}: corpus callosum.}
    \begin{tabularx}{\textwidth}{ c c *{9}{Y}}
		\toprule
		\textbf{Split} &
		\textbf{Condition} &
		\textbf{MRI} & 
		\textcolor{red}{\textbf{WM}} & \textcolor{ForestGreen}{\textbf{Vent}} &
		\textcolor{orange}{\textbf{Cer}} & \textcolor{Goldenrod}{\textbf{ECSF}} &
		\textcolor{Cyan}{\textbf{CGM}} & \textcolor{Purple}{\textbf{DGM}} &
		\textcolor{Gray}{\textbf{BS}} & \textcolor{blue}{\textbf{CC}}\\
		\midrule
		Training & Atlas &
		18 & 18 & 18 & 18 & 18 & 18 & 18 & 18 & 18\\
		Training & Control &
		116 & 116 & 116 & 116 & 54 & 0 & 0 & 0 & 18\\
		Training & Spina Bifida & 
		30 &  30 &  30 &  30 & 0  & 0  & 0  & 0  & 0 \\
	    \midrule
	    Testing & Control & 34 & 34 & 34 & 34 & 34 & 15 & 15 & 15 & 0\\
		Testing & Spina Bifida & 66 & 66 & 66 & 66 & 66 & 25 & 25 & 25 & 41\\
	    \bottomrule
	\end{tabularx}
    \label{tab:data}
\end{table}

\paragraph{Training Data for Fully Supervised Learning.}
$18$ fully-annotated control fetal brain MRI from a spatio-temporal fetal brain MRI atlas~\cite{gholipour2017normative}.

\paragraph{Training Data for Partially Supervised Learning.}
$18$ fully-annotated volumes from the fully-supervised dataset, combined with
$146$ partially annotated fetal brain MRI from a private dataset.
The segmentations available for those 3D MRI are detailed in Table~\ref{tab:data}.

\paragraph{Multi-centric Testing Data.}
The testing dataset contains a total of 
$100$ fetal brain 3D MRI.
This includes $60$ volumes from \uzl{} and $40$ volumes from the publicly available FeTA dataset~\cite{payette2021automatic}.
The segmentations available for those 3D MRI are detailed in Table~\ref{tab:data}.
The 3D MRI of the FeTA dataset come from a different center than the training data.
Automatic brain masks for the FeTA data were obtained using atlas affine registration with a normal fetal brain and a spina bifida spatio-temporal atlas~\cite{fidon2021atlas,gholipour2017normative}.

\subsection{Brain Tumor multi-parametric MRI Dataset}

\begin{figure}[bt]
    \centering
    \includegraphics[width=\linewidth]{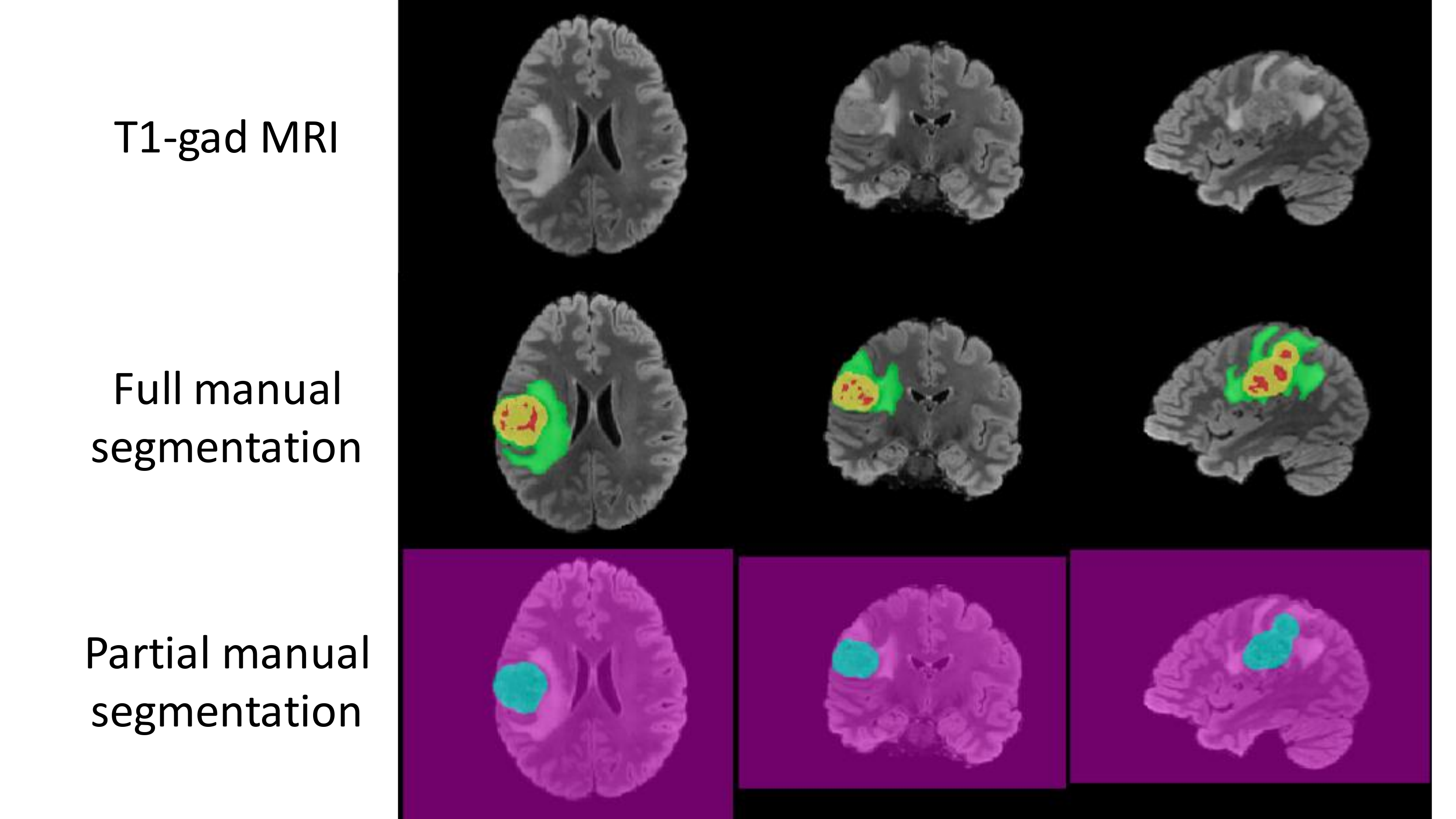}
    \caption{
    Illustration of the partial annotations used for experiments on brain tumor segmentation.
    In blue is the tumor core, a label-set containing the enhancing tumor (yellow) and non-enhancing tumor (red).
    In purple is another label-set that contains the peritumoral edema (green) and the background.
    }
    \label{fig:brats_task}
\end{figure}

We have used the BraTS 2021 training dataset\footnote{\url{https://www.synapse.org/\#!Synapse:syn25829067/wiki/610865}}~\cite{baid2021rsna} in our experiments.
The dataset contains $1251$ cases, and the same four MRI sequences (T1, ceT1, T2, and FLAIR) for all cases, corresponding to patients with either a high-grade Gliomas~\cite{bakas2017HGG} or a low-grade Gliomas~\cite{bakas2017LGG}.
All the cases were manually segmented for peritumoral edema, enhancing tumor, and non-enhancing tumor core using the same labeling protocol~\cite{menze2014multimodal,bakas2018identifying,bakas2017advancing,baid2021rsna}.
For each case, the four MRI sequences are available after co-registration to the same anatomical template, interpolation to $1$mm isotropic resolution, and skull stripping~\cite{menze2014multimodal}.

\paragraph{Partial Labels for Brain Tumor Segmentation.}

The $1215$ studies of BraTS 2021 training dataset are all fully-segmented.
To perform an ablation study on the ratio of partially and fully segmented samples, we have created partial segmentation for all of them using two label-set: 
the tumor core containing the enhancing tumor and the non-enhancing tumor,
and a label-set containing the background and the peritumoral edema.
An illustration is given in Fig.~\ref{fig:brats_task}.
Such partial segmentations correspond to the manual segmentations that would be typically available in clinical data used for resection surgery or radiotherapy.

\section{Experiments}\label{sec:experiments}
In this section, we compare three partially supervised methods and one fully supervised method using the fetal brain 3D MRI dataset described in section~\ref{sec:dataset}.
We also perform an ablation study on the ratio of partially segmented and fully segmented cases using the BraTS 2021 dataset as described in section~\ref{sec:dataset}.

\begin{figure}[tb]
    \centering
    \includegraphics[width=\columnwidth]{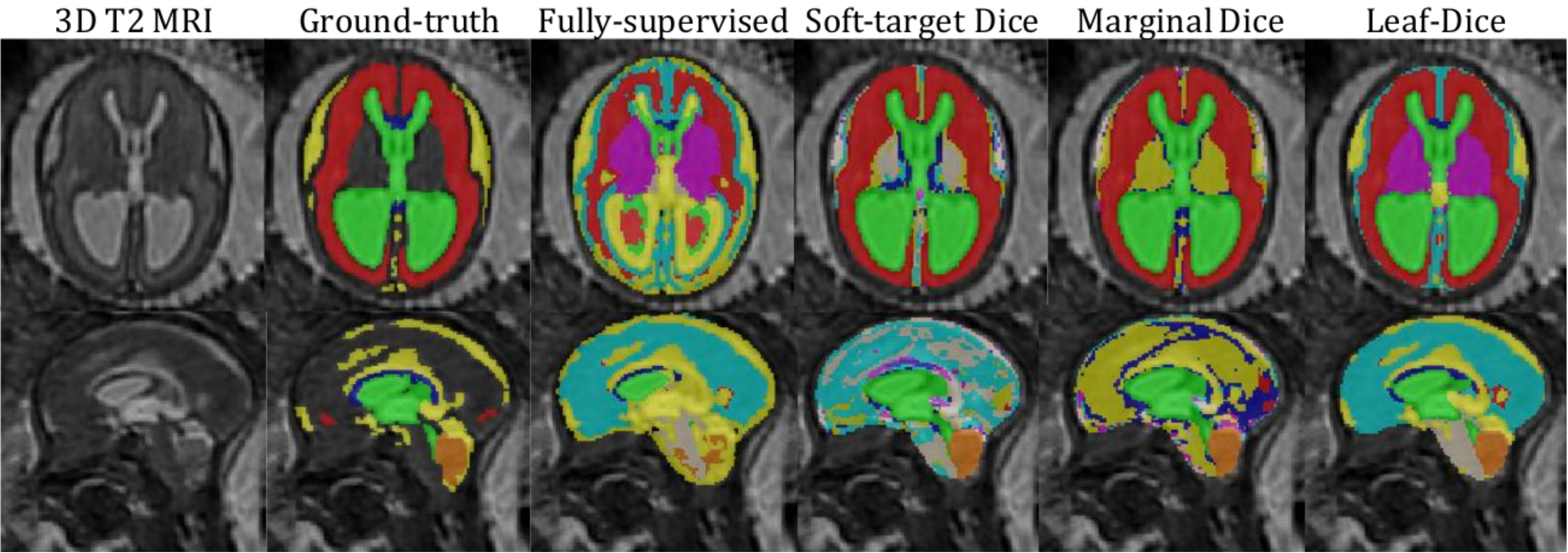}
    \caption{Qualitative comparison on an open spina bifida case.
    Only the proposed Leaf-Dice loss provides satisfactory segmentations for all tissue types.
    }
    \label{fig:qualitative_results}
\end{figure}

\paragraph{Deep Learning Pipeline for Fetal Brain Segmentation.}
We use a 3D U-Net~\cite{cciccek20163d} architecture with $5$ levels, 
% $30$ channels at its lowest level, 
leaky ReLU and instance normalization~\cite{ulyanov2016instance}.
For training and inference, the entire volumes were used as input of the 3D U-Net after padding to $144\times160\times144$ voxels.
% which was chosen to fit the largest fetal brain volume in our dataset.
% 
For each method of Table~\ref{tab:models_results}, an ensemble of $10$ 3D U-Net is used.
Using an ensemble of models 
% is the state of the art in image segmentation.
% and it 
makes the comparison of segmentation results across methods less sensitive to the random initialization of each network.
Each 3D U-Net is trained using a random split of the training dataset into $90\%$ for training and $10\%$ for early stopping.
During training, we used a batch size of $3$, a learning rate of $0.001$, and Adam~\cite{kingma2014adam} optimizer with default hyperpameter values. 
The learning rate was tuned for the fully supervised learning baseline.
Random right/left flip, 
random scaling, 
gamma intensity augmentation,
contrast augmentation,
and additive Gaussian noise
were used as data augmentation during training.
Our code for the label-set loss functions and the deep learning pipeline are publicly available\footnote{\url{https://github.com/LucasFidon/label-set-loss-functions}}$^{,}$\footnote{\url{https://github.com/LucasFidon/fetal-brain-segmentation-partial-supervision-miccai21}}.

\paragraph{Deep Learning Pipeline for Brain Tumor Segmentation.}
We have also used pipeline based on 3D U-Net for the experiments on brain tumor segmentation.
The details of the deep learning pipeline can be found in our previous work~\cite{fidon2021generalized}.

For the ablation study on BraTS, we perform an ablation study on the ratio of partially segmented and fully segmented cases. 
We compare training with fully-supervised learning using only the fully-segmented cases and training with partially-supervised learning using both fully-segmented and partially-segmented cases.
For fully-supervised learning we use the sum of the Dice loss and the cross entropy loss.
For partially-supervised learning we use the sum of the marginalized Dice loss and the marginalized cross entropy.
It is worth noting that the leaf-Dice loss cannot be used in this case because all voxels are annotated using label-set for partially-segmented cases. 
This is because with the proposed definition of the leaf-Dice loss function, the loss is uniformly equal to $1$ and its gradient to $0$ when all voxels are annotated using label-sets containing more than one element.

The BraTS 2021 dataset was split into $5$ folds and the same folds were used for all the experiments on brain tumor segmentation.
Each configuration is evaluated using 5-fold cross validation on the $1251$ cases of the BraTS 2021 dataset.

\paragraph{Hardware.}
For training we used NVIDIA Tesla V100 GPUs.
Training each model took from one to two days.
For inference, we used a NVIDIA GTX 1070 GPU.
The inference for a 3D MRI takes between one and two minutes.
It is worth noting that the proposed loss functions are not the computational bottleneck during training and have no influence on the inference time.

\paragraph{Specificities for each Loss Function for Fetal Brain Segmentation.}
Baseline 1 is trained using fully supervised learning,
the mean class Dice loss~\cite{fidon2017generalised,milletari2016v}, referred as $\mathcal{L}_{Dice}$, and the training dataset for fully supervised learning of Section~\ref{sec:dataset} (18 training volumes only).
The three other methods are trained using partially supervised learning and the training dataset for partially supervised learning of Section~\ref{sec:dataset}.
Baseline 2 is trained using the soft-target Dice loss function defined as 
$\mathcal{L}_{Dice}(\textbf{p}, \Psi_0(\textbf{g}))$, where $\Psi_0$ is defined in \eqref{eq:psi0}. 
Note that 
Baseline 2
%this soft-target Dice loss 
does not satisfy the
label-set axiom~\eqref{eq:axiom}.
%axiom of label-set loss functions in Section~\ref{sec:axiom}.
% 
Baseline 3 is trained using the marginal Dice loss~\cite{shi2021marginal}. 
Our method is trained using the proposed Leaf-Dice loss defined in \eqref{eq:ls_dice}. 
The loss functions of Baseline~3 and our method satisfy the axiom of label-set loss functions \eqref{eq:axiom}.

\paragraph{Statistical Analysis.}
We used the two-sided Wilcoxon signed-rank test.
Differences in mean values are considered significant when $p < 0.01$.

\begin{table}[t!]
	\caption{\textbf{Evaluation on the multi-centric testing fetal brain dataset (100 3D MRIs).}}
	We report mean (standard deviation) for the Dice score (DSC) in percentages and the Hausdorff distance at $95\%$ (HD95) in millimeters for the eight tissue types.
    \underline{Methods underlined} are trained using partially supervised learning and can therefore take advantage of the training volumes that are partially annotated ($146$ volumes) while the baseline trained with fully-supervised learning can only used the fully-annotated volumes ($18$ volumes).
    The same testing images and annotations are used for all methods for evaluation.
    \textbf{Loss functions in bold} satisfy the axiom of label-set loss functions.
    Best mean values are in bold.
    Mean values significantly better with $p<0.01$ (resp. worse) than the ones achieved by the fully-supervised learning baseline are marked with a $*$ (resp. a $\dag$).
	\begin{tabularx}{\columnwidth}{c *{9}{Y}}
	    \toprule
		\textbf{Model} & \textbf{Metric} & 
		\textcolor{red}{\textbf{WM}} & \textcolor{ForestGreen}{\textbf{Vent}} &
		\textcolor{orange}{\textbf{Cer}} & \textcolor{Goldenrod}{\textbf{ECSF}} &
		\textcolor{Cyan}{\textbf{CGM}} & \textcolor{Purple}{\textbf{DGM}} &
		\textcolor{Gray}{\textbf{BS}} & \textcolor{blue}{\textbf{CC}}\\
		\midrule
		Baseline 1 & DSC & 76.4 (12.5) & 72.1 (18.8) & 67.3 (28.6) & 63.9 (31.3) & 47.3 (10.9) & $\textbf{72.7}$ (8.7) & 56.0 (27.7) & 51.6 (10.5)\\
		Fully-Supervised & HD95 & 3.3 (1.2) & 4.8 (3.7) & 4.9 (5.2) & 7.3 (7.8) & 5.3 (0.9) & $\textbf{5.8}$ (2.1) & 9.1 (8.2) & 5.1 (3.1)\\
		\cmidrule(lr){1-10}
		\underline{Baseline 2} & DSC & $89.5^*$ (6.5) & $87.5^*$ (9.6) & $87.2^*$ (10.3) & $37.7^{\dag}$ (34.2) & $31.4^{\dag}$ (14.8) & $18.2^{\dag}$ (20.5) & $20.2^{\dag}$ (13.1) & $12.0^{\dag}$ (10.8)\\
		Soft-target Dice & HD95 & $1.7^*$ (1.3) & $1.6^*$ (2.2) & $2.2^*$ (4.5) & 7.1 (6.8) & $6.2^{\dag}$ (2.0) & $24.2^{\dag}$ (9.0) & $33.9^{\dag}$ (5.9) & $29.3^{\dag}$ (8.0)\\
		\cmidrule(lr){1-10}
		\underline{Baseline 3} & DSC & $89.6^*$ (6.7) & $87.7^*$ (10.4) & $87.6^*$ (9.5) & $66.6^*$ (27.7) & 43.9 (15.1) & $37.8^{\dag}$ (11.3) & $39.4^{\dag}$ (16.8) & $11.1^{\dag}$ (12.3)\\
		\textbf{Marginal Dice} & HD95 & $1.7^*$ (1.3) & $1.6^*$ (2.2) & $2.4^*$ (5.4) & $6.2^*$ (7.6) & $4.4^{*}$ (1.4) & $26.7^{\dag}$ (6.7) & $33.4^{\dag}$ (6.1) & $28.7^{\dag}$ (6.3)\\
		\cmidrule(lr){1-10}
		\underline{Ours} & DSC & $\textbf{91.5}^*$ (6.7) & $\textbf{90.7}^*$ (8.9) & $\textbf{89.6}^*$ (10.1) & $\textbf{75.3}^*$ (24.9) & $\textbf{56.6}^*$ (14.3) & 71.4 (8.6) & $\textbf{61.5}^*$ (21.7) & $\textbf{62.0}^*$ (10.9)\\
		\textbf{Leaf-Dice} & HD95 & $\textbf{1.5}^*$ (1.1) & $\textbf{1.4}^*$ (2.0) & $\textbf{1.7}^*$ (1.8) & $\textbf{5.4}^*$ (8.3)& $\textbf{3.9}^*$ (1.3) & $7.3^{\dag}$ (2.3) & $\textbf{7.9}$ (4.0) & $\textbf{2.9}^*$ (1.5)\\
        \bottomrule
	\end{tabularx}
	\label{tab:models_results}
\end{table}

\begin{figure}[bt]
    \centering
    \includegraphics[width=\linewidth]{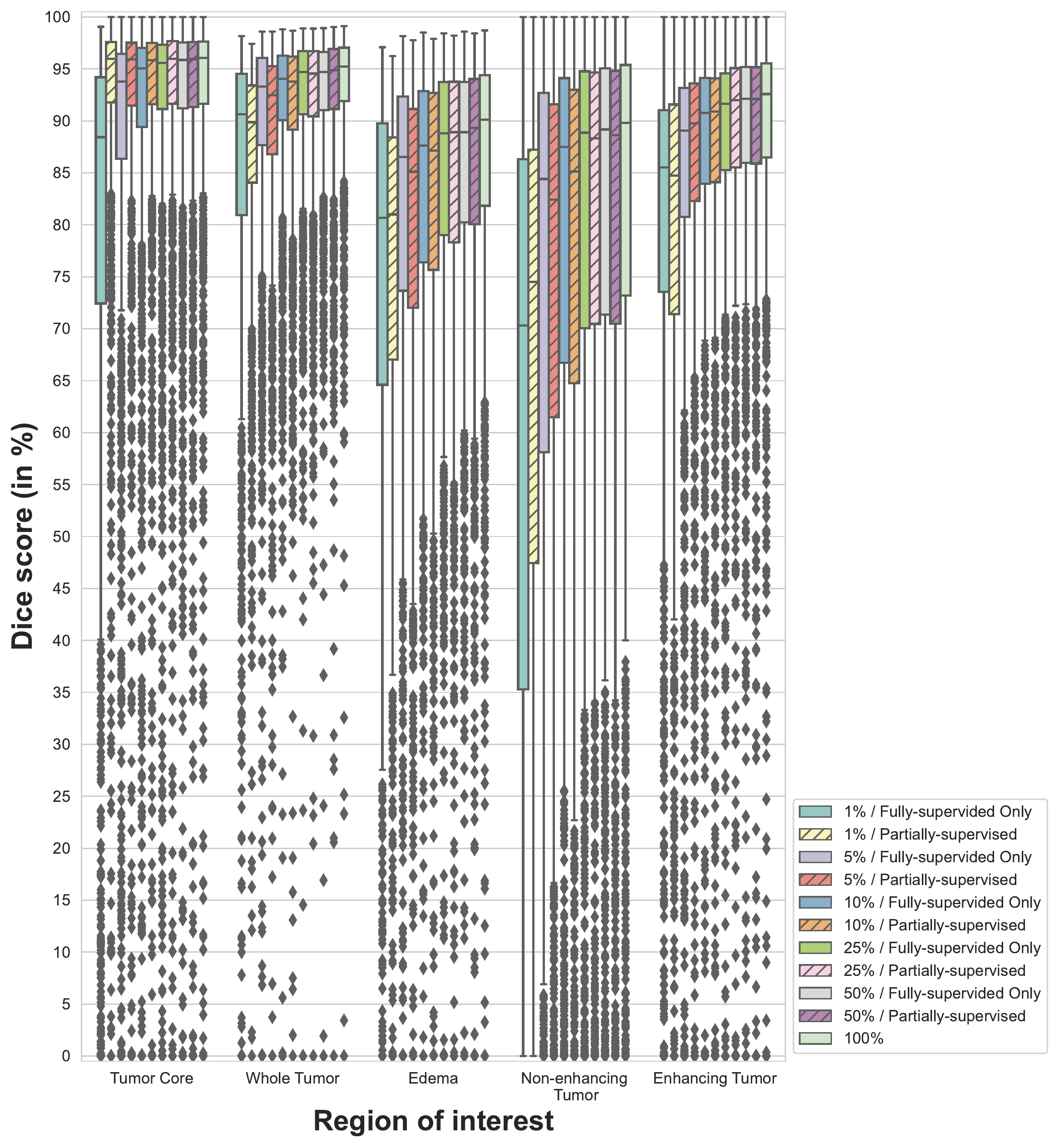}
    \caption{
    \textbf{Dice score evaluation for the ablation study on the ratio of partially segmented and fully segmented cases using the BraTS 2021 training dataset.}
    The \textbf{percentages} in the legend correspond to the percentage of fully-segmented cases used during training. The rest of the segmentation of the training dataset are converted into partial segmentations as described in Fig.~\ref{fig:brats_task}.
    \textbf{Fully-supervised learning} (no hatches) corresponds to training with standard Dice loss plus cross entropy using only the fully segmented data. In this case the partially segmented cases are not used.
    \textbf{Partially-supervised learning} (hatches) corresponds to training with both fully-segmented and partially-segmented cases using the marginalized Dice loss plus the marginalized cross entropy.
    Each box plot results from 5-fold cross validation on the $1251$ cases.
    Box limits are the first quartiles and third quartiles. The central ticks are the median values. The whiskers extend the boxes to show the rest of the distribution, except for points that are determined to be outliers.
    Outliers are data points outside the range median $\pm 1.5\times$ interquartile range.
    }
    \label{fig:brats_partialsup_dice}
\end{figure}

\begin{figure}[bt]
    \centering
    \includegraphics[width=\linewidth]{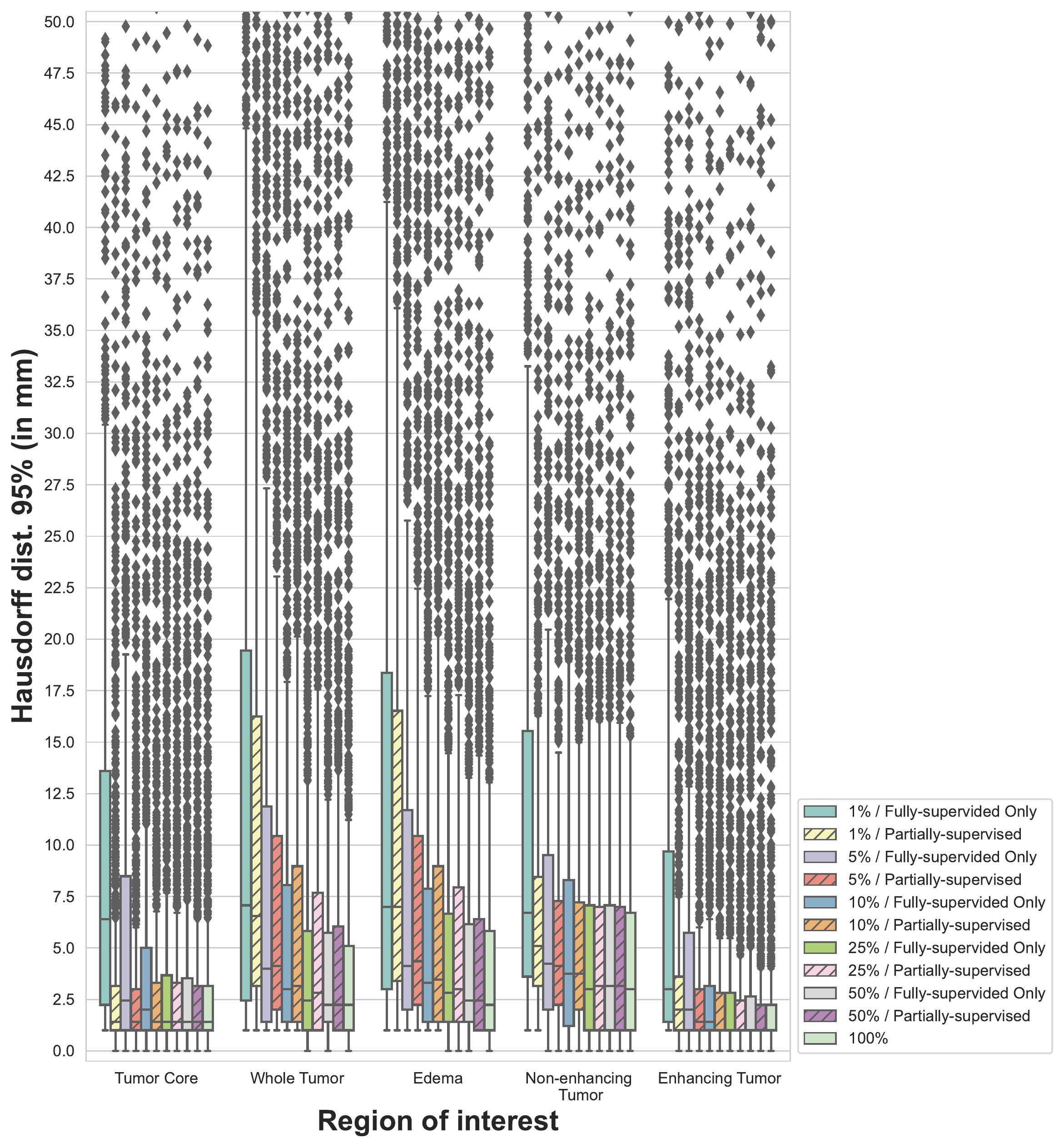}
    \caption{
    \textbf{Hausdorff distance at percentile 95\% evaluation for the ablation study on the ratio of partially segmented and fully segmented cases using the BraTS 2021 training dataset.}
    The \textbf{percentages} in the legend correspond to the percentage of fully-segmented cases used during training. The rest of the segmentation of the training dataset are converted into partial segmentations as described in Fig.~\ref{fig:brats_task}.
    \textbf{Fully-supervised learning} (no hatches) corresponds to training with standard Dice loss plus cross entropy using only the fully segmented data. In this case the partially segmented cases are not used.
    \textbf{Partially-supervised learning} (hatches) corresponds to training with both fully-segmented and partially-segmented cases using the marginalized Dice loss plus the marginalized cross entropy.
    Each box plot results from 5-fold cross validation on the $1251$ cases.
    Box limits are the first quartiles and third quartiles. The central ticks are the median values. The whiskers extend the boxes to show the rest of the distribution, except for points that are determined to be outliers.
    Outliers are data points outside the range median $\pm 1.5\times$ interquartile range.
    Outliers higher than $50$ mm are not displayed.
    }
    \label{fig:brats_partialsup_hd}
\end{figure}

\paragraph{Comparison of Fully Supervised and Partially Supervised Learning.}
The quantitative evaluation for fetal brain MRI segmentation can be found in Table~\ref{tab:models_results}.
The partially supervised learning methods of Table~\ref{tab:models_results} all perform significantly better than the fully supervised learning baseline in terms of Dice score and Hausdorff distance on the tissue types for which annotations are available for all the training data of Table~\ref{tab:data}, i.e. white matter, ventricles, and cerebellum.
Some tissue types segmentations are scarce in the partially supervised training dataset as can be seen in Table~\ref{tab:data}, i.e. cortical gray matter, deep gray matter, brainstem, and corpus callosum.
This makes it challenging for partially supervised methods to learn to segment those tissue types.
Only Leaf-Dice achieves similar or better segmentation results than the fully supervised baseline for the scarce tissue types, except in terms of Hausdorff distance for the deep gray matter.
The extra-ventricular CSF is an intermediate case with almost half the data annotated.
The Leaf-Dice and the Marginalized Dice significantly outperforms the fully supervised baseline for all metrics for the extra-ventricular CSF, and the soft-target Dice performs significantly worse than this baseline.

The quantitative evaluation for brain tumor segmentation in MRI can be found in Fig.~\ref{fig:brats_partialsup_dice} for the Dice score and Fig.~\ref{fig:brats_partialsup_hd} for the Hausdorff distance.
Here we compare partially-supervised learning and fully-supervised learning for different percentages of fully segmented cases in the training dataset: $1\%$, $5\%$, $10\%$, $25\%$, $50\%$, and $100\%$ (fully segmented dataset).

The tumor core is one of the label-set and is therefore always segmented for every cases independently to the percentage
However, in theory, only the partially-supervised method using a label-set loss, exploits all of them all the time.
We observe that the segmentation performance of the partially-supervised method is consistently optimal in terms of Dice scores and Hausdorff distances for all percentages.
This suggests that the labels-set loss function allows the 3D U-Nets to learn to segment the tumor core to the same level of accuracy using partial and full segmentations.
For $1\%$, $5\%$ and $10\%$ the partially-supervised method outperforms the fully-supervised method for the tumor core with a large margin for the two metrics.
For higher percentages, the partially-supervised method achieves similar or slightly better metric values than the fully-supervised method for the tumor core.

For $1\%$ and $5\%$, the partially-supervised method also outperforms the fully-supervised method for all the other regions in terms of Hausdorff distance, in particular at the queue of the Hausdorff distance distribution with the highest values.
For the Dice score it is less clear.
For $1\%$ the tail of the Dice score distribution shows higher values for the partially-supervised method for whole tumor, edema, and non-enhancing tumor but not for enhancing tumor.
Overall, this suggests that when many more partially segmented cases than fully-segmented cases are available for training, partially-supervised learning with a label-set loss function can help improving the robustness for the regions not annotated in the partially-segmented data.
 
For percentages strictly higher than $5$, the partially-supervised learning method achieve similar or lower segmentation performance for the regions not segmented in the partially-segmented data for the two metrics.
This is in contrast to the higher segmentation performance for the tumor core reported above for the partially supervised method.
We hypothesize this might be because the label-set loss implicitly put a higher weight on the tumor core as compared to the other regions and that the number of additional partially segmented cases is not enough to compensate this weighting effect in this case.

It also worth noting, that there are little differences of performance between percentages $100$, $50$, and $25$ for the fully-supervised method for the two metrics.
In other words, using only a quarter of the training dataset with fully-supervised learning already leads to segmentation performances similar to using the full training dataset.
In the appendix~\ref{appendix:partial-brats} we have performed experience with a larger 3D U-Net with $4$ times more trainable parameters using the full training dataset with fully-supervised learning.
We found no improvements after increasing the capacity of the 3D U-Net.
This suggests we may have reached a stage for BraTS where more training data does not lead to much better segmentations even though many outliers remain present.

\paragraph{Comparison of Loss Functions for Partially Supervised Learning.}
The Leaf-Dice performs significantly better than the soft-target Dice for all tissue types and all metrics.
The Marginal Dice performs significantly better than the soft-target Dice in terms of Dice score for extra-ventricular CSF, cortical gray matter, deep gray matter, and brainstem.
Since the soft-target Dice loss is the only loss that does not satisfy the proposed axiom for label-set loss functions, this suggests label-set loss functions satisfying our axiom perform better in practice.

In addition, our Leaf-Dice performs significantly better than the Marginal Dice~\cite{shi2021marginal} for all metrics and all tissue types.
The qualitative results of Fig.~\ref{fig:qualitative_results} also suggest that Leaf-Dice performs better than the other approaches.
% for the tissue types for which only a few training data are annotated.
The Marginal Dice loss and the soft-target Dice loss are obtained by adapting the existing Dice loss function for fully-annotated volumes to partially-annotated volumes. This is not the case of the leaf-Dice that has been proposed specifically for partially-annotated volumes.
This suggests that using a converted fully supervised loss function, as proposed in section~\ref{sec:conversion} and in previous work~\cite{fang2020multi,roulet2019joint,shi2021marginal}, 
% might not the best approach for partially supervised learning.
may be outperformed by dedicated generalised label-set loss functions.

\paragraph{Outliers and worst-case performance in the brain tumor segmentation experiment.}
    In this section, we study the consistency of the outliers across methods for brain tumor segmentation.
    The goal is to address the question: are the cases with poor segmentations always the same when using fully supervised learning or partially supervised learning and when using different ratio of the manual annotations in the BraTS 2021 dataset?
    
    For simplicity, let us define as outliers the brain tumor segmentations for which the Dice score for at least one of the ROIs is equal to $0$.
    
    We have computed the intersection over union of the set of outliers for every method compared to training on the full dataset with all the annotations (the "$100\%$" model in Figure 4.4). This metric measures the overlap of the set of outliers across models.
    
    We find that all the intersection over unions of the sets of outliers are between $50\%$ and $71\%$.
    If the threshold of the Dice score used to define outliers is chosen to be "Dice score $\leq 25\%$", all the intersection over unions of the sets of outliers are between $55\%$ and $82\%$.
    And for "Dice score $\leq 50\%$", all the intersection over unions of the sets of outliers are between $68\%$ and $87\%$.
    
    The outliers are therefore highly consistent across methods.
    In particular, it implies that the majority of the outliers cannot be attributed to the stochasticity of the training method.
    
    There are several possible causes for those recurrent outliers.
    This could be a limitation the nnU-Net deep learning pipeline~\cite{isensee2021nnu} that we use for all models. Only the annotations and the loss vary across the models that we have compared.
    The recurrent outliers could also be due to suboptimal manual annotations.
    A third possible cause, would be that the recurrent outliers correspond to isolated samples out of the training domain distribution.
    Assessing the later for BraTS 2021 is not possible because other than the volumes and the manual segmentations, no center or patient information are publicly available.
    The problem of robustness to underrepresented and out-of-distribution samples is studied in \Chapref{chap:dro} and \Chapref{chap:twai} of this thesis.

\section{Conclusion}
In this chapter, we present the first axiomatic definition of label-set loss functions for training a deep learning model with partially segmented images.
We propose a generalization of the Dice loss, Leaf-Dice, that complies with our axiom for the common case of missing labels that were not manually segmented.
We prove that loss functions that were proposed in the literature for partially supervised learning satisfy the proposed axiom.
In addition, we prove that there is one and only one way to convert a loss function for fully segmented images into a label-set loss function for partially segmented images. 

We propose the first application of partially supervised learning to fetal brain 3D MRI segmentation and brain tumor segmentation in MRI.
Our experiments illustrate the advantage of using partially segmented images in addition to fully segmented images for the two segmentation tasks. 
The comparison of our Leaf-Dice to three baselines suggests that label-set loss functions that satisfy our axiom perform significantly better for fetal brain 3D MRI segmentation.

\chapter[Distributionally Robust Deep Learning]{Distributionally Robust \\Deep Learning using \\Hardness Weighted Sampling}
\label{chap:dro}
\minitoc
\begin{center}
	\begin{minipage}[b]{0.9\linewidth}
		\small
		\textbf{Foreword\,}
		This chapter is to a large extent an \emph{in extenso} reproduction of \cite{fidon2020distributionally} and \cite{fidon2021distributionally}.
		Limiting failures of machine learning systems is of paramount importance for safety-critical applications.
        In order to improve the robustness of machine learning systems,
        Distributionally Robust Optimization (DRO) has been proposed as a generalization of Empirical Risk Minimization (ERM).
        However, its use in deep learning has been severely restricted due to the relative inefficiency of the optimizers available for DRO in comparison to the wide-spread variants of Stochastic Gradient Descent (SGD) optimizers for ERM.
        We propose SGD with hardness weighted sampling, a principled and efficient optimization method for DRO in machine learning that is particularly suited in the context of deep learning.
        Similar to a hard example mining strategy in practice, the proposed algorithm is straightforward to implement and computationally as efficient as SGD-based optimizers used for deep learning, requiring minimal overhead computation.
        In contrast to typical ad hoc hard mining approaches, we prove the convergence of our DRO algorithm for over-parameterized deep learning networks with $\relu$ activation and finite number of layers and parameters.
        Our experiments on fetal brain 3D MRI segmentation and brain tumor segmentation in MRI demonstrate the feasibility and the usefulness of our approach.
        Using our hardness weighted sampling for training a state-of-the-art deep learning pipeline leads to improved robustness to anatomical variabilities in automatic fetal brain 3D MRI segmentation using deep learning and to improved robustness to the image protocol variations in brain tumor segmentation.
	    Our code is available at~\url{https://github.com/LucasFidon/HardnessWeightedSampler}.
	\end{minipage}
\end{center}

\section{Introduction on Distributionally Robust Deep Learning}
\begin{figure}[tb!]
    \centering
    \includegraphics[width=\textwidth]{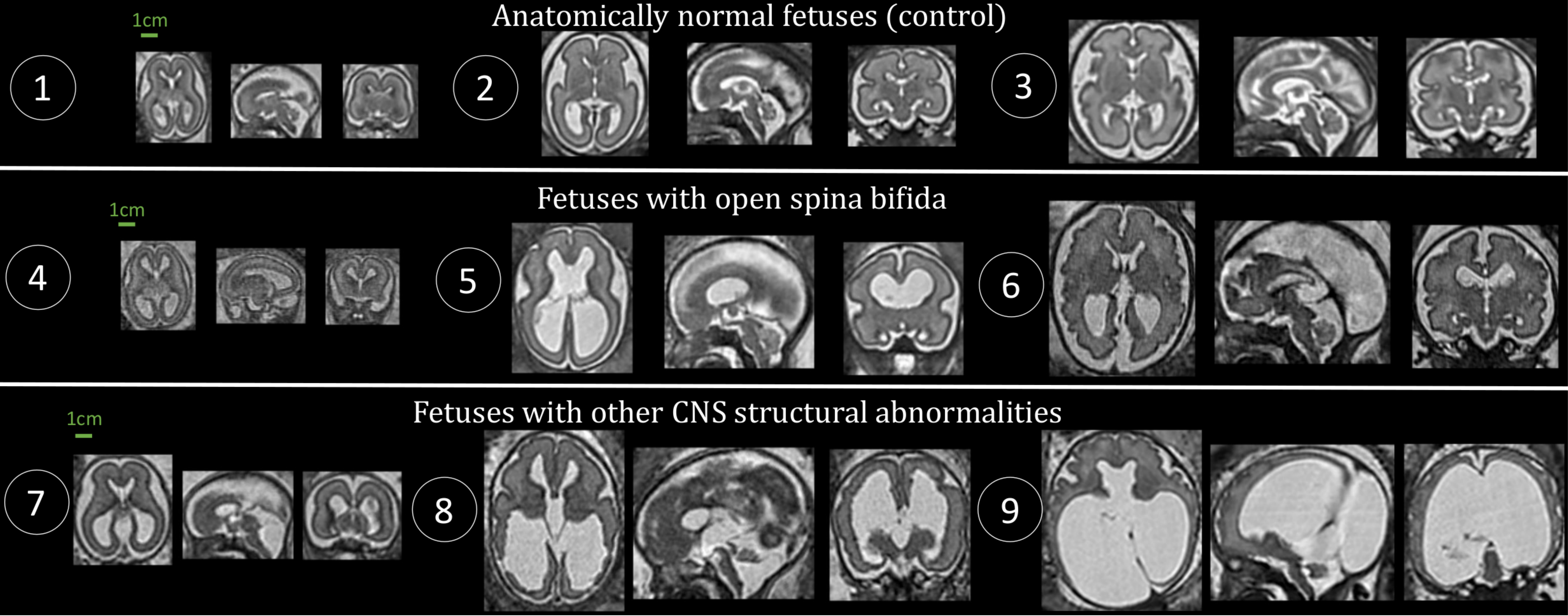}
    \caption{Illustration of the anatomical variability in fetal brain across gestational ages and diagnostics.
    1: Control (22 weeks);  % AUS00215
    2: Control (26 weeks);  % AUS00008
    3: Control (29 weeks);  % AUS00113
    4: Spina bifida (19 weeks);
    5: Spina bifida (26 weeks);
    6: Spina bifida (32 weeks);
    7: Dandy-walker malformation with corpus callosum abnormality (23 weeks);
    8: Dandy-walker malformation with ventriculomegaly and periventricular nodular heterotopia (27 weeks);
    9: Aqueductal stenosis (34 weeks).
    }
    \label{fig:anatomy_variability}
\end{figure}

As discussed in \Chapref{chap:intro},
datasets used to train deep neural networks typically contain some underrepresented subsets of cases.
These cases are not specifically dealt with by the training algorithms currently used for deep neural networks.
This problem has been referred to as hidden stratification~\cite{oakden2020hidden}.
Hidden stratification has been shown to lead to deep learning models with good average performance but poor performance on underrepresented but clinically relevant subsets of the population~\cite{larrazabal2020gender,oakden2020hidden,puyol2021fairness}.
In \Figref{fig:anatomy_variability} we give an example of hidden stratification in fetal brain MRI. 
The presence of abnormalities associated with diseases with low prevalence~\cite{aertsen2019reliability} exacerbates the anatomical variability of the fetal brain between 18 weeks and 38 weeks of gestation.

While uncovering the issue,
the study of~\cite{oakden2020hidden} does not study the cause or propose a method to mitigate this problem.
In addition, the work of~\cite{oakden2020hidden} is limited to classification.
In standard deep learning pipelines, this hidden stratification is ignored and the model is trained to minimize the mean per-example loss, which corresponds to the standard Empirical Risk Minimization (ERM) problem.
As a result, models trained with ERM are more likely to underperform on those examples from the underrepresented subdomains, seen as \textit{hard examples}.
This may lead to \textit{unfair} AI systems~\cite{larrazabal2020gender,puyol2021fairness}.
For example,
state-of-the-art deep learning models for brain tumor segmentation (currently trained using ERM) underperform for cases with confounding effects, such as low grade gliomas, despite achieving good average and median performance~\cite{bakas2018identifying}.
For safety-critical systems, such as those used in healthcare, this greatly limits their usage as ethics guidelines of regulators such as~\cite{ethics} require AI systems to be technically robust and fair prior to their deployment in hospitals.

Distributionally Robust Optimization (DRO) is a robust generalization of ERM that has been introduced in convex machine learning 
%to allow
to model the uncertainty in the training data distribution~\cite{chouzenoux2019general,duchi2016statistics,namkoong2016stochastic,rafique2018non}.
Instead of minimizing the mean per-example loss on the training dataset, DRO seeks to optimize for the hardest \emph{weighted} empirical training data distribution around the (uniform) empirical training data distribution.
This suggests a link between DRO and Hard Example Mining.
However, DRO as a generalization of ERM for machine learning still lacks optimization methods that are principled and computationally as efficient as SGD in the non-convex setting of deep learning.
Previously proposed principled optimization methods for DRO consist in alternating between approximate maximization and minimization steps~\cite{jin2019minmax,lin2019gradient,rafique2018non}.
However, they differ from SGD methods for ERM by the introduction of additional hyperparameters for the optimizer such as a second learning rate and a ratio between the number of minimization and maximization steps.
This makes DRO difficult to use as a drop-in replacement for ERM in practice.

In contrast, efficient weighted sampling methods, including Hard Example Mining~\cite{chang2017active,loshchilov2015online,shrivastava2016training} and weighted sampling~\cite{berger2018adaptive,puyol2021fairness}, have been empirically shown to mitigate class imbalance issues and to improve deep embedding learning~\cite{harwood2017smart,suh2019stochastic,wu2017sampling}.
However, even though these works typically start from an ERM formulation, it is not clear how those heuristics formally relate to ERM in theory.
This suggests that bridging the gap between DRO and weighted sampling methods could lead to a principled Hard Example Mining approach, or conversely to more efficient optimization methods for DRO in deep learning.

Given an efficient solver for the inner maximization problem in DRO, DRO could be addressed by maintaining a solution of the inner maximization problem and using a minimization scheme akin to the standard ERM but over an adaptively weighted empirical distribution.
However, even in the case where a closed-form solution is available for the inner maximization problem, it would require performing a forward pass over the entire training dataset at each iteration. This cannot be done efficiently for large datasets.
This suggests identifying an approximate, but practically usable, solution for the inner maximization problem based on a closed-form solution.

From a theoretical perspective, analysis of previous optimization methods for non-convex DRO~\cite{jin2019minmax,lin2019gradient,rafique2018non} made the assumption that the model is either smooth or weakly-convex, but none of those properties are true for deep neural networks with $\relu$ activation functions that are typically used.

In this chapter, we propose SGD with \textit{hardness weighted sampling}, a novel, principled optimization method for training deep neural networks with DRO and inspired by Hard Example Mining, that is computationally as efficient as SGD for ERM.
Compared to SGD, our method only requires introducing an additional $\softmax$ layer and maintaining a stale per-example loss vector to compute sampling probabilities over the training data.
In the context of the hardness weighted sampler, \emph{stale} means that the loss values in the loss vector are not always up-to-date with the current values of the weights of the network.
In other words, we update the loss value of a training image in the loss vector only when this image is in the training batch rather than updating it during every training iteration.
This is critical to making the hardness sampling method computationally efficient.
Indeed, updating the loss values for the entire training set at every iteration requires an additional computational complexity of O(n), while updating only for the images in the batch requires no additional computation since the loss values for the batch are already computed for the update of the model's weights.

This work is an extension of our previous preliminary work~\cite{fidon2021distributionally} in which we applied the proposed \textit{hardnes weighted sampler} to distributionally robust fetal brain 3D MRI segmentation and studied the link between DRO and the minimization of percentiles of the per-example loss.
In this extension, we formally introduce our \textit{hardness weighted sampler} and we generalize recent results in the convergence theory of SGD with ERM and over-parameterized deep learning networks with $\relu$ activation functions~\cite{allen2018convergence,allen-zhu19a,cao2019generalization,zou2019improved} to our SGD with hardness weighted sampling for DRO.
This is, to the best of our knowledge, the first convergence result for deep learning networks with $\relu$ trained with DRO.
We also formally link DRO in our method with Hard Example Mining.
As a result, our method can be seen as a principled Hard Example Mining approach.
In terms of experiments, we have extended the evaluation on fetal brain 3D MRI with $69$ additional fetal brain 3D MRIs. We have also added experiments on brain tumor segmentations and experiments on image classification with MNIST as a toy example.
We show that our method outperforms plain SGD in the case of class imbalance, and improves the robustness of a state-of-the-art deep learning pipeline for fetal brain segentation and brain tumor segmentation.
We evaluate the proposed methodology for the automatic segmentation of white matter, ventricles, and cerebellum based on fetal brain 3D T2w MRI.
We used a total of $437$ fetal brain 3D MRIs including anatomically normal fetuses, fetuses with spina bifida aperta, and fetuses with other central nervous system pathologies for gestational ages ranging from $19$ weeks to $40$ weeks.
Our empirical results suggest that the proposed training method based on distributionally robust optimization leads to better percentiles values for abnormal fetuses.
In addition, qualitative results shows that distributionally robust optimization allows to reduce the number of clinically relevant failures of nnU-Net.
For brain tumor segmentation our DRO-based method allows reducing the interquartile range of the Dice scores of $2\%$ for the segmentation of the enhancing tumor and the tumor core regions.

\section{Related Works}

% \paragraph{Group-distributionally robust deep learning:}
An optimization method for group-DRO was proposed in~\cite{sagawa2019distributionally}.
In contrast to the formulation of DRO that we study in this paper, their method requires additional labels allowing to identify the underrepresented group in the training dataset.
However, those labels may not be available or may even be impossible to obtain in most applications.
\cite{sagawa2019distributionally} show that, when associated with strong regularization of the weights of the network, their group DRO method can tackle spurious correlations that are known a priori in some classification problems.
It is worth noting that, in contrast, no regularization was necessary in our experiments with MNIST.

Biases of convolutional neural networks applied to medical image classification and segmentation has been studied in the literature.
State-of-the-art deep neural networks for brain tumor segmentation underperform for cases with confounding effects, such as low grade gliomas~\cite{bakas2018identifying}.
It has been shown that scans coming from $15$ different studies can be re-assigned with $73.3\%$ accuracy to their source using a random forest classifier~\cite{wachinger2019quantifying}.
A state-of-the-art deep neural networks for the diagnosis of $14$ thoracic diseases using X-ray trained on a dataset with a gender bias underperform on X-ray of female patients~\cite{larrazabal2020gender}.
And a state-of-the-art deep learning pipeline for cardiac MRI segmentation was found to underperform when evaluated on racial groups that were underrepresented in the training dataset~\cite{puyol2021fairness}.
To mitigate this problem, \cite{puyol2021fairness} proposed to use a stratified batch sampling approach during training that shares similarities with the group-DRO approach mentioned above~\cite{sagawa2019distributionally}.
In contrast to our hardness weighted sampler, their stratified batch sampling approach requires additional labels, such as the racial group, that may not be available for training data. In addition, they do not study the formal relationship between the use of their stratified batch sampling approach and the training optimization problem.

In this chapter, we focus on DRO with a $\phi$-divergence~\cite{csiszar2004information}.
In this case, the data distributions that are considered in the DRO problem~\eqref{eq:dro} are restricted to sharing the support of the empirical training distribution. In other words, the weights assigned to the training data can change, but the training data itself remains unchanged.
Another popular formulation is DRO with a Wasserstein distance~\cite{chouzenoux2019general,duchi2016statistics,sinha2017certifying,staib2017distributionally}.
In contrast to $\phi$-divergences, using a Wasserstein distance in DRO seeks to apply small data augmentation to the training data to make the deep learning model robust to small deformation of the data, but the sampling weights of the training data distribution typically remains unchanged.
In this sense, DRO with a $\phi$-divergence and DRO with a Wasserstein distance can be considered as orthogonal endeavours.
While we show that DRO with $\phi$-divergence can be seen as a principled Hard Exemple Mining method, it has been shown that DRO with a Wasserstein distance can be seen as a principled adversarial training method~\cite{sinha2017certifying,staib2017distributionally}.

Among the other methods recently proposed for DRO, just train twice~\cite{liu2021just} proposes a method for classification that requires only to label the images of the validation dataset as hard examples.
    The method consists in first training for the classification task at hand an identification deep neural network using standard ERM, second identifying the training samples misclassified by the identification network as hard samples, and finally train a new model for the classification task with a higher and fixed sampling weight for the hard samples.
    The group labels of the validation dataset are used to tune the four DRO-specific hyperparameters of the identification network.
    Those hyperparameters consist of one weight for the hard samples, and the number of epochs, the learning rate and the weight decay for the identification model.
    In contrast, our method uses dynamic weights that are updated at each training iteration, does not require additional group labelling on the validation dataset, and requires to tune only one hyperparameter specific to DRO.
    In addition, just train twice is specific to classification problems.

% TODO (TAKEN FROM REBUTTAL)
    Multiplicative weighting and weighted sampling affect the optimization dynamic in different ways.
    Multiplicative weighting corresponds to adapting the learning rate, i.e. the size of the gradient descent step, but the number of gradient evaluation remains uniform across the training dataset.
    In contrast, with weighted sampling the learning rate is not affected and instead gradient evaluation is performed more frequently for the training samples with a higher weight.
    When importance sampling is used, both multiplicative weighting and multiplicative sampling are used.
The effect of \textit{multiplicative weighting} during training, rather than \textit{weighted sampling} used in our algorithm, has been studied empirically by
\cite{byrd2019effect} for image classification.
They find that the effect of multiplicative weighting vanishes over training for classification tasks in which we can achieve zero loss on the training dataset.
However, \textit{multiplicative weighting} and \textit{weighted sampling} affect the optimization dynamic in different ways. This may explain why we did not observe this vanishing effect in our experiments on classification and segmentation.
Previous work have also studied empirical and convergence results 
of DRO for linear models~\cite{hu2018does}.

\section{Methods}
\subsection{Supervised Deep Learning with Distributionally Robust Optimization}
% For ease of presentation, we focus on the supervised machine learning setting.

Standard training procedures in machine learning are based on Empirical Risk Minimization (ERM)~\cite{bottou2018optimization}.
% 
% A predictor $h: \vx \mapsto \vy$, such as a deep neural network, is trained using a training dataset $\{(\vx_i, \vy_i)\}_{i=1}^n$ to perform well \textit{on average} on a task
% %T
% for which the performance is measured on a per-example basis by a smooth criteria $\cL$.
% %
% Optionally, $\cL$ can contain a parameter regularization term.
% 
For a neural network $h$ with parameters $\vtheta$, a per-example loss $\cL$, and a training dataset $\left\{(\rx_i, \ry_i)\right\}_{i=1}^n$, where $\rx_i$ are the inputs and $\ry_i$ are the labels, 
the ERM problem corresponds to
% , defined as the non-convex optimization problem
\begin{equation}
    \label{eq:erm_intro}
    \min_{\vtheta} 
    \left\{
    \E_{(\rx,\ry) \sim \textbf{p}_{\textrm{train}}} \left[\cL\left(h(\rx; \vtheta), \ry\right)\right]
    = \frac{1}{n} \sum_{i=1}^n \cL \left(h(\rx_i;\vtheta), \ry_i\right)
    \right\}
\end{equation}
where $\textbf{p}_{\textrm{train}}$ is the empirical uniform distribution on the training dataset.
This may include the use of a finite or infinite number of data augmentations.
For our theoretical results, we suppose that $\textbf{p}_{\textrm{train}}$ contains a finite number of examples.
In general, $\textbf{p}_{\textrm{train}}$ can contain an infinite number of data due to data augmentation~\cite{chapelle2000vicinal}. This case is discussed in section~\ref{sec:algo_DRO}.
%The extension of our \Algref{alg:1} to an infinite number of data augmentations using importance sampling is presented in section~\ref{sec:algo_DRO}.
%
Optionally, $\cL$ can contain a parameter regularization term that is only a function of $\vtheta$.

The ERM training formulation assumes that $\textbf{p}_{\textrm{train}}$ is an unbiased approximation of the true data distribution.
However, this is generally impossible in domains such as medical image computing. This makes models trained with ERM at risk of underperforming on images from parts of the data distribution that are underrepresented in the training dataset.

In contrast, Distributionally Robust Optimization (DRO) is a family of generalization of ERM in which the uncertainty in the training data distribution is modelled by minimizing the worst-case expected loss over an \textit{uncertainty set} of training data distributions~\cite{rahimian2019distributionally}.

In this paper, we consider training deep neural networks with DRO based on a $\phi$-divergence.
We denote 
$\Delta_n := \left\{\left(p_i\right)_{i=1}^n \in [0,1]^n \,\, | \,\,
\sum_{i=1}^n p_i = 1\right\}$
the set of empirical training data probabilities vectors under consideration (i.e. the uncertainty set).
The different probabilities vectors in $\Delta_n$ correspond to all the possible weighting of the training dataset. Every $\textbf{p}$ in $\Delta_n$ gives a weight to each training examples but keep the examples the same.
% The comparison of different DRO formulations is out of the scope of this paper.
% 
We use the following definition of $\phi$-divergence in the remainder of the paper.
%
% Let $\vtheta$ be the set of parameters of the predictor $h(\,.\,;\vtheta): \vx \mapsto \vy$ we want to train, and $\vh: \vtheta \mapsto \left(h(\vx_i; \vtheta)\right)_{i=1}^n$ be the vector of inferred outputs from the training data.
% %
% We also denote $\cL(\vh(\vtheta))=\left(\cL(h(\vx_i;\vtheta), \vy_i)\right)_{i=1}^n$.
%
\begin{definition}[$\phi$-divergence]
\label{def:phi_divergence}
Let $\phi : \sR_+ \rightarrow \sR$ 
% such that $\forall z \in \sR_+,\, \phi(z) \geq \phi(1)=0$.
be two times continuously differentiable on $[0, n]$, $\rho$-strongly convex on $[0, n]$,
     i.e.
     $
     \exists \rho > 0,
     \forall z, z' \in [0,n],\,\, \phi(z') \geq \phi(z) + \phi'(z)(z' - z) + \frac{\rho}{2}(z - z')^2,
     $
     and satisfying
     $\forall z \in \sR,\,\, \phi(z) \geq \phi(1)=0,\,\, \phi'(1) = 0$.\\
The $\phi$-divergence $D_{\phi}$ is defined as,
for all $p=(p_i)_{i=1}^n, q=(q_i)_{i=1}^n \in \Delta_n$,
\begin{equation}
    D_{\phi}\left(q \Vert p\right)
    = \sum_{i=1}^n p_i \phi\left(\frac{q_i}{p_i}\right)
\end{equation}
\end{definition}
We refer to our example~\ref{ex:softmax} on page~\pageref{ex:softmax} to highlight that the KL divergence is indeed a $\phi$-divergence.

The DRO problem for which we propose an optimizer for training deep neural networks can be formally defined as
\begin{equation}
        \label{eq:dro}
        \min_{\vtheta}\,
        \left\{
        R\left(\vL(h(\vtheta))\right) :=
        \max_{q \in \Delta_n}
        \left(
        \E_{(\rx,\ry) \sim q} \left[\cL\left(h(\rx; \vtheta), \ry\right)\right]
        - \frac{1}{\de} D_{\phi}\left(q \Vert \textbf{p}_{\textrm{train}}\right)
        \right)
        \right\}
\end{equation}
where $\textbf{p}_{\textrm{train}}$ is the uniform empirical distribution, and $\de > 0$ an hyperparameter.
The choice of $\de$ and $\phi$ controls how the unknown training data distribution $q$ is allowed to differ from $\textbf{p}_{\textrm{train}}$.
Here and thereafter, we use the notation 
$\vL(h(\vtheta)) := \left(\cL(h(\rx_i;\vtheta), \ry_i)\right)_{i=1}^n$
to refer to the vector of loss values of the $n$ training samples for the value $\vtheta$ of the parameters of the neural network $h$.

In the remainder of the paper, we will refer to $R$ as the \textit{distributionally robust loss}.
\subsection{Hardness Weighted Sampling for Distributionally Robust Deep Learning}

In the case where $h$ is a non-convex predictor (such as a deep neural network), existing optimization methods for the DRO problem \eqref{eq:dro} alternate between approximate minimization and maximization steps \cite{jin2019minmax,lin2019gradient,rafique2018non}, requiring the introduction of additional hyperparameters compared to SGD. However, these are difficult to tune in practice and convergence has not been proven for non-smooth deep neural networks such as those with $\relu$ activation functions.

In this section, we present an SGD-like optimization method for training a deep learning model $h$ with the DRO problem \eqref{eq:dro}.
We first highlight, in Section~\ref{sec:sampling_approach_DRO}, mathematical properties that allow us to link DRO with stochastic gradient descent (SGD) combined with an adaptive sampling that we refer to as \textit{hardness weighted sampling}.
In Section~\ref{sec:algo_DRO}, we present our \Algref{alg:1} for distributionally robust deep learning.
Then, in Section~\ref{sec:theoretical_results}, we present theoretical convergence results for our hardness weighted sampling.

%%%%%%%%%%%%%%%%%%%%%%%%%%%%%%%%%%%%%%%%%%%%%%%%%%%%%%%%%%%%%%%%%%%%%%%%%%
\subsubsection{A sampling approach to Distributionally Robust Optimization}\label{sec:sampling_approach_DRO}
The goal of this subsection is to show that a stochastic approximation of the gradient of the \textit{distributionally robust loss} can be obtained by using a weighted sampler.
This result is a first step towards our \Algref{alg:1} for efficient training with the \textit{distributionally robust loss} presented in the next subsection.

Our analysis of the properties of the \textit{distributionally robust loss} $R$ relies on the Fenchel duality~\cite{moreau1965proximite} and the notion of Fenchel conjugate~\cite{fenchel1949conjugate}.
\begin{definition}[Fenchel Conjugate Function]
    \label{def:fenchel_conjugate}
    Let $f: \sR^m \rightarrow \sR \cup \{+\infty\}$ be a proper function. The Fenchel conjugate of $f$ is defined as
    $\forall \vv \in \sR^m,\,\, f^*(\vv)= \max_{\vx \in \sR^m} \langle \vv, \vx\rangle - f(\vx)$.
    % \begin{equation}
    %     \forall \vv \in \sR^m,\,\, f^*(\vv)= \max_{\vx \in \sR^m} \langle \vv, \vx\rangle - f(\vx)
    % \end{equation}
\end{definition}

To reformulate $R$ as an unconstrained optimization problem over $\sR^n$ (rather than constraining it to $\Delta_n$), we define
\begin{equation}
    \label{eq:G_def}
    \begin{aligned}
    \forall \textbf{p} \in \sR^n,\quad 
        % G:\,\,&\sR^n \rightarrow \sR\\
             G(\textbf{p}) = \frac{1}{\de} D_{\phi}(\textbf{p} \Vert \textbf{p}_{\textrm{train}}) + \delta_{\Delta_n}(\textbf{p})
    \end{aligned}
\end{equation}
where $\delta_{\Delta_n}$ is the characteristic function of the closed convex set $\Delta_n$, i.e.
\begin{equation}
    \forall \textbf{p} \in \sR^n,\quad \delta_{\Delta_n}(\textbf{p})=\left\{
\begin{array}{cl}
    0 & \text{if } \textbf{p} \in \Delta_n \\
    +\infty & \text{otherwise}
\end{array}
\right.
\end{equation}
The distributionally robust loss $R$ in \eqref{eq:dro} can now be rewritten using the Fenchel conjugate function $G^*$ of $G$.
This allows us to obtain regularity properties for $R$.
\begin{lemma}[Regularity of $R$]
\label{lemma:R_property}
    If $\phi$ satisfies Definition \ref{def:phi_divergence} 
    (i.e. can be used for a $\phi$-divergence), then $G$ and $R$ satisfy the following:
    % , and $G: p \mapsto \frac{1}{\de} D_{\phi}(p\Vert \ptrain) + \delta_{\Delta_n}(p)$
    \begin{equation}
        \label{eq:strong_convexity_G}
        G \text{ is} \left(\frac{n\rho}{\de}\right)\text{-strongly convex}
    \end{equation}
    \begin{equation}
        \label{eq:link_R_and_G}
        \forall \vtheta, \quad 
        R(\vL(h(\vtheta)))
        =
        \max_{\textbf{q} \in \sR^n} 
            \Big(
            \langle \vL(h(\vtheta)), \textbf{q}\rangle - G(\textbf{q})
            \Big)
        = G^*\left(\vL(h(\vtheta))\right)
    \end{equation}
    % And
    \begin{equation}
        \label{eq:gradient_Lip_R}
        R \text{ is } \left(\frac{\de}{n\rho}\right)\text{-gradient Lipschitz continuous.}
    \end{equation}
\end{lemma}
Equation \eqref{eq:link_R_and_G} follows from Definition \ref{def:fenchel_conjugate}. Proofs of \eqref{eq:strong_convexity_G} and \eqref{eq:gradient_Lip_R} can be found in Appendix~\ref{s:regularity_R}.
According to \eqref{eq:strong_convexity_G}, the optimization problem \eqref{eq:link_R_and_G} is strictly convex and admits a unique solution in $\Delta_n$, which we denote as
\begin{equation}
    \label{eq:hardness_weighted_proba}
    \bar{\textbf{p}}(\vL(h(\vtheta))) = \argmax_{\textbf{q} \in \sR^n} 
        \left(
        \langle \vL(h(\vtheta)), \textbf{q}\rangle - G(\textbf{q})
        \right)
\end{equation}

Thanks to those properties, we can now show the following lemma that is essential for the theoretical foundation of our \Algref{alg:1}.
Equation \eqref{eq:p} states that the gradient of the distributionally robust loss $R$ is a weighted sum of the the gradients of the per-example losses (i.e. the gradients computed by the backpropagation algorithm in deep learning)
with the weights given by the empirical distribution $\bar{\textbf{p}}(\vL(\vh(\vtheta)))$.
We further show that straightforward analytical formulas exist for $\bar{\textbf{p}}$, and give an example of such probability distribution for the Kullback-Leibler (KL) divergence.
\begin{lemma}[Stochastic Gradient of the Distributionally Robust Loss]
\label{lemma:robust_loss_sg}
For all $\vtheta$, we have
\begin{equation}
    \label{eq:p}
    \begin{aligned}
        \nabla_\vtheta (R \circ \vL \circ ~h)(\vtheta) 
        & = \E_{(\rx,\ry) \sim \bar{\textbf{p}}(\vL(h(\vtheta)))}\left[\nabla_\vtheta \cL\left(h(\rx; \vtheta), \ry\right)\right]
    \end{aligned}
\end{equation}
\end{lemma}
The proof is found in Appendix~\ref{s:robust_loss_sg}.
We now provide a closed-form formula for $\bar{\textbf{p}}$ given $\cL(h(\vtheta))$ for the KL divergence as the choice of $\phi$-divergence. 
% Closed-form formulas exist for other $\phi$-divergences, e.g. the Pearson $\chi^2$ divergence.
%
\begin{example}
    \label{ex:softmax}
    % 1. 
    For $\phi: z \mapsto z\log(z) - z + 1$, $D_{\phi}$ is the Kullback-Leibler (KL) divergence:
    \begin{equation}
        D_{\phi}(\textbf{q} \Vert \textbf{p}) = \KL (\textbf{q} \Vert \textbf{p}) = \sum_{i=1}^n q_i\log\left(\frac{q_i}{p_i}\right)
    \end{equation}
    In this case, we have (see Appendix~\ref{s:proof_softmax} for a proof)
    % $
    \begin{equation}\label{eq:KL_softmax}
        \bar{\textbf{p}}(\vL(h(\vtheta))) 
        = \softmax\left(\de \vL(h(\vtheta))\right)
    \end{equation}
    % $.
    % 2. For $\phi: z \mapsto (z-1)^2$, $D_{\phi}$ is the Pearson $\chi^2$ divergence:
    % \begin{equation}
    %     D_{\phi}(q \Vert p) = \chi^2 (q \Vert p) = \sum_{i=1}^n \frac{\left(q_i - p_i\right)^2}{p_i}
    % \end{equation}
    % In this case, we have
    % \begin{equation}
    %     \forall i,\quad \bar{p}_i(\cL(h(\vtheta))) 
    %     = \relu\left(
    %         \frac{1}{n}
    %         \left(
    %         1 +
    %         \frac{\de}{2}\left(\cL(h(\vtheta))_i - \frac{1}{n}\sum_{j=1}^n \cL(h(\vtheta))_j\right)
    %         \right)
    %     \right)
    % \end{equation}
\end{example}

%%%%%%%%%%%%%%%%%%%%%%%%%%%%%%%%%%%%%%%%%%%%%%%%%%%%%%%%%%%%%%%%%%%%%%%%%%%
\subsubsection{Proposed Efficient Algorithm for Distributionally Robust Deep Learning}\label{sec:algo_DRO}
We now describe our algorithm for training deep neural networks with DRO using our hardness weighted sampling.

% ALGO
\begin{algorithm}[htb!]
\caption{Training procedure for DRO with Hardness Weighted Sampling.
% our proposed online optimizer for training deep neural networks with Distributionally Robust Optimization (DRO).
}
\label{alg:1}
\begin{algorithmic}[1]
\Require{$\left\{(\rx_i, \ry_i)\right\}_{i=1}^n$: training dataset with $n>0$ the number of training samples.}
\Require{$b\in \{1,\ldots, n\}$: batch size.}
\Require{$\cL$: (any) smooth per-example loss 
function.}
\Require{$\de > 0$: robustness parameter defining the distributionally robust optimization problem.}
\Require{$\vtheta_{0}$: initial parameter vector for the model $h$ to train.}
\Require{$\vL_{init}$: initial stale per-example loss values vector.}
\State{$t \leftarrow 0$ (initialize the time step)}
\State{$\vL \leftarrow \vL_{init}$ (initialize the vector stale loss values)}
\While{$\vtheta_t$ has not converged}
    % Sampling
    \State{$\vp_t \leftarrow \softmax(\de \vL)$
    }\Comment{online estimation of the \textit{hardness weights}}
    
    \State{$I \sim \vp_t$
    }\Comment{hardness weighted sampling}
    
    % Importance sampling weight
    \If{importance sampling is not used}
    \State{$\forall i \in I,\,\, w_i = 1$}
    \Else{}
    \State{$\forall i \in I,\,\,
    w_{i} \leftarrow 
    \exp\left(\de(\cL(h(\rx_i;\vtheta), \ry_i) - L_{i})\right)
    % \approx \frac{p_{t+1, i}}{p_{t,i}}
    $}\Comment{importance sampling weights}
    \State{$\forall i \in I,\,\,
    w_{i} \leftarrow \textup{clip}\left(w_i, [w_{min}, w_{max}]\right)$
    }\Comment{clip the weights for stability}
    % Params update
    \EndIf
    
    % Loss estimates update
    \State{$\forall i \in I,\,\, 
    L_i \leftarrow \cL(h(\rx_i;\vtheta), \ry_i)$
    }\Comment{update the vector of stale loss values}
    
    \State{$\vg_t \leftarrow \frac{1}{b}\sum_{i \in I}
    w_i
    \nabla_{\vtheta}\cL(h(\rx_i;\vtheta_{t}), \ry_i)$ 
    % (SGD direction; change for other directions like Adam)
    }
    
    \State{$\vtheta_{t+1} \leftarrow \vtheta_{t} 
    - \eta \,\vg_t$}\Comment{gradient descent step}
\EndWhile
\State{\textbf{Output:} $\vtheta_t$}
\end{algorithmic}
\end{algorithm}

Equation \eqref{eq:p} implies that $\nabla_\vtheta \cL(h(\rx_i;\vtheta), \ry_i)$ is an unbiased estimator of the gradient of the distributionally robust loss gradient when $i$ is sampled with respect to $\bar{\textbf{p}}(\vL(h(\vtheta)))$.
This suggests that the distributionally robust loss can be minimized efficiently by SGD by sampling mini-batches with respect to $\bar{\textbf{p}}(\vL(h(\vtheta)))$ at each iteration.
However, even though closed-form formulas were provided in Example~\ref{ex:softmax} for $\bar{\textbf{p}}$,
evaluating exactly $\vL(h(\vtheta))$, i.e. doing one forward pass on the whole training dataset at each iteration, is computationally prohibitive for large training datasets.

In practice, we propose to use a stale version of the vector of per-example loss values
% $\cL(h(\vtheta))$ 
by maintaining an online history of the loss values of the examples seen during training $\left(\cL(h(\rx_i;\vtheta^{(t_i)}), \ry_i)\right)_{i=1}^n$,
where for all $i$, $t_i$ is the last iteration at which the per-example loss of example $i$ has been computed.
Using the Kullback-Leibler divergence as $\phi$-divergence, this leads to the SGD with hardness weighted sampling algorithm proposed in \Algref{alg:1}.

When data augmentation is used, an infinite number of training examples is virtually available.
In this case, we keep one stale loss value per example irrespective of any augmentation as an approximation of the loss for this example under any augmentation.

Importance sampling is often used when sampling with respect to a desired distribution cannot be done exactly~\cite{kahn1953methods}.
In \Algref{alg:1}, an up-to-date estimation of the per-example losses (or equivalently the hardness weights) in a batch is only available \emph{after} sampling and evaluation through the network. Importance sampling can be used to compensate for the difference between the initial and the updated stale losses within this batch.
We propose to use importance sampling in steps 9-10 of \Algref{alg:1} and highlight that this is especially useful to deal with data augmentation. Indeed, in this case, the stale losses for the examples in the batch are expected to be less accurate as they were estimated under a different augmentation.
For efficiency, we use the following approximation
$
w_i = \frac{p_i^{new}}{p_i^{old}} \approx \exp\left(\de(\cL(h(\rx_i;\vtheta), \ry_i) - L_{i})\right)
$
where we have neglected the typically small change in the denominator of the $\softmax$. More details are given in Appendix~\ref{appendix:importance_sampling}.
% To complement the hardness weighted sampler for examples, we propose to optionally use importance sampling (steps 9-10 in \Algref{alg:1}).
% 
To tackle the typical instabilities that can arise when using importance sampling~\cite{owen2000safe}, the importance weights are clipped.
% (steps 10 in \Algref{alg:1}).

% In contrast to alternate min-max optimization methods, our SGD with an adaptive sampling strategy is similar in practice to the SGD-based optimizers used by the vast majority of deep learning practitioners.
%
Compared to standard SGD-based training optimizers for the mean loss, our algorithm requires only an additional $\softmax$ operation per iteration and to store an additional vector of scalars of size $n$ (number of training examples), thereby making it well suited for deep learning applications.

%%%%%%%%%%%%%%%%%%%%%%%%%%%%%%%%%%%%%%%%%%%%%%%%%%%%%%%%%%%%%%%%%%%%%%%%%%
\subsection{Overview of Theoretical Results}\label{sec:theoretical_results}
In this section, we present convergence guarantees for \Algref{alg:1} in the framework of over-parameterized deep learning.
We further demonstrate properties of our hardness weighted sampling that allow to clarify its link with Hard Example Mining and with the minimization of percentiles of the per-sample loss on the training data distribution.

\subsubsection{Convergence of SGD with Hardness Weighted Sampling for Over-parameterized Deep Neural Networks with \texorpdfstring{$\relu$}{$\relu$}}\label{s:convergence_main_text}
% \paragraph{Convergence of SGD with Hardness Weighted Sampling for Over-parameterized DNN}\label{s:convergence_main_text}
%
Convergence results for over-parameterized deep learning have recently been proposed in~\cite{allen-zhu19a}.
Their work gives convergence guarantees for deep neural networks $h$ with any activation functions (including $\relu$), and with any (finite) number of layers $L$ and parameters $m$, under the assumption that $m$ is large enough.
In our work, we extend the convergence theory developed by~\cite{allen-zhu19a} for ERM and SGD to DRO using the proposed SGD with hardness weighted sampling and stale per-example loss vector (as stated in~\Algref{alg:1}).
The proof in Appendix~\ref{s:proof_stale_loss_history} deals with the challenges raised by the non-linearity of $R$ with respect to the per-sample stale loss and the non-uniform dynamic sampling used in \Algref{alg:1}.

\begin{theorem}[Convergence of \Algref{alg:1} for neural networks with $\relu$]
\label{th:convergence_dro}
Let $\cL$ be a smooth per-example loss function, $b \in \{1,\ldots,n\}$ be the batch size, and $\epsilon > 0$.
If the number of parameters $m$ is large enough, and the learning rate is small enough, then, with high probability over the randomness of the initialization and the mini-batches, \Algref{alg:1} (without importance sampling) guarantees 
$\norm{\nabla_{\vtheta} (R\circ \vL \circ h)(\vtheta)} \leq \epsilon$ after a finite number of iterations.
\end{theorem}

A detailed description of the assumption for this theorem is described in Appendix~\ref{th:conv_sgd_stale_loss_history} and its proof can be found in Appendix~\ref{s:proof_stale_loss_history}.
Our proof does not cover the case where importance sampling is used. However, our empirical results suggest that convergence guarantees still hold with importance sampling.

%%%%%%%%%%%%%%%%%%%%%%%%%%%%%%%%%%%%%%%%%%%%%%%%%%%%%%%%%%%%%%%%%%%%%%%%%%%
\subsubsection{Link between Hardness Weighted Sampling and Hard Example Mining}
% \paragraph{Link between Hardness Weighted Sampling and Hard Example Mining}
%
In this section, we discuss the relationship between the proposed hardness weighted sampling for DRO and Hard Example Mining.
The following result shows that using the proposed \textit{hardness weighted sampler} the hard training examples, those training examples with relatively high values of the loss, are sampled with higher probability.

\begin{theorem}
\label{th:hard_example_mining}
Let a $\phi$-divergence that satisfies Definition~\ref{def:phi_divergence}, and $\textbf{L} = \left(L_i\right)_{i=1}^n \in \sR^n$ a vector of loss values for the examples $\{\rx_1, \ldots, \rx_n\}$.
The proposed hardness weighted sampling probabilities vector $\bar{\textbf{p}}\left(\textbf{L}\right) = \left(\bar{p}_i\left(\textbf{L}\right)\right)_{i=1}^n$
defined as in \eqref{eq:hardness_weighted_proba} verifies:
\begin{enumerate}
    \item For all $i \in \{1, \ldots, n\}$, $\bar{p}_i$ is an increasing function of $L_i$.
    \item For all $i \in \{1, \ldots, n\}$, $\bar{p}_i$ is an non-increasing function of any $L_j$ for $j \neq i$.
\end{enumerate}
\end{theorem}
See Appendix~\ref{s:hard_example_mining} for the proof.
The second part of Theorem \ref{th:hard_example_mining} implies that as the loss of an example diminishes, the sampling probabilities of all the other examples increase.
As a result, the proposed SGD with hardness weighted sampling balances exploitation (i.e. sampling the identified \textit{hard examples}) and exploration (i.e. sampling any example to keep the record of \textit{hard examples} up to date).
Heuristics to enforce this trade-off are often used in Hard Example Mining methods~\cite{berger2018adaptive,harwood2017smart,wu2017sampling}.

%%%%%%%%%%%%%%%%%%%%%%%%%%%%%%%%%%%%%%%%%%%%%%%%%%%%%%%%%%%%%%%%%%%%%%%%%%%
\subsubsection{Link between DRO and the Minimization of a Loss Percentile}
In this section, we show that the DRO problem \eqref{eq:dro} using the KL divergence is equivalent to a relaxation of the minimization of the per-example loss percentile shown thereafter in equation \eqref{eq:perc}.

Instead of the average per-example loss \eqref{eq:erm_intro}, for robustness,
%algorithms,
one might be more interested in minimizing
the
%$\alpha^{\textrm{th}}$
percentile
$l_{\alpha}$ at $\alpha$ (e.g. 5\%)
of the per-example loss function.
Formally, this corresponds to the minimization problem
\begin{equation}
    \label{eq:perc}
        \min_{\vtheta,\, l_{\alpha}} \quad l_{\alpha} \qquad
        \textrm{such that}
        \qquad  
        p_{\textrm{train}}\left(
            \cL \left(h(\rx;\vtheta), \ry\right) \geq l_{\alpha}
            \right) \leq \alpha
\end{equation}
where $p_{\textrm{train}}$ is the empirical distribution defined by the training dataset.
% 
%For example,
In other words,
if $\alpha=0.05$, the optimal $l_{\alpha}^*(\vtheta)$ of \eqref{eq:perc} for a given value set of parameters $\vtheta$ is the value of the loss such that the per-example loss function is worse than $l_{\alpha}^*(\vtheta)$ $5\%$ of the time.
As a result, training the deep neural network using \eqref{eq:perc} corresponds to minimizing the percentile of the per-example loss function $l_{\alpha}^*(\vtheta)$.

Unfortunately, the minimization problem \eqref{eq:perc} cannot be solved directly using stochastic gradient descent to train a deep neural network.
We now propose a tractable upper bound for $l_{\alpha}^*(\vtheta)$ and show that it can be solved in practice using distributionally robust optimization.

The Chernoff bound~\cite{chernoff1952measure} applied to the per-example loss function and the empirical training data distribution states that for all $l_{\alpha}$ and $\de>0$
\begin{equation}
    p_{\textrm{train}}\left(
            \cL \left(h(\rx;\vtheta), \ry\right) \geq l_{\alpha}
            \right) 
    \leq 
        \frac{\exp\left(-\de l_{\alpha}\right)}{n} 
        \sum_{i=1}^n \exp\left(\de \cL \left(h(\vx_i;\vtheta), \vy_i\right)\right)
\end{equation}
To link this inequality to the minimization problem \eqref{eq:perc}, we set $\de>0$ and
\begin{align}
    % \alpha &= \frac{\exp\left(-\de \hat{l}_{\alpha}(\vtheta)\right)}{n} 
    %     \sum_{i=1}^n \exp\left(\de \cL \left(h(\vx_i;\vtheta), \vy_i\right)\right)\\
    % \iff 
    \hat{l}_{\alpha}(\vtheta) &= \frac{1}{\de} \log\left(
        \frac{1}{\alpha n}
        \sum_{i=1}^n \exp\left(\de \cL \left(h(\vx_i;\vtheta), \vy_i\right)\right)
    \right)
\end{align}
In this case, we have
\begin{equation}
    p_{\textrm{train}}\left(
            \cL \left(h(\rx;\vtheta), \ry\right) \geq \hat{l}_{\alpha}(\vtheta)
            \right) 
    \leq \alpha =
        \frac{\exp\left(-\de \hat{l}_{\alpha}(\vtheta)\right)}{n} 
        \sum_{i=1}^n \exp\left(\de \cL \left(h(\vx_i;\vtheta), \vy_i\right)\right)
\end{equation}
$\hat{l}_{\alpha}(\vtheta)$ is therefore an upper bound for the optimal $l^*_{\alpha}(\vtheta)$ in equation \eqref{eq:perc}, independently to the value of $\vtheta$.
Equation \eqref{eq:perc} can therefore be relaxed by 
% $\min_{\vtheta} \hat{l}_{\alpha}(\vtheta)$ translating as
\begin{equation}
    \label{eq:expvar}
    \min_{\vtheta} \frac{1}{\de} \log\left(
        \sum_{i=1}^n \exp\left(\de \cL \left(h(\vx_i;\vtheta), \vy_i\right)\right)
    \right)
    %- \frac{1}{\de} \log\left(\alpha n\right)
\end{equation}
where $\de>0$ is a hyperparameter,
and where
the term $\frac{1}{\de} \log\left(\frac{1}{\alpha n}\right)$ was dropped as being independent of $\vtheta$.
% 
%The variable $\alpha$ is substituted by $\de$. 
While in \eqref{eq:expvar}, $\alpha$ does not appear in the optimization problem directly anymore, $\de$ essentially acts as a substitute for $\alpha$.
The higher the value of $\de$, the higher weights the per-example losses with a high value will have in \eqref{eq:expvar}.

We give a proof in Appendix~\ref{s:proof_dro_and_percentile}
that \eqref{eq:expvar} is equivalent to solving the following DRO problem
\begin{equation}
    \label{eq:dro2}
    \min_{\vtheta}\, \max_{\vq \in \Delta_n}
        \left(
        \sum_{i=1}^n q_i \cL\left(h(\vx_i; \vtheta), \vy_i\right)
        - \frac{1}{\de} D_{KL}\left(\vq\, \biggr\Vert\, 
        p_{\textrm{train}}
        % \frac{1}{n}\mathbf{1}
        \right)
        \right)
\end{equation}
This is a special case of the DRO problem \eqref{eq:dro} where $\phi$ is chosen as the KL-divergence and it corresponds to the setting of \Algref{alg:1}.

\section{Experiments}
In this section, we experiment with the proposed \textit{hardness weighted sampler} for DRO as implemented in the proposed \Algref{alg:1}.
In the subsection~\ref{s:mnist}, we give a toy example with the task of automatic classification of digits in the case where the digit $3$ is underrepresented in the training dataset.
And in subsection~\ref{s:medical_image_segmentation}, we report the results of our experiments on two medical image segmentation tasks: fetal brain segmentation using 3D MRI, and brain tumor segmentation using 3D MRI.

\subsection{Toy Example: MNIST Classification with a Class Imbalance}\label{s:mnist}
The goal of this subsection is to illustrate key benefits of training a deep neural network using
DRO in comparison to ERM when a part of the sample distribution is underrepresented in the training dataset.
We take the MNIST dataset~\cite{lecun1998mnist} as a toy example, in which the task is to automatically classify images representing digits between $0$ and $9$.
In addition, we verify the ability of our \Algref{alg:1} to train a deep neural network for DRO and illustrates the behaviour of SGD with hardness weighted sampling for different values of $\de$.

\paragraph*{Material:}
We create a bias between training and testing data distribution of MNIST~\cite{lecun1998mnist} by keeping only $1\%$ of the digits $3$ in the training dataset, while the testing dataset remains unchanged.

For our experiments on MNIST, we used a Wide Residual Network (WRN)~\cite{zagoruyko2016wide}.
The family of WRN models has proved to be very efficient and flexible, achieving state-of-the-art accuracy on several dataset.
More specifically, we used WRN-$16$-$1$~\cite[section 2.3]{zagoruyko2016wide}.
For the optimization we used a learning rate of $0.01$.
No momentum or weight decay were used.
No data augmentation was used.
For DRO no importance sampling was used.
We used a GPU NVIDIA GeForce GTX 1070 with 8GB of memory for the experiments on MNIST.

\begin{figure}[bth!]
    \centering
    \includegraphics[width=0.98\linewidth]{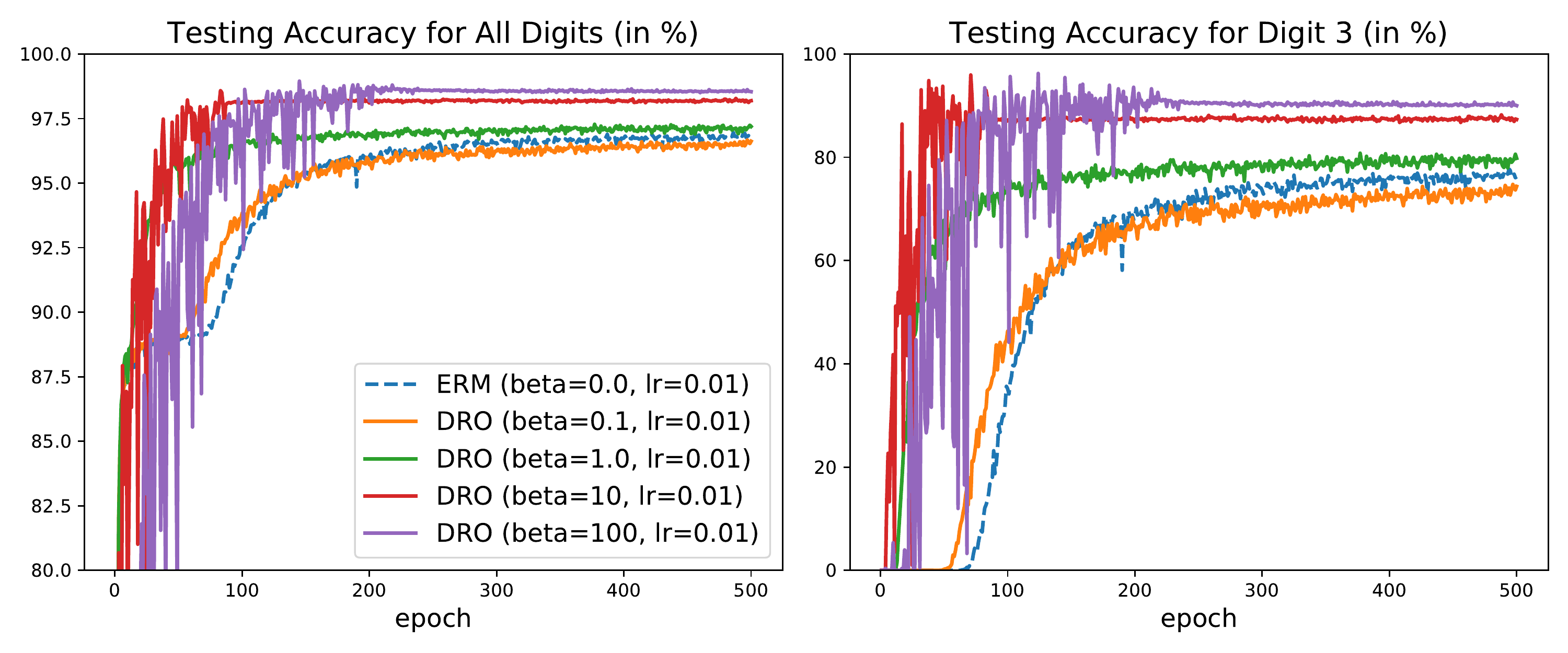}
    \includegraphics[width=0.98\linewidth,trim=0cm 0cm 1cm 0cm]{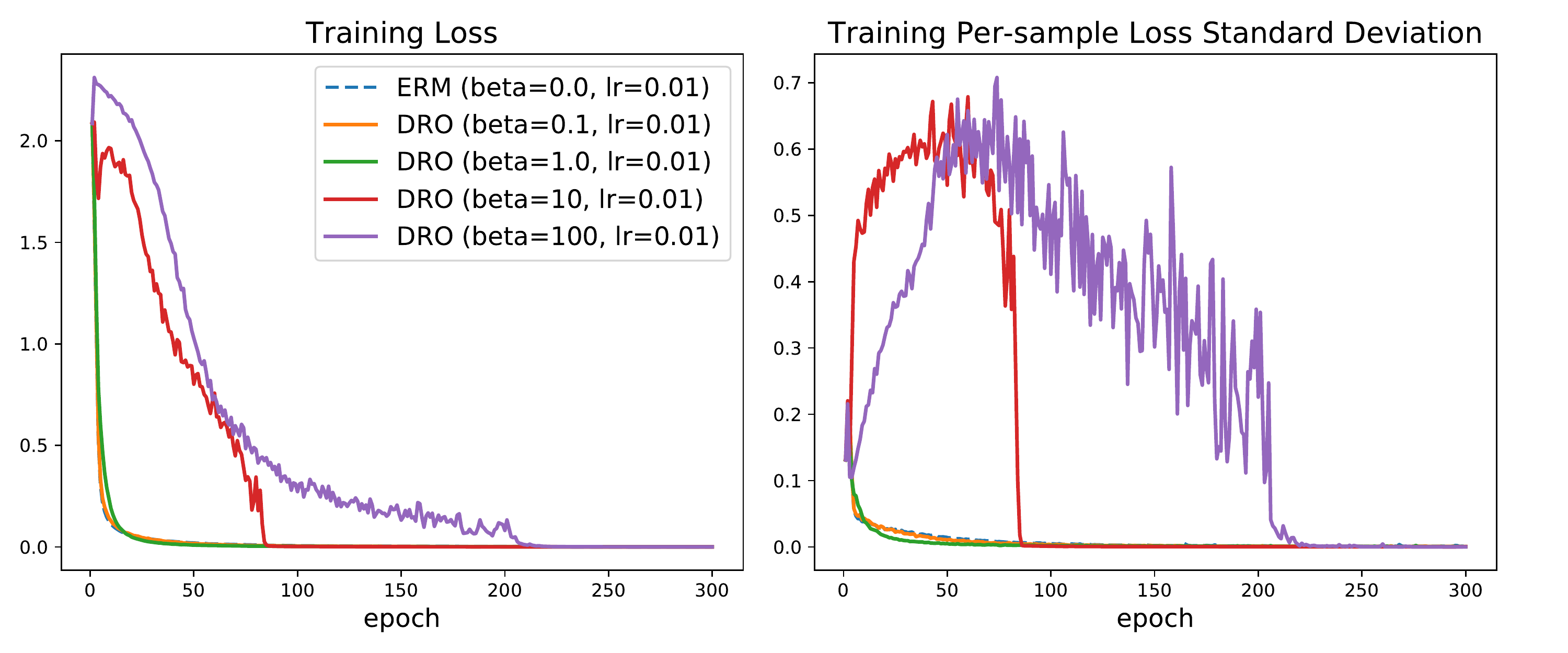}
    \caption{\label{fig:mnist_1percent_all}
        \textbf{Experiments on MNIST.}
        We compare the learning curves at testing (top panels) and at training (bottom panels) for ERM with SGD (\textcolor{blue}{blue}) and DRO with our SGD with hardness weighted sampling for different values of $\de$
        (\textcolor{orange}{$\de=0.1$}, \textcolor{green}{$\de=1$}, \textcolor{red}{$\de=10$},
        \textcolor{violet}{$\de=100$}).
        The models are trained on an imbalanced MNIST dataset (only $1\%$ of the digits $3$ kept for training) and evaluated on the original MNIST testing dataset.
        }
\end{figure}

\paragraph*{Results:}
Our experiment suggests that DRO and ERM lead to different optima.
Indeed, DRO for $\beta=10$ outperforms ERM by more than $15\%$ of accuracy on the underrepresented class, as illustrated in \Figref{fig:mnist_1percent_all}.
This suggests that DRO is more robust than ERM to domain gaps between the training and the testing dataset.
In addition, \Figref{fig:mnist_1percent_all} suggests that DRO with our SGD with hardness weighted sampling can converge faster than ERM with SGD.

Furthermore, the variations of learning curves with $\de$ shown in \Figref{fig:mnist_1percent_all} are consistent with our theoretical insight.
As $\de$ decreases to $0$, the learning curve of DRO with our \Algref{alg:1} converges to the learning curve of ERM with SGD.

For large values of $\de$ (here $\de \geq 10$), instabilities appear before convergence in the \textbf{testing learning curves}, as illustrated in the top panels of \Figref{fig:mnist_1percent_all}.
However, the bottom left panel of \Figref{fig:mnist_1percent_all} shows that the \textbf{training loss curves} for $\de \geq 10$ were stable there. 
We also observe that during iterations where instabilities appear on the \textbf{testing set}, the standard deviation of the per-example loss on the \textbf{training set} is relatively high (i.e. the hardness weighted probability is further away from the uniform distribution).
This suggests that the apparent instabilities on the \textbf{testing set} are related to differences between the distributionally robust loss and the mean loss.
\subsection{Medical Image Segmentation}\label{s:medical_image_segmentation}

In this section, we illustrate the application of \Algref{alg:1} to improve the robustness of deep learning methods for medical image segmentation.
We first discuss the specificities of applying the proposed \textit{hardness weighted sampling} to medical image segmentation in relation to the use of patch-based sampling.
We evaluated the proposed method on two applications:
fetal brain 3D MRI segmentation using 
a subset of the dataset described in \Chapref{chap:fetaldataset},
% the FeTA dataset and a private dataset,
and brain tumor multi-sequence MRI segmentation using the BraTS 2019 dataset~\cite{bakas2017HGG,bakas2017LGG}.

\subsubsection{Hardness Weighted Sampler with Large Images}\label{sec:patch_sampling}
In medical image segmentation, the image used as input of the deep neural network are typically large 3D volumes.
For this reason, state-of-the-art deep learning pipelines use patch-based sampling rather than full-volume sampling during training with ERM ~\cite{isensee2021nnu} as described in subsection~\ref{sec:material}.

This raised the question of what is the training distribution $p_{\textrm{train}}$ in the ERM \eqref{eq:erm_intro} and DRO \eqref{eq:dro} optimization problems.
Here, since the patches are large enough to cover most of the brains, we consider that patches are good approximation of the whole volumes and $p_{\textrm{train}}$ is the distribution of the full volumes.
Therefore, in the hardness weighted sampler of \Algref{alg:1}, we have only one weight per full volume.

In the case the full volumes are too large to be well covered by the patches, one can divide each full volume into a finite number of subvolumes prior to training.
For example, for chest CT, one can divide the volumes into left and right lungs~\cite{tilborghs2020comparative}.

\subsubsection{Material}\label{sec:material}

\paragraph{Fetal Brain Dataset.}
\begin{table}[th!]
	\centering
	\caption{
	\textbf{Training and Testing Fetal Drain 3D MRI Dataset Details.}
	Other Abn: brain structural abnormalities other than spina bifida.
	There is no overlap of subjects between training and testing.
	}
	\begin{tabularx}{\textwidth}{ *{5}{Y}}
		\toprule
		Train/Test & Origin & Condition & Volumes & Gestational age (in weeks)\\
		\midrule
		Training & Atlas& Control & 18 & [21,\,38]\\
		Training & FeTA & Control & 5 & [22,\,28]\\
		Training & UHL & Control & 116 & [20,\,35]\\
		Training & UHL & Spina Bifida & 28 & [22,\,34]\\
		Training & UHL & Other Abn & 10 & [23,\,35]\\
		\midrule
		Testing & FeTA & Control & 31 & [20,\,34]\\
		Testing & FeTA & Spina Bifida & 38 & [21,\,31]\\
		Testing & FeTA & Other Abn & 16 & [20,\,34]\\
		Testing & UHL & Control & 76 & [22,\,37]\\
		Testing & UHL and MUV & Spina Bifida & 74 & [19,\,35]\\
		Testing & UHL & Other Abn & 25 & [21, 40]\\
	\bottomrule
	\end{tabularx}
	\label{tab:data_fetal_dro}
\end{table}
The fetal brain MRI dataset used in this chapter is a subset of the dataset described in \Chapref{chap:fetaldataset}.
In this paragraph, we only specify which fetal brain 3D MRI were used and refer to \Chapref{chap:fetaldataset} for the description of the data acquisition and processing.
A summary of the data used in this chapter is given in Table \ref{tab:data_fetal_dro}.

We used the 18 control fetal brain 3D MRIs of the spatio-temporal fetal brain atlas\footnote{\url{http://crl.med.harvard.edu/research/fetal_brain_atlas/}}~\cite{gholipour2017normative} for gestational ages ranging from $21$ weeks to $38$ weeks.
We also used $80$ volumes from the publicly available FeTA MICCAI challenge dataset\footnote{DOI: 10.7303/syn25649159}~\cite{payette2021automatic}
and the $10$ 3D MRIs from the testing set of the first release of the FeTA dataset for which manual segmentations are not publicly available.
% 
% For those 3D MRIs, manual segmentations and corrections of the segmentations were performed by authors MA and LF to reduce the variability against the published segmentation guidelines that was released with the FeTA dataset~\cite{payette2021automatic}.
% % 
% Part of those corrections were performed as part of our previous work~\cite{fidon2021label,fidon2021partial} and are publicly available\footnote{DOI: 10.5281/zenodo.5148611}.
% %
% Brain masks for the FeTA data were obtained via affine registration using two fetal brain atlases\footnote{DOI: 10.7303/syn25887675}~\cite{fidon2021atlas,gholipour2017normative}.
% 
In addition, we used $329$ 3D MRIs from a private dataset.
All images in the private dataset were part of routine clinical care and were acquired at University Hospital Leuven (UHL) and Medical University of Vienna (MUW)
due to congenital malformations seen on ultrasound.
In total, 
$102$ cases with spina bifida aperta,
$35$ cases with other central nervous system pathologies,
and 
$192$ cases with other malformations, though with normal brain, and referred as controls,
were included.
The gestational age at MRI ranged from $19$ weeks to $40$ weeks.
Some of those 3D MRIs and their manual segmentations were used in previous studies~\cite{emam2021longitudinal,fidon2021atlas,fidon2021label,mufti2021cortical}.

\paragraph{Brain Tumor Dataset.}
We have used the BraTS 2019 dataset because it is the last edition of the BraTS challenge for which information about the image acquisition center is available at the time of writing.
The dataset contains the same four MRI sequences (T1, ceT1, T2, and FLAIR) for 448 cases, corresponding to patients with either a high-grade Gliomas or a low-grade Gliomas.
All the cases were manually segmented for peritumoral edema, enhancing tumor, and non-enhancing tumor core using the same labeling protocol~\cite{menze2014multimodal,bakas2018identifying,bakas2017advancing}.
We split the 323 cases of the BraTS 2019 \textit{training} dataset into 268 for training and 67 for validation.
In addition, the BraTS 2019 \textit{validation} dataset that contains 125 cases was used for testing.

\paragraph{Deep Learning Pipeline.}
The deep learning pipeline used was based on nnU-Net~\cite{isensee2021nnu}, which is a generic deep learning pipeline for medical image segmentation, that has been shown to outperform other deep learning pipelines on 23 public datasets without the need to manually tune the loss function or the deep neural network architecture.
Specifically, we used nnU-Net version 2 in 3D-full-resolution mode which is the recommended mode for isotropic 3D MRI data and the code is publicly available at~\url{https://github.com/MIC-DKFZ/nnUNet}.

Like most deep learning pipelines in the literature, nnU-Net is based on ERM.
For clarity, in the following we will sometimes refer to the unmodified nnU-Net as nnU-Net-ERM.

The meta-parameters used for the deep learning pipeline used were determined automatically using the heuristics developed in nnU-Net~\cite{isensee2021nnu}.
The 3D CNN selected for the brain tumor data is based on 3D U-Net~\cite{cciccek20163d} with 5 (resp. 6) levels for fetal brain segmentation (resp. brain tumor segmentation) and 32 features after the first convolution that are multiplied by 2 at each level with a maximum set at 320.
The 3D CNN uses leaky $\relu$ activation, instance normalization~\cite{ulyanov2016instance}, max-pooling downsampling operations and linear upsampling with learnable parameters.
In addition, the network is trained using the addition of the mean Dice loss and the cross entropy, and deep supervision~\cite{lee2015deeply}.
The default optimization step is SGD with a momentum of $0.99$ and Nesterov update, a batch size of 4 (resp. 2) for fetal brain segmentation (resp. brain tumor segmentation), and a decreasing learning rate defined for each epoch $t$ as
\[
\eta_t = 0.01 \times \left(1 - \frac{t}{t_{max}}\right)^{0.9}
\]
where $t_{max}$ is the maximum number of epochs fixed as $1000$.
Note that in nnU-Net, one epoch is defined as equal to 250 batches, irrespective of the size of the training dataset.
A patch size of $96 \times 112 \times 96$ (resp. $128 \times 192 \times 128$) was selected for fetal brain segmentation (resp. brain tumor segmentation), which is not sufficient to fit the whole brain of all the cases. As a result, a patch-based approach is used as often in medical image segmentation applications.
A large number of data augmentation methods are used: random cropping of a patch, random zoom, gamma intensity augmentation, multiplicative brightness, random rotations, random mirroring along all axes, contrast augmentation, additive Gaussian noise, Gaussian blurring and simulation of low resolution.
nnU-Net automatically splits the training data into 5 folds $80\%$ training/$20\%$ validation.
For the experiments on brain tumor segmentation, only the networks corresponding to the first fold were trained.
For the experiments on fetal brain segmentation, 5 models were trained, one for each fold, and the predicted class probability maps of the 5 models are averaged at inference to improve robustness~\cite{isensee2021nnu}.
GPUs NVIDIA Tesla V100-SXM2 with 16GB of memory were used for the experiments.
Training each network took from 4 to 6 days.

Our only modifications of the nnU-Net pipeline is the addition of our hardness weighted sampling when "DRO" is indicated and for some experiments we modified the optimization update rule as indicated in Table~\ref{tab:models_results_dro}.
Our implementation of the nnU-Net-DRO training procedure is publicly available at \url{https://github.com/LucasFidon/HardnessWeightedSampler}.
If "ERM" is indicated and nothing is indicated about the optimization update rule, it means that nnU-Net~\cite{isensee2021nnu} is used without any modification.

\paragraph{Hyper-parameters of the Hardness Weighted Sampler.}
For brain tumor segmentation, we tried the values $\{10, 100, 1000\}$ of $\de$ with or without importance sampling. Using $\de=100$ with importance sampling lead to the best mean dice score on the validation split of the training dataset.
For fetal brain segmentation, we tried only $\beta=100$ with importance sampling.
When importance sampling is used, the clipping values $w_{min}=0.1$ and $w_{max}=10$ are always used.
No other values of $w_{max}$ and $w_{min}$ have been tested.
% 
% Our implementation of the hardness weighted sampler is publicly available at \url{https://github.com/LucasFidon/HardnessWeightedSampler}.

\paragraph{Metrics.}
We evaluate the quality of the automatic segmentations using the Dice score~\cite{dice1945measures,fidon2017generalised}.
We are particularly interested in measuring the statistical risk of the results as a way to evaluate the robustness of the different methods.

In the BraTS challenge, this is usually measured using the interquartile range (IQR) which is the difference between the percentiles at $75\%$ and $25\%$ of the the metric values~\cite{bakas2018identifying}.
We therefore reported the mean, the median and the IQR of the Dice score in Table~\ref{tab:models_results_dro_brats}.
For fetal brain segmentation, in addition to the mean, median, and IQR, we also report the percentiles of the Dice score at $25\%$, $10\%$, and $5\%$.
In Table~\ref{tab:models_results_dro}, we report those quantities for the Dice scores of the three tissue types white matter, intra-axial CSF, and cerebellum.

For each method, nnU-Net is trained 5 times using different train/validation splits and different random initializations.
The 5 same splits, computed randomly, are used for the two methods.
The results for fetal brain 3D MRI segmentation in Table~\ref{tab:models_results_dro} are for the ensemble of the 5 3D U-Nets.
Ensembling is known to increase the robustness of deep learning methods for segmentation~\cite{isensee2021nnu}.
It also makes the evaluation less sensitive to the random initialization and to the stochastic optimization.

\begin{table}[th!]
	\centering
	\caption{\textbf{Evaluation of Distribution Robustness with Respect to the Pathology (260 3D MRIs).}
	\textbf{nnU-Net-ERM} is the unmodified nnU-Net pipeline~\cite{isensee2021nnu} in which Empirical Risk Minimization (ERM) is used.
	\textbf{nnU-Net-DRO} is the nnU-Net pipeline modified to use the proposed \textit{hardness weighted sampler} and in which Distributionally Robust Optimization (DRO) is therefore used.
	\textcolor{red}{\bf WM}: White matter, 
	\textcolor{ForestGreen}{\bf In-CSF}: Intra-axial CSF, 
	\textcolor{blue}{\bf Cer}: Cerebellum.
	IQR: interquartile range,
	$\textbf{p}_{X}$: $X^{\textrm{th}}$ percentile of the Dice score distribution in percentage.
    Best values are in bold.
	}
	\begin{tabularx}{\textwidth}{c c c *{6}{Y}}
		\toprule
        \multicolumn{1}{c}{} & \multicolumn{1}{c}{} & \multicolumn{1}{c}{}
        & \multicolumn{6}{c}{Dice Score ($\%$)} \\
        \cmidrule(lr){4-9} 
		\multicolumn{1}{c}{\bf Method} & \multicolumn{1}{c}{\bf CNS} & \multicolumn{1}{c}{\bf ROI} & 
		Mean & Median & IQR & $\textbf{p}_{25}$ & $\textbf{p}_{10}$ & $\textbf{p}_5$ \\ 
		\midrule
		(baseline) & \textbf{Controls} &
		        \textcolor{red}{\bf WM} & 
		        $\bf94.4$ & 95.2 & \bf 2.8 & \bf 93.3 & \bf 91.5 & \bf 90.6 \\
		nnU-Net-ERM & (107 volumes) & \textcolor{ForestGreen}{\bf In-CSF} &
		        90.3 & 92.4 & 6.4 & 87.8 & 80.7 & 79.0 \\
		        & & \textcolor{blue}{\bf Cer} &
		        \bf 95.7 & 97.0 & 3.4 & \bf 94.2 & 91.3 & \bf 90.4 \\
	        \cmidrule(lr){2-9}
	        & \textbf{Spina Bifida} &
		        \textcolor{red}{\bf WM} & 
		        89.6 & 92.1 & 4.1 & 89.5 & 80.6 & 73.8 \\
		        & (112 volumes)  & \textcolor{ForestGreen}{\bf In-CSF} &
		        91.4 & 93.9 & \bf 6.4 & 89.6 & \bf 86.9 & \bf 83.7 \\
		        & & \textcolor{blue}{\bf Cer} &
		        76.8 & 87.8 & 11.1 & 80.4 & 15.8 & \bf 0.0 \\
		    \cmidrule(lr){2-9}
	        & \textbf{Other Abn.} &
		        \textcolor{red}{\bf WM} & 
		        90.3 & \bf 92.6 & \bf 4.6 & 90.1 & 88.0 & 71.6 \\
		        & (41 volumes) & \textcolor{ForestGreen}{\bf In-CSF} &
		        87.4 & 87.9 & 10.4 & 82.7 & 77.7 & 75.9 \\
		        & & \textcolor{blue}{\bf Cer} &
		        90.4 & 92.8 & \bf5.4 & \bf90.7 & \bf87.5 & 81.4 \\
	\cmidrule(lr){1-9}
        (ours) & 
		    \textbf{Controls} &
		        \textcolor{red}{\bf WM} & 
		        \bf 94.4 & \bf95.3 & 3.0 & 93.2 & 91.1 & 90.1 \\
		nnU-Net-DRO & (107 volumes) & \textcolor{ForestGreen}{\bf In-CSF} &
		        \bf 90.4 & \bf 92.7 & \bf 6.2 & \bf 87.9 & \bf 81.7 & \bf 79.1 \\
		        & & \textcolor{blue}{\bf Cer} &
		        \bf 95.7 & \bf 97.1 & \bf 3.3 & \bf 94.2 & \bf 91.4 & 90.1 \\
	        \cmidrule(lr){2-9}
	        & \textbf{Spina Bifida} &
		        \textcolor{red}{\bf WM} & 
		        $\bf 90.1$ & \bf 92.2 & \bf 4.0 & \bf 89.9 & \bf 81.6 & \bf 74.8\\
		        & (112 volumes)  & \textcolor{ForestGreen}{\bf In-CSF} &
		        $\bf 91.6$ & \bf 94.1 & \bf 6.4 & \bf 90.0 & 86.7 & 83.6 \\
		        & & \textcolor{blue}{\bf Cer} &
		        $\bf 77.8$ & \bf 87.9 & \bf 9.7 & \bf 82.0& \bf 43.3 & \bf 0.0 \\
		    \cmidrule(lr){2-9}
	        & \textbf{Other Abn.} &
		        \textcolor{red}{\bf WM} & 
		        \bf 90.4 & \bf 92.6 & 4.7 & \bf 90.2 & \bf 88.2 & \bf 73.5 \\
		        & (41 volumes) & \textcolor{ForestGreen}{\bf In-CSF} &
		        \bf 87.9 & \bf 88.1 & \bf 9.5 & \bf 83.3 & \bf 80.4 & \bf 77.7 \\
		        & & \textcolor{blue}{\bf Cer} &
		        \bf 91.3 & \bf 93.0 & 5.5 & \bf 90.7 & \bf 87.5 & \bf 82.7\\
	\bottomrule
	\end{tabularx}
	\label{tab:models_results_dro}
\end{table}

\begin{table}[th!]
	\centering
	\caption{\textbf{Dice Score Evaluation on the BraTS 2019 Online Validation Set (125 cases).}
	Metrics were computed using the BraTS online evaluation platform (\url{https://ipp.cbica.upenn.edu/}).
	ERM: Empirical Risk Minimization,
	DRO: Distributionally Robust Optimization,
	SGD: plain SGD (no momentum used),
	Nesterov: SGD with Nesterov momentum,
	IQR: Interquartile range.
	The best values are in bold.
	}
	\begin{tabularx}{\textwidth}{c c *{9}{Y}}
		\toprule
        \textbf{Optim.}
        & \textbf{Optim.}
        & \multicolumn{3}{c}{Enhancing Tumor} 
        & \multicolumn{3}{c}{Whole Tumor}
        & \multicolumn{3}{c}{Tumor Core}\\
    \cmidrule(lr){3-5} \cmidrule(lr){6-8} \cmidrule(lr){9-11}
		\textbf{problem} & \textbf{update} & 
		Mean & Median & IQR & Mean & Median & IQR & Mean & Median & IQR\\
	\midrule
	    ERM & SGD
	     & 71.3 & 86.0 & 20.9
	     & 90.4 & 92.3 & 6.1
	     & 80.5 & 88.8 & 17.5\\
	\midrule
	    DRO & SGD
	     & 72.3 & 87.2 & 19.1
	     & 90.5 & \bf92.6 & 6.0
	     & 82.1 & 89.7 & 15.2\\
	\midrule
		ERM & Nesterov
		 & 73.0 & 87.1 & 15.6
		 & \bf90.7 & \bf92.6 & \bf5.4
		 & 83.9 & \bf90.5 & 14.3\\
	\cmidrule(lr){1-11}
        DRO & Nesterov
        % ($\de=100$, IS)
         & \bf74.5 & \bf87.3 & \bf13.8
		 & 90.6 & \bf92.6 & 5.9
		 & \bf84.1 & 90.0 & \bf12.5\\
	\bottomrule
	\end{tabularx}
	\label{tab:models_results_dro_brats}
\end{table}

\begin{figure}[ht!]
    \setlength{\lineskip}{0pt}
    \centering
    \includegraphics[width=\textwidth]{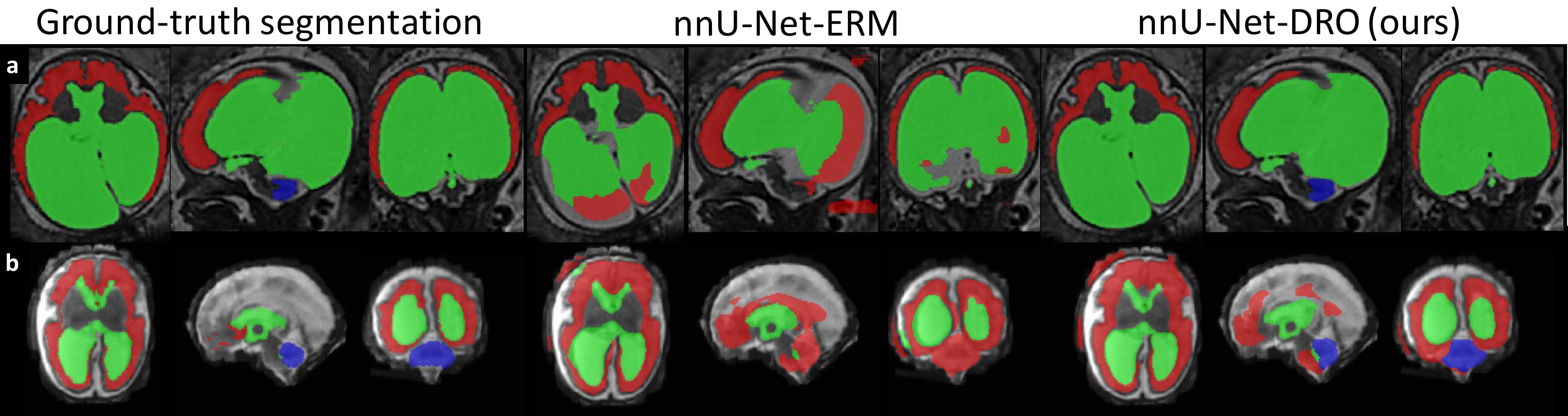}
    \includegraphics[width=\textwidth,trim=0cm 0cm 0cm 1.5cm,clip]{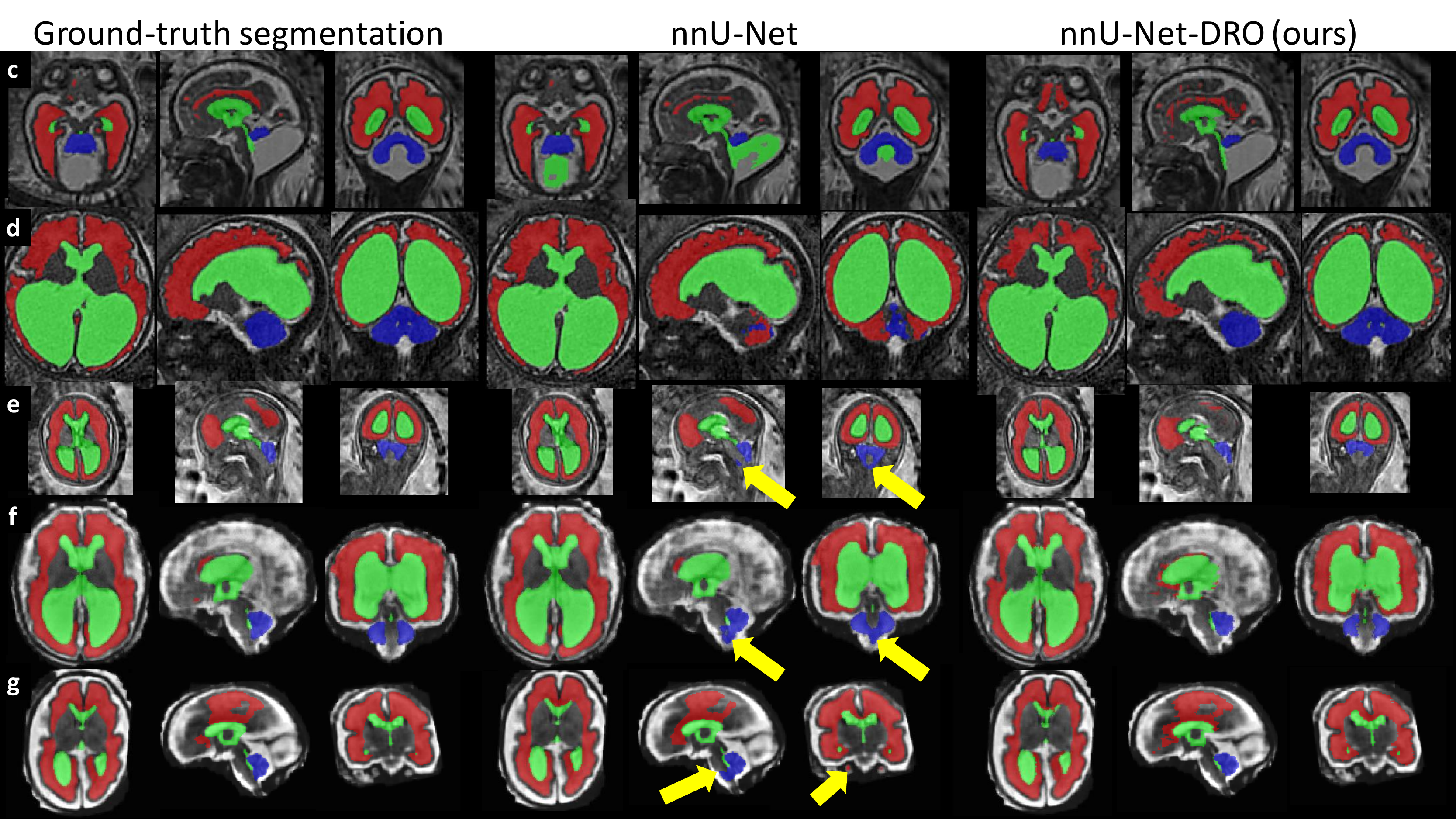}
    \caption{\textbf{Qualitative Results for Fetal Brain 3D MRI Segmentation using DRO.}
    a) Fetus with aqueductal stenosis (34 weeks).
    b) Fetus with spina bifida aperta (27 weeks).
    c) Fetus with Blake's pouch cyst (29 weeks).
    d) Fetus with tuberous sclerosis complex (34 weeks).
    e) Fetus with spina bifida aperta (22 weeks).
    f) Fetus with spina bifida aperta (31 weeks).
    g) Fetus with spina bifida aperta (28 weeks).
    For cases a) and b), nnU-Net-ERM~\cite{isensee2021nnu} misses completely the cerebellum and achieves poor segmentation for the white matter and the ventricles.
    % This is not the case for nnU-Net-DRO.
    % % 
    % Our nnU-Net-DRO achieves satisfactory segmentation for the cerebellum for the two cases, and for all tissue types for the aqueductal stenosis case.
    % 
    For case c), a large part of the Blake's pouch cyst is wrongly included in the ventricular system segmentation by nnU-Net-ERM. This is not the case for the proposed nnU-Net-DRO.
    For case d), nnU-Net-ERM fails to segment the cerebellum correctly and a large part of the cerebellum is segmented as part of the white matter. In contrast, our nnU-Net-DRO correctly segment cerebellum and white matter for this case.
    For cases e) f) and g), we have added yellow arrows pointing to large parts of the brainstem that nnU-Net-ERM wrongly included in the cerebellum segmentation. 
    nnU-Net-DRO does not make this mistake.
    We emphasise that the segmentation of the cerebellum for spina bifida aperta is essential for studying and evaluating the effect of surgery in-utero.
    }
    \label{fig:res_qualitative_fetal}
\end{figure}

\begin{figure}[th!]
    \centering
    \includegraphics[width=\linewidth]{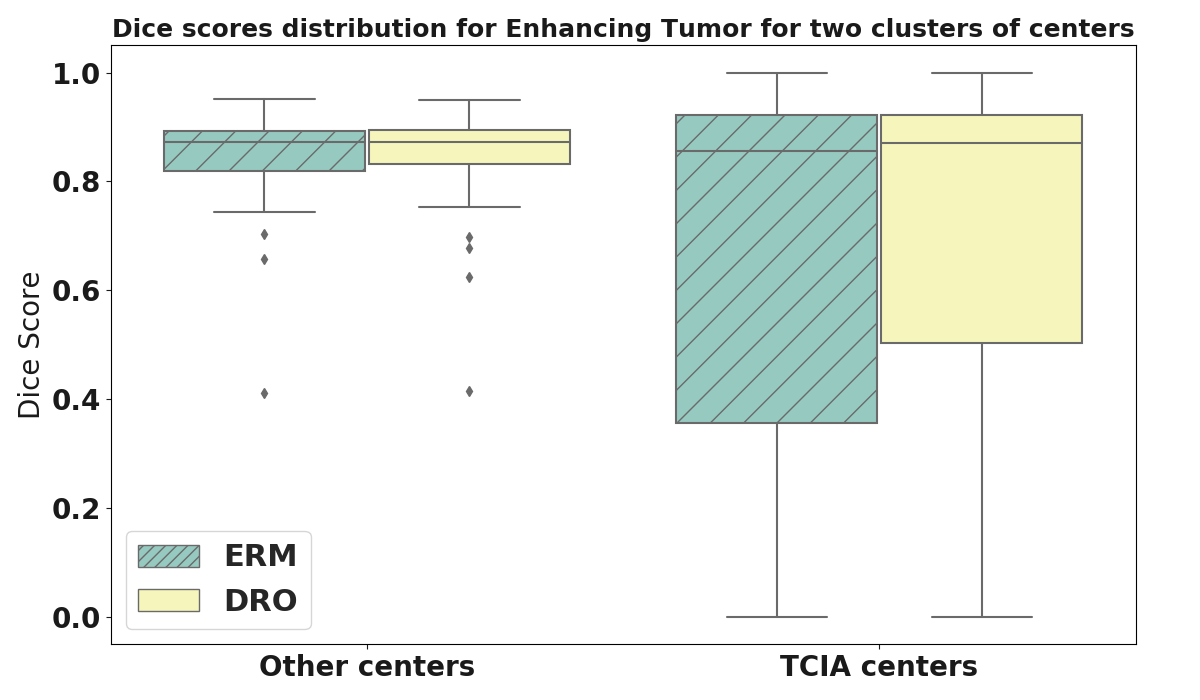}
    \caption{\label{fig:difference_centers}
        \textbf{Dice scores distribution on the BraTS 2019 validation dataset for cases from a center of TCIA (76 cases) and cases from other centers (49 cases).}
        This shows that the lower interquartile range of DRO for the enhancing tumor comes specifically from a lower number of poor segmentations on cases coming from The Cancer Imaging Archive (TCIA).
        This suggests that DRO can deal with some of the confounding biases present in the training dataset, and lead to a model that is more fair.
        }
\end{figure}

\paragraph*{Results.}
The quantitative comparison of nnU-Net-ERM and nnU-Net-DRO on fetal brain 3D MRI segmentation for the three different central nervous system conditions control, spina bifida, and other abnormalities can be found in Table~\ref{tab:models_results_dro}.

For spina bifida and other brain abnormalities, the proposed nnU-Net-DRO achieves same or higher mean Dice scores than nnU-Net-ERM~\cite{isensee2021nnu} with $+0.5$ percentage points (pp) for white matter and $+1$pp for the cerebellum of spina bifida cases and $+0.9$pp for the cerebellum for other abnormalities.
In addition, nnU-Net-DRO achieves comparable (at most $0.1$pp of difference) or lower IQR than nnU-Net-ERM with $-1.4$pp for the cerebellum of spina bifida cases and $-0.9$pp for the intra-axial CSF of cases with other abnormalities.
For controls, the mean, median, and IQR of the Dice scores of nnU-Net-DRO and nnU-Net-ERM differ by less than $0.2$pp for the three tissue types.
This suggests that nnU-Net-DRO is more robust to anatomical variabilities associated with abnormal brains, while retaining the same segmentation performance on neurotypical cases.

In terms of median Dice score, nnU-Net-DRO and nnU-Net-ERM differ by less than $0.3$pp for all tissue types and conditions.
Therefore the differences in terms of mean Dice scores mentioned above are not due to improved segmentation in the middle of the Dice score performance distribution.

The comparison of the percentiles at $25\%$, $10\%$, and $5\%$ of the Dice score allows us to compare methods at the tail of the Dice scores distribution where segmentation methods reach their worst-case performance.
For spina bifida, nnU-Net-DRO achieves higher values of percentiles than nnU-Net-ERM for the white matter ($+1.0$pp for $\textbf{p}_{10}$ and $+1.0$pp for $\textbf{p}_{5}$), 
and for the cerebellum ($+1.6$pp for $\textbf{p}_{25}$ and $+27.5$pp for $\textbf{p}_{10}$).
And for other brain abnormalities, nnU-Net-DRO achieves higher values of percentiles than nnU-Net-ERM for the white matter ($+1.9$pp for $\textbf{p}_{5}$), 
for the intra-axial CSF ($+0.6$pp for $\textbf{p}_{25}$, $+2.3$pp for $\textbf{p}_{10}$ and $+1.8$pp for $\textbf{p}_{5}$), 
and for the cerebellum ($+1.3$pp for $\textbf{p}_{5}$).
All the other percentile values differ by less than $0.5$pp of Dice score between the two methods.
This suggests that nnU-Net-DRO achieves better worst case performance than nnU-Net-ERM for abnormal cases.
However, both methods have a percentile at $5\%$ of the Dice score equal to $0$ for the cerebellum of spina bifida cases. This indicates that both methods completely miss the cerebellum for spina bifida cases in $5\%$ of the cases.

As can be seen in the qualitative results of \Figref{fig:res_qualitative_fetal}, there are cases for which nnU-Net-ERM predicts an empty cerebellum segmentation while nnU-Net-DRO achieves satisfactory cerebellum segmentation.
There were no cases for which the converse was true.
However, there were also spina bifida cases for which both methods failed to predict the cerebellum.
Robust segmentation of the cerebellum for spina bifida is particularly relevant for the evaluation of fetal brain surgery for spina bifida aperta~\cite{aertsen2019reliability,danzer2020fetal,sacco2019fetal}.
All the spina bifida 3D MRIs with missing cerebellum in the automatic segmentations were 3D MRIs from the FeTA dataset~\cite{payette2021automatic} and represented brains of fetuses with spina bifida before they were operated on.
The cerebellum is more difficult to detect using MRI before surgery as compared to early or late after surgery~\cite{aertsen2019reliability,danzer2007fetal}.
No 3D MRI with the combination of those two factors were present in the training datatset (Table.~\ref{tab:data_fetal_dro}).
This might explain why DRO did not help improving the segmentation quality for those cases.
DRO aims at improving the performance on subgroups that were underrepresented in the training dataset, not subgroups that were not represented at all.

In Table~\ref{tab:models_results_dro}, it is worth noting that overall the Dice score values decrease for the white matter and the cerebellum between controls and spina bifida and abnormal cases.
It was expected due to the higher anatomical variability in pathological cases.
However, the Dice score values for the ventricular system tend to be higher for spina bifida cases than for controls.
This can be attributed to the large proportion of spina bifida cases with enlarged ventricles because the Dice score values tend to be higher for larger regions of interest.

For our experiments on brain tumor segmentation, Table~\ref{tab:models_results_dro_brats} summarizes the performance of training nnU-Net using ERM or using DRO.
Here, we experiment with two SGD-based optimizers.
For both ERM and DRO, the optimization update rule used was either plain SGD without momentum (SGD), or SGD with a Nesterov momentum equal to $0.99$ (Nesterov).
Especially, for the latter, this implies that step 12 of \Algref{alg:1} is modified to use SGD with Nesterov momentum. It was also the case for our experiments on fetal brain 3D MRI segmentation.
For DRO, the results presented here are for $\de=100$ and using importance sampling (step 6 of \Algref{alg:1}).

As illustrated in Table~\ref{tab:models_results_dro_brats}, for both ERM and DRO, the use of SGD with Nesterov momentum outperforms plain-SGD for all metrics and all regions of interest.
This result was expected for ERM, for which it is common practice in the deep learning literature to use SGD with a momentum.
Our results here suggest that the benefit of using a momentum with SGD is retained for DRO.

For both optimizers, DRO outperforms ERM in terms of IQR for the enhancing tumor and the tumor core by approximately $2$pp of Dice score, and in terms of mean Dice score for the enhancing tumor by $1$pp for the plain-SGD and $1.5$pp for SGD with Nesterov momentum.
For plain-SGD, DRO also outpermforms ERM in terms of mean Dice score for the tumor core by $1.6$pp.
The IQR is the global statistic used in the BraTS challenge to measure the level of robustness of a method~\cite{bakas2018identifying}.
In addition, \Figref{fig:difference_centers} shows that the lower IQR of DRO for the enhancing tumor comes specifically from a lower number of poor segmentations on cases coming from The Cancer Imaging Archive (TCIA).
This suggests that DRO can deal with some of the confounding biases present in the training dataset, and lead to a model that is more fair with respect to the acquisition center of the MRI.

Since the same improvements are observed independently of the optimization update rule used.
This suggests that in practice \Algref{alg:1} still converges when a momentum is used, even if Theorem~\ref{th:convergence_dro} was only demonstrated to hold for plain-SGD.

The value $\de=100$ and the use of importance sampling was selected based on the mean Dice score on the validation split of the training dataset.
Results for $\de \in \{10, 100, 1000\}$ with Nesterov momentum and with or without importance sampling can be found in Appendix~\ref{s:more_results_brats} Table~\ref{tab:models_all_results}.
The tendency described previously still holds true for the enhancing tumor for $\de$ equal to $10$ or $100$ with and without importance sampling.
The mean Dice score is improved by $0.4$pp to $2.3$pp
and the IQR is reduced by $1.3$pp to $2.3$pp for the four DRO models as compared to the ERM model.
For the tumor core with $\de=100$ mean and IQR are improved over ERM with and without importance sampling.
However, for $\de=10$ with importance sampling there was a loss of performance as compared to ERM for the whole tumor.
This problem was not observed with $\de=10$ without importance sampling. For the other models with $\de$ equal to $10$ or $100$ similar Dice score performance similar to the one ERM was observed for the whole tumor.
This suggests that overall the use of ERM or DRO does not affect the segmentation performance of the whole tumor.
One possible explanation of this is that Dice scores for the whole tumor are already high for almost all cases when ERM is used with a low IQR.
In addition, DRO and the \textit{hardness weighted sampler} are sensitive to the loss function, here the mean-class Dice loss plus cross entropy loss.
In the case of brain tumor segmentation, we hypothesise that the loss function is more sensitive to the segmentation performance for the tumor core and the enhancing tumor than for the whole tumor.

When $\de$ becomes too large ($\de=1000$) a decrease of the mean and median Dice score for all regions is observed as compared to ERM.
In this case, DRO tends towards the maximization of the worst-case example only which appears to be unstable using our \Algref{alg:1}.
    In contrast, when $\beta$ is too small, DRO tends towards ERM.
    The value of $\beta$ controls the number of training samples with a sampling probability that is high enough to allow those samples to be sampled regularly during training.
    This leads to the question: what is a good range of values for $\beta$?
    We hypothesize that good values of $\beta$ are of the order of the inverse of the standard deviation of the vector of per-volume (stale) losses during the training epochs that precede convergence.
	It corresponds, up to one order of magnitude, to the values we have found in practice. However, it is difficult to conclude on a rule for setting $\beta$ from those observation.
	In theory, one can show that the sampling probabilities in our hardness weighted sampler are independent to the mean of the vector of per-volume (stale) losses.
	It is easy to see it for the softmax and one can prove it in the general case with any $\phi$-divergence (according to our definition of $\phi$-divergence).
	It is therefore intuitive to think about the inverse of the standard deviation of the vector of losses.
	It is possible to prove mathematically that the sampling probabilities is not invariant to this variance.
	However, it remains unclear how to characterize rigorously what is a good value for $\beta$.
	Maybe using a schedule for $\beta$ during training could lead to improve results but this would introduce more hyperparameters and we have not investigated it.

For all values of $\de$ the use of importance sampling, as described in steps 6-8 of \Algref{alg:1}, improves the IQR of the Dice scores for the enhancing tumor and the tumor core.
We therefore recommend to use \Algref{alg:1} with importance sampling.

\paragraph*{Efficiency of \Algref{alg:1}:}
The main additional computational cost is \Algref{alg:1} is due to the hardness weighted sampling in steps 4 and 5 that is dependent on the number $n$ of training examples.
We have computed that steps 4 and 5 together take less than $1$ second for up to $n=10^8$ using a batch size of $1$ and the function \texttt{random.choice} of Numpy version $1.21.1$.
For our brain tumor segmentation training set of n=268 volumes and a batch size of 2, the additional memory usage of Algorithm 1 is only 2144 bytes of memory (one float array of size n) and the additional computational time is approximately $10^{-4}$ seconds per iteration using the Python library numpy, i.e. approximately $0.005 \%$ of the total duration of an iteration.
The size of the training dataset for fetal brain 3D MRI segmentation being lower, the additional memory usage and the additional computational time are even lower than for brain tumor segmentation.
\section{Discussion and Conclusion}

In this chapter, we have shown that efficient training of deep neural networks with Distributionally Robust Optimization (DRO) with a $\phi$-divergence is possible.

The proposed \textit{hardness weighted sampler} for training a deep neural network with Stochastic Gradient Descent (SGD) for DRO is as straightforward to implement, and as computationally efficient as SGD for Empirical Risk Minimization (ERM).
It can be used for deep neural networks with any activation function (including $\relu$), and with any per-example loss function.
We have shown that the proposed approach can formally be described as a principled Hard Example Mining strategy (Theorem~\ref{th:hard_example_mining}) and is related to minimizing the percentile of the per-example loss distribution \eqref{eq:perc}.
In addition, we prove the convergence of our method for over-parameterized deep neural networks (Theorem~\ref{th:convergence_dro}).
Thereby, extending the convergence theory of deep learning of~\cite{allen-zhu19a}.
This is, to the best of our knowledge, the first convergence result for training a deep neural network based on DRO.

In practice, we have shown that our hardness weighted sampling method can be easily integrated in a state-of-the-art deep learning framework for medical image segmentation.
Interestingly, the proposed algorithm remains stable when SGD with momentum is used.

The high anatomical variability of the developing fetal brain across gestational ages and pathologies hampers the robustness of deep neural networks trained by maximizing the average per-volume performance.
Specifically, it limits the generalization of deep neural networks to abnormal cases for which few cases are available during training.
In this paper, we propose to mitigate this problem by training deep neural networks using Distributionally Robust Optimization (DRO) with the proposed hardness weighted sampling.
We have validated the proposed training method on a multi-centric dataset of $437$ fetal brain T2w 3D MRIs with various diagnostics.
nnU-Net trained with DRO achieved improved segmentation results for pathological cases as compared to the unmodified nnU-Net, while achieving similar segmentation performance for the neurotypical cases.
Those results suggest that nnU-Net trained with DRO is more robust to anatomical variabilities than the original nnU-Net that is trained with ERM.
In addition, we have performed experiments on the open-source multiclass brain tumor segmentation dataset BraTS~\cite{bakas2018identifying}.
Our results on BraTS suggests that DRO can help improving the robustness of deep neural network for segmentation to variations in the acquisition protocol of the images.

However, we have also found in our experiments that all deep learning models, either trained with ERM or DRO, failed in some cases.
For example, the models evaluated all missed the cerebellum in at least $5\%$ of the spina bifida aperta cases.
As a result, while our results do suggest that DRO with our method can improve the robustness of deep neural networks for segmentation, they also show that DRO alone with our method does not provide a guarantee of robustness.
DRO with a $\phi$-divergence reweights the examples in the training dataset but cannot account for subsets of the true distribution that are not represented at all in the training dataset.

% SAMPLING WITH MANY EXAMPLES
We have shown that the additional computational cost of the proposed hardness weighted sampling is small enough to be negligible in practice and requires less than one second for up to $n=10^8$ examples.
The proposed \Algref{alg:1} is therefore as computationally efficient as state-of-the-art deep learning pipeline for medical image segmentation.
% 
% However, it is unlikely that the stale loss
% 
However, when data augmentation is used, an infinite number of training examples is virtually available.
We mitigate this problem using importance sampling and only one probability per non-augmented example.
We found that importance sampling led to improved segmentation results.

We have also illustrated in our experiments that reporting the mean and standard deviation of the Dice score is not enough to evaluate the robustness of deep neural networks for medical image segmentation.
A stratification of the evaluation is required to assess for which subgroups of the population and for which image protocols a deep learning model for segmentation can be safely used.
In addition, not all improvements of the mean and standard deviation of the Dice score are equally relevant as they can result from improvements of either the best or the worst segmentation cases.
Regarding the robustness of automatic segmentation methods across various conditions, one is interested in improvements of segmentation metrics in the tail of the distribution that corresponds to the worst segmentation cases.
% 
% Reporting only the mean and the standard deviation is therefore not enough.
% 
To this end, one can report the interquartile range (IQR) and measures of risk such as percentiles.

\chapter[Spina Bifida Fetal Brain MRI Atlas]{A Spatio-temporal Atlas of the Developing Fetal Brain with Spina Bifida Aperta}
\label{chap:atlas}
\minitoc
\begin{center}
	\begin{minipage}[b]{0.9\linewidth}
		\small
		\textbf{Foreword\,}
		This chapter is to a large extent an \emph{in extenso} reproduction of \cite{fidon2021atlas}.
		The fetal brain atlas computed in this chapter plays a central role in the proposed implementation of our trustworthy AI framework for fetal brain MRI segmentation described in \Chapref{chap:twai}.
		\begin{itemize}
		    \item The annotation protocol was developed by Elizabeth Viola under the supervision of Michael Aertsen and myself.
		    \item The manual annotation of the landmarks on the 3D MRI has been performed by Elizabeth Viola.
		    \item The intra-rater variability study fir the anatomical landmarks has been performed by Elizabeth Viola.
		    \item The method for computing the atlas has been developed and implemented by myself.
		    \item 3D super-resolution and reconstruction of the fetal brain MRIs has been performed by myself using the library \texttt{NiftyMIC}.
		\end{itemize}
	\end{minipage}
\end{center}

% Content
\section{Introduction}
% Spina bifida aperta (SBA) is the most prevalent fetal brain defect with approximately five per 10,000 live births in Europe \cite{khoshnood2015long}. 
% % 
% It occurs when the neural tube fails to close in the first four weeks after conception. 
% % 
% Most cases of SBA are accompanied by severe anatomical brain abnormalities~\cite{pollenus2020impact}
% with enlargement of the ventricles and a type II Chiari malformation being most prevalent.
% % 
% The Chiari malformation type II is characterized by a small posterior fossa and hindbrain herniation in which the medulla, cerebellum, and fourth ventricule are displaced caudally into the spinal canal~\cite{naidich1980computed}.
% % 
% The corpus callosum of fetuses with SBA is also abnormal~\cite{kunpalin2021incidence,pollenus2020impact}
% % 
% and has been found to be significantly smaller for fetuses with SBA than for normal fetuses~\cite{crawley2014structure,dennis2016white,kunpalin2021incidence}.
% % 
% SBA fetuses have also smaller hippocampus~\cite{treble2015prospective}, 
% % 
% abnormal cortical thickness and gyrification~\cite{mufti2021cortical,treble2013functional},
% % 
% and smaller deep gray matter volume and total brain volume~\cite{hasan2008quantitative,mandell2015volumetric}.
% % 
% For all those reasons 
As discussed in \Chapref{chap:fetaldataset},
the anatomy of the brain of fetuses with SBA differs from the normal fetal brain anatomy.
In addition, the mechanisms underlying those anatomical brain abnormalities remain incompletely understood~\cite{danzer2020fetal}.

% Due to its severity, spina bifida aperta often requires in-utero fetal surgery or treatment after birth by means of a ventriculoperitoneal shunt in 80\% of infants. Anatomical biomarkers based on Magnetic Resonance Imaging (MRI) have shown great promise in quantifying the abnormalities due to spina bifida aperta and supporting patient-management decisions \cite{ma}.

Brain atlases are used to study common trends and variations in the brain anatomy of a population.
They provide a model of a population of brain magnetic resonance images (MRIs) that represents the average brain anatomy of a population,
allow the comparison of measurements in a cohort study, and can be used for the automatic segmentation of brain MRI~\cite{dittrich2014spatio,habas2010spatiotemporal,makropoulos2018review,serag2012multi}.
Atlases can also be used to measure variability in the brain anatomy of an individual as compared to the whole population~\cite{dittrich2014spatio}.
Age and disease specific atlases allow a more accurate model of specific populations of human brains to be obtained~\cite{evans2012brain}.

Previous work on fetal brain atlases has focused on age-specific atlases by proposing various spatio-temporal fetal brain MRI atlases~\cite{dittrich2014spatio,dittrich2011learning,gholipour2017normative,habas2010spatiotemporal,serag2012multi,zhan2013spatial,wu2021age}.
% blabla about previous atlas methods
A spatio-temporal atlas does not consist in only one average volume, but instead consists in a collection of age-specific average volumes.
This allows the development of the fetal brain anatomy to be modelled.
However, existing studies have only used brain MRIs of fetuses with a normal brain development, except for one study that combined fetuses with a normal brain and fetuses with lissencephaly in the same atlas~\cite{dittrich2014spatio}.
In particular, no fetal brain atlas for the developing fetal brain with SBA has been proposed in the literature.

In this chapter, we propose the first spatio-temporal fetal brain MRI atlas for SBA.
Our atlas covers all the weeks of gestation between $21$ weeks and $34$ weeks.
This range of gestational ages is of particular interest for SBA because it starts before the time at which in-utero surgery for SBA is currently performed~\cite{danzer2020fetal} and covers most of the time until birth.
The atlas is computed using $90$ fetal brain MRIs from $37$ fetuses with SBA.
% as part of a retrospective study. 
% 
We hypothesise that the high variability of the brain anatomy in SBA is one of the main challenges in adapting methods developed for normal fetal brain atlases for SBA.
To tackle this issue, we propose a semi-automatic method for the computation of the proposed fetal brain MRI atlas for SBA.
We propose a protocol for the annotation of $11$ anatomical landmarks in fetal brain 3D MRI of fetuses.
Those anatomical landmarks are used for two things in our pipeline.
% for the computation of the atlas.
% 
The anatomical landmarks are used firstly to initialize the computation of the atlas using a weighted generalized Procrustes method
and secondly to regularize the non-linear image registration of fetal brain 3D MRIs to the atlas.

We performed an intra-rater variability evaluation for the proposed landmarks using a subset of $31$ 3D MRIs from our cohort.
Based on this evaluation, $4$ anatomical landmarks were excluded and $7$ were selected to help for the computation of the spatio-temporal atlas.
In addition, we evaluated the automatic fetal brain segmentations computed using the proposed atlas for SBA on $40$ fetal brain 3D MRIs of the publicly available FeTA dataset~\cite{payette2021automatic}. It contains $15$ MRIs of normal fetuses and $25$ MRIs of fetuses with SBA.
We compared the automatic segmentations computed using our SBA atlas to the segmentations computed using a state-of-the-art normal fetal brain MRI atlas~\cite{gholipour2017normative}.
We have found that the proposed SBA atlas outperforms the normal fetal brain atlas on cases with SBA.
The proposed spatio-temporal fetal brain MRI atlas for SBA is made publicly available  \href{https://doi.org/10.7303/syn25910198}{here}.

\newpage
\section{Materials}

\begin{figure}[t]
    \centering
    \includegraphics[width=0.7\linewidth]{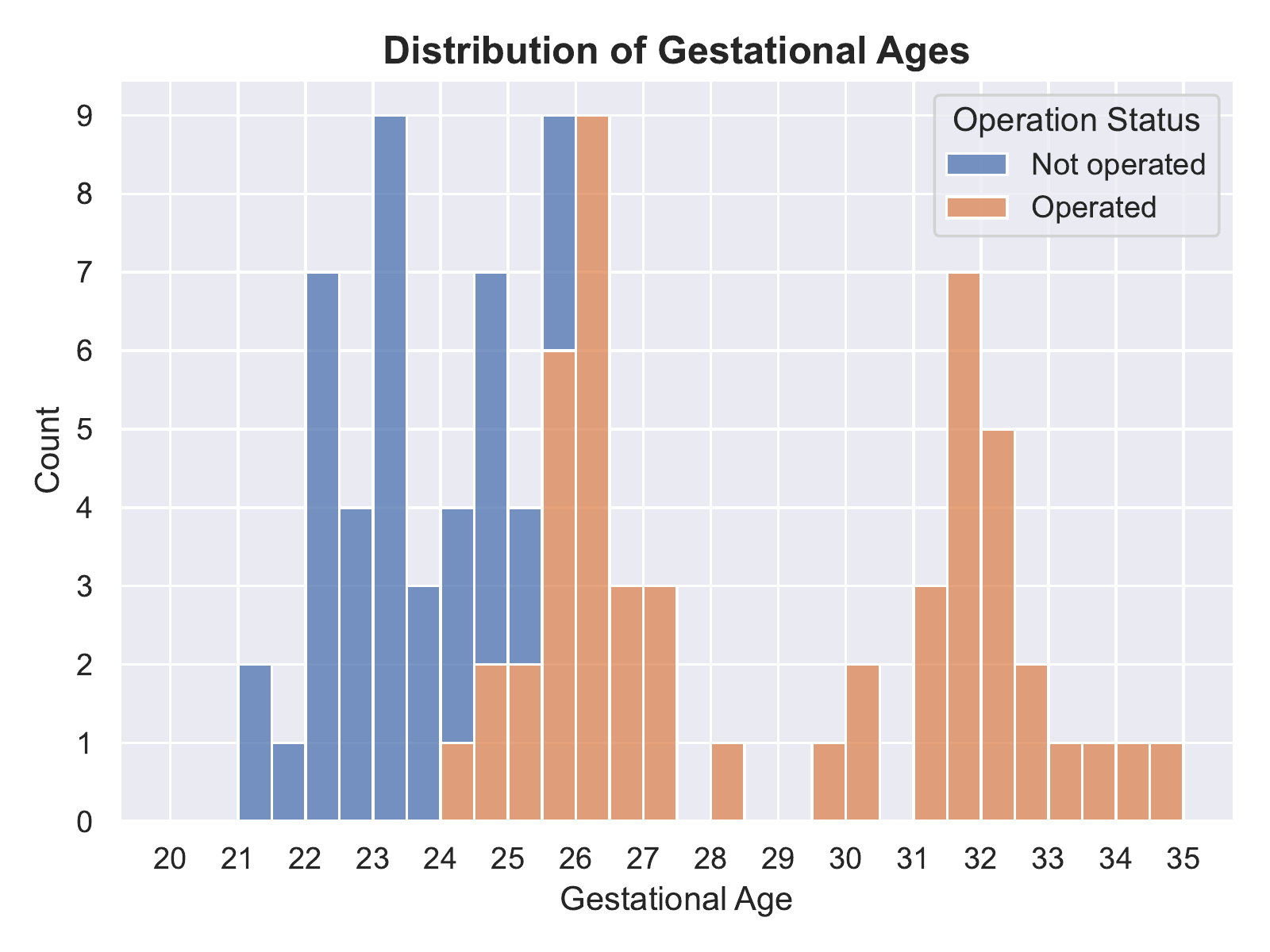}
    \caption{\textbf{Distribution of gestational ages for operated (fetal surgery) and non-operated fetal brains.}
    The dataset used to compute the atlas contains $39$ magnetic resonance imaging (MRI) examinations of non-operated fetuses and $51$ MRI examinations of operated fetuses.
    }
    \label{fig:ga}
\end{figure}

In this section, we describe the fetal brain MRI data used to compute the atlas and for the evaluation of automatic segmentations obtained using the atlas.
This is a subset of the dataset described in \Chapref{chap:fetaldataset}.
We will therefore refer to \Chapref{chap:fetaldataset} whenever possible regarding details about the data and processing that were used.

% \subsection{Ethics statement}
% The MRI data were automatically pseudonymized  using the \texttt{GIFT-Cloud} data sharing platform~\cite{doel2017gift} prior to using them for research.
% % 
% \\At University Hospitals Leuven, ethical approval to use the data for research was given by the Ethics Committee University Hospitals Leuven (ethical approval S63598).
% % 
% A retrospective study does not fall under the Belgian law of May 7, 2004 regarding experiments on the human person.
% % 
% However, given the use of potentially identifying MRIs in the study, the requirements set forth in the EU Regulation 2016/679 (General Data Protection Regulation, GDPR) must be met. The sponsor of this study is University Hospitals Leuven, and University Hospitals Leuven maintains "public interest" as the legal basis for data processing. 
% % 
% Article 14 of the GDPR mentions the information obligation of the data controller (= sponsor of the study) to the data subject whose personal data are collected. An information obligation is therefore sufficient according to GDPR, and informed consent is not legally required for the use of the MRIs for illustrative purposes. All snapshots of fetal MRIs used in our figures are based on MRI acquired at Leuven.
% % 
% \\At University College London Hospital (UCLH) the study was approved by the Caldicott guardian at UCLH and patient consent was not required as these images were acquired for clinical purposes and the data used retrospectively.

\paragraph{Spina bifida aperta cohort used to compute the spatio-temporal atlas}\label{material:cohort}

% The fetal brain MRI data used in this work are part of a retrospective study.
% 
A total of $90$ fetal brain MRI examinations from $37$ fetuses were used in this work.
% 
% 
% All the MRI examinations were performed as part of clinical routine following abnormal findings during ultrasound examination.
% % 
% All the fetuses in this cohort were diagnosed with spina bifida aperta at fetal ultrasound examinations.
% % 
% MRI scans were acquired at two surgical centers, University Hospitals Leuven and UCLH
% (see \textit{Underlying data}).
% % 
% For each study, at least three orthogonal T2-weighted HASTE series of the fetal brain were collected on a $1.5$T scanner using an echo time of $133$ms, a repetition time of $1000$ms, with no slice overlap nor gap, pixel size $0.39$mm to $1.48$mm, and slice thickness $2.50$mm to $4.40$mm.
% % 
% A radiologist attended all the acquisitions for quality control.
% % 
The dataset contains longitudinal MRI examinations with up to $5$ examinations per fetus.
In addition, $51$ of the MRI examinations were performed after open fetal surgery performed before $26$ weeks of gestation, to close the spina bifida aperta defect.
The distribution of gestational ages for MRI examinations and whether they were done before or after surgery can be found in Figure~\ref{fig:ga}.

Details about the MRI data acquisition can be found in \Chapref{chap:fetaldataset}.

\paragraph{Fetal brain 3D MRI used for the evaluation of automatic segmentation}\label{material:evaluation}
For the evaluation of automatic fetal brain segmentation we have used the publicly available \href{https://www.synapse.org/#!Synapse:syn23747212/wiki/608434}{FeTA} dataset~\cite{payette2021automatic} (first release).
The FeTA dataset contains $40$ reconstructed 3D MRI, including $15$ MRIs of fetuses with a normal brain and $25$ MRIs of fetuses with spina bifida aperta.
For all the 3D MRI, segmentations are available for seven tissue types:
white matter, ventricular system, cerebellum, extra-axial cerebrospinal fluid, cortical gray matter, deep gray matter, and brainstem.
More details about those data can be found in \Chapref{chap:fetaldataset}.

% The $40$ 3D MRIs and original segmentations (as provided with the FeTA dataset) were inspected by two paediatric radiologists within our institutions, MA and PD, with more than 8 years of experience in segmenting fetal brains.
% % 
% Corrections of the segmentations were performed~\cite{fidon2021partial,fidon2021distributionally} to reduce the variability against the published segmentation guidelines that was released with the FeTA dataset~\cite{payette2021automatic}.
% % 
% Two volumes of spina bifida aperta cases were excluded because the poor quality of the 3D reconstruction (\texttt{sub-feta007} and \texttt{sub-feta009}) did not allow to segment them reliably for the seven tissue types.

\paragraph{Spatio-temporal atlas for the normal developing fetal brain}\label{material:control_atlas}
For comparison to a spatio-temporal atlas of the normal developing fetal brain, we have used the publicly available spatio-temporal \href{http://crl.med.harvard.edu/research/fetal_brain_atlas/}{fetal brain atlas}~\cite{gholipour2017normative}.
This atlas contains $18$ 3D MRI of average normal fetal brain for gestational ages ranging from $21$ weeks to $38$ weeks.

\section{Atlas Computation Method}

\begin{figure}[t!]
    \centering
    \includegraphics[width=\linewidth]{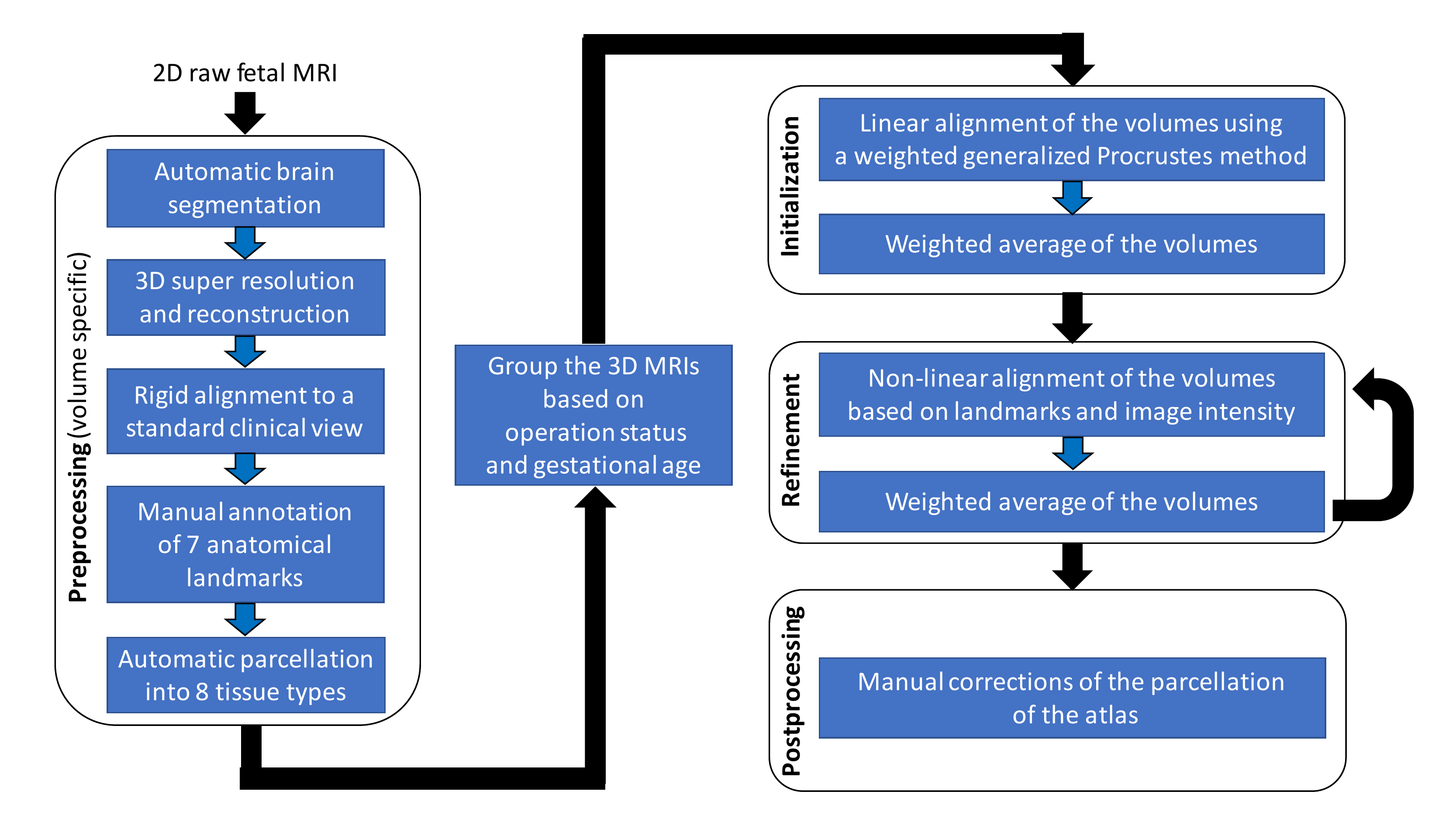}
    \caption{
    \textbf{Overview of the spatio-temporal atlas construction pipeline. MRI: magnetic resonance imaging.}
    }
    \label{fig:overview}
\end{figure}

In this section, we describe our pipeline for computing the spina bifida aperta (SBA) spatio-temporal fetal brain atlas.
An overview of the pipeline can be found in Figure~\ref{fig:overview}.

\subsection{Data Preprocessing}

In this subsection, we give details about the preprocessing steps as can be found in Figure~\ref{fig:overview}.

\subsubsection{Automatic brain segmentation}\label{sec:brain_mask}
One of the main challenges in fetal brain MRI is the motion of the fetus.
To tackle this issue, MRI sequences used for fetal MRI are designed to produce multiple stacks of 2D slices rather than a 3D image.
Original 2D slices typically have lower resolution, suffers from motion between neighboring slices, motion artefact, and suboptimal cross-section~\cite{ebner2020automated}.

Automatic segmentation of the fetal brain in the raw 2D MRI are obtained using a deep learning-based method~\cite{martaisbi2021}.
Those brain masks are an input required by the 3D super resolution and reconstruction algorithm described below.
A public implementation of the deep learning pipeline \texttt{MONAIfbs}~\cite{martaisbi2021}, used in this study to obtain the brain masks, can be found at \href{https://github.com/gift-surg/MONAIfbs}{here} (main git branch, commit $bcab52a$).

\subsubsection{3D super resolution and reconstruction}
% One of the main challenge of fetal brain MRI is the motion of the fetus.
% % 
% To tackle this issue, MR sequences used for fetal MRI are designed to produce multiple stacks of 2D slices rather than a 3D image.
% % 
% Original 2D slices typically have lower resolution, suffers from motion between neighboring slices, motion artefact, and suboptimal cross-section.
 
We use a 3D super resolution and reconstruction algorithm to improve the resolution, and remove motion between neighboring slices and motion artefacts present in the original 2D slices~\cite{ebner2020automated}.
The output of the 3D super resolution and reconstruction algorithm~\cite{ebner2020automated} is a reconstructed 3D MRI of the fetal brain with an isotropic image resolution of $0.8$ mm.
We hypothesize that the reconstructed 3D MRI facilitates the manual delineation and annotation of the fetal brain structures as compared to the original 2D slices.

We used a state-of-the-art 3D super resolution and reconstruction algorithm~\cite{ebner2020automated} publicly available in the \href{https://github.com/gift-surg/NiftyMIC}{\texttt{NiftyMIC}} pipeline version $0.8$ with Python $3.8$.
The original 2D MRI slices were also corrected for bias field in the \href{https://github.com/gift-surg/NiftyMIC}{\texttt{NiftyMIC}} pipeline version $0.8$ using a N4 bias field correction step as implemented in \href{https://simpleitk.org}{SimpleITK} version $1.2.4$.
The 3D super resolution and reconstruction algorithm~\cite{ebner2020automated} also combines the brain masks obtained in section~\secref{sec:brain_mask}. This results in a 3D brain mask for the 3D reconstructed MRI that is computed fully-automatically.

\subsubsection{Rigid alignment to a standard clinical view}
The 3D reconstructed MRI were rigidly aligned to a time-point volume of the control fetal brain 4D atlas~\cite{gholipour2017normative} as implemented in \href{https://github.com/gift-surg/NiftyMIC}{\texttt{NiftyMIC}}~\cite{ebner2020automated} version $0.8$.
All the 3D reconstructed MRIs are therefore aligned to a standard clinical view in which the axes are aligned with the axial, sagittal, and coronal planes of the fetal brain. This facilitates the manual delineation and annotation of the fetal brain structures.
The target time-point in the control 4D atlas is chosen based on the brain volume computed using the automatic 3D brain mask.

\subsubsection{Anatomical 
landmarks}\label{sec:landmarks}

\begin{figure}[t]
    \centering
    % trim: left bottom right top
    \includegraphics[width=\linewidth,trim=1cm 4.5cm 1cm 3.3cm, clip]{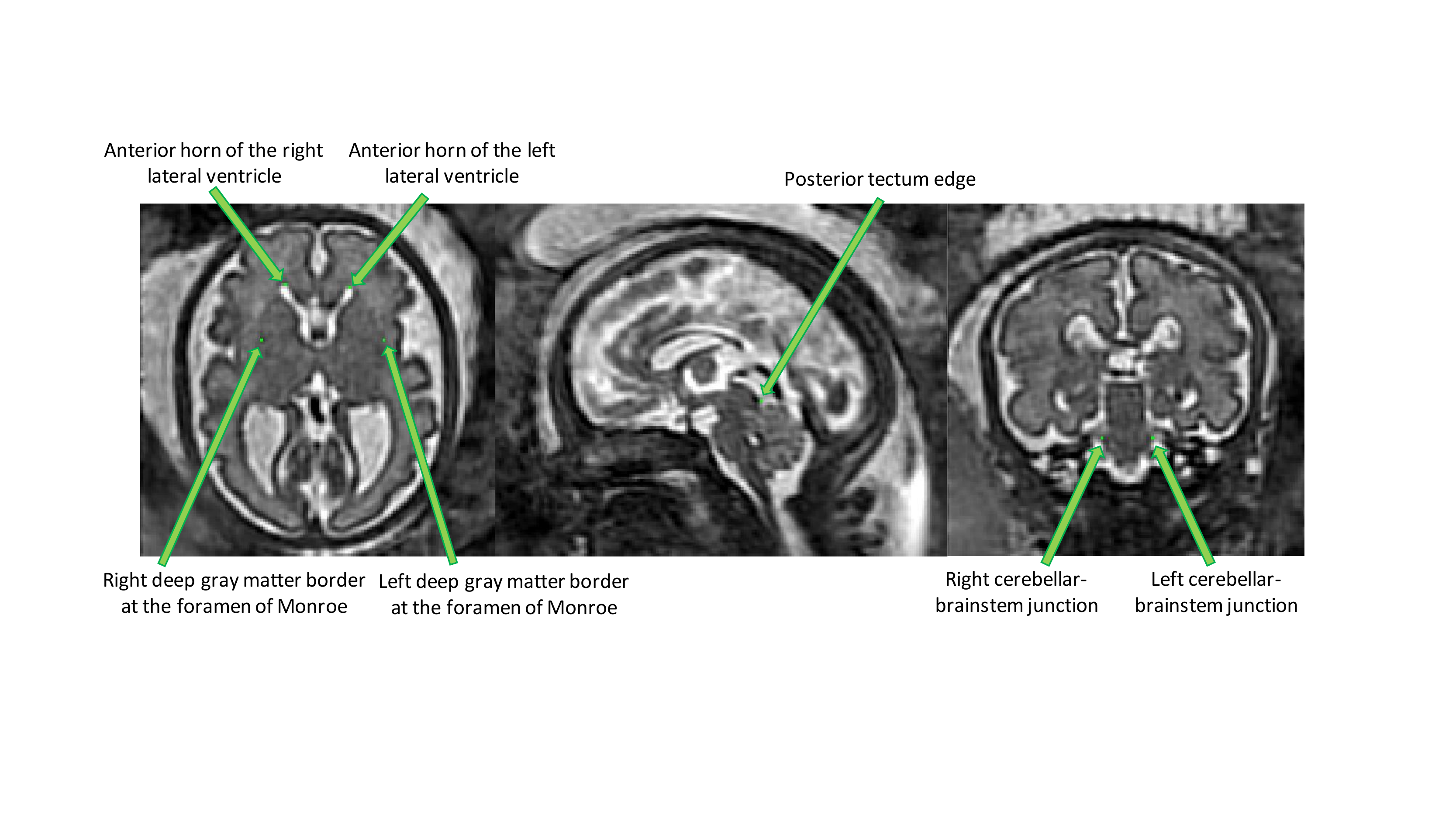}
    \caption{\textbf{Overview of the proposed anatomical landmarks.} 
    Those landmarks were annotated for all the 3D reconstructed magnetic resonance imaging (MRI). 
    They aim at improving the accuracy and the robustness of the image registration steps.}
    \label{fig:lmks_overview}
\end{figure}

Seven anatomical landmarks were manually annotated to regularize and improve the accuracy of the image registration steps used in the computation of the spina bifida 4D atlas.
Details can be found in section~\secref{sec:atlas_method}.

The anatomical landmarks that were selected are:
the right and left anterior horn of the lateral ventricles,
the posterior tectum plate,
the right and left junctions between the cerebellum and the braistem,
and the right and left deep gray matter border at the foramen of Monro.
An illustration of those anatomical landmarks can be found in Figure~\ref{fig:lmks_overview}.

Those landmarks include anatomical structures that have been reported to be reliably identifiable in the fetal MRI clinical research literature~\cite{aertsen2019reliability,garel2001fetal,geerdink2012essential}.
Another selection criteria was to choose landmarks that are spread over the fetal brain anatomy to efficiently support image registration.
Our proposed annotation protocol can be found in \secref{annexe:protocol}.

The manual annotations of the $90$ 3D reconstructed MRIs were performed.
%by author EV.
% 
Manual annotations of landmarks were performed using the software ITK-SNAP~\cite{yushkevich2016itk} version $3.8.0$.
The annotation of one volume took $12$ min on average.
It is worth noting, that landmarks can be missing, especially for fetal MRIs before $26$ weeks of gestation.

The intra-rater reliability for the anatomical landmarks has been evaluated, as described in Section~\secref{sec:irr}. 
The proposed anatomical landmarks protocol also included 
the right and left deep gray matter border at the anterior cavum septi pellucidi line
and the right and left deep gray matter border at the posterior cavum septi pellucidi line.
However, those landmarks were found to be unreliable and often missing due to the high variation in shape of the cavum septi pellucidi.
For this reason, those landmarks were not used for the computation of the atlas but they are present in the annotation protocol.
Details can be found in Section~\secref{sec:irr}.

\subsubsection{Age and operation status specific groups of 3D reconstructed MRIs}\label{sec:group}
The 3D reconstructed MRIs were grouped with respect to their operation status and their gestational age.
Each group of 3D reconstructed MRIs went through the atlas construction pipeline described in section~\secref{sec:atlas_method} and lead to the computation of a unique volume of our spatio-temporal atlas.

SBA surgery affects the evolution of the fetal brain anatomy~\cite{aertsen2019reliability,danzer2020fetal,mufti2021cortical}.
Therefore, we have chosen to separate the 3D reconstructed MRIs of operated and non-operated fetuses.
A group either contains only 3D reconstructed MRIs of fetuses that have been operated for SBA in-utero, or contains only 3D reconstructed MRIs of fetuses that have not been operated.

Each group is assigned with a gestational age ranging from $21$ weeks to $34$ weeks.
Volumes are included in a group only if the gestational age at the time of the acquisition is within $9$ days of the gestational age of the group.
The description of the cohort used can be found in section~\secref{material:cohort} and the distribution of gestation ages can be found in Figure~\ref{fig:ga}.
As can be seen in Figure~\ref{fig:sb_atlas_not_operated} and Figure~\ref{fig:sb_atlas_operated}, groups for non-operated fetuses covers the gestational ages from $21$ weeks to $25$ weeks and groups for operated fetuses covers gestational ages from $25$ weeks to $34$ weeks.

% \paragraph{Exclusion criteria:}
A group is excluded if it contains less than three 3D reconstructed MRIs.
In addition, we excluded a group if it did not include both 3D reconstructed MRIs with gestational ages higher and lower than the gestational age of the group.
This avoids, for example, to have a group for non-operated fetuses at $26$ weeks of gestation that would contain only MRIs at gestational ages $25$ weeks or less.
\\
\textbf{Data augmentation:}
We used right-left flipping as a data augmentation to synthetically increase the amount of volumes in each group.
This encourages the atlas to be symmetrical with respect to the central sagittal plane.
Right-left flipping has been used in several previous studies on brain MRI atlases~\cite{fonov2011unbiased,grabner2006symmetric}.
Imposing symmetry between right and left hemispheres of the atlas volumes aims at reducing potential biases in the cohort used to compute the atlas.
In addition, it allows to use the atlas for the study of asymmetry between right and left hemispheres~\cite{grabner2006symmetric}.
Asymmetry between brain hemispheres for normal fetuses has been described as well as the role of hemispheric asymmetry in isolated corpus callosum agenesis.~\cite{glatter2020improved,kasprian2011prenatal}.
To the best of our knowledge, hemispheric asymmetry has not been studied yet in SBA.

\subsection{Atlas Construction}\label{sec:atlas_method}
In this section we describe the different steps for the computation of the spina bifida atlas as can be seen in the Initialization and Refinement boxes of the pipeline overview in Figure~\ref{fig:overview}.

\subsubsection{Weighted average of the volumes}\label{sec:weighted_average}
In this section, we describe the method to average the intensity of 3D reconstructed MRIs after spatial alignment.
As described in section~\secref{sec:group}, data are grouped with respect to their operation status and gestational age.
After aligning spatially all the 3D reconstructed MRIs of a group, we average their image intensity to obtain an average fetal brain MRI for the group.

\textbf{Weighted average:}
To reflect the gestational age associated with each group, we used a time-weighted average.
The weight for the volume $i$ is defined using a Gaussian kernel as follow
\begin{equation}
    w_{i} = \frac{1}{\sqrt{2 \pi} \sigma} 
                \exp{\left(-\frac{1}{2} \left(\frac{GA_i - GA_{target}}{\sigma}\right)^2\right)}
    % \left\{
    %     \begin{tabular}{c c}
    %             0 & if $|GA_i - GA_{target}| > 3 \sigma$ \\
    %             $
    %             \frac{1}{\sqrt{2 \pi} \sigma} 
    %             \exp{\left(-\frac{1}{2} \left(\frac{GA_i - GA_{target}}{\sigma}\right)^2\right)}
    %             $
    %             & otherwise
    %     \end{tabular}
    % \right.
\end{equation}
where $GA_{target}$ is the gestational age of the group and $GA_{i}$ is the gestational age of volume $i$.
The standard deviation value is set to $\sigma=3$ days. 
% This value was chosen so that $2 \sigma$ approximately covers the duration of a week.

In addition, we average each image and its symmetric by right-left flipping to impose to the average volume to be exactly symmetric with respect to the central sagittal plane.
This is performed in addition to the data augmentation described in section~\secref{sec:group}.

Formally, let $\{I_i\}_{i=1}^N$ be a set of $N$ co-registered 3D reconstructed MRIs to average. The weighted average is computed as
\begin{equation}
    I_{average} = \frac{1}{2N} \sum_{i=1}^N w_i \left(I_i + S(I_i)\right)
\end{equation}
where $S$ is the operator that computes the symmetric of a volume with respect to the central sagittal plane.

\textbf{Preprocessing:}
Before averaging, we transform the intensity of each volume linearly to set the mean (resp. the standard deviation) of the image intensity inside the brain mask to $2000$ (resp. $500$).
Those values were set to approximate the intensity profile of a spatio-temporal fetal brain atlas of normal fetuses~\cite{gholipour2017normative}.

\subsubsection{Weighted generalized Procrustes}\label{sec:procrustes}
In this section, we describe the optimization method that we used for the joint initial linear alignment of the volumes in a group of 3D reconstructed MRIs. 
This method is based on a weighted generalized Procrustes method and uses only the anatomical landmarks.
Especially, note that the image intensity is not used.

Generalized Procrustes methods~\cite{gower1975generalized} aims at matching simultaneously $n$ configurations of landmarks using linear spatial transformations.
% 
% The method is named after Procrustes, a thief and innkeeper in the Greek mythology who made his victims fit in his bed by stretching or cutting off their body.
% % 
% For an introduction to generalized Procrustes methods we refer the reader to~\cite{gower2010procrustes}.
% 
Generalized Procrustes methods (without constraints) can be defined as optimization problems of the form~\cite{gower1975generalized,gower2010procrustes}
\begin{equation}
    \label{eq:generalized_procrustes}
    \min_{\{M_i, t_i\}} \frac{1}{2} \sum_{i=1}^n \sum_{k=1}^K \norm{
        M_i x_{i,k} + t_i - \frac{1}{n} \sum_{j=1}^n \left(M_j x_{j,k} + t_j\right)
    }^2
\end{equation}
where $n$ is the number of samples, $K$ is the number of landmarks, $x_{i,k}$ is the vector of coordinates for the landmark $k$ of sample $i$, $t_i$ is the translation for the sample $i$, and $M_i$ is the linear transformation for the sample $i$.
In this work we restrict the linear transformations $M_i$ to be anisotropic scaling transformations.

However, for the computation of the spina bifida atlas we have to take into account that landmarks can be missing for some samples.
We also would like to weight differently the samples based on their gestational age alike what is done for the weighted average of the 3D reconstructed MRIs in section~\secref{sec:weighted_average}.

In this work, we introduce weights in the generalized Procrustes methods. A weight of zeros represents a missing landmark for a sample.
The proposed weighted generalized Procrustes method corresponds to the optimization problem
\begin{equation}
    \label{eq:weighted_generalized_procrustes}
    \min_{\{M_i, t_i\}} \frac{1}{2} \sum_{i=1}^n \sum_{k=1}^K w_{i,k} \norm{
        M_i x_{i,k} + t_i - \frac{\sum_{j=1}^n w_{j,k} \left(M_j x_{j,k} + t_j\right)}{\sum_{j=1}^n w_{j,k}}
    }^2
\end{equation}
where $w_{i,k} \geq 0$ is the weight for the landmark $k$ of sample $i$.
For landmark $k$, sample $i$ of gestational age $GA_{i}$, and the target gestational age $GA_{target}$, we propose to define the weight $w_{i,k}$ as
\begin{equation}
    w_{i,k} = \left\{
        \begin{tabular}{l l}%\rowcolors{2}{white}{white}
                0 & if landmark k is missing for sample i \\
                0 & if $|GA_i - GA_{target}| > 3 \sigma$ \\
                $
                \frac{1}{\sqrt{2 \pi} \sigma} 
                \exp{\left(-\frac{1}{2} \left(\frac{GA_i - GA_{target}}{\sigma}\right)^2\right)}
                $
                & otherwise
        \end{tabular}
    \right.
\end{equation}
The standard deviation value is $\sigma=3$ days.

We assume that every landmark was annotated at least once in each group.
As a result, $\forall k,\, \sum_{j=1}^n w_{j,k} > 0$ and the fractions used in \eqref{eq:weighted_generalized_procrustes} are well defined.

In general, the optimization problem \eqref{eq:generalized_procrustes} admits an infinity of solutions, including the trivial solution that send all the landmarks to the origin.
To tackle this issue, constraints on the size of the system are added~\cite{gower1975generalized,gower2010procrustes}.
The optimization problem \eqref{eq:weighted_generalized_procrustes} suffers from the same under-specification problem.
We therefore choose to constrain the center of mass of the barycenter of the system and the size of the system because it is the most intuitive approach.
This leads to the optimization problem
\begin{equation}
\label{eq:weighted_generalized_procrustes_with_constraints}
\begin{split}
    \min_{\{M_i, t_i\}, \{g_k\}}\,\, & 
    \frac{1}{2} \sum_{i=1}^n \sum_{k=1}^K w_{i,k} \norm{M_i x_{i,k} + t_i - g_k}^2\\
    \textup{s.t.}\quad & 
    \frac{1}{K}\sum_{k=1}^K g_k = 
    \frac{1}{K}\sum_{k=1}\frac{\sum_{j=1}^n w_{j,k} x_{j,k}}{\sum_{j=1}^n w_{j,k}}\\
    \textup{and}\quad
    & \frac{1}{K}\sum_{k=1}^K \norm{g_k - \frac{1}{K}\sum_{l=1}^K g_l}^2 = 
    \frac{1}{K}\sum_{k=1}\norm{
        \frac{\sum_{j=1}^n w_{j,k} x_{j,k}}{\sum_{j=1}^n w_{j,k}} - 
        \frac{1}{K}\sum_{l=1}\frac{\sum_{j=1}^n w_{j,l} x_{j,l}}{\sum_{j=1}^n w_{j,l}}
    }^2
\end{split}
\end{equation}
This optimization problem can be solved efficiently using an alternating least squares approach~\cite{gower2010procrustes}.

\subsubsection{Non-linear image registration}\label{sec:non_linear_reg}
In this section, we describe the non-linear image registration method that we used for the refinement step of the 4D atlas as can be seen in Figure~\ref{fig:overview}.
In the refinement step, intermediate atlas MRI volumes have already been computed for all time points.
The goal of this step is to improve the image sharpness of the intermediate atlas MRI volumes by registering all the 3D reconstructed MRIs to the intermediate MRI volumes and computing new weighted average volumes using the method described in section~\secref{sec:weighted_average}.

We used \href{https://github.com/KCL-BMEIS/niftyreg}{\texttt{NiftyReg}}~\cite{modat2010fast} to perform non-linear image registration using image intensity and the anatomical landmarks.
The non-linear image registration optimization problem is the following

\begin{equation}
\label{eq:reg}
    \min_{\Theta} \,\, \mathcal{L}(I_{subject},\, I_{atlas},\, \phi(\Theta)) + R(\Theta)
\end{equation}
where $I_{subject}$ is the 3D reconstructed MRI to be aligned to the 3D atlas time point $I_{atlas}$ and $\phi(\Theta)$ is a spatial transformation parameterized by cubic B-splines of parameters $\Theta$.

The regularization term $R$ is a linear combination of the bending energy~\cite{ashburner2007non} (BE) and the linear energy~\cite{ashburner2007non} (LE) regularization functions applied to $\phi(\Theta)$
\begin{equation}
\label{eq:regularization}
\begin{split}
    R(\Theta) = &
    \alpha_{BE} BE(\phi(\Theta)) 
    + \alpha_{LE} LE(\phi(\Theta))
\end{split}
\end{equation}
with $\alpha_{BE}=0.1$ and $\alpha_{LE}=0.3$.
More details about the methodology used to tune image registration parameters can be found below.

The data term $\mathcal{L}$ is a linear combination of the local normalized cross correlation (LNCC)~\cite{cachier2003iconic} and the squared euclidean distances between the landmarks positions
\begin{equation}
\label{eq:data_term}
    \begin{aligned}
        \mathcal{L}(I_{subject},\, I_{atlas},\, \phi(\Theta)) =&
        \alpha_{LNCC} LNCC(I_{subject},\, I_{atlas} \circ \phi(\Theta))\\
        & + \alpha_{LMKS} \sum_{k \in \Omega_{LMKS}} \norm{\phi(\Theta)(x^{subject}_k) - x^{atlas}_k}^2
    \end{aligned}
\end{equation}
where $\Omega_{LMKS}$ is the set of landmarks that are present for both $I_{subject}$ and $I_{atlas}$, $\alpha_{LMKS}=0.001$ and $\alpha_{LNCC}=(1 - \alpha_{LMKS})(1 - \alpha_{BE} - \alpha_{LE})$ as implemented in \texttt{NiftyReg}~\cite{modat2010fast}.
The standard deviation of the Gaussian kernel of the LNCC was set to $6$ mm.
More details about the methodology used to tune image registration parameters can be found below.

\textbf{Implementation details:}
Registrations that solve the optimization problem \eqref{eq:reg} were computed using the publicly available code for \href{https://github.com/KCL-BMEIS/niftyreg}{\texttt{NiftyReg}} ~\cite{modat2010fast}.
We used the latest version of the code on the master branch (git commit \textit{99d584e}).

The transformation $\phi$ in \eqref{eq:reg} is parameterized by cubic B-Splines of order $3$ with a grid spacing equal to $3$ mm.
\texttt{NiftyReg}~\cite{modat2010fast} uses a pyramidal approach to solve \eqref{eq:reg}.
We used $3$ levels of pyramid which is the default value in \texttt{NiftyReg}.
The brain mask were used to mask the voxels outside the brain.

The transformation $\phi$ in \eqref{eq:reg} was initialized with an affine transformation.
The affine transformation was computed using a symmetric block-matching approach~\cite{modat2014global} based on image intensities and the brain masks.
The implementation of the affine image registration method is included in \texttt{NiftyReg}.

\textbf{Parameters tuning:}
The parameters $\alpha_{BE}$, $\alpha_{LE}$, $\alpha_{LMKS}$, and the standard deviation of the Gaussian kernel of the LNCC of equations \eqref{eq:regularization} and \eqref{eq:data_term} were tuned using a grid search.
The other parameters of the image registration were not tuned.
The values of $\alpha_{BE}$ were $\{0.001, 0.01, 0.03, 0.1, 0.3\}$,
the values of $\alpha_{LE}$ were $\{0.01, 0.03, 0.1\}$,
the values of $\alpha_{LMKS}$ were $\{0.0003, 0.001, 0.003\}$,
and the values for the standard deviation of the LNCC were $\{1, 2, 4, 6, 8\}$.
We also tried to use the normalized mutual information (NMI) in place of the LNCC. There are no additional hyper-parameters related to NMI.

We selected the best set of parameter values using a subset of $22$ pairs of 3D reconstructed MRIs covering the range of gestational ages available.
The selection criteria was the mean of the Dice scores for the white matter, the ventricular system, and the cerebellum between volumes after non-linear registration.
Details about the segmentation protocol can be found in section~\secref{sec:segmentation}.

It is worth noting that the gradients of the different terms of the objective function in \eqref{eq:reg} have different scales.
Therefore, comparing the contribution of the different terms based on their weights is misleading.
Our parameter tuning protocol suggests that all the terms of the objective function are important to obtain optimal image registration results.

\subsection{Semi-automatic Segmentation of the Atlas}\label{sec:segmentation}
In this section, we describe the semi-automatic method that was used to obtain the segmentation for the proposed spatio-temporal atlas for SBA.

The fetal brains were divided into a total of eight tissue types:
white matter (excluding the corpus callosum), ventricular system with the cavum septi pellucidi and cavum vergae, cerebellum, extra-axial cerebrospinal fluid, cortical gray matter, deep gray matter, brainstem, and corpus callosum.
A visualization of the segmentations of those tissue types can be found in Figure~\ref{fig:sb_atlas_not_operated} and Figure~\ref{fig:sb_atlas_operated}.
The annotation protocol follows the annotation guidelines of the FeTA dataset~\cite{payette2021automatic}.
In addition, the corpus callosum was also delineated.

Automatic 3D tissue types probability maps were obtained using a deep learning pipeline trained using partially supervised learning~\cite{fidon2021partial}.
An ensemble of ten deep neural networks trained using the Leaf-Dice loss~\cite{fidon2021partial} has been used.
The code and the pre-trained networks used for the automatic segmentation are available \href{https://github.com/LucasFidon/fetal-brain-segmentation-partial-supervision-miccai21}{here}.
An average 3D tissue types probability maps for the atlas was obtained using a weighted average method analogous to the one described in section~\secref{sec:weighted_average} for the 3D reconstructed MRIs.
Formally, let $\{P_i\}_{i=1}^N$ be a set of $N$ co-registered 3D tissue types probability maps to average. The weighted average is computed as
\begin{equation}
    P_{average} = \frac{1}{2N} \sum_{i=1}^N w_i \left(P_i + S(P_i)\right)
\end{equation}
where $S$ is the operator that computes the symmetric of a volume with respect to the central sagittal and the weights $w_i$ are defined as in section~\secref{sec:weighted_average}.
An initial segmentation of the atlas was obtained using the tissue types of maximum probability for each voxel.

The initial segmentations of the spatio-temporal atlas were quality controlled and corrected when necessary by authors LF and MA, a paediatric radiologist specialized in fetal brain anatomy with eight years of experience in segmenting fetal brain MRIs.
Manual segmentations were performed using the software ITK-SNAP~\cite{yushkevich2016itk} version $3.8.0$.

\section{Annotation Protocol of Anatomical Landmarks for Fetuses with Spina Bifida Aperta}\label{annexe:protocol}

In this section, protocols designed for the selection of imaging landmarks in MRI images of fetal brains with spina bifida aperta (SBA) are outlined. This is aimed to improve the accuracy of image registration. A total of 11 anatomical landmarks per study have been selected for initial assessment. Four in each cerebral hemisphere and three in the posterior fossa.

The first seven landmarks described below were found to be sufficiently reliable.
The last four landmarks involving the cavum septi pellucidi were found to be insufficiently reliable.

\begin{figure}
    \centering
    \begin{subfigure}[t]{0.4\textwidth}
        \centering
        \includegraphics[width=\linewidth]{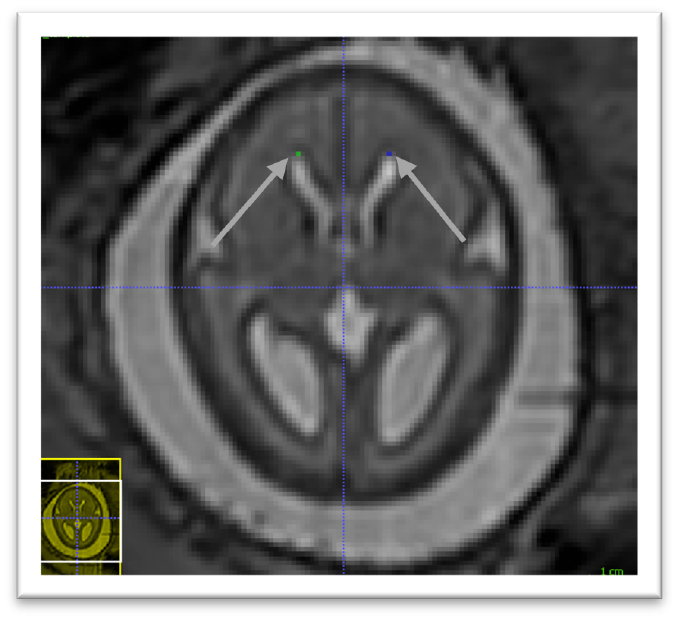}
        % \caption{Axial view.} \label{fig:LV}
    \end{subfigure}
    \hspace{1em}
    \begin{subfigure}[t]{0.4\textwidth}
        \centering
        \includegraphics[width=\linewidth]{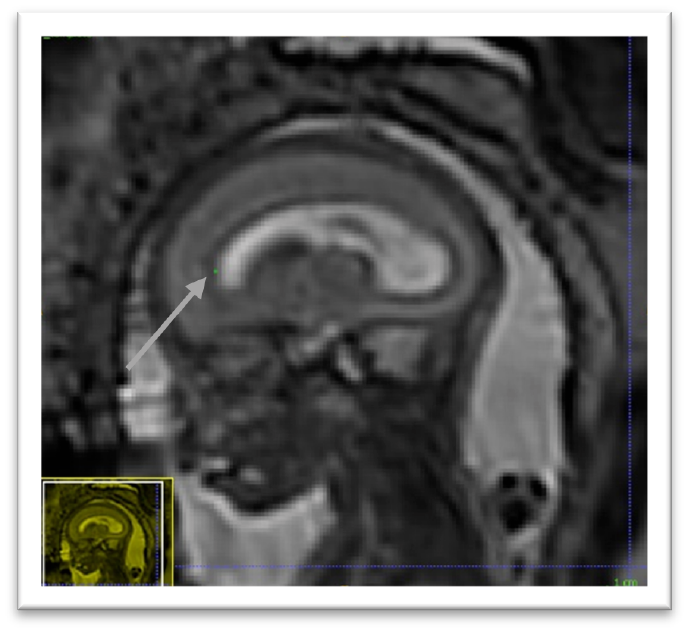}
        % \caption{Sagittal view.} \label{fig:LV2}
    \end{subfigure}
    \caption{Anterior horn of the right lateral ventricle (green) and anterior horn of the left lateral ventricle (blue).}
\end{figure}

\subsection{Anterior horn of the right lateral ventricle}

In the axial plane identify the right lateral ventricle. 
Use the view in the sagittal plane to select the most anterior slice reached by the ventricle. 
When this slice is not unique, which occurs when the anterior border of the ventricle is flattened, select the slice at the centre.
With the smallest possible marker the most anterior border. The border is considered as the brighter intensity value of the two lines of intensity values showing the greatest difference.

\subsection{Anterior horn of the left lateral ventricle}
In the axial plane identify the Left Lateral Ventricle. 
Use the view in the sagittal plane to select the most anterior slice reached by the ventricle. 
When this slice is not unique, which occurs when the anterior border of the ventricle is flattened, select the slice at the centre.  
With the smallest possible marker the most anterior border. The border is considered as the brighter intensity value of the two lines of intensity values showing the greatest difference.

\begin{figure}
    \centering
    \begin{subfigure}{0.36\textwidth}
        \centering
        \includegraphics[width=\linewidth]{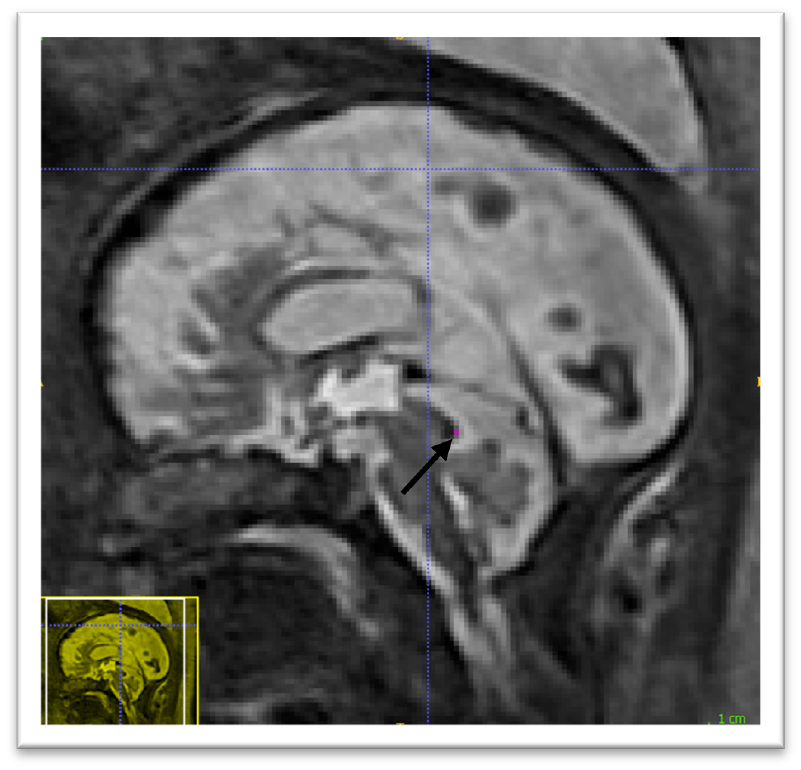}
        % \caption{Sagittal view.} \label{fig:PT}
    \end{subfigure}
    \hspace{1em}
    \begin{subfigure}{0.4\textwidth}
        \centering
        \includegraphics[width=\linewidth]{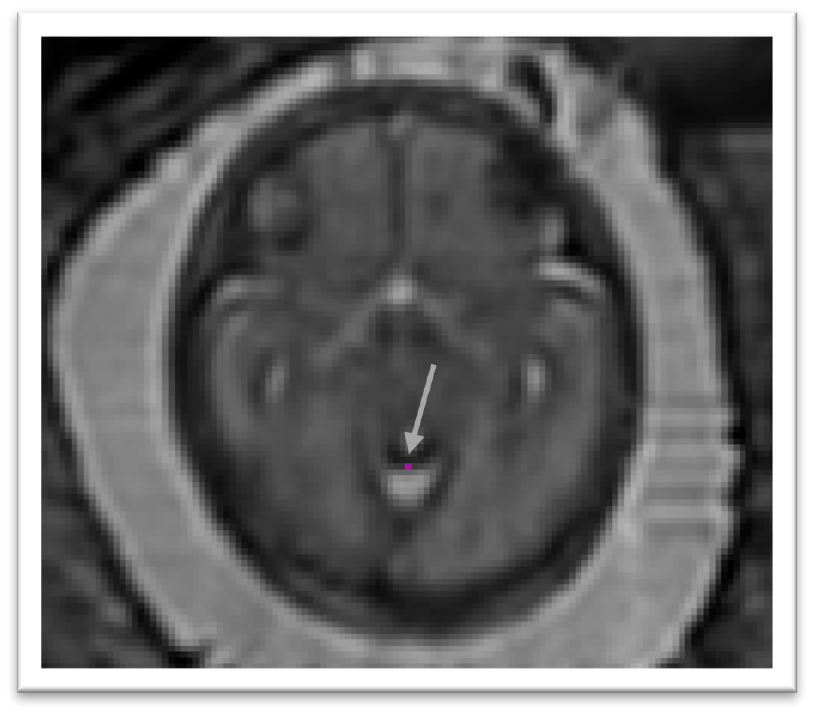}
        % \caption{Axial view.} \label{fig:PT2}
    \end{subfigure}
    \caption{Posterior tectum plate (pink).}
\end{figure}
\subsection{Posterior tectum plate}
Using the sagittal and axial planes locate the tectum. In the axial plane select the midline sagittal slice. Confirm using the sagittal plane that the axial slice is viewing the most prominent part of the tectum. Using the smallest marker select the most posterior point of the tectum tissue. This considered to be the lower intensity value of the two intensity values at the posterior peak showing the greatest difference.

\begin{figure}
    \centering
    \includegraphics[width=0.35\linewidth]{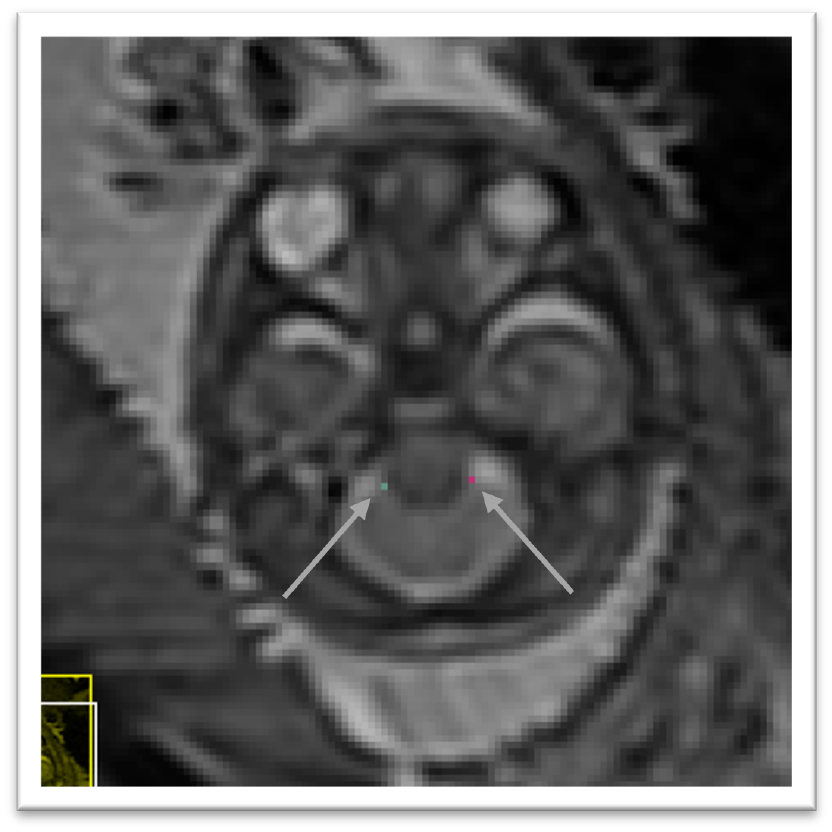}
    \caption{Right cerebellar-brainstem junction (turquoise) and left cerebellar-brainstem junction (pink).}
    \label{fig:my_label}
\end{figure}
\subsection{Left cerebellar-brainstem junction}
In the axial view we locate the cerebellum and select the slice with the greatest cerebellar width, preferably where the posterior fossa also is seen at its greatest width. The brainstem is found just anterior to the cerebellum and directly meets with the cerebellum along its posterior borders. In this area, we select with the smallest possible marker the most anterior point where the cerebellum and brainstem meet on the left side. The marker should be within cerebellar tissue as oppose to the tissue of the brainstem.

\subsection{Right cerebellar-brainstem junction}
In the axial view we locate the cerebellum and select the slice with the greatest cerebellar width, preferably where the posterior fossa also is seen at its greatest width. The brainstem is found just anterior to the cerebellum and directly meets with the cerebellum along its posterior borders. In this area, we select with the smallest possible marker the most anterior point where the cerebellum and brainstem meet on the right side. The marker should be within cerebellar tissue as oppose to the tissue of the brainstem.

\begin{figure}
    \centering
    \begin{subfigure}[t]{0.3\textwidth}
        \centering
        \includegraphics[width=\linewidth]{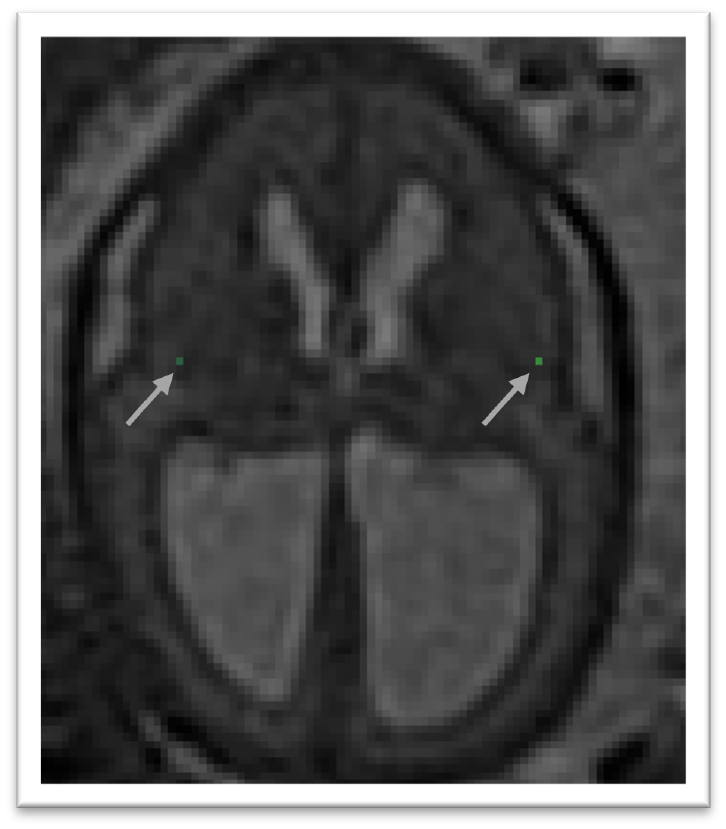}
        % \caption{Axial view.} \label{fig:FOM}
    \end{subfigure}
    \hspace{1em}
    \begin{subfigure}[t]{0.3\textwidth}
        \centering
        \includegraphics[width=\linewidth]{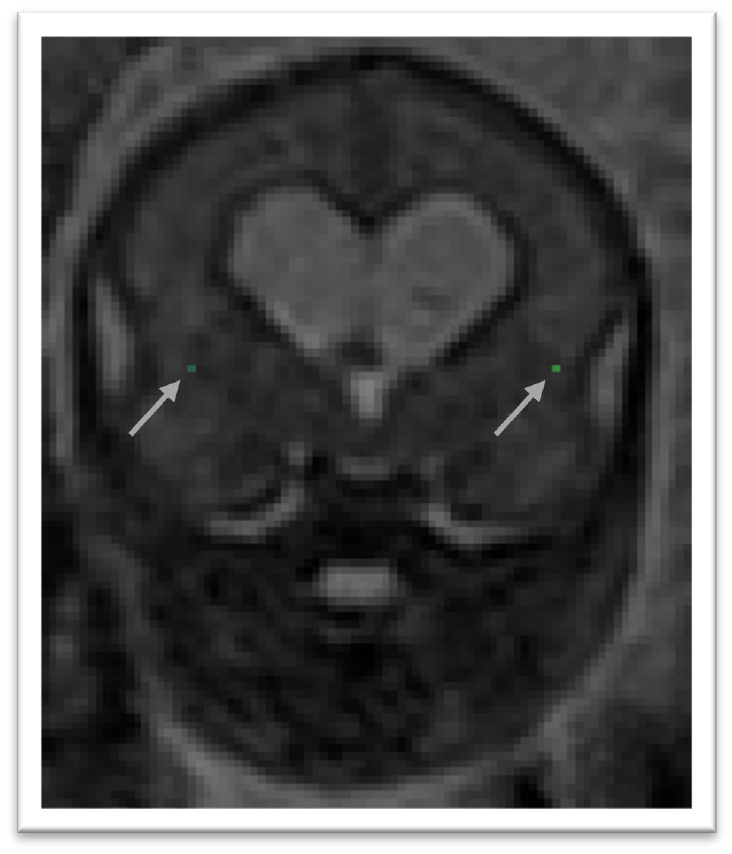}
        % \caption{Coronal view.} \label{fig:FOM2}
    \end{subfigure}
    \caption{Left deep grey matter border at foramen of Monro (dark olive) and right deep grey matter border at foramen of Monro (lime green).}
\end{figure}
\subsection{Left deep grey border at foramen of Monro}
In the axial view locate the foramen of Monro or the interventricular foramen. The paired foramina connect the lateral ventricles to the third ventricle. The point where the foramina lead into the third ventricle, a horseshoe or trough shaped border is formed anteriorly. If not visible in this way, it can also be observed in the coronal view connecting the anterior horns of the lateral ventricle to the third ventricle.  Select the mid-sagittal slice and trace a horizontal line left across from this border. The correct position of the line is considered as the row of brighter intensity value of the two rows of intensity values showing the greatest contrast. The edge of the deep grey matter on the left side which should be visible forming a darker grey arch from the left anterior horn to the left posterior horn of the lateral ventricles. Using the smallest possible marker, mark the edge of the deep grey matter where it intersects with the line.

\subsection{Right deep grey border at foramen of Monro}
In the axial view locate the foramen of Monro or the interventricular foramen. The paired foramina connect the lateral ventricles to the third ventricle. The point where the foramina lead into the third ventricle, a horseshoe or trough shaped border is formed anteriorly. If not visible in this way, it can also be observed in the coronal view connecting the anterior horns of the lateral ventricle to the third ventricle.  Select the mid-sagittal slice and trace a horizontal line right across from this border. The correct position of the line is considered as the row of brighter intensity value of the two rows of intensity values showing the greatest contrast. The edge of the deep grey matter on the right side which should be visible forming a darker grey arch from the right anterior horn to the right posterior horn of the lateral ventricles. Using the smallest possible marker, mark the edge of the deep grey matter where it intersects with the line.

\begin{figure}
    \centering
    \begin{subfigure}[t]{0.32\textwidth}
        \centering
        \includegraphics[width=\linewidth]{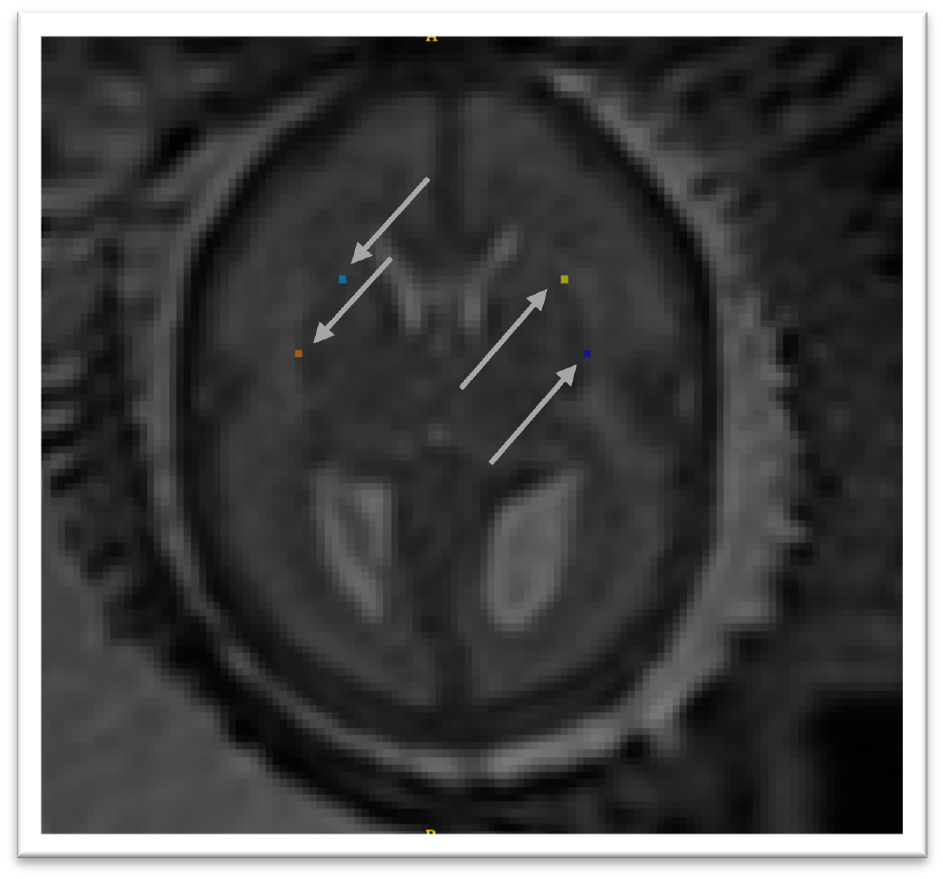}
        \caption{Left deep grey border at anterior CSP line (yellow), right deep grey border at the anterior CSP line (light blue), left deep grey border at posterior CSP line (dark blue), right deep grey border at the posterior CSP line (orange).} \label{fig:CSP}
    \end{subfigure}
    \hfill
    \begin{subfigure}[t]{0.32\textwidth}
        \centering
        \includegraphics[width=\linewidth]{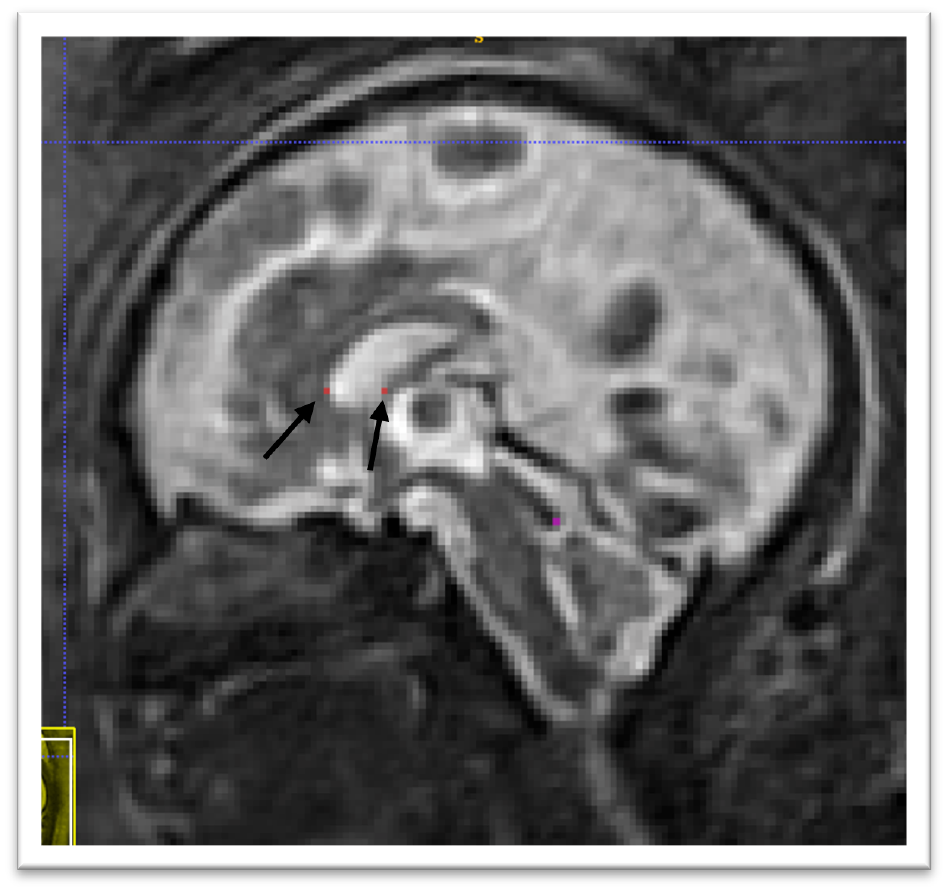}
        \caption{Sagittal view of the position of the horizontal lines used to guide the marking of the deep grey borders at CSP (red).} \label{fig:CSP2}
    \end{subfigure}
    \hfill
    \begin{subfigure}[t]{0.27\textwidth}
        \centering
        \includegraphics[width=\linewidth]{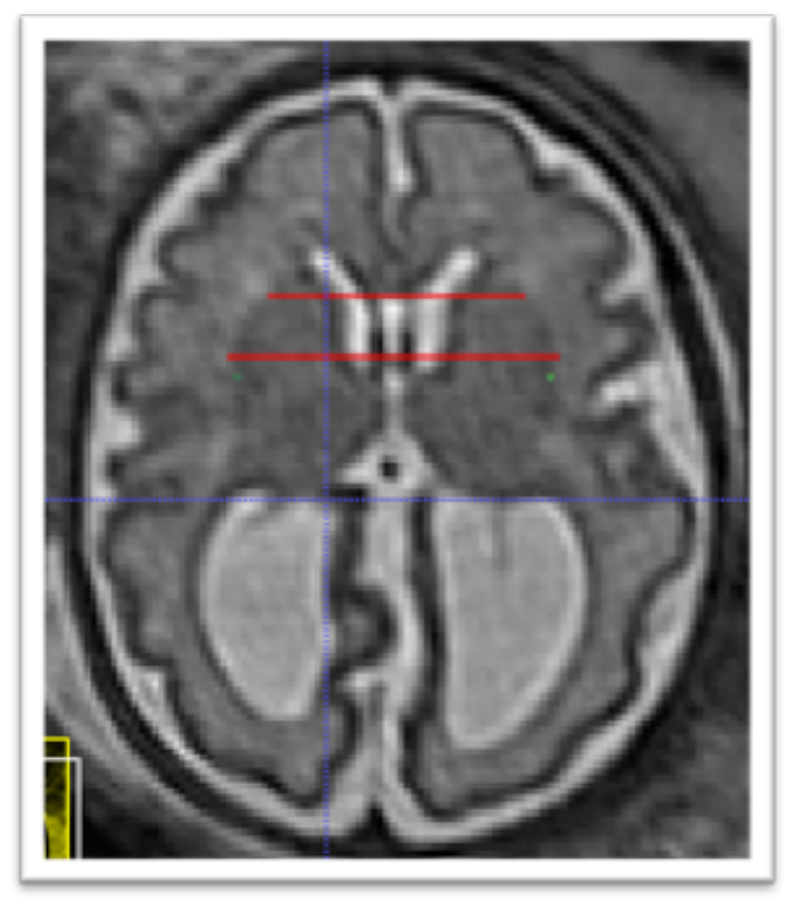}
        \caption{Axial view of the position of the horizontal lines used to guide the marking of the deep grey borders at CSP (red).} \label{fig:CSP3}
    \end{subfigure}
    \caption{Deep grey matter border with respect to the cavum septi pellucidi (CSP).}
\end{figure}
\subsection{Left deep grey border at anterior cavum septi pellucidi line}
In the axial view locate the cavum septi pellucidi (CSP), a cavity in the fetal brain, the leaflets of the septum pellucidum are located between the anterior horns of the lateral ventricles. Select the slice in which the anterior wall of the cavity is found most anteriorly. If there is significant abnormality in this structure it may be helpful to use the sagittal plane to assist in defining this area. Trace a horizontal line left across from the anterior wall of the cavum septi pellucidi. The correct position of the line is considered as the row of brighter intensity value of the two rows of intensity values showing the greatest contrast. The edge of the deep grey matter on the left side forms a darker arch from the left anterior horn to the left posterior horn of the lateral ventricles. Using the smallest possible marker mark the edge of the deep grey matter where it intersects with that line.

\subsection{Right deep grey border at the anterior cavum septi pellucidi line}
In the axial view locate the cavum septi pellucidi (CSP), a cavity in the fetal brain, the leaflets of the septum pellucidum are located between the anterior horns of the lateral ventricles. Select the slice in which the anterior wall of the cavity is found most anteriorly. If there is significant abnormality in this structure it may be helpful to use the sagittal plane to assist in defining this area. Trace a horizontal line right across from the anterior wall of the cavum septi pellucidi. The correct position of the line is considered as the row of brighter intensity value of the two rows of intensity values showing the greatest contrast. The edge of the deep grey matter on the right side forms a darker arch from the right anterior horn to the right posterior horn of the lateral ventricles. Using the smallest possible marker mark the edge of the deep grey matter where it intersects with that line.

\subsection{Left deep grey border at posterior cavum septi pellucidi line}
In the axial view locate the cavum septi pellucidi, a cavity in the fetal brain, the leaflets of the septum pellucidum are located between the anterior horns of the lateral ventricles. Select the slice in which the anterior wall of the cavity is found most anteriorly. If there is significant abnormality in this structure it may be helpful to use the sagittal plane to assist in defining this area. At this level trace a horizontal line left across from the posterior wall of the cavum septi pellucidi. The correct position of the line is considered as the row of brighter intensity value of the two rows of intensity values showing the greatest contrast. The edge of the deep grey matter on the left side forms a darker arch from the left anterior horn to the left posterior horn of the lateral ventricles. Using the smallest possible marker mark the edge of the deep grey matter where it intersects with that line.

\subsection{Right deep grey border at the posterior cavum septi pellucidi line}
In the axial view locate the cavum septi pellucidi, a cavity in the fetal brain, the leaflets of the septum pellucidum are located between the anterior horns of the lateral ventricles. Select the slice in which the anterior wall of the cavity is found most anteriorly. In this slice trace a horizontal line right across from the posterior wall of the cavum septi pellucidi. The correct position of the line is considered as the row of brighter intensity value of the two rows of intensity values showing the greatest contrast. The edge of the deep grey matter on the right side forms a darker arch from the right anterior horn to the right posterior horn of the lateral ventricles. Using the smallest possible marker mark the edge of the deep grey matter where it intersects with that line.

\begin{figure}[t]
    \centering
    \includegraphics[width=\linewidth]{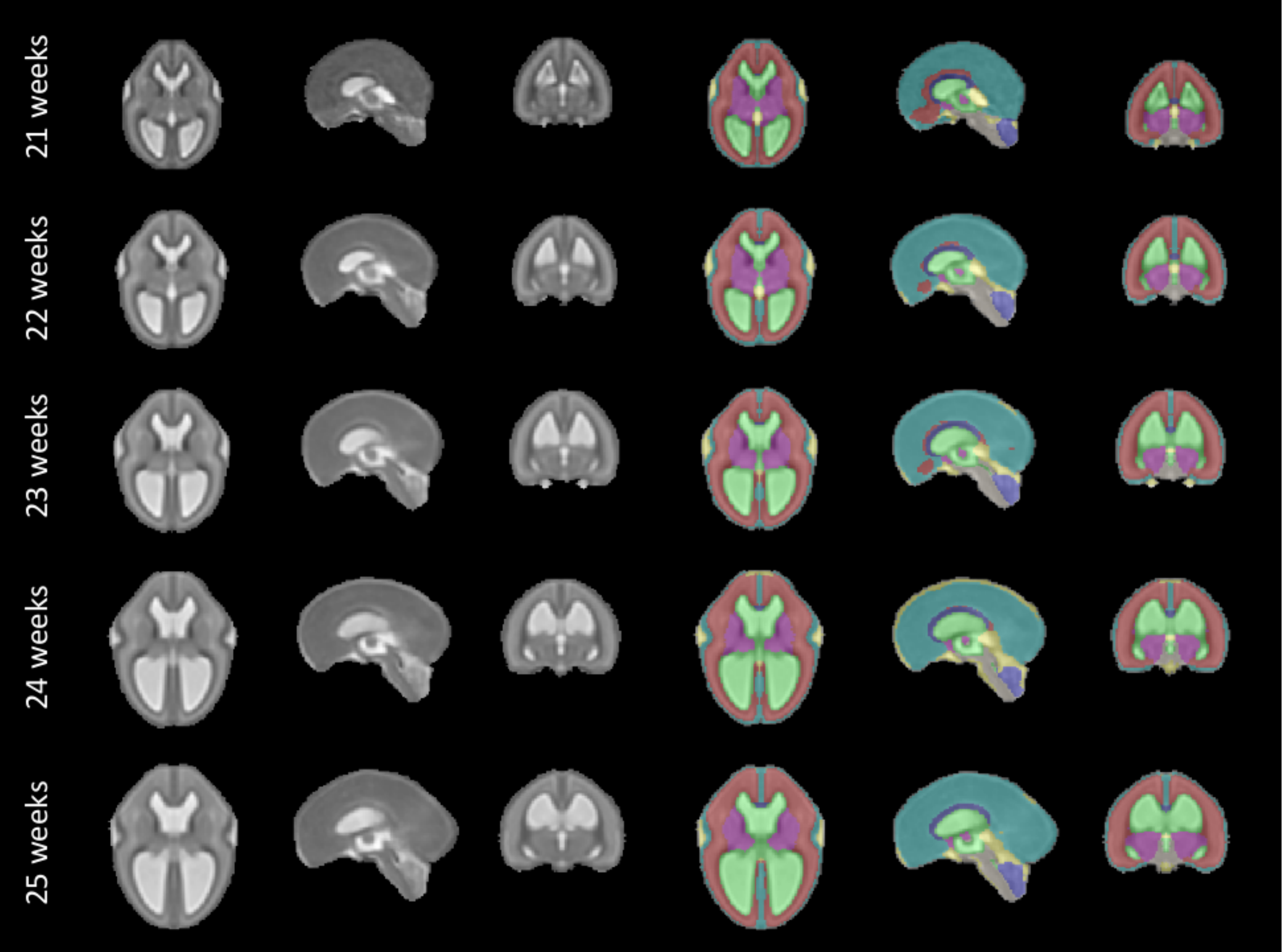}
    \caption{\textbf{Our spatio-temporal atlas for spina bifida aperta - Part I (not operated).}
    Publicly available \href{https://doi.org/10.7303/syn25910198}{here}.}
    \label{fig:sb_atlas_not_operated}
\end{figure}

\begin{figure}
    \centering
    \includegraphics[width=0.93\linewidth]{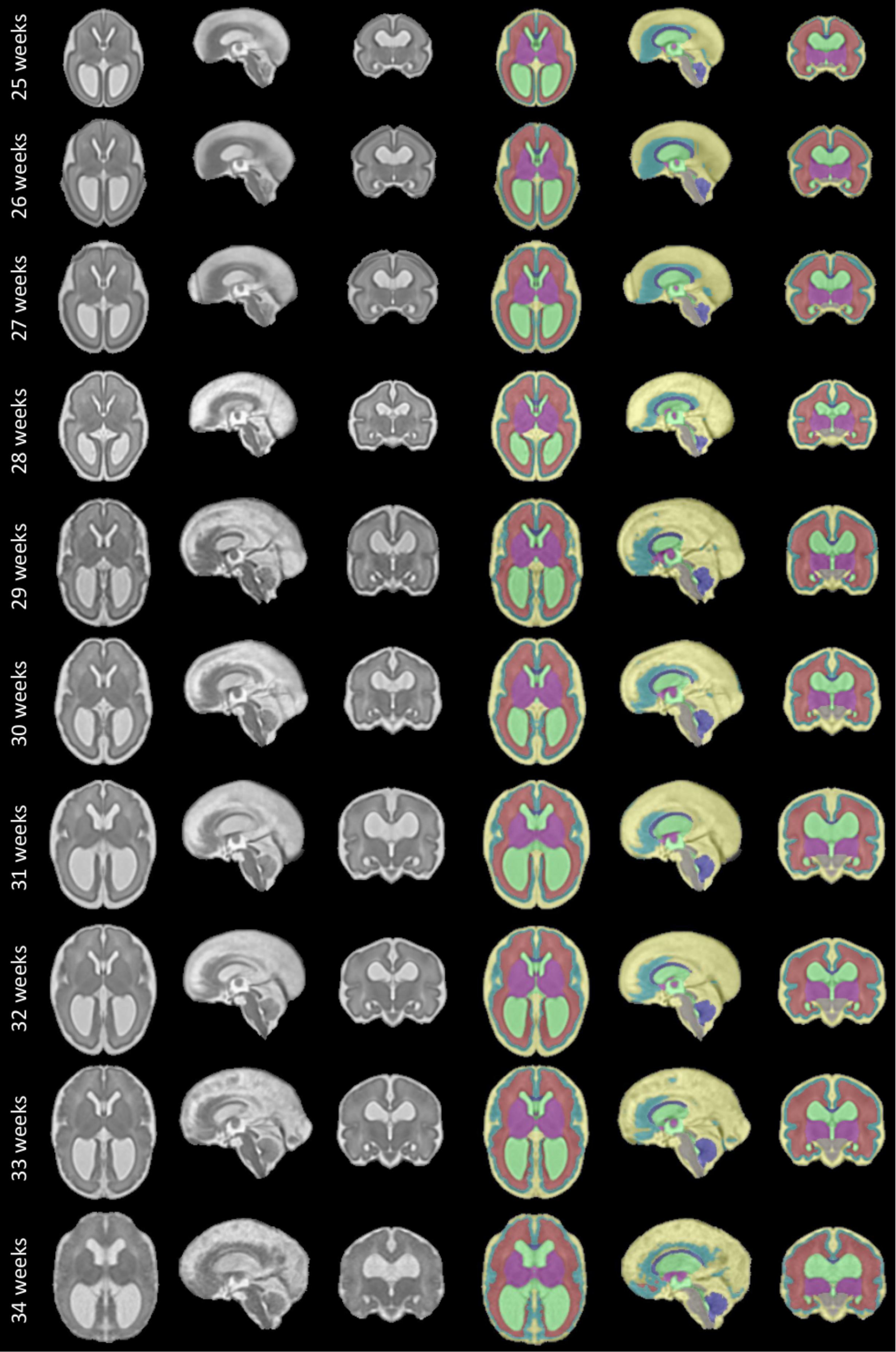}
    \caption{\textbf{Spatio-temporal atlas for spina bifida aperta - Part II (operated).}
    Publicly available \href{https://doi.org/10.7303/syn25910198}{here}.}
    \label{fig:sb_atlas_operated}
\end{figure}

\newpage
\section{Results}

\subsection{Intra-rater Variability for the Annotation of the Anatomical Landmarks}\label{sec:irr}

\begin{table}[t]
	\caption{\textbf{Evaluation of the reliability of the landmarks.}
	We report the estimated percentiles of distances in millimeters between first and second marking for each proposed landmarks.
    $P_{75}$: $75$th percentile of distances in millimeters.
    $P_{80}$: $80$th percentile of distances in millimeters.
    $P_{95}$: $95$th percentile of distances in millimeters.
    Our reliability score is defined in section~\ref{sec:irr}.
	\textbf{LALV}: Anterior Horn of the Left Lateral Ventricle,
	\textbf{RALV}: Anterior Horn of the Right Lateral Ventricle,
	\textbf{PTP}: Posterior Tectum Plate,
	\textbf{LCB}: Left Cerebellar Brainstem Junction,
	\textbf{RCB}: Right Cerebellar Brainstem Junction,
	\textbf{LFOM}: Left Deep Grey Border at Foramen of Monro,
	\textbf{RFOM}: Right Deep Grey Border at Foramen of Monro,
	\textbf{LACSP}: Left Deep Grey Border at Anterior Cavum Septi Pellucidi line,
	\textbf{RACSP}: Right Deep Grey Border at Anterior Cavum Septi Pellucidi line,
	\textbf{LPCSP}: Left Deep Grey Border at Posterior Cavum Septi Pellucidi line,
	\textbf{RPCSP}: Right Deep Grey Border at the Posterior Cavum Septi Pellucidi line.
	}
	\begin{tabularx}{\columnwidth}{cc*{4}{Y}}
	    \toprule
		\textbf{Landmark} & \textbf{Ratio of Missing} (\%) & $P_{75}$ (mm) & $P_{80}$ (mm) & $P_{95}$ (mm) & \textbf{Reliability}\\
		\midrule
		LALV & 0    & 1.73 & 1.95 & 3.02 & Good\\
	    RALV & 0    & 1.70 & 1.91 & 2.96 & Good\\
	    PTP  & 3 & 1.15 & 1.29 & 2.00 & Excellent\\
	    LCB  & 0    & 1.70 & 1.90 & 2.95 & Good\\
	    RCB  & 0    & 1.78 & 2.00 & 3.10 & Good\\
	    LFOM & 3 & 2.83 & 3.17 & 4.91 & Poor\\
	    RFOM & 0    & 2.50 & 2.81 & 4.35 & Satisfactory\\
	    LACSP& 16 & 2.74 & 3.07 & 4.77 & Poor\\
	    RACSP& 29 & 2.59 & 2.91 & 4.51 & Satisfactory\\
	    LPCSP& 16 & 3.35 & 3.76 & 5.83 & Poor\\
	    RPCSP& 16 & 3.12 & 3.50 & 5.43 & Poor\\
        \bottomrule
	\end{tabularx}
	\label{tab:irr}
\end{table}

\begin{figure}[t]
    \centering
    \includegraphics[width=\linewidth]{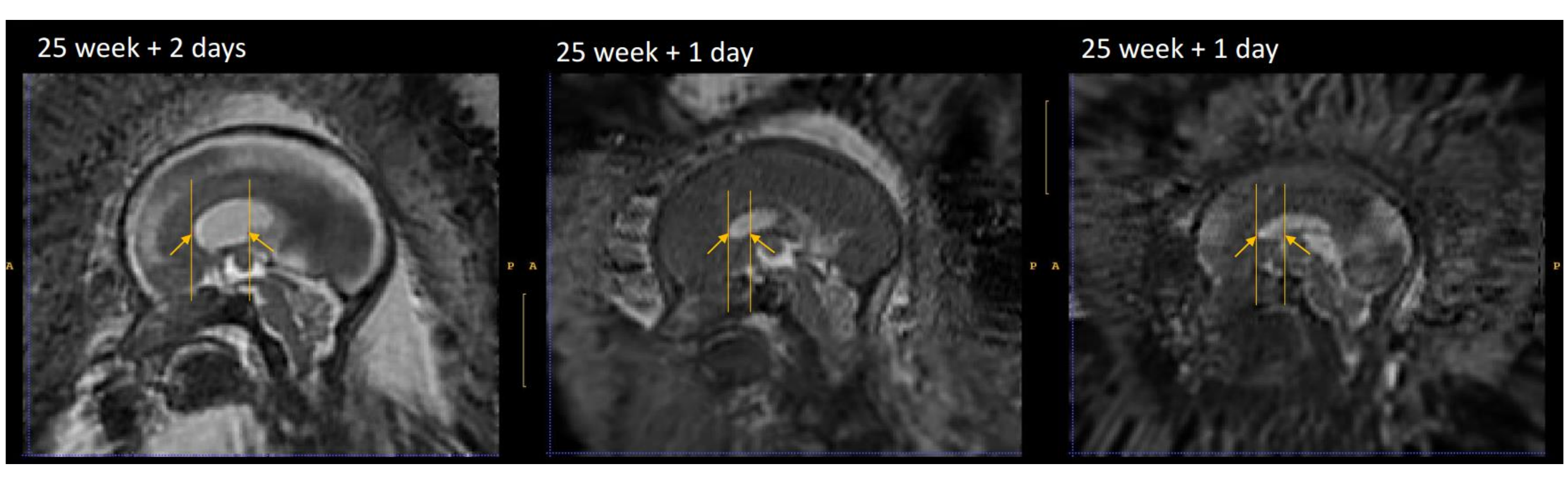}
    \caption{Cavum septi pellucidi (CSP) variation fetuses with 25 weeks of gestation. Yellow arrows indicate the anterior and posterior borders of the CSP as defined by the landmark localisation protocol. 
    This visualisation illustrates the disparity between volumes in terms of shape and size of the CSP.
    % and therefore reference points anterior CSP line and posterior CSP line (yellow lines and arrows) as defined in the protocol.
    }
    \label{fig:csp}
\end{figure}

To assess intra-rater variability, a subset of $31$ 3D reconstructed MRIs, selected at random, were marked two times by the same rater, EV. 
The mean gestational age was $26.2$ weeks and the range of gestational ages in the reliability set was $22-34$ weeks.
Those statistics closely match the one of the full cohort as described in section~\secref{material:cohort} (the mean gestational age is $26.1$ weeks and the range is $21-35$ weeks for the full dataset).
The two ratings were performed with an interval of at least three weeks to mitigate the bias caused by observer recollection.
A landmark was marked absent when the anatomical position described by the protocol was not found within the volume.

The two landmark placements are said to be in agreement if the second landmark placement is inside a $3\times 3\times 3$ voxel cube where the original placement is the central voxel.
When $95\%$ of the second landmarks fall within this radius, the landmark is considered ‘Excellent’ in terms of intra-rater reliability,
when $80\%$ of are in agreement, intra-rater reliability is considered ‘Good’, 
where $75\%$ fall within the radius of agreement intra-rater reliability is considered ‘Satisfactory’.
For landmarks with a probability of agreement of less than $75\%$, the reliability is considered ‘Poor’.
The probabilities that pairs of landmarks are in agreement is estimated based on the assumption that the distribution of distances between first and second marks is Gaussian.

\subsection{Automatic Segmentation of Fetal Brain 3D MRIs}

\begin{table}[t]
	\caption{\textbf{Evaluation of automatic fetal brain segmentation.}
	We report mean (standard deviation) for the Dice score (DSC) in percentages and the Hausdorff distance at $95\%$ (HD95) in millimeters for the seven tissue types.
    The spina bifida atlas is the atlas contributed in this chapter.
    The normal atlas is a spatio-temporal atlas computed from 3D MRIs of fetuses with a normal brain~\cite{gholipour2017normative}.
	\textbf{WM}: white matter,
	\textbf{Vent}: ventricular system,
	\textbf{Cer}: cerebellum,
	\textbf{CSF}: cerebrospinal fluid,
	\textbf{ECSF}: extra-axial CSF,
\textbf{CGM}: cortical gray matter,
	\textbf{DGM}: deep gray matter,
	\textbf{BS}: brainstem.
    % 
    % Best mean values for the evaluation on the cohort of spina bifida aperta cases are in bold.
    }
	\begin{tabularx}{\columnwidth}{ccc*{7}{Y}}
	    \toprule
		\textbf{Atlas} & \textbf{Cohort} & \textbf{Metric} & 
		\bf WM & \bf Vent & \bf Cer & \bf ECSF & \bf CGM & \bf DGM & \bf BS\\
		\midrule
		Normal & Normal 
		& DSC & 87.1 (2.3) & 83.8 (4.4) & 86.9 (3.7) & 86.7 (2.3) & 67.9 (5.3) & 82.7 (3.2) & 81.7 (3.2)\\
		& & HD95 & 1.9 (0.6) & 1.2 (0.2) & 1.6 (0.4) & 1.2 (0.4) & 1.5 (0.6) & 2.7 (0.7) & 2.5 (0.5)\\
		\midrule[0.3pt]\midrule[0.3pt]
		Normal & Spina Bifida 
		& DSC & 69.4 (16.4) & 79.9 (8.0) & 50.2 (31.3) & 49.6 (35.7) & 41.7 (21.6) & 69.6 (14.5)& 62.2 (21.2)\\
		& & HD95 & 4.3 (2.2) & 3.5 (2.8) & 5.8 (4.2) & 10.3 (9.8) & 4.1 (2.6) & 4.3 (2.5) & 3.9 (2.7)\\
		\midrule[0.3pt]
		Spina Bifida & Spina Bifida 
		& DSC & 80.6 (6.4) & 84.0 (10.1) & 69.8 (15.6) & 54.9 (26.3) & 48.8 (15.9)& 77.2 (5.1) & 71.2 (10.9)\\
		& & HD95 & 3.4 (1.6) & 2.0 (1.4) & 2.7 (0.8) & 9.6 (9.1) & 3.2 (1.7) & 3.0 (0.7) & 3.1 (1.2)\\
        \bottomrule
	\end{tabularx}
	\label{tab:seg_results}
\end{table}

In this section, we compare the automatic segmentations obtained either using an 
atlas of normal fetal brains~\cite{gholipour2017normative}
or using the proposed atlas for spina bifida aperta (SBA).
The quantitative evaluation can be found in Table~\ref{tab:seg_results}.

We studied the automatic segmentation of fetal brain 3D MRIs into seven tissue types.
Fetal brain 3D MRIs from the FeTA dataset~\cite{payette2021automatic} were used for the evaluation.
More details about the dataset used for the evaluation can be found in section~\secref{material:evaluation}.

The automatic segmentations are obtained in two steps: first a volume of the atlas, chosen based on the gestational age, is registered to each fetal brain 3D MRI, and second, after registration, the segmentation of the atlas is propagated.
Non-linear image registration is implemented as described in section~\secref{sec:non_linear_reg}.
In particular, we used the same hyper-parameter values.
The automatic segmentations for the corpus callosum and the white matter were merged into white matter, since the corpus callosum is part of the white matter segmentation in the FeTA dataset.
 
Automatic segmentations for the SBA cases are computed using either a normal fetal brain atlas~\cite{gholipour2017normative} or our SBA fetal brain atlas as can be seen in the last four rows of Table~\ref{tab:seg_results}.
In addition, we have also compute automatic segmentations for the normal brain cases using the normal fetal brain atlas~\cite{gholipour2017normative} as can be seen in the first two rows of Table~\ref{tab:seg_results}. 
The evaluation was performed for each tissue type using the Dice score~\cite{dice1945measures,fidon2017generalised} and the Hausdorff distance at percentile $95$~\cite{hausdorff1991set}.

\section{Discussion}\label{sec:discusion}
The proposed spatio-temporal atlas for spina bifida aperta (SBA) is illustrated in Figure~\ref{fig:sb_atlas_not_operated} and Figure~\ref{fig:sb_atlas_operated} (see \textit{Data availability} \cite{fidon_lucas_2021_5524312} and \textit{Software availability} for full atlas).
As described in section~\secref{material:cohort}, the cohort used to compute this atlas contains longitudinal data.
This longitudinal dataset of 90 MRIs might be less representation of the whole SBA population than a dataset of 90 MRI that would contain only singletons.
However, the use of longitudinal data adds some implicit temporal consistency in the atlas.

The landmarks in the ventricles, the posterior tectum plate, and at the junction of the cerebellar and the brainstem were all found to be reliable enough in terms of distance between successive marks by the same rater as can be seen in Table~\ref{tab:irr}.
In addition, those anatomical landmarks were always present, except for the posterior tectum plate that was missing for one reconstructed 3D MRI.
However, the landmarks in the deep gray were almost all found to be poorly reliable in terms of distance between successive marks by the same rater.
One can group the landmarks in the deep gray matter into two groups:
the landmarks based on the foramen of Monro,
and the landmarks based on the cavum septi pellucidi.
The landmarks based on the foramen of Monro were almost always present.
This is in contrast with the landmarks based on the cavum septi pellucidi that were missing up to $29\%$ of the time.
In Figure~\ref{fig:csp}, we give an illustration of the anatomical variability of the cavum septi pellucidi in fetuses with SBA.
This suggests that the position of landmarks based on the cavum septi pellucidi can vary widely from one subject to the other.
As a result, we choose to use the two landmarks based on the foramen of Monro for the computation of the atlas, but to excluded the four landmarks based on the cavum sceptum pellucidum.

The evaluation of automatic segmentation of fetal brain 3D MRIs in Table~\ref{tab:seg_results} suggests that using the proposed atlas for SBA leads to more accurate segmentation of SBA cases than a normal fetal brain atlas.
The proposed atlas for SBA outperforms the normal fetal brain atlas in terms of mean Dice scores and mean Hausdorff distances for all tissue types.
The proposed atlas also leads to lower standard deviations of Dice scores and Hausdorff distances for all tissue types except for the ventricular system.
This suggests that automatic segmentation using image registration of an atlas is more robust for SBA when a SBA atlas is used.

In addition, when comparing automatic segmentations of normal fetuses and fetuses with SBA obtained using a normal fetal brain atlas we found a decrease of segmentation accuracy in terms of Dice scores and hausdorff distances for all tissue types.
For the cerebellum, the mean Dice score decreased from $86.9\%$ for normal fetuses to $50.2\%$ for fetuses with SBA.
This can be attributed to the Chiari malformation type II which is found in most SBA cases~\cite{pollenus2020impact}.
The decrease of mean Dice score and the increase of mean Hausdorff distance for the extra-axial cerebrospinal fluid (CSF) can be attributed to the quasi absence of extra-axial CSF in fetuses with SBA.

\section{Limitations}
In this chapter, we have used MRIs of operated and non-operated fetuses,
ie that have or have not undergone fetal surgery to close the spina bifida aperta (SBA) defect in utero.
\textit{In-utero} fetal surgery is currently recommended to be performed prior to $26$ weeks of gestation.
The surgery has been found to influence the evolution of the fetal brain anatomy starting within one week after the operation~\cite{aertsen2019reliability}.
Therefore, a normative atlas for SBA should be computed using only MRIs of non-operated fetuses.
This limitation of our work is however due to the clinical data used.
To make this limitation clear we have separated the atlas into two parts as illustrated in Figure~\ref{fig:sb_atlas_not_operated} and Figure~\ref{fig:sb_atlas_operated}.
This separations is also reflected in the data structure chosen to share the atlas, as detailed in the next section \cite{fidon_lucas_2021_5524312}.

In Figure~\ref{fig:ga}, it is worth noting that relatively little cases are available in the range of gestational ages $27-31$ weeks.
As a result, the proposed atlas might be less representative of the SBA population in this range of gestational ages.
In particular, this might explain why the ventricle size does not appear to increase linearly for those gestational ages as can be seen in Figure~\ref{fig:sb_atlas_operated}.

\section{Data Availability}

Zenodo: A Spatio-temporal Atlas of the Developing Fetal Brain with Spina Bifida Aperta. \url{https://doi.org/10.5281/zenodo.5524312} \cite{fidon_lucas_2021_5524312}.

The project contains $15$ folders, each corresponding to a unique volume of our spatio-temporal fetal brain atlas, as illustrated in Figure~\ref{fig:sb_atlas_not_operated} and Figure~\ref{fig:sb_atlas_operated}, and contains four nifti files:
\begin{itemize}
	\item srr.nii.gz (average 3D reconstructed MRI).
	\item mask.nii.gz (3D brain mask).
	\item parcellation.nii.gz (3D segmentation of the fetal brain into 8 tissue types as described in section~\secref{sec:segmentation}).
	\item lmks.nii.gz (annotations for the 7 anatomical landmarks described is section~\secref{sec:landmarks}).
\end{itemize}

Data are available under the terms of the \href{https://creativecommons.org/publicdomain/zero/1.0/}{Creative Commons Zero "No rights reserved" data waiver} (CC0 1.0 Public domain dedication). Codes and scripts  are available under the terms of the \href{https://opensource.org/licenses/BSD-3-Clause}{BSD-3-Clause} license.

Alternatively, it is possible to download A Spatio-temporal Atlas of the Developing Fetal Brain with Spina Bifida Aperta on Synapse:
\url{https://doi.org/10.7303/syn25887675}. It is necessary to create a synapse account to be able to download the data.

\section{Software Availability}
Source code available from: \url{https://github.com/LucasFidon/spina-bifida-MRI-atlas}
\\Archived source code at the time of publication: \url{https://doi.org/10.5281/zenodo.5524312} \cite{fidon_lucas_2021_5524312}
\\License: \href{https://opensource.org/licenses/BSD-3-Clause}{BSD-3-Clause}

\section{Conclusion}
In this chapter, we propose the first spatio-temporal fetal brain MRI atlas for spina bifida aperta (SBA).

We propose a semi-automatic pipeline for the computation of spatio-temporal fetal brain atlas.
Our pipeline relies on four main components:
\begin{itemize}
    \item \texttt{\href{https://github.com/gift-surg/MONAIfbs}{MONAIfbs}}~\cite{martaisbi2021}, an automatic method for fetal brain extraction in 2D fetal MRIs.
    \item \texttt{\href{https://github.com/gift-surg/NiftyMIC}{NiftyMIC}}~\cite{ebner2020automated}, a 3D super resolution and reconstruction algorithm that allows to obtain isotropic and motion-free volumetric MRI of the fetal brain.
    \item A proposed protocol for the annotation of 7 anatomical landmarks in 3D reconstructed fetal brain MRIs.
    \item A proposed weighted generalize Procrustes method for an unbiased initialization of the atlas based on the anatomical landmarks.
\end{itemize}

We find that the proposed atlas outperforms a state-of-the-art fetal brain atlas for the automatic segmentation of brain 3D MRIs of fetuses with SBA.
This suggests that the proposed atlas for SBA provides a better anatomical prior about the peri-surgical SBA brain.
We hypothesise that this atlas could also help improving fetal brain MRI segmentation methods that lacks such prior, such as segmentation methods based on deep learning~\cite{fidon2021partial}.
We investigate this research question in \Chapref{chap:twai}.

Recently, deep learning-based methods for computing brain atlases that can be conditioned on variables such as the gestational age have been proposed~\cite{dalca2019learning,dey2021generative}.
In particular, some of those methods attempt to address the problem of robustness of registration and atlas construction algorithms to missing spatial correspondences~\cite{bone2020learning}.
Such missing correspondences are critical in the case for spina bifida aperta.
Applying those methods to spina bifida atlases construction has not yet been explored and is an interesting research direction for future work.

\chapter[A Dempster-Shafer approach to Trustworthy AI]{A Dempster-Shafer approach to Trustworthy AI with application to Fetal Brain MRI Segmentation}
\label{chap:twai}
\minitoc
\begin{center}
	\begin{minipage}[b]{0.9\linewidth}
		\small
		\textbf{Foreword\,}
		This chapter is adapted from \cite{fidon2022trustworthy}.
		Deep learning models for medical image    segmentation can fail unexpectedly and spectacularly for pathological cases and for images acquired at different centers than those used for training, with labeling errors that violate expert knowledge about the anatomy and the intensity distribution of the regions to be segmented.
        Such errors undermine the trustworthiness of deep learning models developed for medical image segmentation.
        Mechanisms with a fallback method for detecting and correcting such failures are essential for safely translating this technology into clinics and are likely to be a requirement of future regulations on artificial intelligence (AI). 
        Here, we propose a principled trustworthy AI theoretical framework and a practical system that can augment any backbone AI system using a fallback method and a fail-safe mechanism based on Dempster-Shafer theory. Our approach relies on an actionable definition of trustworthy AI.
        Our method automatically discards the voxel-level labeling predicted by the backbone AI that are likely to violate expert knowledge and relies on a fallback atlas-based segmentation method for those voxels.
        We demonstrate the effectiveness of the proposed trustworthy AI approach on the largest reported annotated dataset of fetal T2w MRI consisting of $540$ manually annotated fetal brain 3D MRIs with neurotypical or abnormal brain development and acquired from $13$ sources of data across $6$ countries.
        We show that our trustworthy AI method improves the robustness of a state-of-the-art backbone AI for fetal brain MRI segmentation on MRIs acquired across various centers and for fetuses with various brain abnormalities.
	\end{minipage}
\end{center}

% Content
\section{Introduction on Trustworthy AI using a Fallback}

% START FROM AI TRUST
% Automatic segmentation of medical images is needed for personalized medicine and to study anatomical development in healthy populations as well as populations with a pathology.
% % 
% Artificial Intelligence (AI) algorithms for medical image segmentation can reach super-human accuracy on average~\cite{isensee2021nnu} and yet most radiologists do not trust them~\cite{allen20212020,cabitza2019biases}.
% % 
% This is partly because, for some cases, AI algorithms fail spectacularly with errors that violate expert knowledge about the segmentation task when the AI was applied across imaging protocol and anatomical pathologies~\cite{allen20212020,fidon2021distributionally,gonzalez2021detecting} (Fig.\ref{fig:twai-overview}b).
% % 
% This sense of distrust is exacerbated by the current lack of clear
% fit-for-purpose
% regulatory requirements for AI-based medical image software~\cite{van2021artificial}.
% % paving the way for their deployment in clinical practice in the near future.

As discussed in \Chapref{chap:intro},
the legal framework for the deployment in clinics of AI tools for medical segmentation is likely to soon become more stringent once the European Union has proposed its Artificial Intelligence Act to regulate AI
and AI trust is at the core of this proposal~\cite{ethics,regulations2021}.
% by requiring AI to be trustworthy prior to their deployment.
% 
Trust is an attitude of the human user and, in particular, AI trust is context-dependent.
In contrast, AI trustworthiness is an attribute of the AI system~\cite{jacovi2021formalizing,hoffman2017taxonomy}.
It is worth noting that a human user can trust an AI system that is not trustworthy and conversely.
% 
%  WHY DOES TRUST MATTER?
% 
Guidelines for trustworthy AI claim that AI trustworthiness must precede trust in the deployment of AI systems to avoid miscalibration of the human trust with respect to the trustworthiness of an AI system~\cite{ethics}.
In Psychology, trust of humans in AI can be defined as the belief of the human user that the AI system will satisfy the criteria of a set of contracts of trust.
This contract-based definition of human-AI trust reflects the plurality and the context-dependency of human-AI trust. In particular, the user may trust an AI system for one population or one type of scanner but not trust it for others.
An AI system is trustworthy to a contract of trust if it can maintain this contract
in all situations within the contract scope.
The EU ethics guidelines for trustworthy AI, that upheld the AI Act, advocate that
\enquote{AI systems should have safeguards that enable a \textit{fallback} plan in case of problems}~\cite{ethics}.
We argue that those safeguards should implement a \textit{fail-safe} mechanism in relation with a collection of contracts of trust so as to improve the trustworthiness of the overall system.

\begin{figure}[tp]
    \centering
    \includegraphics[width=\linewidth]{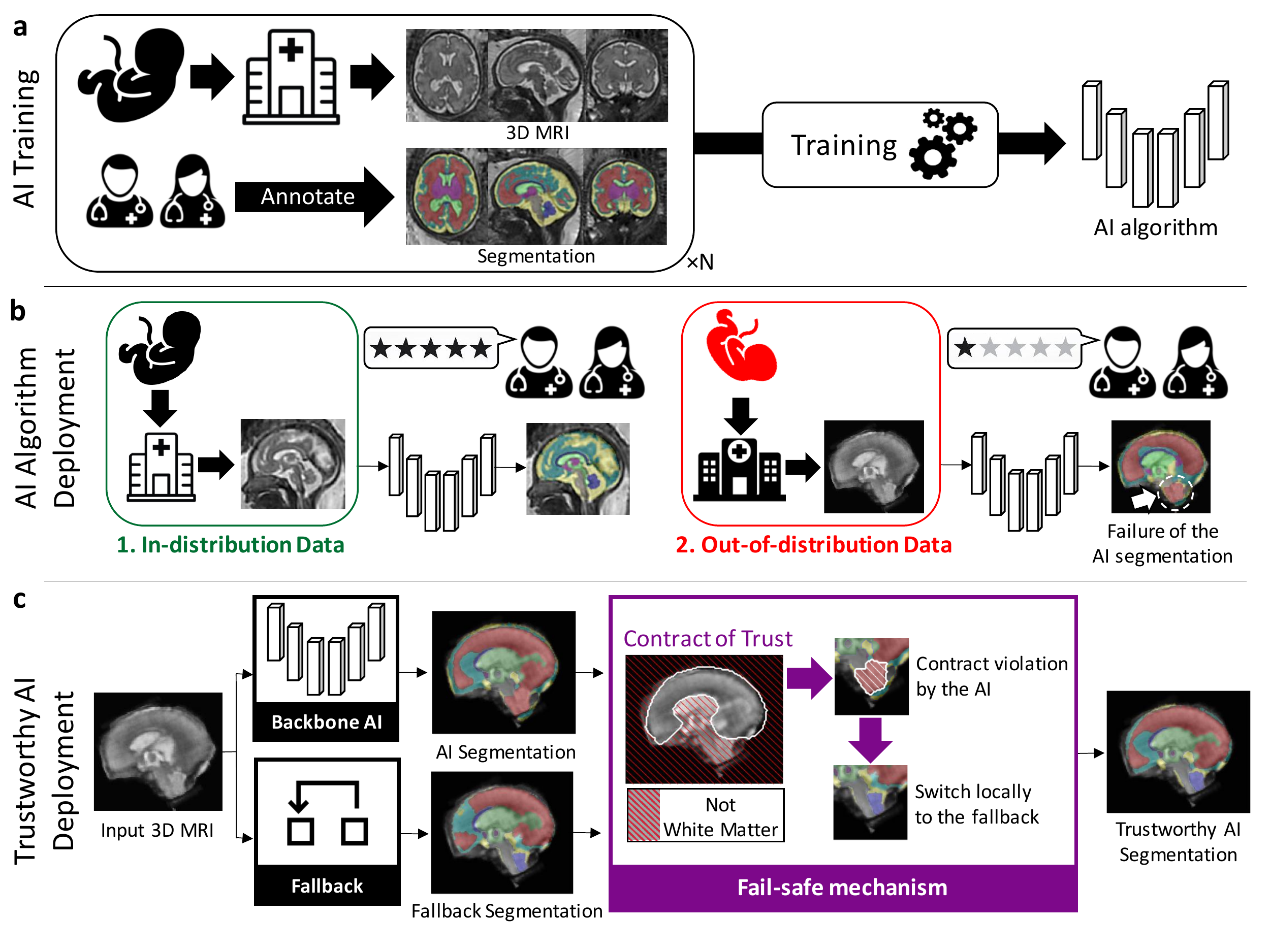}
    \caption{
    \textbf{Schematics of our principled method for trustworthy AI applied to medical image segmentation.}
    \textbf{a.} Deep neural networks for medical image segmentation (AI algorithm) are typically trained on images from a limited number acquisition centers. 
    This is usually not sufficient to cover all the anatomical variability.
    \textbf{b.} When the trained AI algorithm is deployed, it will typically give satisfactory accuracy for images acquired with the same protocol as training images and with a health condition represented in the training dataset (left).
    However, the AI algorithm might fail with errors that are not anatomically plausible, for images acquired with a slightly different protocol as training images and/or representing anatomy underrepresented in the training dataset (right).
    \textbf{c.} Schematic of the proposed trustworthy AI algorithm.
    The backbone AI segmentation algorithm is coupled with a \emph{fallback} segmentation algorithm.
    Experts knowledge about the anatomy is modelled using Dempster-Shafer theory.
    When part of the AI segmentation is found to contradict expert knowledge for a voxel, our trustworthy AI algorithm automatically switches to the fallback segmentation for this voxel.
    }
    \label{fig:twai-overview}
\end{figure}

In this chapter, we propose the first trustworthy AI framework with a fail-safe and a fallback
for medical image segmentation (see \nameref{sec:TWAI}).
The proposed framework consists of three main components: 
first, a backbone AI algorithm, that can be any AI algorithm for the task at hand,
second, a fallback segmentation algorithm, that can be any safe segmentation algorithm but potentially less precise than the backbone AI algorithm,
and third, a fail-safe method that automatically detects local conflicts between the backbone AI algorithm prediction and the contracts of trust and switches to the fallback algorithm in case of conflicts.
This is illustrated for fetal brain 3D MRI segmentation in Fig.~\ref{fig:twai-overview}.
The proposed principled fail-safe method is based on Dempster-Shafer (DS) theory.
DS theory allows to model partial information about the task, which is typically the case for expert knowledge.
For example, in human brain anatomy, one piece of expert knowledge is that the cerebellum is located in the lower back part of the brain.
This gives us information only about the segmentation of the cerebellum while a segmentation algorithms will typically compute segmentation for many other tissue types in addition to the cerebellum.
The Dempster's rule of combination of DS theory is an efficient mathematical tool to combine independent sources of information that discards contradictions among the sources (see \nameref{sec:DS}).
In our framework, the AI-based segmentation and each expert knowledge are treated as independent sources of information and the Dempster's rule of combination is employed to act as the fail-safe (see \nameref{sec:fail-safe}).

To demonstrate the applicability of the developed trustworthy AI framework,
we propose one implementation for fetal brain segmentation in MRI.
For the backbone AI model, we used the state-of-the-art deep learning-based segmentation pipeline nnU-Net~\cite{isensee2021nnu} (see \nameref{sec:nnunet}).
For the fallback model we used a registration-based segmentation method inspired by the state-of-the-art multi-atlas method GIF~\cite{cardoso2015geodesic} (see \nameref{sec:fallback}).
We also show that our fail-safe formulation is flexible enough to model both spatial location-based (see \nameref{sec:anatomical_contract}) and intensity-based (see \nameref{sec:intensity_contract}) contracts of trust about the regions of interest to be segmented (see \nameref{sec:multi_contracts}).
Spatial location-based fail-safes are implemented using the masks of the regions of interest computed by the registration-based fallback algorithm.
However, in the fail-safe, the masks are interpreted differently.
In this case, the mask of a region R is used only to exclude labeling voxels outside of the mask as belonging to R. 
This is illustrated in Fig.~\ref{fig:twai-overview}c.
Inspired by the margins used for safety in radiation therapy planning to account for spatial registration errors~\cite{niyazi2016estro,andres2020dosimetry}, 
we first add spatial margins to the fallback masks before excluding the labels seen as anatomically unlikely according to the dilated fallback masks.
While allowing the masks to overlap, this helps preventing miscoverage of the regions of interest that is the only source of error in this formulation of the fail-safe.

We evaluated the proposed trustworthy AI method on fetal brain segmentation into eight tissue types using 3D T2w MRI.
The segmentation of fetal brain MRI is essential to study normal and abnormal fetal brain development.
In the future, reliable analysis and evaluation of fetal brain structures could also support diagnosis of central nervous system pathology, patient selection for fetal surgery, evaluation and prediction of outcome, hence also parental counselling. 
In particular, fetal brain 3D T2w MRI segmentation presents multiple challenges for trustworthy AI~\cite{fidon2021distributionally}.
There are variations in T2w MRI protocols used across clinical centers and there is a spectacular variation of the fetal brain anatomy across gestational ages and across normal and abnormal fetal brain anatomy.

\FloatBarrier

\section{Materials and Methods}

% DATASET
% DATA (total = 540 volumes)
% TRAINING (191 volumes)
% - 144 Leuven training (116 control + 28 SB)
% - 15 spina bifida atlas
% - 18 Harvard atlas
% - 14 Chinese atlas
% Testing (349 volumes)
% - 76 Leuven control
% - 63 Leuven SB (Fred SB)
% - 35 Leuven ABN
% - 11 Vienna SB
% - 36 FeTA control
% - 36 FeTA SB
% - 16 FeTA ABN
% - 47 UCLH SB (8 Leuven + 27 London + 4 Manchester + 4 Belfast + 2 Cork + 1 New Castle + 1 Liverpool)
% - 29 KCL control

\subsection{Fetal brain MRI dataset}\label{sec:twai-dataset}

\begin{figure}[t]
    \centering
    \includegraphics[width=\linewidth,trim=0cm 0cm 0cm 3.7cm,clip]{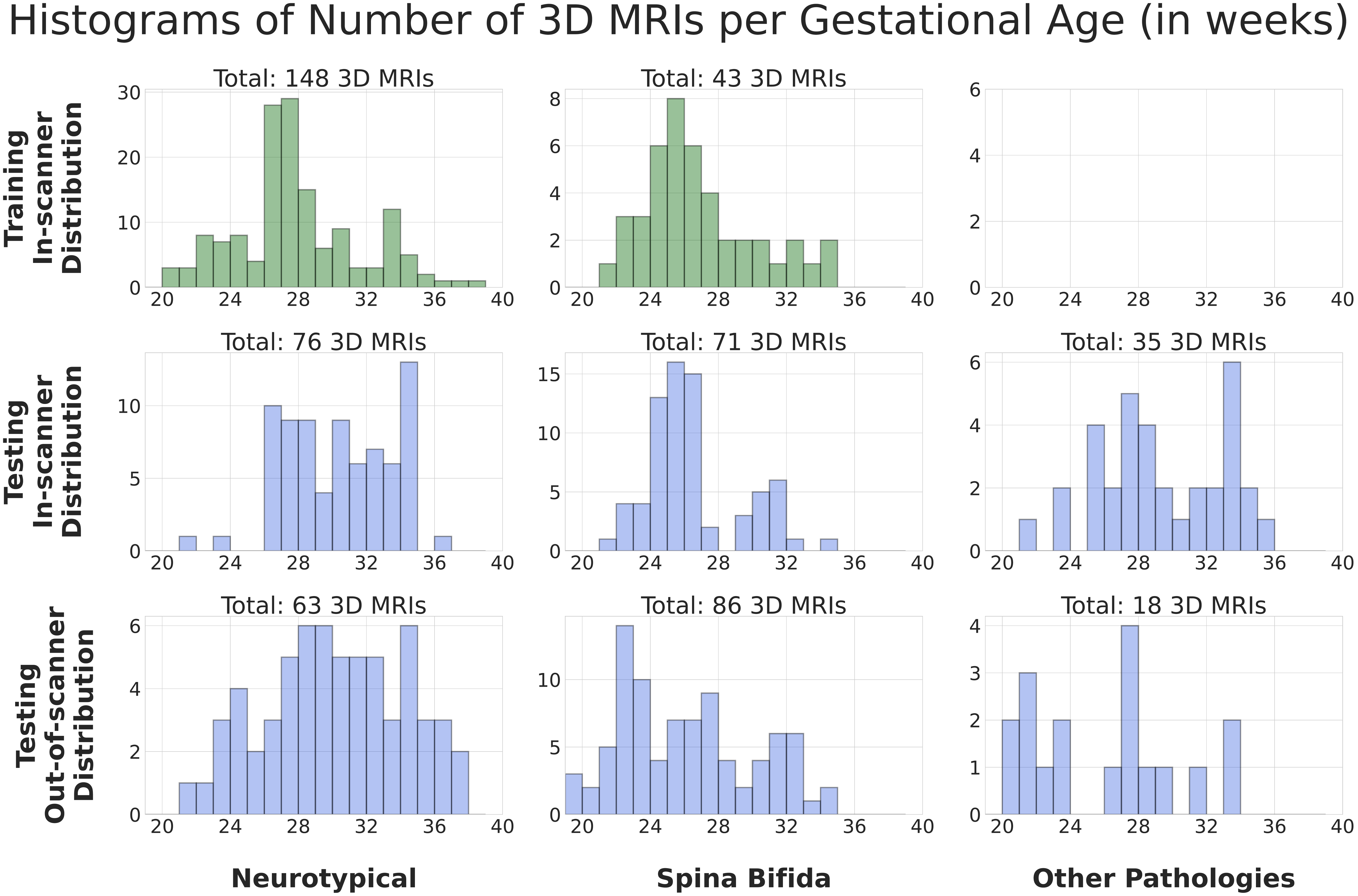}
    \caption{
    \textbf{Composition of the training and testing datasets (total: 540 3D MRIs)}
    \textit{In-scanner distribution} designates the 3D MRIs acquired at the same center as the training data.
    \textit{Out-of-scanner distribution} designates the 3D MRIs acquired at different centers than the training data.
    This is the largest fetal brain MRI dataset reported to date.
    }
    \label{fig:twai-data}
\end{figure}

We have collected a dataset with a total of $540$ fetal brain 3D MRIs with neurotypical or abnormal brain development and from $13$ sources of data across $6$ countries.
The dataset and data processing used are described in \Chapref{chap:fetaldataset}.

The composition of the training and testing datasets is summarized in Fig.\ref{fig:twai-data}.
The training dataset consists of the $47$ volumes from the three fetal brain atlases, $144$ neurotypical cases from \uzlshort{} and $28$ spina bifida cases from \uzlshort{}.
The rest of the data is used for testing ($n=349$).
In the testing dataset, 3D MRIs that were acquired at \uzlshort{} are designated as \textit{in-scanner-distribution} while the data from other acquisition centers are designated as \textit{out-of-scanner-distribution} data.

% % Pre-processing / SRR
% The 3D MRIs with an isotropic image resolution of $0.8$mm have been reconstructed from the stacks of 2D MRI slices acquired at \uzlshort{}, \viennashort{}, \uclhshort{}, \manchestershort{}, \belfastshort{}, \corkshort{}, \newcastleshort{}, and \liverpoolshort{} using the state-of-the-art and publicly available software \href{https://github.com/gift-surg/NiftyMIC/tree/master/data/templates}{\texttt{NiftyMIC}}~\cite{ebner2020automated}.
% % 
% The original 2D MRI slices were also corrected for bias field in the \href{https://github.com/gift-surg/NiftyMIC}{\texttt{NiftyMIC}} pipeline version $0.8$ using a N4 bias field correction step as implemented in \href{https://simpleitk.org}{SimpleITK} version $1.2.4$.
% % 
% Brain masks for those MRIs were computed using \texttt{\href{https://github.com/gift-surg/MONAIfbs}{MONAIfbs}}~\cite{martaisbi2021}, an automatic method for fetal brain extraction in 2D fetal MRIs.
% % 
% The 3D brain masks are reconstructed using the automatic 2D brain masks along with the 3D MRIs in \texttt{NiftyMIC}.

% The 3D MRI reconstructions for the FeTA dataset is described in~\cite{payette2021automatic}.
% % 
% Two volumes of spina bifida cases were excluded from the total FeTA dataset because the poor quality of the 3D MRI reconstruction (\texttt{sub-feta007} and \texttt{sub-feta009}) did not allow to manually segment them reliably for the seven tissue types.
% % 
% The brain masks for those 3D MRIs were computed directly using the 3D MRIs and an atlas-based method as described in our previous work~\cite{fidon2021partial}.

% BACKGROUND HUMAN-AI TRUST
% \input{chapters/twai/humanAITrust}

\subsection*{Mathematical notations}
\begin{itemize}
    \item $\mathbf{C}$: the set of all classes to be segmented
    \item $2^{\mathbf{C}}$: the set of all subsets of $\mathbf{C}$
    \item $\textbf{x}$: a voxel or a pixel
    \item $\Omega$: the set of all voxels or pixels (image domain)
    \item $p$: a probability vector
    \item $m$: a basic probability assignment (BPA) in Dempster-Shafer theory
    \item $\oplus$: Dempster's rule of combination
\end{itemize}

% BACKGROUND DEMPSTER SHAFER
\subsection{Background on Dempster-Shafer theory}\label{sec:DS}
In this section, we give a brief background on Dempster-Shafer theory where we introduce solely the mathematical concepts that will be used for our method:
the basic probability assignment (BPA) and Dempster's rule of combination.

In Dempster-Shafer theory~\cite{shafer1976mathematical}, \textit{basic probability assignments} are a generalization of probabilities that allow to model partial information and to combine different sources of information using Dempster's rule.

Let $\mathbf{C}$ be the set of all classes and $2^{\mathbf{C}}$ the set of all subsets of $\mathbf{C}$.
A \textbf{basic probability assignment (BPA)} on $\mathbf{C}$ is a function $m: 2^{\mathbf{C}} \mapsto [0,\,1]$ that satisfies
\begin{equation}
\left\{
    \begin{aligned}
        m(\emptyset) =& \,0\\
        \sum_{A \subset \mathbf{C}} m(A) = & \,1
    \end{aligned}
\right.
\label{eq:bpa}
\end{equation}
Probabilities on $\mathbf{C}$ are functions $p: \mathbf{C} \mapsto [0,1]$ that satisfy $\sum_{c\in \mathbf{C}}p(c) = 1$.
Probabilities are equivalent to the BPAs that assign non-zeros weights only to the singletons, i.e. the sets $A=\{c\}$ for $c \in \mathbf{C}$.
Indeed, given a probability $p$ the BPA $m^{(p)}$ associated with $p$ is defined as: 
$\forall c \in \mathbf{C},\, m^{(p)}(\{c\})=p(c)$ 
and $\forall A \subset \mathbf{C}$ with $|A| \neq 1$, $m^{(p)}(A)=0$.
Basic probability assignments on $\mathbf{C}$ are therefore more general that probabilities on $\mathbf{C}$.

For $A \subset \mathbf{C}$, $m(A)$ is the probability that our knowledge about the true label is exactly and only: "the true class is one of the classes in $A$".
In particular, it does not imply that $m(B)>0$ for any set $B$ such that $B \subsetneq A$ or $A \subsetneq B$.
This is in contrast to probabilities that can weight only the individual classes.
BPAs allow to represent more precisely than probabilities what we know (and don't know) about the true class of a voxel.
For example, the extreme case where we know nothing about the true class can be represented by the BPA $m$ such that $m(\mathbf{C})=1$.
The best one can do to try representing this case with probabilities is to define a probability $p$ such that $\forall c \in \mathbf{C},\, p(c)= \frac{1}{|\mathbf{C}|}$.
However, this choice of $p$ corresponds to the knowledge that the class distribution is uniformly random which is different from knowing nothing about the true class.

We now introduce the concept of \emph{complete contradiction} between BPAs which we need for Dempster's rule of combination.
Two BPAs on $\mathbf{C}$, $m_1$ and $m_2$, are said to be \textbf{completely contradictory} if 
\begin{equation}
\label{eq:contradiction}
    \sum_{E,F\subset \mathbf{C}|E \cap F = \emptyset} m_1(E)m_2(F) = 1
\end{equation}
Following \eqref{eq:bpa}, $m_1$ and $m_2$ are completely contradictory if and only if one cannot form a pair of overlapping sets of classes $(A,B)$ such that 
$m_1$ commits some belief to $A$, i.e. $m_1(A) > 0$, 
and $m_2$ commits some belief to $B$, i.e. $m_2(B) > 0$.

\subsubsection*{Dempster's rule of combination}
Dempster's rule of combination allows to combine any pair $(m_1, m_2)$ of BPAs on $\mathbf{C}$ that are not completely contradictory using the formula
\begin{equation}
    \forall A \subset \mathbf{C},
    \quad (m_1 \oplus m_2)(A) =
    \left\{
    \begin{array}{cc}
        \frac{\sum_{E,F\subset \mathbf{C}|E \cap F = A} m_1(E)m_2(F)}{1 - \sum_{E,F\subset \mathbf{C}|E \cap F = \emptyset} m_1(E)m_2(F)} & \textup{if}\,\, A \neq \emptyset\\
        0 & \textup{if}\,\, A = \emptyset
    \end{array}
    \right.
    % \frac{\sum_{E \cap F = A} m_1(E)m_2(F)}{1 - \sum_{E \cap F = \emptyset} m_1(E)m_2(F)}
    \label{eq:ds}
\end{equation}
For any pairs $(m_1, m_2)$ of BPAs on $\mathbf{C}$ that are not completely contradictory, $m_1 \oplus m_2$ is also a BPA on $\mathbf{C}$.
In addition, the relation $\oplus$ is symmetrical and associative.

\subsubsection{Special case of combining a probability with a BPA}
One particular case that will be useful for our method is the combination of a probability $p$ on $\mathbf{C}$, which we consider as a BPA, with a generic BPA $m$ on $\mathbf{C}$ using Dempster's rule of combination.

Since $p$ is a probability, for $A\subset\mathbf{C}$, $p(A)$ can be non-zeros only if $A$ is a singleton, i.e. if it exists a class $c \in \mathbf{C}$ such that $A=\{c\}$.
For simplicity, we will therefore use the abusive notation when considering $p$ as a BPA:
$p(c):=p(A)=p(\{c\})$.

The relation of complete contradiction \eqref{eq:contradiction} between $p$ and $m$ can be simplified
\begin{equation}
\label{eq:contradiction_proba}
    \sum_{E,F\subset \mathbf{C}|E \cap F = \emptyset} p(E)m(F)
    = \sum_{c \in \mathbf{C}} \sum_{F \subset \left(\mathbf{C}\setminus \{c\}\right)} p(c)m(F)
    = 1
\end{equation}
% Therefore, if $p$ and $m$ are completely contradictory, for all class $c \in \mathbf{C}$ and all subset $F\subset \mathbf{C}$,

Similarly, if $p$ and $m$ are not completely contradictory, the Dempster's rule between $p$ and $m$ can be simplified.
Let $A \subset \mathbf{C}$, $A\neq \emptyset$, using \eqref{eq:ds} we have
\begin{equation}
    \begin{split}
        (p \oplus m)(A) 
        &= \frac{\sum_{E,F\subset \mathbf{C}|E \cap F = A} p(E)m(F)}{1 - \sum_{E,F\subset \mathbf{C}|E \cap F = \emptyset} p(E)m(F)}\\
        &= \frac{\sum_{c \in \mathbf{C},F\subset \mathbf{C}|\{c\} \cap F = A} p(c)m(F)}{1 - \sum_{c \in \mathbf{C}} \sum_{F \subset \left(\mathbf{C}\setminus \{c\}\right)} p(c)m(F)}
    \end{split}
\end{equation}

We remark that $p\oplus m$ is also a probability on $\mathbf{C}$.
Indeed, let $c\in \mathbf{C}$, it can exist $F \subset \mathbf{C}$ such that $\{c\} \cap F = A$ only if $A$ is the singleton $A=\{c\}$ (we have assumed $A \neq \emptyset$).
Therefore if $A$ is not a singleton, i.e. $|A|>1$, the sum on the numerator is empty and equal to $0$.
As a result, we can write for any class $c \in \mathbf{C}$
\begin{equation}
    \begin{split}
        (p \oplus m)(c) 
        &= \frac{\sum_{c' \in \mathbf{C},F\subset \mathbf{C}|\{c'\} \cap F = \{c\}} p(c')m(F)}{1 - \sum_{c' \in \mathbf{C}} \sum_{F \subset \left(\mathbf{C}\setminus \{c'\}\right)} p(c')m(F)}\\
        &= \frac{p(c)\sum_{F\subset \mathbf{C}|c \in F} m(F)}{1 - \sum_{c' \in \mathbf{C}} \sum_{F \subset \left(\mathbf{C}\setminus \{c'\}\right)} p(c')m(F)}
    \end{split}
    \label{eq:drc_proba}
\end{equation}

% TRUSTWORHTY AI
\subsection{A Dempster-Shafer approach to Trustworthy AI for medical image segmentation}\label{sec:TWAI}

In this section, we present the formulation of our method for trustworthy AI-based medical image segmentation.

Our trustworthy AI segmentation method consists of three main components:
1) a backbone AI segmentation algorithm; 2) a fallback segmentation algorithm; and 3) and 
a fail-safe method that detects area of conflict between the AI algorithm segmentation and the contracts of trust and switches to the fallback algorithm for those regions.
An illustration for the task of fetal brain MRI segmentation is given in Fig.~\ref{fig:overview}.

The AI segmentation algorithm is a high-accuracy segmentor that can be, for example, a state-of-the-art convolutional neural network for medical image segmentation.
The fallback segmentation algorithm is a segmentor that might achieve lower accuracy than the AI, but is superior to the AI for other desirable properties such as robustness.
The AI and fallback segmentation algorithms take as input an image to be segmented and compute for each voxel of the image a probabilities vector with one probability for each class to be segmented.

The fail-safe mechanism aims at detecting erroneous predictions of the AI segmentation algorithm that contradict one of the contracts of trust.
The contracts of trust embed domain knowledge such as "there can't be white matter in this part of the brain" or "hyperintense voxels on T2 fetal brain MRI are always cerebrospinal fluid".
% 
% As a result, the detection mechanism consists of a collection of rules based on domain knowledge.
% % 
% Those rules represent human-AI contracts of trust.
% 
It is worth noting that, in general, those contracts are not sufficient to segment an entire image. 
Most contract will not enforce a specific segmentation but rather impose that the automatic segmentation meets certain constraints.
The fail-safe mechanism is not a segmentation algorithm.
In the context of image segmentation, contract of trusts can only reduces the set of possible classes and reweights the class probabilities of the segmentation of a pixel or voxel.
To implement the fail-safe mechanism, we propose to use a basic probability assignment (BPA) that acts on the backbone AI and the fallback class probabilities using Dempster's rule of combination~\eqref{eq:drc_proba}.
In addition, we assume that the fallback class probabilities never completely contradicts the BPA representing the contracts of trust.
As a result, Dempster's rule of combination can be used to switch automatically between the backbone AI algorithm and the fallback algorithm when the AI class probabilities completely contradict the BPA.
% 
% In contrast to the segmentation algorithms, the fail-safe mechanism is modeled as an algorithm that takes as input an image to be segmented and computes for each voxel of the image a basic probability assignment (BPA) with one mass for each subset of the classes to be segmented.
% % 
% Here, we use BPAs rather than probabilities to represent the contracts of trust because a contract of trust only reduces the set of possible classes and reweights the class probabilities of the segmentation of a pixel or voxel.

Formally, the trustworthy segmentation prediction is defined for an input image $I$ and for all voxel position $\textbf{x}$ as
\begin{equation}
\label{eq:trustworhtyAI}
    p^{\TWAI}_{I, \textbf{x}} = 
    \left(
    (1 - \epsilon) p^{\AI}_{I, \textbf{x}} + \epsilon p^{\fallback}_{I, \textbf{x}}
    \right)
    \oplus m^{\failsafe}_{I, \textbf{x}}
\end{equation}
where $\oplus$ is the Dempster's combination rule \eqref{eq:ds},
$p^{\AI}_{I, \textbf{x}}$ is the class probability prediction of the AI segmentation algorithm for voxel $\textbf{x}$ of image $I$,
$p^{\fallback}_{I, \textbf{x}}$ is the class probability prediction of the fallback segmentation for voxel $\textbf{x}$ of image $I$,
and $m^{\failsafe}_{I, \textbf{x}}$ is the BPA of the fail-safe mechanism for voxel $\textbf{x}$ of image $I$.
The parameter $\epsilon$ is a constant between $0$ and $1$.
% that weights the contribution of $p^{\fallback}_{\textbf{x}}$ as compared to $p^{\AI}_{\textbf{x}}$.

\subsubsection{Toy example: trustworthy traffic lights.}
One contract of trust for a trustworthy traffic light system at a crossing is that green should not be shown in all directions of the crossing at the same time.
To maintain this contract, traffic light controllers may use a fail-safe conflict monitor unit to detect conflicting signals and switch to a fallback light protocol.
One possible fallback is to display flashing warning signal for all traffic lights.

Formally, for the example of two traffic lights at a single-lane passage, the set of classes is all the pairs of color the two traffic light can display at the same time
$\mathbf{C}=\left\{
(c_1,c_2)\,|\,c_1,c_2 \in \{\textup{green},\textup{orange},\textup{red},\textup{flash}\}
\right\}$.

Let $p^{\textrm{backbone}}$ be the probability of the pair of traffic lights for the default light algorithm.
The probability of the fallback algorithm $p^{\fallback}$ is then defined such as $p^{\fallback}(\textup{flash},\textup{flash})=1$.
The contract of trust is $m$ defined as $m^{\textrm{not-all-green}}\left(\mathbf{C}\setminus \{(\textup{green},\textup{green})\}\right)=1$.

Let $\epsilon \in ]0,1]$, the trustworthy light algorithm is given by
\begin{equation}
    p^{\TWAI}=
    \left((1-\epsilon)p^{\textrm{backbone}}+\epsilon p^{\fallback}\right)
    \oplus m^{\textrm{not-all-green}}
\end{equation}
Using \eqref{eq:drc_proba}, we obtain
\begin{equation}
    \begin{split}
        p^{\TWAI}((\textup{green},\textup{green})) &= 0\\
        p^{\TWAI}((\textup{flash},\textup{flash})) 
        &= \frac{\epsilon p^{\fallback}(\textup{flash},\textup{flash})}{1 - (1-\epsilon)p^{\textrm{backbone}}((\textup{green},\textup{green}))}
    \end{split}
\end{equation}
where the amount of conflict \eqref{eq:contradiction_proba} between $p^{\textrm{backbone}}$ and $m^{\textrm{not-all-green}}$ is equal to $p^{\textrm{backbone}}((\textup{green},\textup{green}))$ and the amount of conflict between $p^{\fallback}$ and $m^{\textrm{not-all-green}}$ is equal to $0$.
In the case of complete contradiction between the default algorithm and the contract of trust, i.e. $p^{\textrm{backbone}}((\textup{green},\textup{green}))=1$, the trustworthy algorithm switch completely to the fallback algorithm with $p^{\TWAI}((\textup{flash},\textup{flash}))=p^{\fallback}(\textup{flash},\textup{flash})=1$.

\subsubsection{Fail-safe mechanism: switching between the AI algorithm and the fallback}\label{sec:fail-safe}
In this section, we describe how the the fail-safe mechanism $m^{\failsafe}$ allows to switch between the backbone AI segmentation algorithm and the fallback segmentation algorithm.

In our framework, we assume that the fallback segmentation algorithm always produces segmentation probabilities that do not contradict entirely the BPA of the contracts of trust.
A trivial example of such fallback, is the uniform segmentation algorithm that assigns an equal probability to all the classes to be segmented and for all voxels.
In contrast, we do not make such compatibility assumption for the AI segmentation algorithm.
Not only does this make our approach applicable with any AI segmentation algorithm, but our method also relies on the incompatibility between the AI segmentation algorithm prediction and the contracts of trust to detect failure of the AI segmentation algorithm and to switch to the fallback segmentation algorithm.
Formally, however small $\epsilon$ is, as long at it remains strictly positive ($\epsilon>0$), when $p^{\AI}_{I, \textbf{x}}$ is completely contradictory with $m^{\failsafe}_{I, \textbf{x}}$, we obtain that $p^{\TWAI}_{I, \textbf{x}}$ depends only on $p^{\fallback}_{I, \textbf{x}}$ and not on $p^{\AI}_{I, \textbf{x}}$.
On the contrary, when $p^{\AI}_{I, \textbf{x}}$ is not contradictory with $m^{\failsafe}_{I, \textbf{x}}$, we obtain that $p^{\TWAI}_{I, \textbf{x}}$ depends mainly on $p^{\AI}_{I, \textbf{x}}$ and the contribution of $p^{\fallback}_{I, \textbf{x}}$ is negligible for $\epsilon$ small enough.

This is due to a property of Dempster's rule of combination that can be illustrated by the following example of pairs of BPAs~\cite{Zadeh1979OnTV}.
Let $\epsilon_1,\epsilon_2 \in ]0, 1]$ and let $m_1$ and $m_2$ two BPAs on $\{a,b,c\}$ defined by: 
$m_1(\{a\}) = 1 - \epsilon_1$,
$m_1(\{b\}) = \epsilon_1$,
$m_1(\{c\}) = 0$,
$m_2(\{a\}) = 0$,
$m_2(\{b\}) = \epsilon_2$, and
$m_2(\{c\}) = 1 - \epsilon_2$.
We obtain $(m_1 \oplus m_2)(\{b\}) = 1$, however small $\epsilon_1, \epsilon_2 > 0$ are.

This characteristic of Dempster's rule of combination has sometimes been pointed out as counter-intuitive and as a motivation against the use of this operation~\cite{Zadeh1979OnTV}.
In contrast, we believe this to be a interesting characteristics and we use this property at our advantage to implement the fail-safe mechanism to switch between the AI algorithm and the fallback algorithm.
We recommend to choose a value of $\epsilon$ small compared to $1$. As a result, the fallback class with the highest probability becomes the trustworthy AI prediction highest probability only when the AI highest probabilities prediction have been eliminated by the contracts of trust BPAs under the Dempster's combination rule.
In this sense, the fallback probabilities act as a discounting factor in Dempster Shafer theory~\cite{shafer1976mathematical,smets1993belief}.
Rolf Haenni~\cite{haenni2002alternatives} summarized the motivation for the use of such discounting coefficient by Sherlock Holmes' statement~\cite{doyle2010sign}: 
\enquote{When you have eliminated the impossible, whatever remains, however improbable, must be the truth}.

Here, we consider the case in which the AI algorithm predicted probability $p^{\AI}_{I,\textbf{x}}$ is completely contradictory with $m^{\failsafe}_{I,\textbf{x}}$ for a voxel $\textbf{x}$.
Using \eqref{eq:contradiction_proba}, this implies
\begin{equation*}
\left\{
    \begin{aligned}
        &\sum_{c'\in \mathbf{C}}
        \left(
        \sum_{\mathbf{C}'\subset (\mathbf{C}\setminus \{c'\})}
        p^{\AI}_{I, \textbf{x}}(c')\,m^{\failsafe}_{I,\textbf{x}}\left(\mathbf{C}'\right)
        \right)
        = 1\\
        & \forall c' \in \mathbf{C},\,\forall \mathbf{C}'\subset \mathbf{C}\,|\, c' \in \mathbf{C}',\quad
        p^{\AI}_{I, \textbf{x}}(c')\,m^{\failsafe}_{I,\textbf{x}}\left(\mathbf{C}'\right) = 0
    \end{aligned}
\right.
\end{equation*}
Since by construction of our Trustworthy AI approach,
the fallback algorithm probability $p^{\fallback}_{I,\textbf{x}}$ is \emph{not} completely contradictory with $m^{\failsafe}_{I,\textbf{x}}$, we have that 
$\left((1 - \epsilon) p^{\AI}_{I, \textbf{x}} + \epsilon p^{\fallback}_{I, \textbf{x}}\right)$ is not completely contradictory with $m^{\failsafe}_{I,\textbf{x}}$.
Using Dempster's rule of combination \eqref{eq:drc_proba} we obtain, for every class $c \in \mathbf{C}$
\begin{equation*}
    p^{\TWAI}_{I, \textbf{x}}(c) =
    \frac{
    \sum_{\mathbf{C}'\subset \mathbf{C}\,|\, c \in \mathbf{C}'}
    \left((1 - \epsilon) p^{\AI}_{I, \textbf{x}}(c) + \epsilon p^{\fallback}_{I, \textbf{x}}(c)\right)
    m^{\failsafe}_{I,\textbf{x}}\left(\mathbf{C}'\right)
    }{
    1 - 
    \sum_{c'\in \mathbf{C}}\sum_{\mathbf{C}'\subset (\mathbf{C}\setminus \{c'\})}
    \left((1 - \epsilon) p^{\AI}_{I, \textbf{x}}(c') + \epsilon p^{\fallback}_{I, \textbf{x}}(c')\right)
    m^{\failsafe}_{I,\textbf{x}}\left(\mathbf{C}'\right)
    }
\end{equation*}
Since, in this illustrative case, we assume that $p^{\AI}_{I,\textbf{x}}$ is completely contradictory with $m^{\failsafe}_{I,\textbf{x}}$ and $\epsilon > 0$, $p^{\TWAI}_{I, \textbf{x}}(c)$ becomes
\begin{equation*}
    \begin{aligned}
        p^{\TWAI}_{I, \textbf{x}}(c) 
        &=
        \frac{
        \sum_{\mathbf{C}'\subset \mathbf{C}\,|\, c \in \mathbf{C}'}
        \epsilon p^{\fallback}_{I, \textbf{x}}(c)\,m^{\failsafe}_{I,\textbf{x}}\left(\mathbf{C}'\right)
        }{
        1 - (1-\epsilon) - 
        \sum_{c'\in \mathbf{C}}\sum_{\mathbf{C}'\subset (\mathbf{C}\setminus \{c'\})}
        \epsilon p^{\fallback}_{I, \textbf{x}}(c')\,
        m^{\failsafe}_{I,\textbf{x}}\left(\mathbf{C}'\right)
        }\\
        &=
        \frac{
        \sum_{\mathbf{C}'\subset \mathbf{C}\,|\, c \in \mathbf{C}'}
        p^{\fallback}_{I, \textbf{x}}(c)\,m^{\failsafe}_{I,\textbf{x}}\left(\mathbf{C}'\right)
        }{
        1 - 
        \sum_{c'\in \mathbf{C}}\sum_{\mathbf{C}'\subset (\mathbf{C}\setminus \{c'\})}
        p^{\fallback}_{I, \textbf{x}}(c')\,
        m^{\failsafe}_{I,\textbf{x}}\left(\mathbf{C}'\right)
        }\\
        &=
        \left(
        p^{\fallback}_{I, \textbf{x}} \oplus m^{\failsafe}_{I,\textbf{x}}
        \right)(c)
    \end{aligned}
\end{equation*}
In this last equality, one can observe that, however small $\epsilon>0$ can be, the trustworthy AI prediction for voxel $\textbf{x}$ does not depend anymore on the AI algorithm probability but only on the fallback algorithm probability.
In other words, we have switched totally from the backbone AI algorithm to the fallback algorithm.

\subsubsection{General case with multiple contracts of trust:}\label{sec:multi_contracts}
In general, $m^{\failsafe}_{I, \textbf{x}}$ is a sum of contracts of trust BPAs that are not completely contradictory and $m^{\failsafe}_{I, \textbf{x}}$ can be written as
\begin{equation}
    m^{\failsafe}_{I, \textbf{x}} = \bigoplus_{k=1}^K m^{(k)}_{I, \textbf{x}}
\end{equation}
where each $m^{(k)}_{I, \textbf{x}}$ is a basic probability assignment (BPA), $K$ is the number of BPAs, and $\bigoplus_{k=1}^K$ is the Dempster's rule of combination \eqref{eq:ds} of $K$ BPAs computed in any order.
$\bigoplus_{k=1}^K$ is well defined because Dempster's rule of combination $\oplus$ is associative and because the $m^{(k)}_{I, \textbf{x}}$ are not completely contradictory.
The $m^{(k)}_{I, \textbf{x}}$ represent the contracts of trust in our framework.

Specifically, for medical image segmentation we propose the following trustworthy AI model:
\begin{equation}
    \label{eq:trustworhtyAI-fetal}
    p^{\TWAI}_{I, \textbf{x}} = 
    \left((1 - \epsilon) p^{\AI}_{I, \textbf{x}} + \epsilon p^{\fallback}_{I, \textbf{x}}\right)
    \oplus m^{\anatomy}_{I, \textbf{x}}
    \oplus m^{\intensity}_{I, \textbf{x}}
\end{equation}
where
$m^{\anatomy}_{I, \textbf{x}}$ is the anatomical contract of trust BPA for voxel $\textbf{x}$ of image $I$,
and $m^{\intensity}_{I, \textbf{x}}$ is the intensity contract of trust BPA for voxel $\textbf{x}$ of image $I$.
The definitions and properties of $m^{\anatomy}_{I, \textbf{x}}$ and $m^{\intensity}_{I, \textbf{x}}$ will be derived in following two subsections of this section \nameref{sec:anatomical_contract} and \nameref{sec:intensity_contract}.

\subsubsection{Dempster-Shafer anatomical contract of trust for fetal brain segmentation}\label{sec:anatomical_contract}
In this section, we describe our proposed anatomical prior basic probability assignment (BPA) $m^{\anatomy}$ that is used in our trustworthy AI method \eqref{eq:trustworhtyAI-fetal}.

Our anatomical prior is computed using the segmentations computed using the multi-atlas segmentation algorithm~\cite{iglesias2015multi}.
Atlas-based segmentation algorithms are anatomically-constrained due to the spatial smoothness that is imposed to the spatial transformation used to compute the segmentation.
In practice, this is achieved thanks to the parameterization of the spatial transformation and the regularization loss in the registration optimization problem~\cite{cardoso2015geodesic,fidon2019incompressible,modat2010fast}.
Therefore, if implemented correctly, atlas-based automatic segmentations can inherit from the anatomical prior represented by segmentation atlases.

In terms of contract of trust, every binary segmentation mask corresponding to a specific region of interest in a fetal brain atlas is associated with an anatomical contract of trust.
Indeed, each of those binary segmentation masks represents the anatomy of a given tissue type, for a given gestational age and a given population of fetuses.
The anatomical contracts derived from atlas-based segmentation are therefore context-dependent since the segmentation masks are specific to a class, to a gestational age, and to the population of fetuses that was used to compute the atlas~\cite{fidon2021atlas,gholipour2017normative,wu2021age}.
Since only neurotypical fetal brain atlases~\cite{gholipour2017normative,wu2021age} and a spina bifida fetal brain atlas~\cite{fidon2021atlas} are available in our work, our anatomical contract of trust will hold only for neurotypical and spina bifida fetuses.

\begin{figure}[t]
    \centering
    \includegraphics[width=\linewidth]{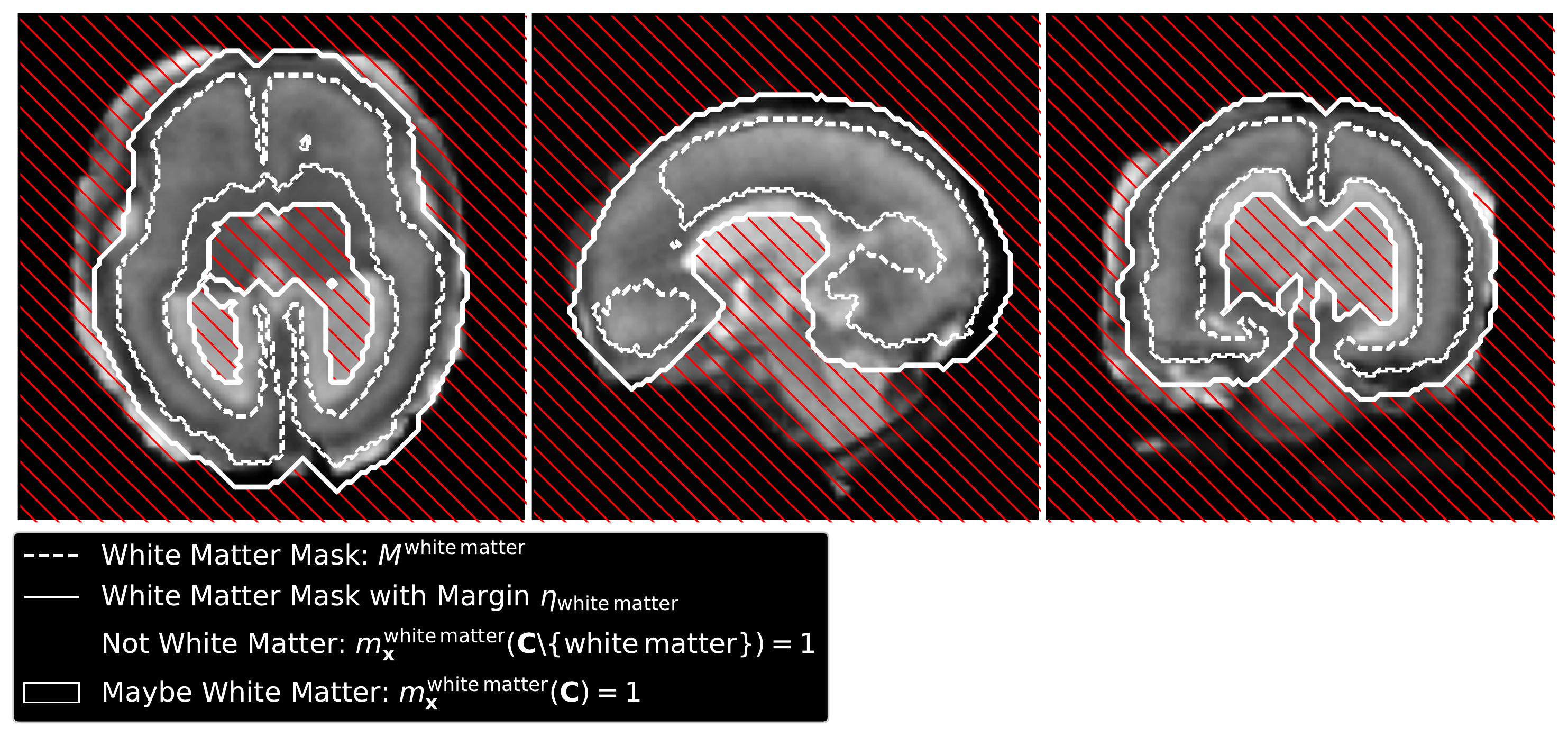}
    \caption{
    \textbf{Implementation of the anatomical BPA for the white matter of a fetus with spina bifida.}
    The white matter basic probability assignment (BPA) is computed by dilating the white matter mask of the fallback algorithm $M^{\mathrm{white\,matter}}$ using the margin $\eta_{\mathrm{white\,matter}}$.
    The margin aims at eliminating the false negative for the mask $M^{\mathrm{white\,matter}}$ and are estimated using the training dataset.
    The white matter BPA imposes that no $\mathrm{white\,matter}$ can be predicted outside the dilated mask.
    No constraint is imposed inside the dilated mask.
    The same approach is applied to all the regions of interest to be segmented.
    }
    \label{fig:margins}
\end{figure}

Due to the spatial smoothness imposed to the spatial transformation, atlas-based automatic segmentations will usually be correct up to a spatial margin.
Therefore, we propose to add spatial margins to the atlas-based segmentation to compute the BPAs of our anatomical contract of trust.
This approach is inspired by the safety margins used in radiotherapy to account for errors including spatial registration errors~\cite{niyazi2016estro}
and is illustrated in Fig.~\ref{fig:margins} for the white matter of a fetal brain 3D MRI.

Formally,
let $M^c$ a 3D (binary) mask from an atlas for class $c \in \mathbf{C}$.
We propose to define the BPA map $m^{(c)}=\left(m^{(c)}_{\textbf{x}}\right)_{\textbf{x}\in \Omega}$ associated with $M^c$ as
\begin{equation}
\label{eq:anatomical_bpa}
\forall \textbf{x},\quad
    \left\{
    \begin{aligned}
        m^{(c)}_{\textbf{x}}(\mathbf{C} \setminus \{c\}) &= 1 - \phi(d(\textbf{x}, M^c))\\
        m^{(c)}_{\textbf{x}}(\mathbf{C}) &=\phi(d(\textbf{x}, M^c))\\
    \end{aligned}
    \right.
\end{equation}
where $d(\textbf{x}, M^c)$ the Euclidean distance from any voxel position vector $\textbf{x}$ to $M^c$,
and $\phi: \mathbb{R}_{+} \xrightarrow{} [0,\, 1]$ with $\phi(0)=1$ and $\phi$ non-increasing.
In the following, we call \textbf{thresholding function} any function $\phi$ that satisfies the properties above.
We give below two examples of thresholding functions.
The first one is based on hard margins, while the second one is based on exponential soft margins.
\begin{equation}
    \begin{split}
        % a)\quad \forall d \geq 0,\quad \phi(d) &= \exp(-\eta d) \\
        1)\quad \forall d \geq 0,\quad \phi(d) &= \left\{ \begin{array}{cc}
            1 & \texttt{if } d \leq \eta \\
            0 & \texttt{otherwise }
        \end{array}\right.\\
        2)\quad \forall d \geq 0,\quad \phi(d) &= \exp\left(-\frac{d}{\eta}\right)
    \end{split}
\end{equation}
In both case, $\eta > 0$ is a hyper-parameter homogeneous to a distance and can be interpreted as a safety margin for the anatomical prior.
The BPA for the first thresholding function (hard margins) can be implemented efficiently without computing explicitly the distance between every voxel $\textbf{x}$ and the mask $M^c$.
An illustration is given in Fig.~\ref{fig:margins} for the white matter.
We describe a method to tune the margins at training time for each class at the end of this section.

With the definition of the BPA $m^{(c)}_{\textbf{x}}$ in \eqref{eq:anatomical_bpa}, we formalize the following belief: far enough from the mask $M^c$ we know for sure that the true class is not $c$, i.e. $m^{(c)}_{\textbf{x}}(\mathbf{C} \setminus \{c\}) = 1$, otherwise we do not know anything for sure regarding class $c$, i.e. $m^{(c)}_{\textbf{x}}(\mathbf{C}) > 0$.

%We now 
%%would like to
%define the anatomical prior BPA $m^{\anatomy}$ used in \eqref{eq:trustworhtyAI-fetal} as the sum for Dempster's rule of combination of all the $m_c$ across the classes.

\textbf{No contradiction between the anatomical contracts of trust:}
Following the assumption of Dempster's rule of combination in \eqref{eq:ds}, we need to make sure that the BPAs $m_c$ defined as in \eqref{eq:anatomical_bpa} are nowhere completely contradictory.

\emph{Proof:}
We show that there is always at least one class that is compatible with the set of anatomical prior.
For this we need to show that for all voxel $\textbf{x}$, there exists $c \in \mathbf{C}$ such that
$
m^{(c)}_{\textbf{x}}\left(\mathbf{C} \setminus \{c\}\right) < 1
\quad
\textup{and}
\quad
m^{(c)}_{\textbf{x}}\left(\mathbf{C}\right) > 0
$.
This holds in our case because the masks $\{M^c\}_{c \in \mathbf{C}}$ form a partition of the set of all the voxels and because, following \eqref{eq:anatomical_bpa}, for the voxels $\textbf{x}$ inside mask $M^c$, $m^{(c)}_{\textbf{x}}(\mathbf{C})=1$ and $m^{(c)}_{\textbf{x}}\left(\mathbf{C} \setminus \{c\}\right)=0$.

\textbf{Formula for the anatomical prior:}
We can now define the anatomical prior BPA
used in \eqref{eq:trustworhtyAI-fetal}
for image $I$ and voxel $\textbf{x}$ as
\begin{equation}
    \label{eq:anatomical_prior}
    m^{\anatomy}_{I, \textbf{x}} = \bigoplus_{c \in \mathbf{C}} m^{(c)}_{\textbf{x}}
\end{equation}
where $m_c$ is the BPA associated to the mask $M^c_I$ for a given thresholding function, and $M^c_I$ is the mask for class $c$ of the segmentation obtained using the multi-atlas fallback segmentation algorithm.

    The Dempster's rule of combination $\oplus$ satisfies the mathematical property of associativity.
    This is why, by convention, we use the symbol $\oplus$ for this relation here.
    In terms of implementation, the associativity implies that one can use a for loop to compute $\bigoplus_{c \in \mathbf{C}} m^{(c)}_{\textbf{x}}$.
    This also makes it possible to parallelize the computation of this quantity with respect to the contracts.
    Therefore, the computation of the composition of contracts scales in $O(log(K))$ where $K$ is the number of contracts.

We prove that for all voxel $\textbf{x}$ and for all subset of classes $\mathbf{C}' \subset \mathbf{C}$,
the anatomical BPA mass that the true label of $\textbf{x}$ is not in $\mathbf{C}'$ is equal to
\begin{equation}
    \label{eq:anatomical_BPA}
    m^{\anatomy}_{I, \textbf{x}}(\mathbf{C}\setminus \mathbf{C}') =
    \prod_{c\in \mathbf{C}}
    \left(
        \delta_{c}(\mathbf{C}')m^{(c)}_{\textbf{x}}(\mathbf{C} \setminus \{c\}) 
        + (1 - \delta_{c}(\mathbf{C}'))m^{(c)}_{\textbf{x}}(\mathbf{C})
    \right)
\end{equation}
where for all $c \in \mathbf{C}$, $\delta_c$ is the Dirac measure associated with $c$ defined as
\begin{equation}
    \forall \mathbf{C}' \subset \mathbf{C}, \quad \delta_c(\mathbf{C}') =
    \left\{
    \begin{array}{cc}
        1 & \textup{if}\,\, c \in \mathbf{C}'\\
        0 & \textup{if}\,\, c \not \in \mathbf{C}'
    \end{array}
    \right.
\end{equation}
The proof of \eqref{eq:anatomical_BPA} can be found in the Appendix.
% Note that this is a special case of discounted coefficient, as used in Dempster-Shafer theory literature~\cite{todo}.

In practice, we are particularly interested in summing the anatomical prior BPA with probabilities using the particular case of Dempster's rule in \eqref{eq:drc_proba}.
Let $\textbf{x}$ a voxel and $p_{I,\textbf{x}}$ a probability on $\mathbf{C}$ for voxel $\textbf{x}$ of image $I$ that is not completely contradictory with $m^{\anatomy}_{I,\textbf{x}}$.
For all $c \in \mathbf{C}$, we can show that
\begin{equation}
    \label{eq:DRC_proba_anatomical_BPA}
    \left(
    p_{I,\textbf{x}} \oplus m^{\anatomy}_{I,\textbf{x}}
    \right)
    \left(c\right) =
    \frac{p_{I,\textbf{x}}(c)m^{(c)}_{\textbf{x}}(\mathbf{C})}{\sum_{c'\in \mathbf{C}}p_{I,\textbf{x}}(c')m^{(c')}_{\textbf{x}}(\mathbf{C})}
\end{equation}
A proof of this equality can be found in the Appendix.
It is worth noting that, due to the specific form of $m^{\anatomy}_{I,\textbf{x}}$ and because $p_{I,\textbf{x}}$ is a probability, the computational cost of $p_{I,\textbf{x}} \oplus m^{\anatomy}_{I,\textbf{x}}$ is $\mathcal{O}(\mathbf{C})$ even though there are $2^{|\mathbf{C}|}$ elements in $2^{\mathbf{C}}$.

Another important remark is that when $p_{I,\textbf{x}}$ is completely contradictory with $m^{\anatomy}_{I,\textbf{x}}$, we have
$\sum_{c'\in \mathbf{C}}p_{I,\textbf{x}}(c')m^{(c')}_{\textbf{x}}(\mathbf{C})=0$.

\textbf{No contradiction between the anatomical prior and the fallback:}
For our trustworthy AI model \eqref{eq:trustworhtyAI-fetal} to be valid, we need to show that the probability
$\left((1 - \epsilon) p^{\AI}_{I, \textbf{x}} + \epsilon p^{\fallback}_{I, \textbf{x}}\right)$, 
is not completely contradictory with the anatomical prior BPA.

\emph{Proof:} Since $\epsilon > 0$, it is sufficient to show that $p^{\fallback}_{I, \textbf{x}}$ is not completely contradictory with $m^{\anatomy}_{I, \textbf{x}}$ for every voxel $\textbf{x}$.
In \eqref{eq:anatomical_bpa} and \eqref{eq:anatomical_prior} we have defined the BPA maps $m_{I,c}$ based on the multi-atlas segmentation which is equal to $p^{\fallback}_{I}$.
Therefore, for any voxel $\textbf{x}$, let $c \in \mathbf{C}$ the class such that $\textbf{x} \in M^c$. 
We have $d(\textbf{x}, M^c)=0$ and therefore $m_{I,\textbf{x}}(\mathbf{C})=\phi(d(\textbf{x}, M^c))=1$.
And we have $p^{\fallback}_{I, \textbf{x}}(c)>0$.
This shows that $p^{\fallback}_{I, \textbf{x}}$ and $m^{\anatomy}_{I, \textbf{x}}$ are not completely contradictory.

\begin{figure}
    \centering
    \includegraphics[width=0.95\linewidth]{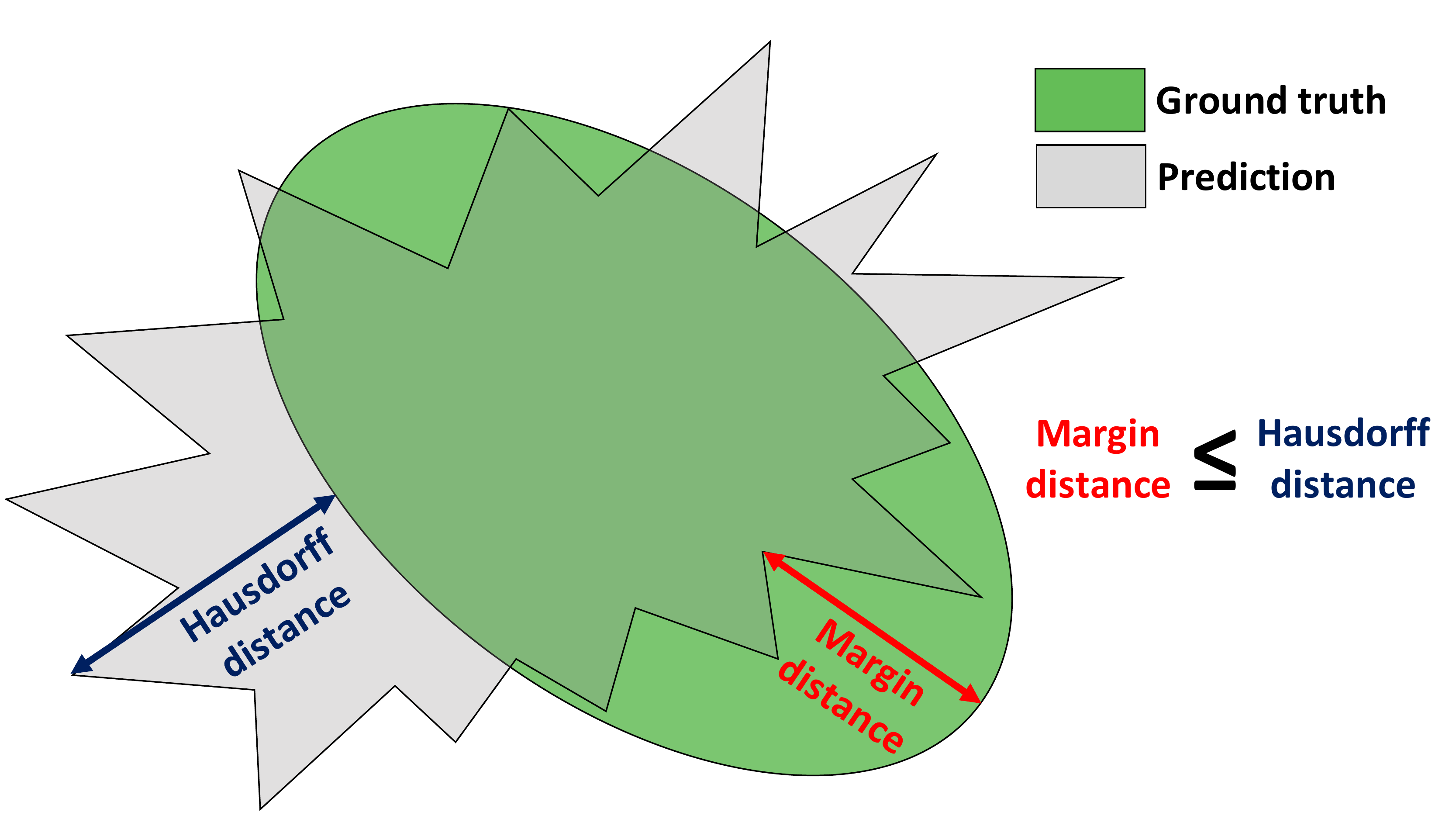}
    \caption{
    \textbf{Illustration of the margin distance.}
    The margin distance is the minimal dilation radius to apply to the predicted binary mask so that it covers entirely the ground-truth binary mask.
    We have proposed to use the margin distance to define our margins used in our definition of the anatomical BPAs.
    }
    \label{fig:margins_definition}
\end{figure}

\textbf{Tuning the margins:}
The margins $\eta$ were tuned for each class and each condition independently using the 3D MRIs of the fold 0 of the training dataset.

In our definition of the anatomical prior BPAs \eqref{eq:anatomical_bpa}, the margins account for false negatives in the multi-atlas segmentation.
The anatomical prior BPAs for a class $c$ will impose to the probabilities of class $c$ to be zeros for every voxel outside of the mask after adding the margins using dilation.
Therefore, we chose the margin for a given class $c$ to be the minimal dilation radius for the dilated mask to cover entirely the true region of class $c$ even if it creates overlaps with other regions.

For this purpose, we propose to use a modified Hausdorff distance, called \textit{margin distance}, that considers only the false negatives.
An illustration is given in Fig.~\ref{fig:margins_definition}.
Let $\textup{HD}_{95}(M_{pred}, M_{gt})$ denotes the Hausdorff distance at $95\%$ of percentile between a predicted binary mask $M_{pred}$ and the ground-truth mask $M_{gt}$.
The margin distance of interest between $M_{pred}$ and $M_{gt}$ is defined as
\begin{equation}
    \textup{HD}^{FN}_{95}(M_{pred}, M_{gt}) = \textup{HD}_{95}(M_{pred}, M_{pred} \cup M_{gt})
\end{equation}

The margin $\eta_{c,cond}$ for class $c\in\mathbf{C}$ and condition $cond$ (neurotypical or spina bifida) is chosen as the $95\%$ percentile value of $\textup{HD}^{FN}_{95}$ on the fold 0 of the training dataset for the given class and condition.

For fetuses with a condition other than neurotypical or spina bifida, we chose
$\eta_{c,other\,pathologies}= \max\{\eta_{c,neurotypical},\,\eta_{c,spina\,bifida}\}$.

\subsubsection{Dempster-Shafer intensity-based contract of trust for fetal brain segmentation}\label{sec:intensity_contract}
In this section, we describe our proposed intensity prior basic probability assignment (BPA) $m^{\intensity}$ that is used in our trustworthy AI method for fetal brain 3D MRI segmentation \eqref{eq:trustworhtyAI-fetal}.

In T2-weighted MRI, which is the 
%modality of MRI data
MRI modality
used in this work, it is known that the hyper-intense voxels inside the brain are highly likely to be part of the cerebrospinal fluid (CSF).
We therefore propose to model this intensity prior about high intensities as a contract of trust.

Regarding hypo-intense voxels, it is unclear how to derive similar prior because even the CSF classes contain hypo-intense voxels, such as the choroid plexus for the intra-axial CSF class and the vein of Galena and straight sinus for the extra-axial CSF class~\cite{payette2021automatic}.
% 
% This is the case for regions between the extra-axial CSF of the two hemispheres in the posterior part of the brain and are included in the extra-axial CSF class~\cite{payette2021automatic}.
% 
It is also worth noting that voxels outside the brain (\emph{background} class) can also be hyper-intense.

\textbf{Formula for the intensity BPA:}
Let $\mathbf{C}_{high}\subset \mathbf{C}$ be the subset of classes that contain all the classes that partition the entire CSF (intra-axial CSF and extra-axial CSF) and the background.
Let $I=\{I_{\textbf{x}}\}_{\textbf{x} \in \Omega}$ be the volume and $\Omega$ the volume domain of a fetal brain 3D MRI.
We propose to fit a Gaussian mixture model (GMM) with two components to the image intensity distribution of $I$.
The two components of parameters $(\mu_{high},\,\sigma_{high})$ and $(\mu_{low},\,\sigma_{low})$ are associated to high and low intensities.
We propose to define the intensity prior BPA for all voxels $\forall \textbf{x}$, up to a normalization factor, as
\begin{equation}
    \left\{
    \begin{aligned}
        m^{\intensity}_{I,\textbf{x}}(\mathbf{C}_{high}) &\propto \frac{1}{\sigma_{high}} \exp \left(
            \frac{1}{2} \left(\frac{I_{\textbf{x}} - \mu_{high}}{\sigma_{high}}\right)^2
        \right)\\
        m^{\intensity}_{I,\textbf{x}}(\mathbf{C}) &\propto \frac{1}{\sigma_{low}} \exp \left(
            \frac{1}{2} \left(\frac{I_{\textbf{x}} - \mu_{low}}{\sigma_{low}}\right)^2
        \right)\\
    \end{aligned}
    \right.
\end{equation}
It is worth noting that $m^{\intensity}_{I,\textbf{x}}(\mathbf{C}) > 0$.
Therefore, no probability will be set to $0$ using the Dempster's rule of combination with $m^{\intensity}$. In other words, $m^{\intensity}$ does not forbid any assignment.
This is in contrast with the anatomical BPAs defined in \nameref{sec:anatomical_contract}.

Let $\textbf{x}$ a voxel and $p_{I,\textbf{x}}$ a probability on $\mathbf{C}$ for voxel $\textbf{x}$ of image $I$.
Since $m^{\intensity}_{I,\textbf{x}}(\mathbf{C}) >0$, $p_{I,\textbf{x}}$ is not completely contradictory with $m^{\intensity}_{I,\textbf{x}}$.
Using Dempster's rule, we have, for all class $c \in \mathbf{C}$
\begin{equation}
    \left(p_{I,\textbf{x}} \oplus m^{\intensity}_{I,\textbf{x}}\right)(c) \propto
    \left\{
    \begin{aligned}
        \left(m^{\intensity}_{I,\textbf{x}}(\mathbf{C}_{high}) + m^{\intensity}_{I,\textbf{x}}(\mathbf{C})\right) p_{I,\textbf{x}}(c) & \,\,\textup{if}\,\, c \in \mathbf{C}_{high}\\
        m^{\intensity}_{I,\textbf{x}}(\mathbf{C})\, p_{I,\textbf{x}}(c) & \,\,\textup{otherwise}\\
    \end{aligned}
    \right.
\end{equation}
This can be interpreted as a soft-thresholding operation
\begin{equation}
    \left(p_{I,\textbf{x}} \oplus m^{\intensity}_{I,\textbf{x}}\right)(c) \propto
    \left\{
    \begin{aligned}
        \left(
        1 + \frac{m^{\intensity}_{I,\textbf{x}}(\mathbf{C}_{high})}{m^{\intensity}_{I,\textbf{x}}(\mathbf{C})}
        \right) p_{I,\textbf{x}}(c) & \,\,\textup{if}\,\, c \in \mathbf{C}_{high}\\
        p_{I,\textbf{x}}(c) & \,\,\textup{otherwise}\\
    \end{aligned}
    \right.
\end{equation}
Thus, only the probabilities for the background and CSF classes in $\mathbf{C}_{high}$ are increased in the case of a voxel $\textbf{x}$ with relatively high intensity.
In particular, the probabilities remain approximately unchanged  for a voxel $\textbf{x}$ with relatively low or medium intensity.
This reflects the fact that the background and CSF classes also contain hypo-intense voxels.
The hyper-intense voxels must be in $\mathbf{C}_{high}$ while we ca not say anything about hypo-intense voxels in general. There are hypo-intense voxels in every class.

\subsubsection{nnU-Net as backbone AI segmentation algorithm}\label{sec:nnunet}
The AI segmentation algorithm used is based on nnU-Net~\cite{isensee2021nnu} which is a state-of-the-art deep learning-based method for medical image segmentation.
We have chosen the nnU-Net deep learning pipeline because it has lead to state-of-the-art results on several segmentation challenge, including the FeTA challenge 2021 for automatic fetal brain 3D MRI segmentation~\cite{fidon2021partial,payette2021automatic}. 
We have used the code available at \url{https://github.com/MIC-DKFZ/nnUNet} without modification
for our backbone AI.

\textbf{Deep learning pipeline:} The nnU-Net pipeline is based on a set of heuristics to automatically select the deep neural network architecture and other training hyper-parameters such as the patch size.
In this work, a 3D U-Net~\cite{cciccek20163d} was selected with one input block, $4$ down-sampling blocks, one bottleneck block, $5$ upsampling blocks, $32$ features in the first level, instance normalization~\cite{ulyanov2016instance}, and the leaky-ReLU activation function with slope $0.01$.
This 3D U-Net has a total of $31.2$M trainable parameters.
The patch size selected is $96 \times 112 \times 96$ voxels.

\textbf{Preprocessing:}
We have used the same pre-processing as in our previous work~\cite{fidon2021partial}.
The 3D MRIs are skull-stripped using the brain mask after applying a dilation operation ($3$ iterations using a structuring element with a square connectivity equal to one) and setting the values outside the dilated brain mask to $0$.
The brain masks are all computed automatically either during the 3D reconstruction for the Leuven data using \texttt{NiftyMIC}~\cite{ebner2020automated,martaisbi2021}, or for the other data using a multi-atlas segmentation method based on affine registration~\cite{modat2014global} and three fetal brain atlases~\cite{fidon2021atlas,gholipour2017normative,wu2021age}.
The intensity values inside the dilated brain mask are clipped to the percentile values at $0.5\%$ and $99.5\%$, and after clipping the intensity values inside the dilated brain mask are normalized to zero mean and unit variance.

\textbf{Training:} the training dataset is split at random into $5$ folds.
In total, five 3D U-Nets are trained with one for each possible combination of $4$ folds for training and $1$ fold for validation.
The AI segmentation algorithm consists of the ensemble of those five 3D U-Nets.
Each 3D U-Net is initialized at random using He initialization~\cite{he2015delving}.
The loss function consists of the sum of the Dice loss and the cross entropy loss. Stochastic gradient descent with Nesterov momentum is used to minimize the empirical mean loss on the training dataset, with batch size $4$, weight decay $3\times 10^{-5}$, initial learning rate $0.01$, deep supervision on $4$ levels, and polynomial learning rate decay with power $0.9$ for a total of $250{,}000$ training iterations.
The data augmentation methods used are: random cropping of a patch, random zoom, gamma intensity augmentation, multiplicative brightness, random rotations, random mirroring along all axes, contrast augmentation, additive Gaussian noise, Gaussian blurring, and simulation of low resolution.
For more implementation details, we refer the interested reader to~\cite{isensee2021nnu} and the nnU-Net \href{https://github.com/MIC-DKFZ/nnUNet}{GitHub page}.

\textbf{Inference:} The probabilistic segmentation prediction of the AI segmentation algorithm is the average of the five probabilistic segmentation prediction of the five 3D U-Nets after training.
In addition, for each 3D U-Net, test-time data augmentation with flip around the $3$ spatial axis is performed.

\subsubsection{Multi-atlas segmentation as fallback segmentation algorithm}\label{sec:fallback}
The fallback segmentation algorithm that we propose to use is based on a multi-atlas segmentation approach.
Multi-atlas segmentation~\cite{iglesias2015multi} is one of the most trustworthy approaches for medical image segmentation in terms of anatomical plausibility.
The multi-atlas segmentation that we use is inspired by the Geodesic Information Flows method (GIF)~\cite{cardoso2015geodesic}, which is a state-of-the-art multi-atlas segmentation algorithm.

In this section, we give details about the three main steps of the multi-atlas segmentation algorithm used.
First, the selection of the atlas volumes to use to compute the automatic segmentation.
Second, the non-linear registration algorithm to propagate each atlas segmentation to the 3D MRI to be segmented.
And third, the fusion method used to combine the propagated segmentations from the atlas volumes. 

\textbf{Atlas volumes selection:}
We used the volumes from two neurotypical fetal brain 3D MRI atlases~\cite{gholipour2017normative,wu2021age} and one spina bifida fetal brain 3D MRI atlas~\cite{fidon2021atlas}.
Let GA be the gestational age rounded to the closest number of weeks of the 3D MRI to be segmented. We select all the atlas volumes with a gestation age in the interval $[\textup{GA} - \Delta \textup{GA},\textup{GA} + \Delta \textup{GA}]$ with $\Delta \textup{GA}=1$ week for the neurotypical fetuses and $\Delta \textup{GA}=3$ for spina bifida fetuses.
This way approximately the same number of atlas volumes are used for neurotypical and spina bifida fetuses.

\textbf{Non-linar registration:}
Our image registration step aims at spatially aligning the selected atlas volumes with the 3D MRI to be segmented.
We used \href{https://github.com/KCL-BMEIS/niftyreg}{\texttt{NiftyReg}}~\cite{modat2010fast} to compute the non-linear image registrations.
The non-linear image registration optimization problem is the following
\begin{equation}
\label{eq:reg-non-linear}
\left\{
    \begin{aligned}
        &\min_{\Theta} \,\, \mathcal{L}(I_{subject},\, I_{atlas},\, \phi(\Theta)) + R(\Theta)\\
        & R(\Theta) = \alpha_{BE} BE(\phi(\Theta)) + \alpha_{LE} LE(\phi(\Theta))
    \end{aligned}
\right.
\end{equation}
where $I_{atlas}$ is the segmented atlas volume to be registered to the 3D reconstructed MRI $I_{subject}$  that we aim to segment, $\phi(\Theta)$ is a spatial transformation parameterized by cubic B-splines of parameters $\Theta$ with a grid size of $4$ mm.
The data term $\mathcal{L}$ is the local normalized cross correlation (LNCC) with the standard deviation of the Gaussian kernel of the LNCC was set to $6$ mm.
The regularization term $R$ is a linear combination of the bending energy (BE) and the linear energy (LE) regularization functions applied to $\phi(\Theta)$ with $\alpha_{BE}=0.1$ and $\alpha_{LE}=0.3$.

Prior to the non-linear registration, the brain mask of $I_{subject}$ was used to mask the voxels outside the brain and $I_{atlas}$ was registered to $I_{subject}$ using an affine transformation.
The affine transformation was computed using a symmetric block-matching approach~\cite{modat2014global} based on image intensities and the brain masks.
The optimization is performed using conjugate gradient descent and a pyramidal approach with $3$ levels~\cite{modat2010fast}.
% 
% The pyramidal optimization aims at making the non-linear registration more robust~\cite{modat2010fast}.
% % 
% In the first (resp. second) level, the images are downsampled two (resp. one) times and the grid space is multiplied by $4$ (resp. $2$).
% % 
% The optimization for each level is stopped either when the loss function \eqref{eq:reg-non-linear} does not decrease anymore or when the maximum number of iterations is reached.
% % 
% The transformation $\phi(\Theta)$ obtained at the end of a level is used as an initialization for the next level.
% 
The hyper-parameters for the non-linear registration were chosen to be the same as in a recent registration pipeline to compute a fetal brain atlas~\cite{fidon2021atlas}.

\textbf{Segmentations fusion:}
Once all the atlas volumes $\{I_{k}\}_{k=1}^K$ and their probabilistic segmentations $\{S_{k}\}_{k=1}^K$ have been registered to the 3D MRI to be segmented $I_{subject}$ using the tranformations $\{\phi_{k}\}_{k=1}^K$, we need to fuse the aligned segmentations $\{S_{k}\circ \phi_k\}_{k=1}^K$ into one segmentation.
This fusion is computed via a voxel-wise weighted average using heat kernels~\cite{cardoso2015geodesic}.

The heat map for atlas $k$ at voxel $\textbf{x}$ is defined as~\cite{cardoso2015geodesic}
\begin{equation}
    w_k(\textbf{x}) = \exp\left(-(D(k, \textbf{x}))^2\right)
\end{equation}
where $D(k, \textbf{x})$ is a surrogate of the morphological similarity between $I_{subject}$ and $I_{k}\circ \phi_k$ at voxel $\textbf{x}$.
The distance $D(k, \textbf{x})$ is the sum of two components
\begin{equation}
    D(k, \textbf{x}) = 
    \alpha L(I_{subject}, I_{k}\circ \phi_k)(\textbf{x})
    + (1 - \alpha) F(\phi_k)(\textbf{x})
\end{equation}
with $\alpha=0.5$, 
$L(I_{subject}, I_{k}\circ \phi_k) = B * \left(I_{subject} - (I_{k}\circ \phi_k)\right)^2$ the local sum of squared differences convoluted (convolution operator $*$) by a B-spline kernel $B$ of order $3$,
and $F(\phi_k)(\textbf{x})$ the Euclidean norm of the displacement field at voxel $\textbf{x}$ (in mm) after removing the low spatial frequencies of $\phi_k$ using a Gaussian kernel with a standard deviation of $20$ mm.
The hyper-parameters chosen are the same as in GIF~\cite{cardoso2015geodesic}.
Before computing $D$, the intensity values of the images are normalized to zero mean and unit variance inside the brain mask.

The multi-atlas segmentation at voxel $\textbf{x}$ is computed using the heat kernels as
\begin{equation}
    S_{multi-atlas}(\textbf{x}) = \frac{\sum_{k=1}^K w_k(\textbf{x}) (S_{k}\circ \phi_k)(\textbf{x})}{\sum_{k=1}^K w_k(\textbf{x})}
\end{equation}

Our implementation is available at \codetwai{}.

\textbf{Hyper-parameters tuning:}
The hyper-parameters that we tuned are $\Delta\textup{GA}$, the selection strategy for the atlas volumes, and the fusion strategy for combining the probabilistic segmentation of the atlas volumes after non-linear registration.
For $\Delta\textup{GA}$ we tried the values $\{0,1,2,3,4\}$.
For the selection strategy we compared the condition-specific strategy described above to using all the atlases irrespective of the condition of the fetus.
And for the fusion strategy we compared the GIF-like fusion strategy described in above to a simple average.

The data used for the selection of the hyper-parameters are the training data of the first fold that was used for the training of the AI segmentation algorithm.
The mean Dice score across all the segmentation classes and the number of volumes to register were used as our selection criteria to find a trade-off between segmentation accuracy and computational time.
The results can be found in the appendix.
The approach selected consists of GIF-like atlas segmentations fusions, condition specific atlas selection and $\Delta \textup{GA}=1$ for the neurotypical condition and $\Delta \textup{GA}=3$ for the spina bifida condition.

\subsection{Human expert scoring method for evaluating the trustworthiness of fetal brain 3D MRI segmentation}\label{sec:scoring_protocol}
The trustworthiness scoring is done for each of the tissue types: white matter, intra-axial CSF, cerebellum, extra-axial CSF, cortical gray matter, deep gray matter, and brainstem.
% and one criteria for the labeling of the hyper-intense voxels as either intra-axial CSF, extra-axial CSF, or background.

The evaluation is performed using a Likert scale ranging from $0$ star to $5$ stars to answer the question \textit{"Is the automatic segmentation of the tissue type X trustworthy?"}:
%\begin{enumerate}[label=\arabic*$\star$]
\begin{description}
    \item[\faStarO\faStarO\faStarO\faStarO\faStarO] Strongly disagree / there are several severe violations of the anatomy that are totally unacceptable
    \item[\faStar\faStarO\faStarO\faStarO\faStarO] Disagree / there is one severe violation of the anatomy that is totally unacceptable
    \item[\faStar\faStar\faStarO\faStarO\faStarO] Moderately disagree / there are violations of the anatomy that make the acceptability of the segmentation questionable
    \item[\faStar\faStar\faStar\faStarO\faStarO] Moderately agree / there are many minor violations of the anatomy that are acceptable
    \item[\faStar\faStar\faStar\faStar\faStarO] Agree / there are a few minor violations of the anatomy that are acceptable
    \item[\faStar\faStar\faStar\faStar\faStar] Strongly agree / perfect fit of the anatomy
\end{description}

This evaluation is performed on $50$ 3D MRIs from the FeTA dataset. We selected at random $20$ neurotypical cases, $20$ spina bifida cases, and $10$ cases with other brain pathologies.

The scoring was performed independently by \numscoring{} individuals or groups of expert raters: 
MA, paediatric radiologist at \uzl{} with several years of experience in manual segmentation of fetal brain MRI;
AJ, MD and group leader at University Children’s Hospital Zurich;
AB, professor of neuroradiology at University Hospital Zurich;
% AE, paediatric radiologist and neuroradiologist at \kcl{};
and by MS and PK jointly, two MDs at \vienna{} (\viennashort{}) with more than $400$ hours of experience in manual segmentation of fetal brain MRI,
under the supervision of 3 experts:
DP, professor of radiology at \viennashort{},
GK, professor of paediatric radiology at \viennashort{},
and IP, neuroradiologist at \viennashort{}.

The human expert raters were given access to the 3D MRIs, the backbone AI segmentation, the fallback segmentation, the trustworthy AI segmentation, and the ground-truth manual segmentation for each case.
The segmentation algorithms are anonymized for the raters by assigning to each of them a number from 1 to 3.
The assignment of numbers to the segmentation methods was performed randomly for each case independently, i.e. we used a different random assignment for each case.

The segmentation protocol that was used for the manual segmentations can be found at \url{http://neuroimaging.ch/feta}.
An empty csv file \texttt{ranking.csv} is made available to each rater to complete.

\section{Code Availability}
The code for our trustworthy AI segmentation algorithm of fetal brain 3D MRI is publicly available at
\codetwai{}.
It included our implementation of the fallback algorithm.
For medical image registration we have used the \texttt{NiftyReg} software that is publicly available at \url{https://github.com/KCL-BMEIS/niftyreg}.
For the backbone AI algorithm we have used the nnU-Net software that is publicly available at \url{https://github.com/MIC-DKFZ/nnUNet}.
nnU-Net is based on the PyTorch deep learning library and we have used PyTorch version $1.10$.

We have used python $3.8.10$ for all experiments.
The plots in Fig.\ref{fig:twai-results},~\ref{fig:twai-dice_GA},~\ref{fig:twai-hausdorff_GA}
and Fig.~\ref{fig:twai-dice_GA_roi},~\ref{fig:twai-hausdorff_GA_roi}, including the confidence intervals, were computed using seaborn version $0.11.2$.
The statistical test were computed using scipy version $1.6.3$.

\section{Results}

\begin{figure}[htb]
    \centering
    \includegraphics[width=\linewidth]{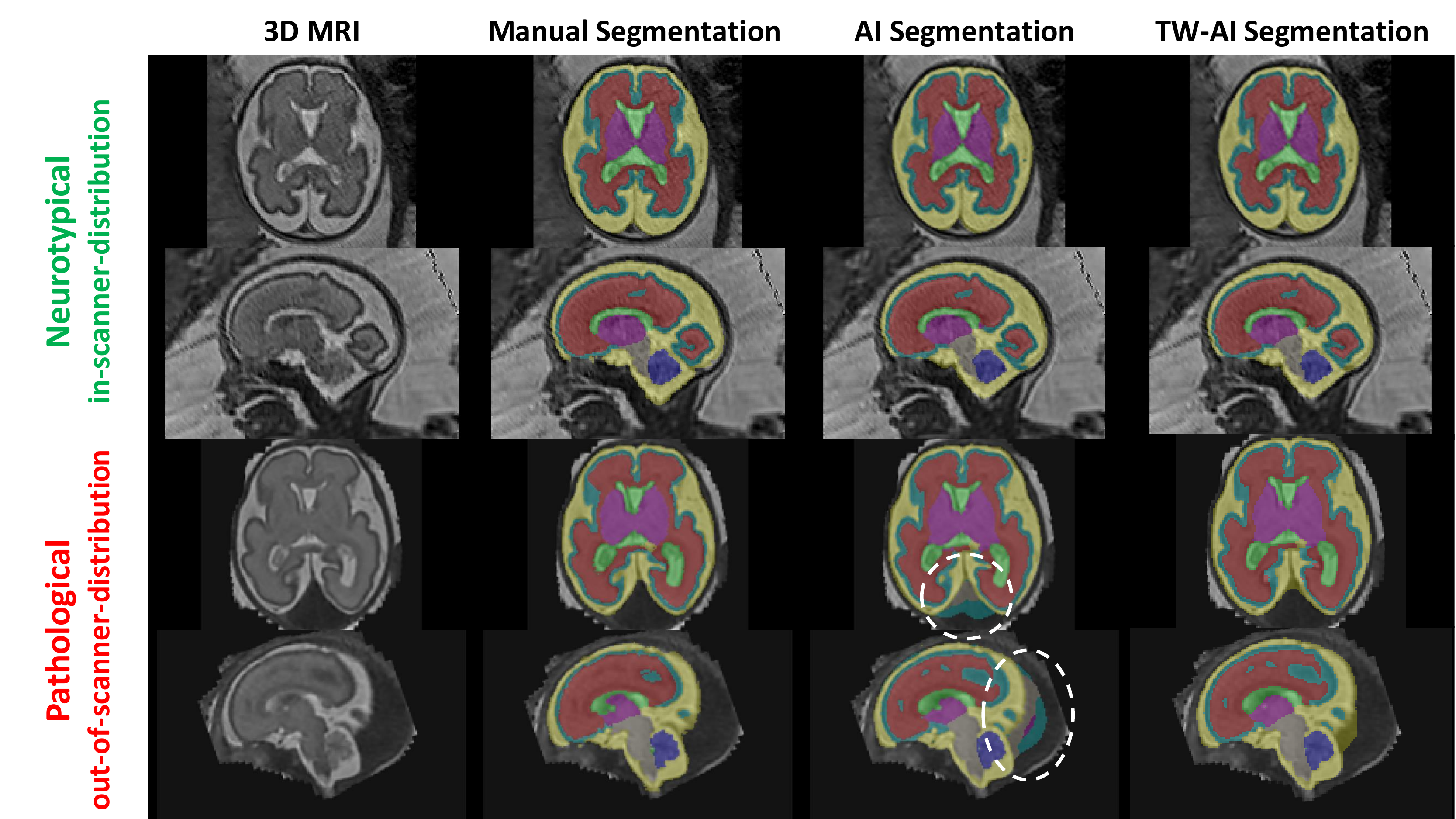}
    \caption{
        Illustration of the improved robustness of the proposed trustworthy AI method (TW-AI) as compared to a state-of-the-art AI method. (Top) 3D MRI of a neurotypical fetus at 28 weeks of gestation acquired at the same center as the training data for the AI. (Bottom) 3D MRI of a fetus with a high-flow dural sinus malformation at 28 weeks of gestation acquired at a different center as the training data for the AI. Severe violations of the anatomy by the AI, but not the TW-AI, are highlighted in white.
    }
    \label{fig:twai-qualitative_results}
\end{figure}

\begin{figure}[htb]
    \centering
    \begin{subfigure}[t]{0.49\linewidth}
        \includegraphics[width=0.98\textwidth,trim=0cm 0cm 0cm 4cm,clip]{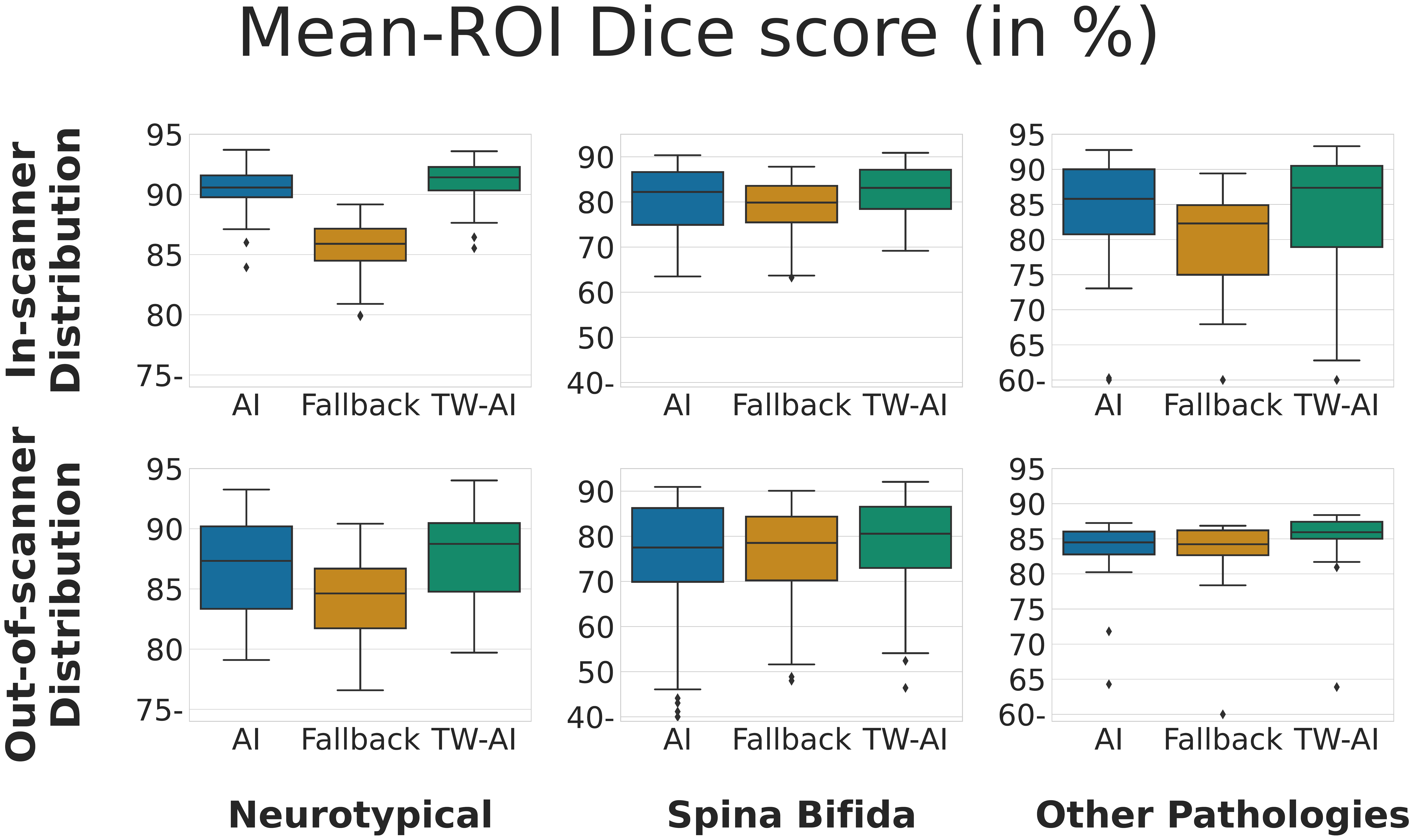}
        \caption{\textbf{Mean-ROI Dice Score (in \%)}}
        \label{fig:twai-dice}
    \end{subfigure}
    \begin{subfigure}[t]{0.49\linewidth}
        \includegraphics[width=0.98\textwidth,trim=0cm 0cm 0cm 4cm,clip]{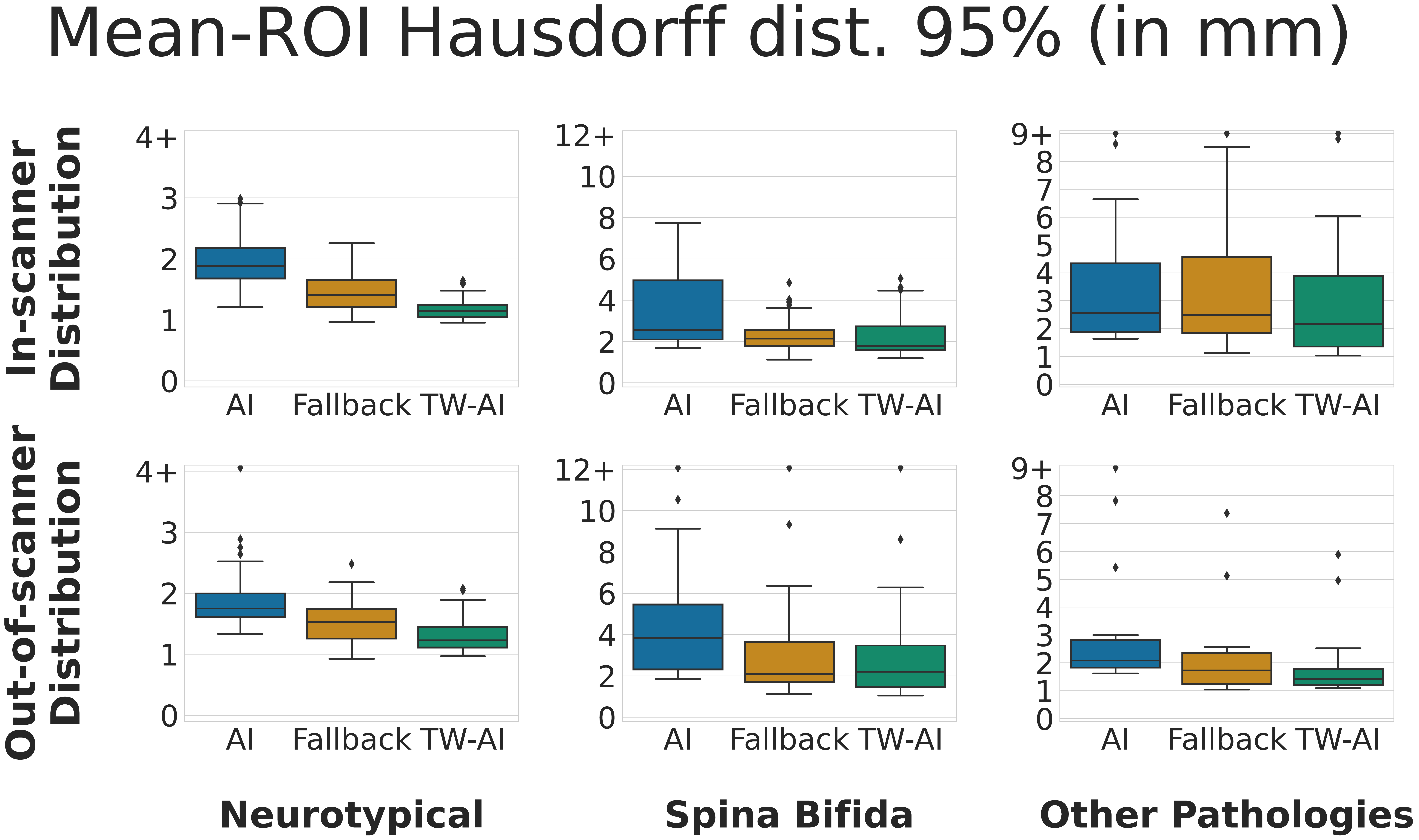}
        \caption{\textbf{Mean-ROI HD95 (in mm)}}
        \label{fig:twai-hausdorff}
    \end{subfigure}
    \hfill
    \begin{subfigure}[t]{0.49\linewidth}
        \includegraphics[width=0.98\textwidth,trim=0cm 0cm 0cm 3cm,clip]{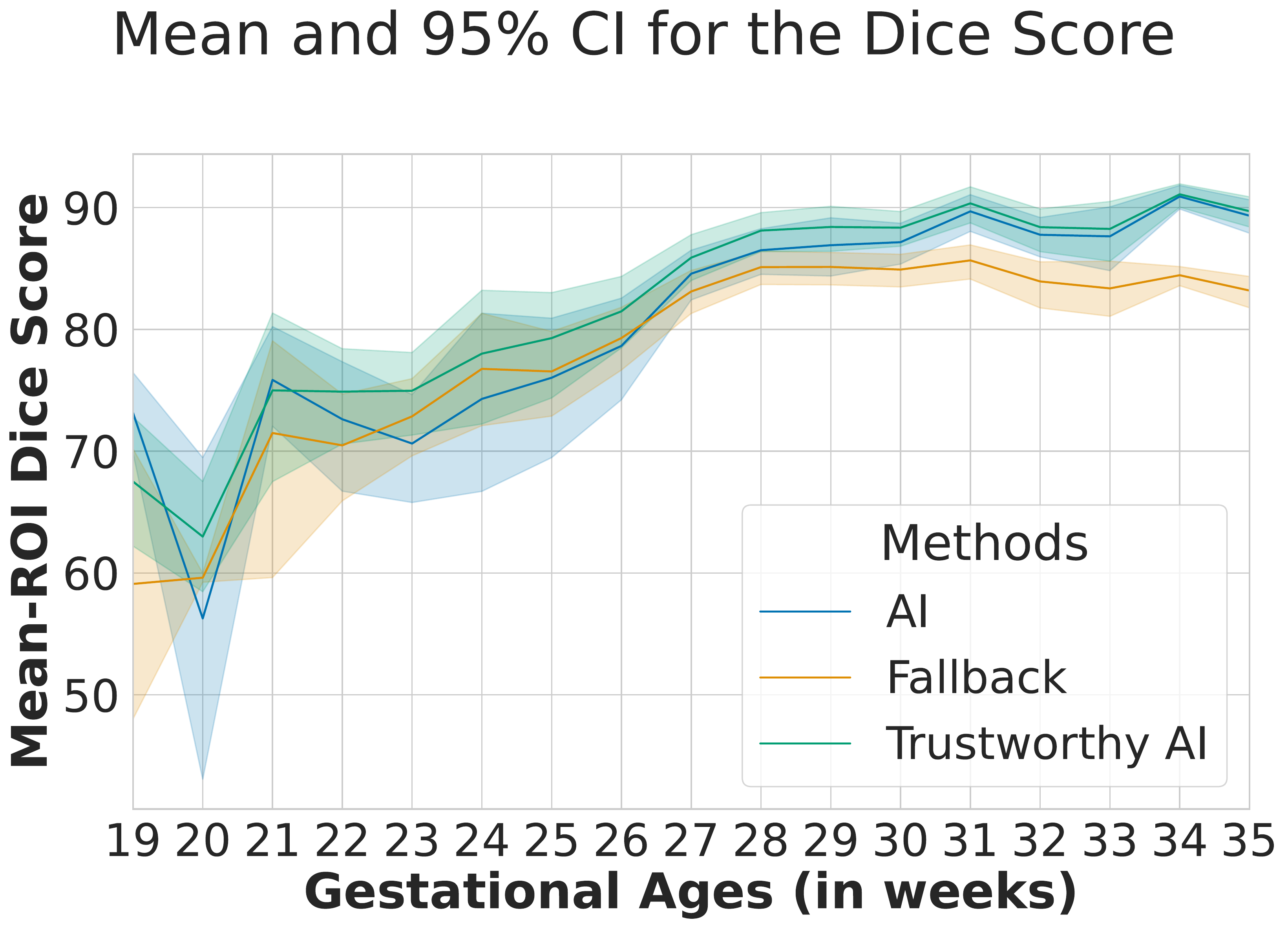}
        \caption{\textbf{Mean and $95\%$ CI of Dice Scores}}
        \label{fig:twai-dice_GA}
    \end{subfigure}
    \begin{subfigure}[t]{0.49\linewidth}
        \includegraphics[width=\textwidth,trim=0cm 0cm 0cm 3cm,clip]{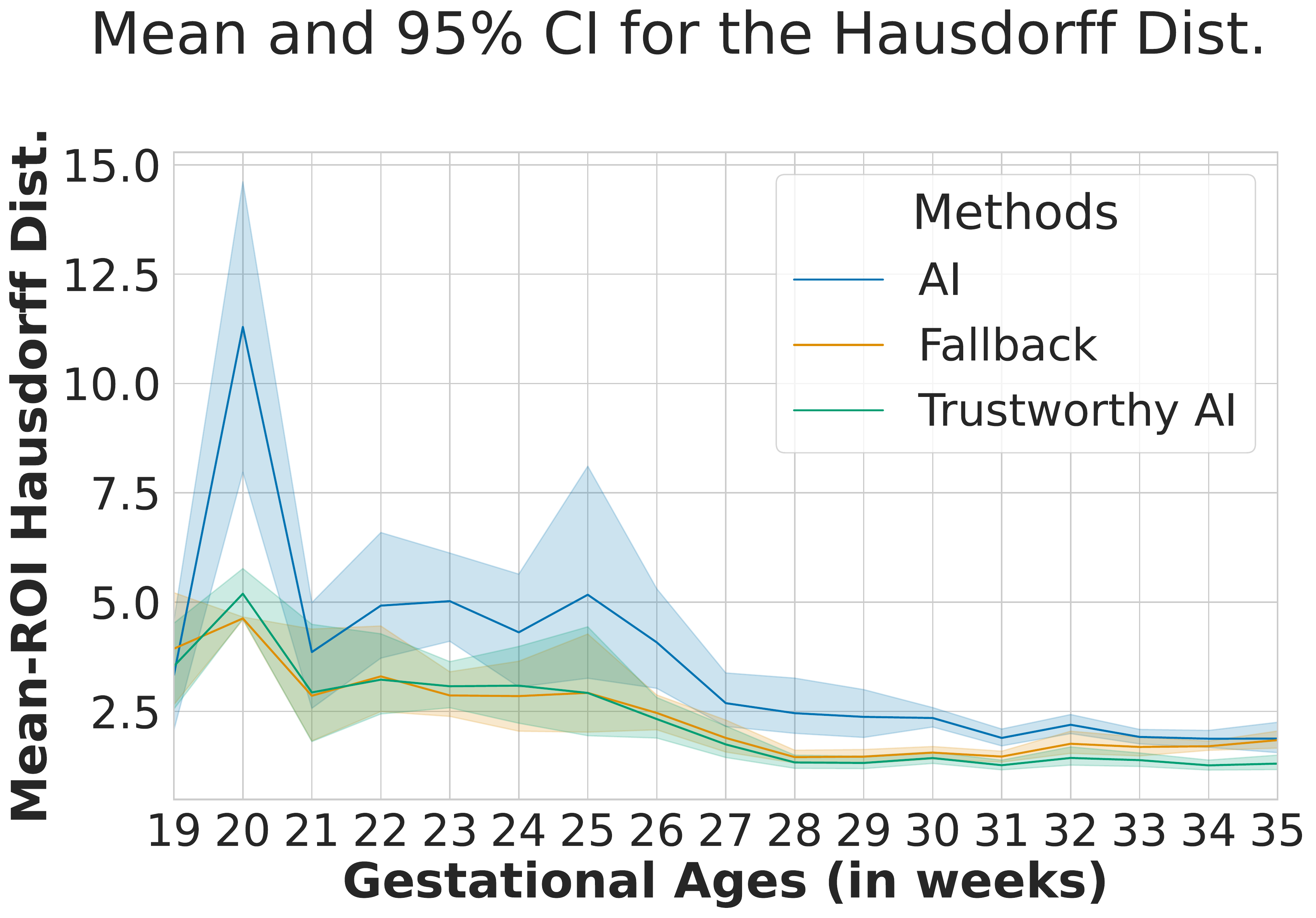}
        \caption{\textbf{Mean and $95\%$ CI of HD95}}
        \label{fig:twai-hausdorff_GA}
    \end{subfigure}
    \hfill
    \begin{subfigure}[t]{0.98\linewidth}
        \includegraphics[width=\textwidth,trim=0cm 0cm 0cm 3cm,clip]{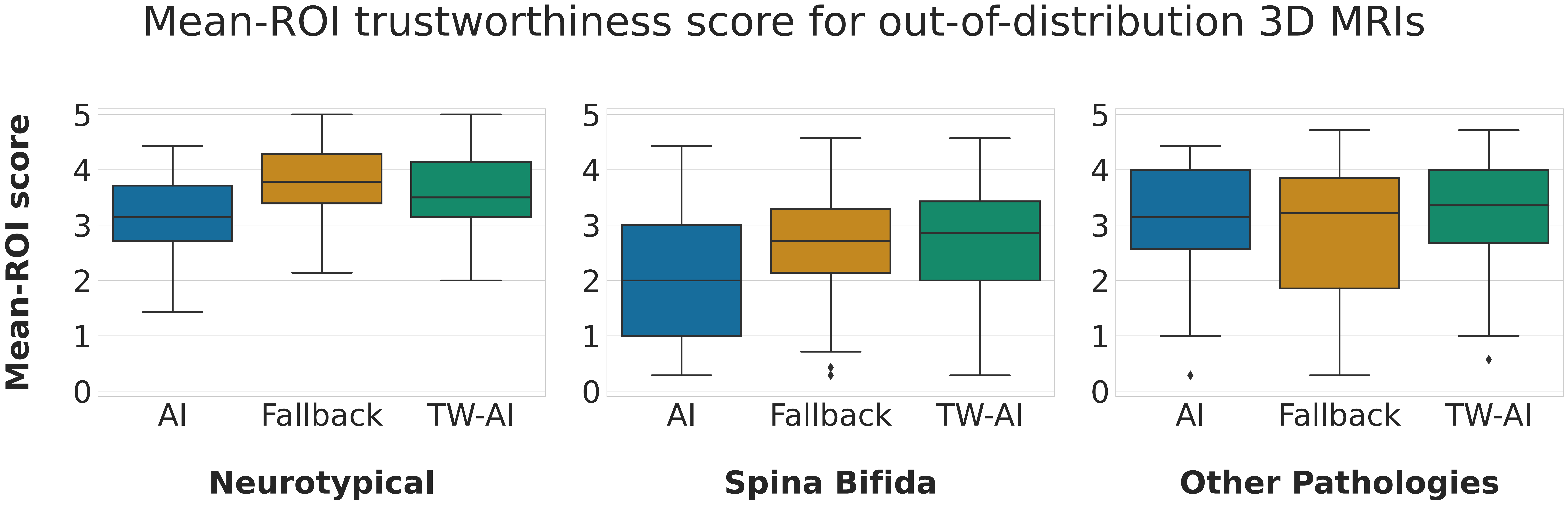}
        \caption{\textbf{Mean-ROI Trustworthiness Scores for out-of-scanner distribution 3D MRIs}}
        \label{fig:twai-scores}
    \end{subfigure}
    \caption{
    \textbf{
    Comparison of the backbone AI, fallback, and trustworthy AI segmentation algorithms.
    }
    In (a) (resp. (b)), we report for each algorithm the distributions of Dice scores (resp. Hausdorff distances at $95\%$ percentile (HD95)) for in-scanner distribution ($n=182$) and out-of-scanner distribution ($n=167$) images and for neurotypical ($n=141$), spina bifida ($n=157$), and other pathologies ($n=51$).
    In (c) (resp. (d)), we report mean and $95\%$ confidence intervals for the Dice scores (resp. HD95) across gestational ages, for neurotypical and spina bifida cases.
    In (e), we report \numscoring{} scoring by a panel of \numraters{} experts of the trustworthiness of the automatic segmentations for a subset of the out-of-scanner distribution testing 3D MRIs ($n=50$).
    Each expert was asked to score from $0$ (totally unacceptable) to 5 (perfect fit) the trustworthiness of each ROI. The scores displayed here are averaged across ROIs.
    Results per ROI can be found in the appendix (Fig.~\ref{fig:twai-dice_roi},~\ref{fig:twai-hausdorff_roi},~\ref{fig:twai-dice_GA_roi},~\ref{fig:twai-hausdorff_GA_roi},~\ref{fig:twai-scores_roi}).
    Box limits are the first quartiles and third quartiles. The central ticks are the median values. The whiskers extend the boxes to show the rest of the distribution, except for points that are determined to be outliers.
    Outliers are data points outside the range median $\pm 1.5\times$ interquartile range.
    }
    \label{fig:twai-results}
\end{figure}

\paragraph{Evaluation on a large multi-center fetal brain T2w MRI dataset.}
To effectively evaluate the performance of the proposed trustworthy AI framework as a suitable method to improve the trustworthiness of a backbone AI model using a fallback model, we have selected the task of fetal brain segmentation in 3D MRI.
This task is clinically relevant and is characterized by large image protocol variations and large anatomical variations.

Deep learning-based AI methods for fetal brain MRI segmentation have recently defined state-of-the-art segmentation performance~\cite{fetit2020deep,fidon2021label,fidon2021partial,hong2020fetal,khalili2019automatic,li2021cas,payette2021automatic,zhao2022automated}, gradually replacing image registration-based segmentation methods~\cite{makropoulos2018review} in the literature.
Most previous work on deep learning for fetal brain MRI segmentation trained and evaluated their models using only MRIs of healthy fetuses or only MRIs acquired at one center.
However, 
the segmentation performance of deep learning methods typically degrades when images from a different center or a different scanner vendor as the one used for training are used or when evaluating the segmentation performance on abnormal anatomy~\cite{alis2021inter,kamraoui2022deeplesionbrain,maartensson2020reliability,oakden2020hidden,perone2019unsupervised,redko2019advances}.
One study has reported such issues for fetal brain MRI segmentation~\cite{fidon2021distributionally}.

Thus, we have compared the proposed trustworthy AI algorithm to the backbone AI algorithm only~\cite{isensee2021nnu} and the fallback algorithm only, consisting of a registration-based segmentation method~\cite{cardoso2015geodesic}.
We have used a large multi-centric fetal brain 3D MRI dataset that consists of a total of $540$ fetal brain 3D MRIs with neurotypical or abnormal brain development, with gestational ages ranging from $19$ weeks to $40$ weeks, and with MRIs acquired at $13$ hospitals across $6$ countries.
% and $3$ continents.
% 
% Our trustworthy AI algorithms aims at effectively combining the backbone AI and the fallback methods.
% 
The task consists of segmenting automatically a fetal brain 3D MRI into eight clinically relevant tissue types:
the corpus callosum, the white matter, the cortical gray matter, the deep gray matter, the cerebellum, the brainstem, the intra-axial cerebrospinal fluid (CSF), and the extra-axial CSF.

\paragraph{Stratified evaluation across brain conditions and acquisition centers.}

The evaluation of AI-based segmentation algorithms has shown that the performance of deep learning models can vary widely across clinically relevant populations and across data acquisition protocols~\cite{fidon2021distributionally} (Fig.~\ref{fig:twai-overview}b, Fig.~\ref{fig:twai-qualitative_results}).

Therefore, we performed a stratified comparison of the backbone AI algorithm, the fallback algorithm, and the trustworthy AI algorithm across two groups of acquisition centers and three groups of brain conditions.
The composition of the dataset for each group is summarized in Fig.~\ref{fig:twai-data} and detailed in \nameref{sec:dataset}.
The acquisition centers were split into two groups, that we called \textit{in-scanner distribution} and \textit{out-of-scanner distribution}, depending if 3D MRIs acquired at a given center were present in the training dataset or not.
Four out of thirteen data sources were used for the training of the backbone AI algorithm.
In addition, the 3D MRIs were also separated based on the underlying brain condition of the fetus.
The first group, \textit{neurotypical}, contains the fetuses diagnosed by radiologists with a normal brain development using ultrasound and MRI.
The second group, \textit{spina bifida}, contains the fetuses with a condition called spina bifida aperta. We use the term \textit{spina bifida} for short in this work.
Cases of spina bifida aperta are typically accompanied by severe anatomical brain abnormalities~\cite{fidon2021atlas,pollenus2020impact}
with a type II Chiari malformation and an enlargement of the ventricles being most prevalent.
The Chiari malformation type II is characterized by a small posterior fossa and hindbrain herniation in which the medulla, cerebellum, and fourth ventricule are displaced caudally into the direction of the spinal canal~\cite{naidich1980computed}.
The third group, \textit{other pathologies}, contains fetuses with various pathologies other than spina bifida and causing an abnormal brain development, such as 
corpus callosum agenesis and dysgenesis,
intracranial hemorrhage and cyst,
aqueductal stenosis,
and Dandy-Walker malformation.
Those other pathologies were not present in the training dataset of the backbone AI algorithm and spatio-temporal atlases are not available for the fallback and the fail-safe algorithm.
Hence, the testing 3D MRIs classified as other pathologies allow us to measure the segmentation performance of the trustworthy AI algorithm outside of the domain covered by the anatomical contracts of trust.

% TODO: add detailed numbers
Figures~\ref{fig:twai-dice},\ref{fig:twai-hausdorff} show the results of the overall stratified evaluation in terms of Dice score and Hausdorff distances at $95\%$ percentile.
The detailed results per region of interest can be found in the appendix (Fig.~\ref{fig:twai-dice_roi},\ref{fig:twai-hausdorff_roi}).

\paragraph{Scoring of trustworthiness by a panel of radiologists.}
The Dice score and the Hausdorff distance are the two most standard metrics used for measuring the quality of automatic segmentations~\cite{bakas2018identifying}.
However, those two metrics do not directly measure the trustworthiness of segmentation algorithms~\cite{kofler2021we}.
Therefore, we have also conducted an evaluation of the trustworthiness of the automatic segmentations as perceived by radiologists.
We have asked a panel of \numraters{} experts from \numscoring{} different hospitals to score the trustworthiness of automatic segmentations from $0$ (totally unacceptable) to $5$ (perfect fit) for each region of interest and for the three segmentation algorithms.
Independent scoring were performed by raters at different hospitals.
% Each hospital provided one independent scoring.
% 
The scoring protocol and details about the panel of experts can be found in section \nameref{sec:scoring_protocol}.
The scoring was performed for the same 3D MRIs by all radiologists.
We have used a subset of $50$ 3D MRIs from the out-of-distribution group of the testing dataset consisting of $20$ neurotypical fetuses, $20$ spina bifida fetuses, and $10$ fetuses with other abnormalities.
Those cases were selected per condition at random among the 3D MRIs of the publicly available FeTA dataset~\cite{payette2021automatic}.
We have chosen to restrict this analysis to fetal brain 3D MRIs from the FeTA dataset, a subset of the \textit{out-of-scanner distribution} group, 
because those data are publicly available which facilitated collaborations across centers and because rating of the quality of the 3D MRIs had been performed in previous work~\cite{payette2021automatic}.
The \textit{out-of-scanner distribution} group is the most relevant group for the evaluation of trustworthiness because this corresponds to the situation in which AI algorithms generalization is the most challenging and clinically relevant.
In addition, this allows us to share our segmentations and scores publicly, thereby improving the reproducibility of our results.
The overall scoring results can be found in Fig.~\ref{fig:twai-scores} and the detailed results per region of interest can be found in the appendix (Fig.~\ref{fig:twai-scores_roi}).

Expert raters reported that most of the 3D MRIs were of good quality.
However, they noticed that the algorithms were very dependent on the quality of the 3D MRIs they were based on for the \textit{spina bifida} group.
We found a positive correlation between the mean-class trustworthiness scores and the quality of the 3D MRI for the \textit{spina bifida} group (Pearson $r=0.43$).
There was no correlation between scores and 3D MRI quality for the \textit{neurotypical} group (Pearson $r=-0.1$) and the 3D MRIs of the \textit{other pathologies} were all of high quality.
The 3D MRI quality scores used to compute the correlations above were performed by the authors of the FeTA dataset~\cite{payette2021automatic}.
% This corroborates previous work on segmentation using the data used for the scoringto~\cite{payette2021aumatic}.
% 
In addition, the more structurally abnormal the brains were due to the pathologies, the more difficult it was to compare the algorithms. In the case of the Chiari malformations, this applies in particular to the cerebellum and brainstem.

\paragraph{Stratified evaluation across gestational ages.}

The anatomy and the size of the fetal brain change significantly from $19$ weeks of gestation until term for both neurotypical fetuses~\cite{gholipour2017normative} and fetuses with spina bifida~\cite{fidon2021atlas}.
This age-related variability is a challenge for segmentation algorithms for fetal brain MRI~\cite{fidon2021distributionally}.
 
We analysed the performance of the proposed trustworthy AI algorithm for fetal brain segmentation as a function of the gestational age and compared it to the backbone AI algorithm based on deep learning~\cite{isensee2021nnu} and the fallback algorithm based on image registration~\cite{cardoso2015geodesic}.
We grouped the fetuses with neurotypical or spina bifida condition with the same gestational age rounded to the closest week.
The mean and the confidence intervals at $95\%$ for the overall performance in terms of Dice sore (resp. Hausdorff distance) across regions of interest can be found in Fig.~\ref{fig:twai-dice_GA} (resp. Fig.~\ref{fig:twai-hausdorff_GA}).
The detailed results per region of interest can be found in the appendix (Fig.~\ref{fig:twai-dice_GA_roi},\ref{fig:twai-hausdorff_GA_roi}).

Our results show how the backbone-AI algorithm, based on deep learning, and the fallback algorithm, based on image registration, are complementary.
Overall, the backbone-AI algorithm achieves higher Dice scores than the fallback algorithm, while the fallback achieved lower Hausdorff distances than the backbone-AI method (Fig.~\ref{fig:twai-results}).
Our proposed trustworthy AI algorithm successfully combines the backbone AI and the fallback algorithms.
It achieves higher or similar segmentation performance than those two algorithms in terms of established segmentation quality metrics such as the Dice score and the Hausdorff distance across all gestational ages for \textit{neurotypical} and \textit{spina bifida}.
Using a Wilcoxon signed-rank test, we found that the trustworthy AI method significantly outperforms the backbone AI in terms of both Dice score and Hausdorff distance for every group ($p<0.05$),
except for the Dice score of the in-scanner distribution \textit{other pathologies} group for which there is no statistical difference between the trustworthy AI and AI methods.
The trustworthy AI method also significantly outperforms the fallback for the two metrics for all groups,
except for the Hausdorff distances of the out-of-scanner-distribution \textit{spina bifida} group for which there is no statistical difference between the trustworthy AI and fallback methods.
This result is further supported by the trustworthiness scores of expert paediatric radiologists specialized in fetal brain anatomy.
The trustworthy AI method outperformed the AI and performed on a par with the fallback for the \textit{neurotypical} and \textit{spina bifida} groups and the trustworthy AI method outperformed the fallback and performed on a par with the AI method for the \textit{other pathologies} group.

\FloatBarrier

\section{Discussion}
% MAIN RESULTS

\paragraph{A principled and practical trustworthy AI method.}
We have mathematically formalized a method for trustworthy AI with a fallback based on Dempster-Shafer theory.
For application to fetal brain MRI segmentation, we have shown that our trustworthy AI method can be implemented using anatomy-based and intensity-based priors.
We have proposed to interpret those priors as contracts of trust in Human-AI trust theory.
Altogether, we showed that our principled trustworthy AI method improves the robustness and the trustworthiness of a state-of-the-art 
AI algorithm for fetal brain 3D MRI segmentation.

\paragraph{AI-based and registration-based algorithms are complementary.}
AI-based algorithms and registration-based algorithms have different error patterns.
In several situations we have found that the registration-based method tends to achieve better segmentation performance in terms of Hausdorff distance as compared to the AI-based method while the AI-based method achieved better segmentation performance in terms of Dice score.
We have found that the segmentation performance of the fallback algorithm decreases less than for the backbone AI algorithm, when comparing out-of-scanner distribution to in-scanner distribution for neurotypical and spina bifida fetal brain 3D MRIs.
In our scoring of trustworthiness on out-of-scanner distribution data, we have also found that
% , overall, 
the fallback algorithm outperformed the backbone AI algorithm for neurotypical and spina bifida cases (Fig.~\ref{fig:twai-scores}).
We think this is because the anatomical prior used by registration-based segmentation methods prevents mislabelling voxels far from the real anatomy.
In contrast, AI-based method are unconstrained and such errors can occur.
This is what we observe for the out-of-distribution cases displayed in Fig.~\ref{fig:twai-overview}c,~\ref{fig:twai-qualitative_results}.
Our proposed fail-safe method uses the registration-based segmentation with added margins with the aim to automatically detect and discard such errors that were found to occur more often for AI-based approach than for registration-based approach.

\paragraph{The contracts of trust hold for sub-populations covered by brain atlases.}
Our implementation of trustworthy AI for fetal brain segmentation depends on the availability of spatio-temporal segmentation atlases of the fetal brain in 3D MRI.
While such atlases currently exist for neurotypical fetal brain~\cite{gholipour2017normative,wu2021age} and fetuses with spina bifida~\cite{fidon2021atlas}, it is not the case for other fetal brain pathologies.
Therefore, our contracts of trust are not expected to hold for the group \textit{other pathologies}.
This illustrates how AI trustworthiness is context-dependent.
We found that the \textit{other pathologies} group is the only one for which radiologists associated the fallback method, based solely on the atlases, with a lower trustworthy scores than the backbone AI algorithm (Fig.~\ref{fig:twai-scores}).
Surprisingly, we found that the trustworthy AI algorithm still performs better or on a par with the backbone AI algorithm for the \textit{other pathologies} group.
We associate this with the use of our margins and to the proposed voxel intensity prior for the cerebrospinal fluid that are specific to the trustworthy AI algorithm.
For the \textit{other pathologies} group, we used the margin values estimated for spina bifida. 
Our group \textit{other pathologies} gathers diverse rare developmental diseases associated with different variations of the fetal brain anatomy.
However, due to the low number of examinations available per pathology, grouping them was necessary for evaluation purposes.
This introduces biases when comparing the segmentation performance for the \textit{other pathologies} groups associated with in-scanner and out-of-scanner distribution.
In particular, some of the fetuses with \textit{other pathologies} in the in-scanner distribution had very severe brain anatomical abnormalities, such as acqueductal stenosis with large supratentorial ventricles and caudal discplacement of the cerebellum or intracranial hemorrhage with parenchymal destruction and ventriculomegaly.
In contrast, the one in the out-of-scanner distribution have milder brain abnormalities, such as moderate ventriculomegaly, and there were no cases with parenchymal destruction.
This explains why, for this condition, we observe more outliers with low Dice scores and high Hausdorff distances for the backbone AI algorithm for in-scanner-distribution as compared to out-of-distribution 3D MRIs (Fig.~\ref{fig:twai-dice},\ref{fig:twai-hausdorff}).
This is also the only group for which the trustworthy AI method does not significantly outperform the AI method in terms of Dice score (Fig.~\ref{fig:twai-dice_roi}a).

% RESULTS - DETAILS
The two histograms of gestational ages for the training spina bifida 3D MRIs and the in-scanner-distribution testing spina bifida 3D MRIs are not uniform and have the same shape (see Fig.~\ref{fig:twai-data}).
In contrast the histogram of gestational ages for the out-of-scanner-distribution testing spina bifida is more uniform.
This might partly explain the degradation of Dice scores and Hausdorff distances, with the appearance of outliers, between in-scanner-distribution and out-of-scanner-distribution for the backbone AI algorithm (Fig.~\ref{fig:twai-dice},\ref{fig:twai-hausdorff}).
Training and in-scanner-distribution testing spina bifida MRIs were mostly clinical data acquired at \uzlshort{}.
In this center, MRI of spina bifida are typically performed a few days before and after the surgery that is performed prior to $26$ weeks of gestation.
In addition, a follow-up MRI is sometimes performed one month after the surgery.
This explains the two modes observed in the histograms for those two groups.
In the training data, the use of the spina bifida atlas~\cite{fidon2021atlas}, that has a uniform gestational age distribution, makes the second mode less visible.
Our results suggest the trustworthy AI algorithm is more robust than the AI algorithm to the gestational ages distributional shift between training and testing.

For gestational ages lower than $27$ weeks, the Dice scores and Hausdorff distances degrade for all the algorithms (Fig.~\ref{fig:twai-dice_GA},\ref{fig:twai-hausdorff_GA}).
For the backbone AI algorithm this is surprising given that more MRIs acquired at gestational ages lower than $27$ weeks than higher were present in the training dataset (Fig.~\ref{fig:twai-data}).
Poorer MRI quality, which is typical for younger fetuses, might explain this degradation.
In addition, the ratio of spina bifida over neurotypical examinations is higher for gestational ages lower than $27$ weeks in our dataset.
The abnormal brain anatomy of spina bifida cases leads to more difficult segmentation compared to neurotypical cases.
This is particularly the case for several classes: the cerebellum, the extra-axial cerebrospinal fluid (CSF), the cortical gray matter, and the brainstem (Fig.~\ref{fig:twai-dice_roi},\ref{fig:twai-hausdorff_roi},\ref{fig:twai-dice_GA_roi},\ref{fig:twai-hausdorff_GA_roi},\ref{fig:twai-scores_roi}).
The cerebellum is more difficult to detect using MRI before surgery as compared to early or late after surgery~\cite{aertsen2019reliability,danzer2007fetal}.
This has already been found to affect the segmentation performance of AI-based algorithms in previous work~\cite{fidon2021distributionally}.
For neurotypical fetuses, the extra-axial CSF is present all around the cortex.
However, for fetal brain MRI of spina bifida fetuses with gestational ages of $27$ weeks or less this is often not the case and the extra-axial CSF might be reduced to several small connected components that do not embrace the entire cortex anymore~\cite{fidon2021atlas}.
We hypothesize that the spina bifida atlas does not cover well this variability of the extra-axial CSF~\cite{fidon2021atlas}.
Due to the explicit spatial regularization, medical image registration cannot tackle such differences of topology.
Therefore, using the atlas currently available, the contract of trust for extra-axial CSF does not apply for this group of spina bifida cases.
It can also influence nearby regions, such as the cortical gray matter in this case.
For the fallback algorithm and the trustworthy AI algorithm, a further degradation of the segmentation performance for gestational ages lower than $21$ weeks was expected because the fetal brain atlases used start at $21$ weeks.
For gestational ages of $21$ weeks or higher, the trustworthy AI outperforms either the backbone AI-algorithm or the fallback algorithm and performs better or on a par with the best other algorithms for all regions of interest in terms of Dice score and Hausdorff distance (Fig.~\ref{fig:twai-dice_GA_roi},\ref{fig:twai-hausdorff_GA_roi}).
The confidence intervals are also similar or narrower for the trustworthy AI algorithm than for the other algorithms for gestational ages higher or equal to $21$ weeks.
This illustrates that our contracts of trust improve the robustness of the proposed trustworthy AI algorithm for spina bifida for the range of gestational ages covered by the atlas used~\cite{fidon2021atlas}.

% FUTURE WORK
\paragraph{Future work.}
For this work we have created the largest manually segmented fetal brain MRI dataset to date that consists of $540$ fetal brain 3D MRIs from $13$ acquisition centers.
A recent trend in medical image processing using AI is to gather even larger multi-institutional datasets using methods such as federated learning~\cite{rieke2020future}.
One can hypothesize that, with enough data, the AI algorithm would get more accurate even in the worst case until eventually reaching the same accuracy as the trustworthy AI algorithm in all cases.
However, results of our stratified evaluation suggest that this will require manually annotated 3D MRIs for every scanner acquisition protocol, for every condition, and for every gestational age.
To give an order of magnitude of the required dataset size, if we consider that $10$ 3D MRIs are required for each gestational age from $19$ weeks to $38$ weeks, for each of $10$ conditions and each of $5$ hospitals, we would already need $10,000$ 3D MRIs for both training and testing.
Given the low prevalence of some conditions~\cite{fidon2021distributionally} and the cost of obtaining fully-segmented data, classical supervised learning approaches might not be sufficient.
This rough estimation does not even include important confounding factors such as ethnicity and gender.
Altogether, this suggests that gathering more training data to improve the AI algorithm prior to deployment might not be sufficient to make the AI algorithm alone trustworthy.

The proposed fail-safe mechanism, that is part of our trustworthy AI method, could be used to help improving the backbone AI continuously after its deployment.
An AI incident could be declared when a large part of the AI algorithm prediction was discarded by the \emph{fail-safe mechanism}.
This would allow automatic detection of images to correct and include in priority in the training set to update the backbone AI algorithm.
In addition, reporting such incidents could help to further improve the trust of the user.
In the context of trust, it is important to report such issues even when the incidents were handled correctly using the fallback segmentation algorithm.
In addition, as part of the 
European Union Medical Device Regulations (EU MDR)
Article 87 on \enquote{Reporting of serious incidents and field safety corrective actions}~\cite{mdr},
%European guidelines on medical devices vigilance system~\cite{meddev}
it is a requirement for medical device manufacturers to report device-related incidents.
Previous methods for global segmentation failures detection, i.e. at the image-level, were proposed~\cite{kofler2021robust,robinson2019automated}.
In contrast, our fail-safe mechanism approaches the problem locally, i.e. at the voxel-level, by using an atlas-based anatomical prior and an intensity-based prior.

The margins used in our trustworthy AI segmentation algorithm for fetal brain MRI could also support interactive segmentation.
Instead of providing voxel-level corrections or scribbles, the annotator could interact with the automatic segmentation by manually adapting the margins for its annotation.
After manual adjustment, the voxels outside the margins are automatically marked as correctly labelled while for the voxel inside the margins will be assigned a set of possible labels.
This yields partial annotations that can be exploited to improve the backbone AI method using partially-supervised learning methods~\cite{fidon2021label}.
This use of margins is similar, in terms of user interaction, to the safety margins that are used in clinics for radiation therapy planning~\cite{niyazi2016estro}.

The expert raters also emphasized that some of the most frequent major violations in the cortex layer could be quickly removed manually and that they would have given higher scores to the segmentations if they could interact with them.
This echoes previous work on computational-aided decision making that found that users are more satisfied with imperfect algorithms if they can interact with them~\cite{dietvorst2018overcoming}.
Our findings suggest that allowing interactions would also increase the trust of human users in AI algorithms for medical image segmentation.

% In our trustworthy AI method, 
% Deep learning and registration predictions of uncertainty could allow to adapt $\epsilon$ for each image.
% In fact this already happening implicitly (and voxel-wise) when we use an ensemble of CNNs.
% This is because the CNNs in complete contradiction with the contracts for a given voxel will be discarded. 
% At the same time we still average the results by dividing by N=number of CNNs.
% Therefore the reweighting of DRC gives more weight to the other CNNs and to the fallback.
% In the end it will be like having an ensemble of N-1 CNNs with an epsilon of $\frac{N}{N-1}\epsilon$.
% 
% The proposed population-wise margins represent uncertainty in the atlas-based segmentation due to registration errors.
% More advanced registration uncertainty could allow to personalize the margins to each volume.

% About using more than two models in the trustworthy AI.
% It is worth noting that our backbone AI and fallback algorithms both use ensembling already.

% Limitations:
% \begin{itemize}
%     % \item There are still confounding factors despite our stratified evaluation 
%     \item How to represent contracts like "the cerebellum is made of exactly one connected components" using our model?
%     The problem here is that, in our method, the basic probability assignments depend only on the image, whereas here it depends only on the predicted segmentation.
% \end{itemize}

\chapter{Conclusion}
\label{chap:conclusion}
\minitoc

\section{Conclusion}

Throughout this thesis, it has been shown that current state-of-the-art deep learning methods are not trustworthy.
In \textbf{\Chapref{chap:dro}} and \textbf{\Chapref{chap:twai}} we show that a state-of-the-art deep learning pipeline for fetal brain MRI segmentation can fail spectacularly on MRIs of fetuses with an abnormal brain and/or on MRIs acquired in a different hospital as the MRIs used for training.
In \textbf{\Chapref{chap:dro}}, we found that state-of-the-art deep learning pipelines for fetal brain MRI segmentation and brain tumor segmentation using multi-parametric MRI under-perform on subpopulations that are underrepresented in the training dataset.
In the case of fetal brain MRI segmentation, the quantification of this problem was only possible thanks to the multi-center fetal brain MRI dataset created in \textbf{\Chapref{chap:fetaldataset}} and that contains the highest number of brain MRIs of fetuses with normal and abnormal brain developments reported to date.

Novel deep learning methods have been proposed in this thesis to tackle this problem.
In \textbf{\Chapref{chap:partialsup}}, a method has been proposed to train deep neural networks using partially segmented images.
We have shown empirically that it allows to train deep neural networks for fetal brain MRI segmentation with order of magnitude more MRIs.
It has been observed that training with more data, even partially segmented, leads to improved segmentation accuracy and robustness.
In \textbf{\Chapref{chap:dro}}, a new optimization problem and optimizer for training deep learning is proposed. 
In practice, the new method makes the deep neural network learn more on the training images seen as \textit{hard samples} during the training process.
We show that for fetal brain MRI segmentation, the proposed optimization leads to deep neural network with an improved worst-case testing performance corresponding to brain MRI of subpopulations of fetuses with abnormal development that were underrepresented in the training dataset.
However, as expected, no improvement was observed for subpopulations of fetuses that were not represented at all in the training dataset.
In \textbf{\Chapref{chap:twai}}, we propose a practical implementation of trustworthy AI for fetal brain MRI segmentation.
Our approach combines deep learning-based and atlas-based segmentation methods.
Our experiments show that the proposed method can successfully detect and correct violations of the anatomy in the segmentation prediction of a state-of-the-art deep learning pipeline for fetal brain MRI.
We measured an improved robustness on MRIs acquired in a different hospital as the MRIs used for training and/or MRIs of fetuses with an abnormal brain when the normal or abnormal subpopulations is one of those covered by an existing fetal brain atlas.
Therefore, the fetal brain atlas for spina bifida aperta proposed in \textbf{\Chapref{chap:atlas}} can be used to improve the trustworthiness of deep learning-based methods for the segmentation of brain MRIs of fetuses with spina bifida aperta.

This thesis also presents several theoretical contributions.
In \textbf{\Chapref{chap:partialsup}} an axiomatic construction of the set of loss functions that should be used when using partially segmented images for training a deep neural network has been proposed.
We propose the name \textit{label-set loss functions} for this family of loss functions because they rely on the interpretation of partial segmentation as images segmented using set of labels instead of labels as voxel-level annotation.
We proved a necessary and sufficient condition for a loss function to satisfy the axiom of label-set loss functions.
We propose the leaf-Dice loss and proved that it is a label-set loss function.
Last but not least, we propose a method to convert any existing loss function for fully-supervised learning into a label-set loss function and prove a unicity result regarding the proposed conversion method.
Therefore, there is an infinity of possible label-set loss functions.
In \textbf{\Chapref{chap:dro}}, we propose an adaptive sampling strategy that can be used during the training of a deep neural network with any optimizer based on stochastic gradient descent (SGD).
We call this sampling strategy \textit{harness weighted sampling} because it samples more often those samples, seen as \textit{hard} samples, that were found to lead to a higher value of the loss function previously during training.
For overparameterized deep neural network trained using SGD, we formally proved that using our hardness weighted sampling corresponds to solve a distributionally robust optimization (DRO) problem instead of the classic empirical risk minimization problem.
In addition, we formally link the proposed hardness weighted sampling with hard example mining and we studied the link between DRO and the minimization of percentiles of the loss function on the training dataset.
In \textbf{\Chapref{chap:atlas}}, it has been proved that sparse linear constraints on the parameters of divergence-conforming B-Splines stationary velocity fields (SVFs) lead to being exactly divergence-free at any point of the continuous spatial domain.
When using divergence-conforming B-splines SVFs for incompressible registration this results in velocity fields that are exactly divergence-free and we proved a bound on the numerical incompressibility error for the transformation in the case of an Euler integration.
In \textbf{\Chapref{chap:twai}}, we propose a principled implementation of trustworthy AI with a fallback and a fail-safe mechanism using Dempster-Shafer (DS) theory.
We theoretically show how Dempster's rule of combination allows to automatically switch from the AI to the fallback when the AI segmentation prediction violate anatomical priors.
A compact and computationally-efficient formula for our proposed implementation of trustworthy AI for fetal brain MRI is also proved.

\section{Discussion and Future Work}

The methods proposed in this thesis were found to improve the trustworthiness of deep learning methods applied to fetal brain MRI segmentation.
However, the proposed methods remain insufficient to guarantee robustness and safety of deep learning-based methods in every case.
It is therefore hypothesized that further theoretical contributions in the field of trustworthy AI is needed.

\subsection{Semi-automatic segmentation for trustworthy AI}

Semi-automatic segmentation methods are promising for improving the trustworthiness of segmentation methods.
In \textbf{\Chapref{chap:twai}}, some of the experts conducting the trustworthiness scoring of the segmentations stressed that they would have given a higher scores to some segmentation if they were able to correct them manually even slightly.
In interactive segmentation, experts are allowed to partially correct automatic segmentations to improve them.
Experts interactions are typically partial to limit the interaction time required.
As result, only a small fraction of the voxels are annotated by the expert in this context.
This suggests the label-set loss functions proposed in \textbf{\Chapref{chap:partialsup}} could be extended for this purpose.
Indeed, label-set loss functions allow to model unannotated voxels by assigning them a set of classes. If no information has been given about a voxel, it corresponds to assigning the set of all classes to this voxel.

Label-set loss functions also allow to model voxels marked as \emph{error} by the annotator.
In this case, the set of all classes except the class predicted and marked as incorrect is assigned to this voxel.

One major difference between training with a fixed training set and training in an interactive segmentation framework is the fact that for the latter the manual annotations are related to an automatic segmentation.
There are therefore two sources of information to learn from: the annotations of the expert and the automatic segmentation that has been corrected.
Since, in essence, Dempster-Shafer theory aims at combining information potentially partial but certain and coming from different sources in a safe manner, it appears as a promising mathematical theory for interactive segmentation.
The margins used in our Dempster-Shafer approach trustworthy AI segmentation algorithm 
could also support interactive segmentation.
Instead of providing voxel-level corrections or scribbles, the annotator could interact with the automatic segmentation by manually adapting the margins for its annotation.
After manual adjustment, the voxels outside the margins are automatically marked as correctly labelled while for the voxel inside the margins will be assigned a set of possible labels.
This yields partial annotations that can be exploited to improve the backbone AI method using partially-supervised learning methods such as the label-set loss functions.
This use of margins is similar, in terms of user interaction, to the safety margins that are used in clinics for radiation therapy planning~\cite{niyazi2016estro}.
In addition, domain knowledge, such as the one used in \textbf{\Chapref{chap:twai}}, could be used in conjunction with expert interactions to make the best of every interactions using Dempster-Shafer theory.

\subsection{Future work on fetal brain segmentation}
Regarding fetal brain MRI segmentation specifically,
the reconstructed 3D MRIs used in this thesis are the result of a super-resolution and reconstruction (SRR) post-processing of the 2D MRI slices coming directly from the scanner.
Therefore, one can hypothesize that information lost during this post-processing may be useful to achieving more trustworthy segmentations and segmenting the fetal brain directly in the 2D MRI slices could be beneficial.
It is however worth noting that SRR algorithms also do provide an important piece of information not present in the 2D MRI slices taken independently:
the spatial mapping between the 2D MRI slices and the reconstructed 3D MRI that reflects the non-trivial spatial correspondence between the slices and between the stacks.
In this sense, the output of a SRR algorithm is better represented by the pair of the input 2D MRI slices and the output spatial transformations than by the output reconstructed 3D MRI.
The 2D MRI slices and spatial transformations are complementary information whose corresponding reconstructed 3D MRI is only a summary that is useful for visualization and for volumetric and shape analysis.

Combining segmentations in the 2D MRI slices and in the reconstructed 3D MRI is also a promising avenue of research.
In this context, the spatial transformation computed by the SRR algorithm can be used as an interface between segmentations in the 2D MRI slices and a segmentation in the corresponding reconstructed 3D MRI.
The same way the intensity of voxels in the 3D reconstructed MRI are interpolation of voxels intensity in different 2D MRI slices, the same interpolation can be used to combine the segmentations in the different 2D MRI slices into a combined segmentation of the reconstructed 3D MRI.
This process is already used to compute 3D brain masks in NiftyMIC~\cite{ebner2020automated}.
Conversely, the spatial transformations computed by the SRR can be used to propagate segmentation in the reconstructed 3D MRI back to the 2D MRI slices.
We hypothesize that the manual segmentation of fetal brain MRI is easier and more accurate when performed using the reconstructed 3D MRI.
Therefore, it may be beneficial to perform the manual segmentation of the 2D MRI slices by segmenting manually in the reconstructed 3D MRI.
In contrast to the 2D MRI slices, reconstructed 3D MRI can always be registered to a standard clinical view where the axial, sagittal, and coronal planes are aligned with the axis of the image.
In addition, the reconstructed 3D MRIs have less image artefacts and it offers the annotator the possibility to using the three planes for the annotation task.

\subsection{Impact of the fetal brain MRI segmentation algorithm developed in this thesis}
The impact of the methods for trustworthy fetal brain T2w MRI segmentation developed in this thesis remains to be fully demonstrated.
Recent advances in fetal neuroimaging have permitted more in-depth analysis of the normal fetal brain development but the challenging application of advanced fetal neuroimaging under pathological conditions is still in its infancy.
Some cases of congenital brain defects can be prevented, while the severity of others can be reduced by prenatal interventions including fetal surgery.  
Anticipation of the potential long-term outcomes is also critical for parental counselling in these pathologies.
We expect trustworthy fetal brain MRI segmentation to play a critical role in these applications where quantitative volumetric and shape analysis are currently not fully exploited.
Encouraging preliminary results have already been obtained using the segmentation methods developed in this thesis.
In \cite{mufti2021cortical,mufti2023}, a segmentation algorithm based on \textbf{\Chapref{chap:dro}} and \textbf{\Chapref{chap:twai}} has been used to study cerebral volume, shape and cortical folding quantitatively for fetuses with spina bifida aperta as compared to fetuses with a normal brain development.
In \cite{emam2021longitudinal}, a segmentation algorithm based on \textbf{\Chapref{chap:partialsup}} and \textbf{\Chapref{chap:dro}} has allowed the quantitative analysis of the intra-axial and extra-axial cerobrospinal fluid spaces and to compute the local gyrification index for a population of fetuses with isolated congenital diaphragmatic hernia (CDH) and a population of fetuses with a normal brain development.
It has been found that longitudinal brain development on MRI in CDH fetuses is significant altered compared to normal controls in the third trimester.
This is one of the first evidence that fetuses with CDH have an abnormal prenatal brain development.
In \cite{deprest2022application}, a segmentation algorithm based on \textbf{\Chapref{chap:dro}} has been tested on brain 3D MRIs of fetuses with severe brain anatomical abnormalities.
Segmentation errors were analyzed in detail from a clinical point of view.
Those results guided us for creating the dataset described in \textbf{\Chapref{chap:fetaldataset}} and to develop the trustworthy AI method described in \textbf{\Chapref{chap:twai}}.
To further accelerate discoveries, we have made publicly available a software for automatic fetal brain T2w MRI segmentation at \url{https://github.com/LucasFidon/trustworthy-ai-fetal-brain-segmentation}.
With a unique command line, the software makes it possible to compute automatic segmentation with a backbone state-of-the-art deep learning pipeline trained using the entire dataset described in \textbf{\Chapref{chap:fetaldataset}} and the distributionaly robust optimization method described in \textbf{\Chapref{chap:dro}}.
The full software pipeline also includes the method for trustworthy AI described in \textbf{\Chapref{chap:twai}} on top of the backbone deep learning pipeline.
This software will need to be tested to establish its segmentation accuracy and its limitations.

\backmatter

\bibliography{JNFull,ThesisBibliography}

% Use this to hide the section from the main text in the table of content
% \unhidefromtoc
\newpage

\setcounter{section}{0}
\setcounter{chapter}{1}
\renewcommand{\thechapter}{\Alph{chapter}}%

\chapter[Appendix]{Appendix}
\label{appendix}
\minitoc
\newpage
\section{Appendix of \chapref{chap:fetaldataset}}

\subsection{Lexic of Fetal Neuroanatomy for Engineers}\label{sec:neuroanatomy}

This section contains useful definitions about fetal neuroanatomy.
The terms are defined in alphabetical order.

\paragraph{Brain parenchyma}
The brain parenchyma is the functional tissue in the brain. It's comprised of two types of cells that are used specifically for cognition and controlling the rest of the body. The remaining brain tissue is known as stroma, which is the supportive or structural tissue. Damage or trauma to the brain parenchyma often results in a loss of cognitive ability or even death.

\paragraph{Caudal}
An anatomic term meaning: 1. Pertaining to the tail or the hind part. 2. Situated in or directed toward the tail or hind part. 3. Inferior to another structure, in the sense of being below it.

\paragraph{Cerebellar Herniation Level (CHL)} 
% Brain herniation is a potential deadly side effect of very high pressure within the skull.
This is the level of descent of the cerebellum caudally into the spinal canal past the foramen magnum. This is caused by a small posterior fossa, and the development of hydrocephalus which obstructs the normal cerebrospinal fluid flow around the body.

\paragraph{Cerebrospinal Fluid (CSF)}
The Cerebrospinal Fluid is the colorless fluid found in the brain and the spinal cord.
It is made of $99\%$ of water.

\paragraph{Chiari II hindbrain malformation - hindbrain herniation (HBH)}
The HBH seen in spina bifida is caused by a downward displacement of the cerebellum below the level of the foramen magnum.
It may result in brainstem dysfunction such as apnoea, dysphgagia, torticollis, and spasticity. The 5 year mortality rate in patients born with MMC is $14\%$ which rises to $35\%$ in patients whose Chiari II malformation leads to brainstem dysfunction~\cite{juranek2010anomalous,grant2011morphometric}.
This obstruction can prevent the movement of CSF and results in hydrocephalus.
The Chiari II malformation is characterised by a small posterior fossa with downward displacement of the cerebellum, medulla and fourth ventricule. 
Previous studies suggest that Chiari II malformation can be (manually) measured reliably using an MRI \cite{adzick2011randomized,aertsen2019reliability}.

% \paragraph{Chorioamniotic membrane separation}

\paragraph{Cognitive impairment}
When a person hasdificulty in short, and long-term memory, learning new skills, concentrating or making decisions that affect their everyday life.

\paragraph{Gestional age (GA)}
“Gestational age” (or “menstrual age”) is the time elapsed between the first day of the last normal menstrual period and the day of delivery~\cite{engle2004age}. GA is expressed as completed weeks.
The GA can be calculated from the first day of the last menstrual period otherwise, it can be estimated using an US via a measurement known aas \emph{crown-rump-length}.
It is worth noting that GA is not calculated from the day of conception (more difficult to estimate in practice).

% \paragraph{Gyrification}
% TODO: example with polymicrogyria and periventricular nodular heterotopia.

\paragraph{Hydrocephalus}
Hydrocephalus is a condition in which an accumulation of cerebrospinal fluid (CSF) occurs within the brain. This typically causes increased pressure inside the skull.

% \paragraph{Ischaemic}

\paragraph{Medulla or Medulla oblongata}
The medulla is a long stem-like structure which makes up part of the brainstem. It is anterior and partially inferior to the cerebellum. It is a cone-shaped neuronal mass responsible for autonomic (involuntary) functions ranging from vomiting to sneezing.

\paragraph{Myelomeningocele (MMC)}
Most common type of spina bifida. It is characterised by a fluid-filled sac containing exposed spinal cord and neural tissues.

\paragraph{Posterior Fossa (PF)}
Small region in the back of the skull that houses the cerebellum and brain stem.

% \paragraph{Preterm Premature Rupture Of the Membranes (PPROM)}

% \paragraph{Transverse Cerebellar Diameter (TCD)}

\paragraph{Ventricles}
Brain cavities in the middle of the brain where CSF circulates and is produced.

\paragraph{Ventriculomegaly}
Ventriculomegaly occurs when the ventricular width is larger then the reference ranges.
They can be graded as mild (8-10mm), moderate (10-15mm), and severe ($>$ 15 mm)~\cite{tulipan2015prenatal}.

% \paragraph{Ventriculoperitoneal shunt}
\newpage
\section{Proofs of \chapref{chap:partialsup}}

\subsection{Summary of mathematical notations}

\begin{itemize}
    \item $N$: number of voxels. 
    \item $\mathbf{L}$: set of final labels (e.g. white matter, ventricular system).
    We call the elements of $\mathbf{L}$ \textit{leaf-labels}.
    \item $2^{\mathbf{L}}$: set of subsets of $\mathbf{L}$
    We call the elements of $2^{\mathbf{L}}$ \textit{label-sets}.
    \item $P(\mathbf{L})$: space of probability vectors for leaf-labels.
    \item $\mathds{1}$: indicator function. 
    let A an assertion, $\mathds{1}(A)=1$ if $A$ is true and $\mathds{1}(A)=0$ if $A$ is false.
\end{itemize}

\subsection{Proof of equation \eqref{eq:lemma_general}}
Let $\mathcal{L}:\,\, \mathbf{E} \xrightarrow{} \mathds{R}$, with $\textbf{E}:=P\left(\mathbf{L}\right)^N \times \left(2^{\mathbf{L}}\right)^N$.
%
% We will demonstrate that $\mathcal{L}$ is a label-set loss function if and only if
% $\forall (\textbf{p},\textbf{g})\in \textbf{E},\,\, \mathcal{L}(\textbf{p}, \textbf{g}) = \mathcal{L}(\Phi(\textbf{p}; \textbf{g}), \textbf{g})$.
% 
Let us first remind that we have defined label-set loss functions as the losses $\tilde{\mathcal{L}}$ that satisfy the axiom
\begin{equation}
    \forall \textbf{g} \in \left(2^{\mathbf{L}}\right)^N,\,\forall \textbf{p}, \textbf{q} \in P(\textbf{L})^N,\quad
    \Phi(\textbf{p}; \textbf{g}) =\Phi(\textbf{q}; \textbf{g})
    \implies 
    \tilde{\mathcal{L}}(\textbf{p},\textbf{g}) = \tilde{\mathcal{L}}(\textbf{q}, \textbf{g})
    \label{eq:supp_labels_set_losses}
\end{equation}
% \begin{equation*}
% \begin{split}
%     & \mathcal{L} \,\,\textup{is compatible with} \sim_{\Phi}\\
%     % 
%     \iff &
%     \forall (\textbf{p},\textbf{g}), (\textbf{q}, \textbf{h}) \in \textbf{E},\,
%     (\textbf{p},\textbf{g}) \sim_{\Phi} (\textbf{q}, \textbf{h}) 
%     \implies 
%     \mathcal{L}(\textbf{p},\textbf{g}) = \mathcal{L}(\textbf{q}, \textbf{h})\\
%     % 
%     \iff &
%     \forall (\textbf{p},\textbf{g}), (\textbf{q}, \textbf{h}) \in \textbf{E},\,
%     \left\{
%     \begin{array}{c}
%         \Phi(\textbf{p}; \textbf{g}) =\Phi(\textbf{q}; \textbf{h})  \\
%         \textbf{g} = \textbf{h} 
%     \end{array}
%     \right. 
%     \implies 
%     \mathcal{L}(\textbf{p},\textbf{g}) = \mathcal{L}(\textbf{q}, \textbf{h})\\
%     % 
%     \iff &
%     \forall \textbf{g} \in \left(2^{\mathbf{L}}\right)^N,\,\forall \textbf{p}, \textbf{q} \in P(\textbf{L})^N,\,
%     \Phi(\textbf{p}; \textbf{g}) =\Phi(\textbf{q}; \textbf{g})
%     \implies 
%     \mathcal{L}(\textbf{p},\textbf{g}) = \mathcal{L}(\textbf{q}, \textbf{g})\\
% \end{split}
% \end{equation*}
% Where the first equivalence is the definition of the compatibility of a function with an equivalence relation. 

\subsubsection*{First implication.}
Let us assume that 
$\forall (\textbf{p},\textbf{g})\in \textbf{E},\,\, \mathcal{L}(\textbf{p}, \textbf{g}) = \mathcal{L}(\Phi(\textbf{p}; \textbf{g}), \textbf{g})$.
Let $\textbf{g} \in \left(2^{\mathbf{L}}\right)^N$ and $\textbf{p}, \textbf{q} \in P(\textbf{L})^N$,
such that $\Phi(\textbf{p}; \textbf{g}) =\Phi(\textbf{q}; \textbf{g})$.
We have
\begin{equation*}
    \mathcal{L}(\textbf{p},\textbf{g}) = \mathcal{L}(\Phi(\textbf{p}; \textbf{g}), \textbf{g})
= \mathcal{L}(\Phi(\textbf{q}; \textbf{g}), \textbf{g})
= \mathcal{L}(\textbf{q},\textbf{g}) 
\,\,\blacksquare
\end{equation*}
% 
% where the first and last equalities hold due to our hypothesis on $\mathcal{L}$ for this implication.
% 
% Therefore, $\mathcal{L}$ is a label-set loss function $\blacksquare$

\subsubsection*{Second implication.}
Let us assume that $\mathcal{L}$ is a label-set loss function.
% 
% Therefore
% \begin{equation*}
%     \forall \textbf{g} \in \left(2^{\mathbf{L}}\right)^N,\,
%     \forall \textbf{p}, \textbf{q} \in P(\textbf{L})^N,\,\,
%     \Phi(\textbf{p}; \textbf{g}) =\Phi(\textbf{q}; \textbf{g})
%     \implies 
%     \mathcal{L}(\textbf{p},\textbf{g}) = \mathcal{L}(\textbf{q}, \textbf{g})
% \end{equation*}
% 
We will use the following lemma which is proved at the end of this paragraph. 
\begin{lemma}
For all $\textbf{g} \in \left(2^{\mathbf{L}}\right)^N$,
the function $\Phi(\cdot\,; \textbf{g}): \textbf{p} \mapsto \Phi(\textbf{p}; \textbf{g})$
is idempotent, 
i.e. $\forall \textbf{p}\in P(\mathbf{L})^N,\, 
\Phi(\Phi(\textbf{p}; \textbf{g}); \textbf{g}) = \Phi(\textbf{p}; \textbf{g})$.
\label{lemma:idempotent}
\end{lemma}
Let $(\textbf{p},\textbf{g}) \in \textbf{E}$.
By Lemma~\ref{lemma:idempotent},
$
\Phi(\Phi(\textbf{p}; \textbf{g}); \textbf{g}) = \Phi(\textbf{p}; \textbf{g}).
$
Therefore, by applying \eqref{eq:supp_labels_set_losses} for $\textbf{p}$ and $\textbf{q}=\Phi(\textbf{p}; \textbf{g})$ we eventually obtain
$
\mathcal{L}(\textbf{p},\textbf{g}) = \mathcal{L}(\Phi(\textbf{p}; \textbf{g}), \textbf{g})
$
$\blacksquare$

\subsubsection*{Proof of Lemma~\ref{lemma:idempotent}}
Let $\textbf{g}= (g_i) \in \left(2^{\mathbf{L}}\right)^N$ and $\textbf{p}=(p_{i,c}) \in P(\textbf{L})^N$.
Let us denote $\Phi(\textbf{p};\textbf{g})= (\tilde{p}_{i,c})$
and $\Phi(\Phi(\textbf{p};\textbf{g});\textbf{g}) = (\vardbtilde{p}_{i,c})$.
For all $i, c$,
if $c \not \in g_i$, 
$\vardbtilde{p}_{i,c} = \tilde{p}_{i,c}$,
and if $c \in g_i$,
$
\vardbtilde{p}_{i,c} = \frac{1}{|g_i|}\sum_{c' \in g_i}\tilde{p}_{i,c'} = 
\frac{1}{|g_i|}\sum_{c' \in g_i}\left(\frac{1}{|g_i|}\sum_{c'' \in g_i}p_{i,c''}\right) = \tilde{p}_{i,c}
$
$\blacksquare$

\subsection{Proof that the leaf-Dice is a label-set loss function}
Let $\textbf{g} \in \gset$ be a partial segmentation that takes its value in a partition of $\mathbf{L}$ of the form
$\{\mathbf{L}'\} \cup \left\{\{c\}\,|\,c \in \mathbf{L}\setminus\mathbf{L}'\right\}$ with $\mathbf{L}' \subsetneq \mathbf{L}$
that contains all the labels of the regions of interest that were not manually segmented.
Let $\textbf{p} \in \pset$.

\begin{equation}
\begin{split}
    \mathcal{L}_{Leaf-Dice}(\textbf{p}, \textbf{g}) 
    &=
    1 - 
    \frac{1}{|\mathbf{L}|} 
    \sum_{c \in \mathbf{L}} 
    \frac{2 \sum_i \mathds{1}(g_i =\{c\})\,p_{i,c}}{
    \sum_i \mathds{1}(g_i =\{c\})^{\alpha}
    + \sum_{i} p_{i,c}^{\alpha}
    +\epsilon
    }\\
    &=
    1 - 
    \frac{1}{|\mathbf{L}|} 
    \sum_{c \in \mathbf{L}\setminus\mathbf{L}'} 
    \frac{2 \sum_i \mathds{1}(g_i =\{c\})\,p_{i,c}}{
    \sum_i \mathds{1}(g_i =\{c\})^{\alpha}
    + \sum_{i} p_{i,c}^{\alpha}
    +\epsilon
    }
\end{split}
\label{eq:supp_LSDice}
\end{equation}
because the terms of the sum for the leaf-labels $c \in \mathbf{L}'$ have a numerator equal to $0$.

To prove that $\mathcal{L}_{Leaf-Dice}$ is a label-set loss function it is sufficient to prove that 
$
\mathcal{L}_{Leaf-Dice}(\textbf{p},\textbf{g})=\mathcal{L}_{Leaf-Dice}(\Phi(\textbf{p;\textbf{g}}),\textbf{g}).
$
For all voxel i and leaf-label c
\begin{equation*}
    % \forall i,c,\quad
    \left[\Phi(\textbf{p;\textbf{g}})\right]_{i,c} =
    \left\{
    \begin{array}{cc}
        \frac{1}{|g_i|}\sum_{c' \in g_i} p_{i,c'} & \textup{if}\,\, c \in g_i \\
        p_{i,c} & \textup{if}\,\, c \not \in g_i
    \end{array}
    \right.
    =
    \left\{
    \begin{array}{cc}
        \frac{1}{|\mathbf{L}'|}\sum_{c' \in \mathbf{L}'} p_{i,c'} & \textup{if}\,\, c \in g_i \,\,\textup{and}\,\, g_i = \mathbf{L}' \\
        p_{i,c} & \textup{if}\,\, c \in g_i \,\,\textup{and}\,\, g_i =\{c\}\\
        p_{i,c} & \textup{if}\,\, c \not \in g_i
    \end{array}
    \right.
\end{equation*}
where the second equality comes from the partition.

We observe that $\left[\Phi(\textbf{p;\textbf{g}})\right]_{i,c} \neq p_{i,c}$ only when $c \in \mathbf{L}'$.
Using \eqref{eq:supp_LSDice} we obtain
$
\mathcal{L}_{Leaf-Dice}(\textbf{p},\textbf{g})=\mathcal{L}_{Leaf-Dice}(\Phi(\textbf{p;\textbf{g}}),\textbf{g})\,
\blacksquare
$

\subsection{Proof of equation \eqref{eq:partial_conversion_unique}}
Using previous results and the fact that $\mathcal{L}_{partial}$ is here defined as a converted fully-supervised loss function,
$\mathcal{L}_{partial}$ is a label-set loss function if and only if 
for all 
$
(\textbf{p},\textbf{g}) \in \pset \times \gset
$
\begin{equation*}
    % \forall \textbf{p},\textbf{g},\quad
    \mathcal{L}_{partial}(\textbf{p}, \textbf{g}) 
    = \mathcal{L}_{partial}(\Phi(\textbf{p}; \textbf{g}), \textbf{g})
    = \mathcal{L}_{fully}(\Phi(\textbf{p}; \textbf{g}), \Psi(\textbf{g}))
\end{equation*}
As a result, we only need to prove that if $\mathcal{L}_{partial}$ is a label-set loss function then $\Psi=\Psi_0$.
Let us suppose that $\mathcal{L}_{partial}$ is a label-set loss function.
Therefore, for all 
$
\textbf{g} \in \gset,\,\, 
\mathcal{L}_{fully}(\Phi(\Psi(\textbf{g}); \textbf{g}), \Psi(\textbf{g}))
= \mathcal{L}_{partial}(\Phi(\Psi(\textbf{g}); \textbf{g}), \textbf{g})
= \mathcal{L}_{partial}(\Psi(\textbf{g}), \textbf{g})
= \mathcal{L}_{fully}(\Psi(\textbf{g}), \Psi(\textbf{g})).
$
Using the unicity of the minima of $\mathcal{L}_{fully}$ we can therefore conclude that 
$
\forall \textbf{g} \in \gset,\,\, \Psi(\textbf{g}) = \Phi(\Psi(\textbf{g}); \textbf{g}).
$
For all $\textbf{g},i,c$,
\begin{equation*}
    \left[\Phi(\Psi(\textbf{g});\textbf{g})\right]_{i,c} =
    \left\{
    \begin{array}{cc}
        \frac{1}{|g_i|}\sum_{c' \in g_i} \left[\Psi(\textbf{g})\right]_{i,c'} & \textup{if}\,\, c \in g_i \\
        % &
        \left[\Psi(\textbf{g})\right]_{i,c} & \textup{if}\,\, c \not \in g_i
    \end{array}
    \right.
    =
    \left\{
    \begin{array}{cc}
        \frac{1}{|g_i|} & \textup{if}\,\, c \in g_i \\
        0 & \textup{if}\,\, c \not \in g_i
    \end{array}
    \right.
    = \left[\Psi_0(\textbf{g})\right]_{i,c}
\end{equation*}
where in the second inequality we have used
$
\forall i,c,\,\, [\Psi(\textbf{g})]_{i,c} > 0 \implies c \in g_i\,\, \blacksquare
$

\subsection{Relation to the marginal Dice loss}
Let $P=\{\mathbf{L}_j\}$ a partition of $\mathbf{L}$ and $\textbf{g}\in \gset$ with its values in $P$.
For all $c \in \mathbf{L}$, we denote $j(c)$ the unique index such that $c \in \mathbf{L}_{j(c)}$.

Using the definitions of $\Psi_0$, $\Phi$, and the partition $P$, we obtain
\begin{equation*}
    \begin{split}
        \mathcal{L}_{Dice}\left(\Phi(\textbf{p}; \textbf{g}), \Psi_0(\textbf{g})\right)
        &=
        1 - \frac{1}{|\mathbf{L}|} \sum_{c \in \mathbf{L}} 
        \frac{
            \sum_i
            \Psi_0(\textbf{g})_{i,c}
            \Phi(\textbf{p}; \textbf{g})_{i,c}
            }{
            \sum_i \Psi_0(\textbf{g})_{i,c}
            +
            \sum_i \Phi(\textbf{p}; \textbf{g})_{i,c}
            }\\
        &=
        1 - \frac{1}{|\mathbf{L}|} \sum_{c \in \mathbf{L}} 
        \frac{
            \sum_i
            \left(
            \frac{1}{|\mathbf{L}_{j(c)}|}\mathds{1}\left(g_i = \mathbf{L}_{j(c)}\right)
            \right)
            \left(
            \frac{1}{|\mathbf{L}_{j(c)}|}\sum_{c'\in \mathbf{L}_{j(c)}}p_{i,c'}
            \right)
            }{
            \sum_i
            \frac{1}{|\mathbf{L}_{j(c)}|}\mathds{1}\left(g_i = \mathbf{L}_{j(c)}\right)
            +
            \sum_i
            \frac{1}{|\mathbf{L}_{j(c)}|}\sum_{c'\in \mathbf{L}_{j(c)}}p_{i,c'}
            }\\
        &=
        1 - \frac{1}{|\mathbf{L}|} \sum_{c \in \mathbf{L}} 
        \frac{
            \left(\frac{1}{|\mathbf{L}_{j(c)}|}\right)^2
            \sum_i
            \mathds{1}\left(g_i = \mathbf{L}_{j(c)}\right)
            \left(\sum_{c'\in \mathbf{L}_{j(c)}}p_{i,c'}\right)
            }{
            \frac{1}{|\mathbf{L}_{j(c)}|}
            \left(
            \sum_i \mathds{1}\left(g_i = \mathbf{L}_{j(c)}\right)
            +
            \sum_i \sum_{c'\in \mathbf{L}_{j(c)}}p_{i,c'}
            \right)
            }\\
        &=
        1 - \frac{1}{|\mathbf{L}|} \sum_{c \in \mathbf{L}} 
        \frac{1}{|\mathbf{L}_{j(c)}|}
        \frac{
            \sum_i
            \mathds{1}\left(g_i = \mathbf{L}_{j(c)}\right)
            \left(\sum_{c'\in \mathbf{L}_{j(c)}}p_{i,c'}\right)
            }{
            \sum_i \mathds{1}\left(g_i = \mathbf{L}_{j(c)}\right)
            +
            \sum_i \sum_{c'\in \mathbf{L}_{j(c)}}p_{i,c'}
            }\\
    \end{split}
\end{equation*}
By grouping the terms of the first sum with respect to the label-sets in $P$
\begin{equation*}
    \begin{split}
        \mathcal{L}_{Dice}\left(\Phi(\textbf{p}; \textbf{g}), \Psi_0(\textbf{g})\right)
        &=
        1 - \frac{1}{|\mathbf{L}|} \sum_{\mathbf{L}' \in P} \sum_{c \in \mathbf{L}'} 
        \frac{1}{|\mathbf{L}_{j(c)}|}
        \frac{
            \sum_i
            \mathds{1}\left(g_i = \mathbf{L}_{j(c)}\right)
            \left(\sum_{c'\in \mathbf{L}_{j(c)}}p_{i,c'}\right)
            }{
            \sum_i \mathds{1}\left(g_i = \mathbf{L}_{j(c)}\right)
            +
            \sum_i \sum_{c'\in \mathbf{L}_{j(c)}}p_{i,c'}
            }\\
        &=
        1 - \frac{1}{|\mathbf{L}|} \sum_{\mathbf{L}' \in P} \sum_{c \in \mathbf{L}'} 
        \frac{1}{|\mathbf{L}'|}
        \frac{
            \sum_i
            \mathds{1}\left(g_i = \mathbf{L}'\right)
            \left(\sum_{c'\in \mathbf{L}'}p_{i,c'}\right)
            }{
            \sum_i \mathds{1}\left(g_i = \mathbf{L}'\right)
            +
            \sum_i \sum_{c'\in \mathbf{L}'}p_{i,c'}
            }\\
        &=
        1 - \frac{1}{|\mathbf{L}|} \sum_{\mathbf{L}' \in P}
        \frac{
            \sum_i
            \mathds{1}\left(g_i = \mathbf{L}'\right)
            \left(\sum_{c'\in \mathbf{L}'}p_{i,c'}\right)
            }{
            \sum_i \mathds{1}\left(g_i = \mathbf{L}'\right)
            +
            \sum_i \sum_{c'\in \mathbf{L}'}p_{i,c'}
            }\\
    \end{split}
\end{equation*}
because for all $\mathbf{L}' \in P$ and all $c \in \mathbf{L}'$, $\mathbf{L}_{j(c)}=\mathbf{L}'$.

The right hand-side of the last equality is the marginal Dice loss~\cite{shi2021marginal} up to the multiplicative constant $\frac{1}{|\mathbf{L}|}$ $\blacksquare$

\FloatBarrier
\section{Additional Experiments on BraTS 2021}\label{appendix:partial-brats}

\begin{figure}[!htb]
    \centering
    \begin{subfigure}[t]{0.9\linewidth}
        \includegraphics[width=\linewidth,trim=0cm 1cm 0cm 0cm,clip]{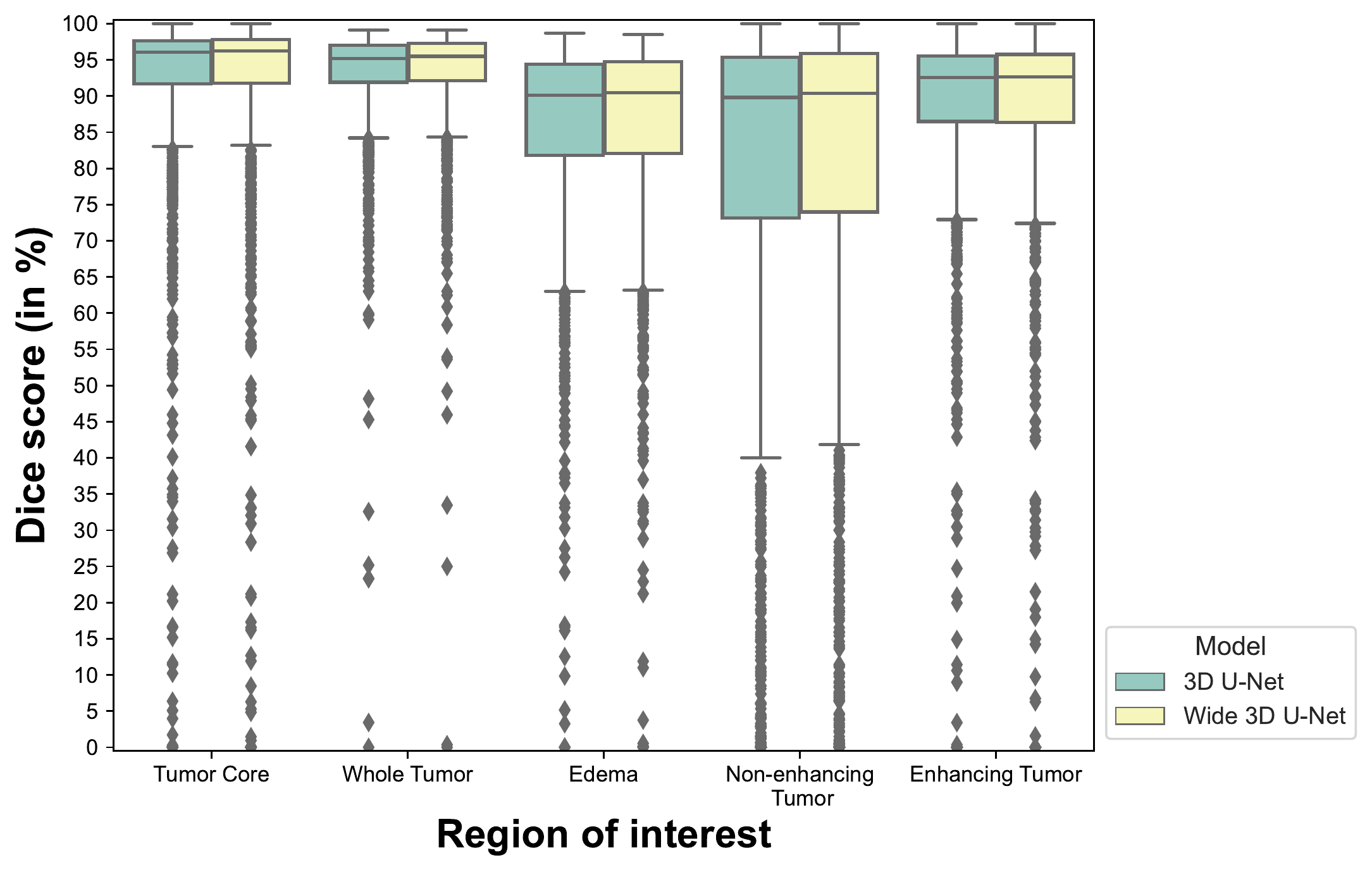}
    \end{subfigure}
    \hfill
    \begin{subfigure}[t]{0.9\linewidth}
        \includegraphics[width=\linewidth,trim=0cm 1cm 0cm 0cm,clip]{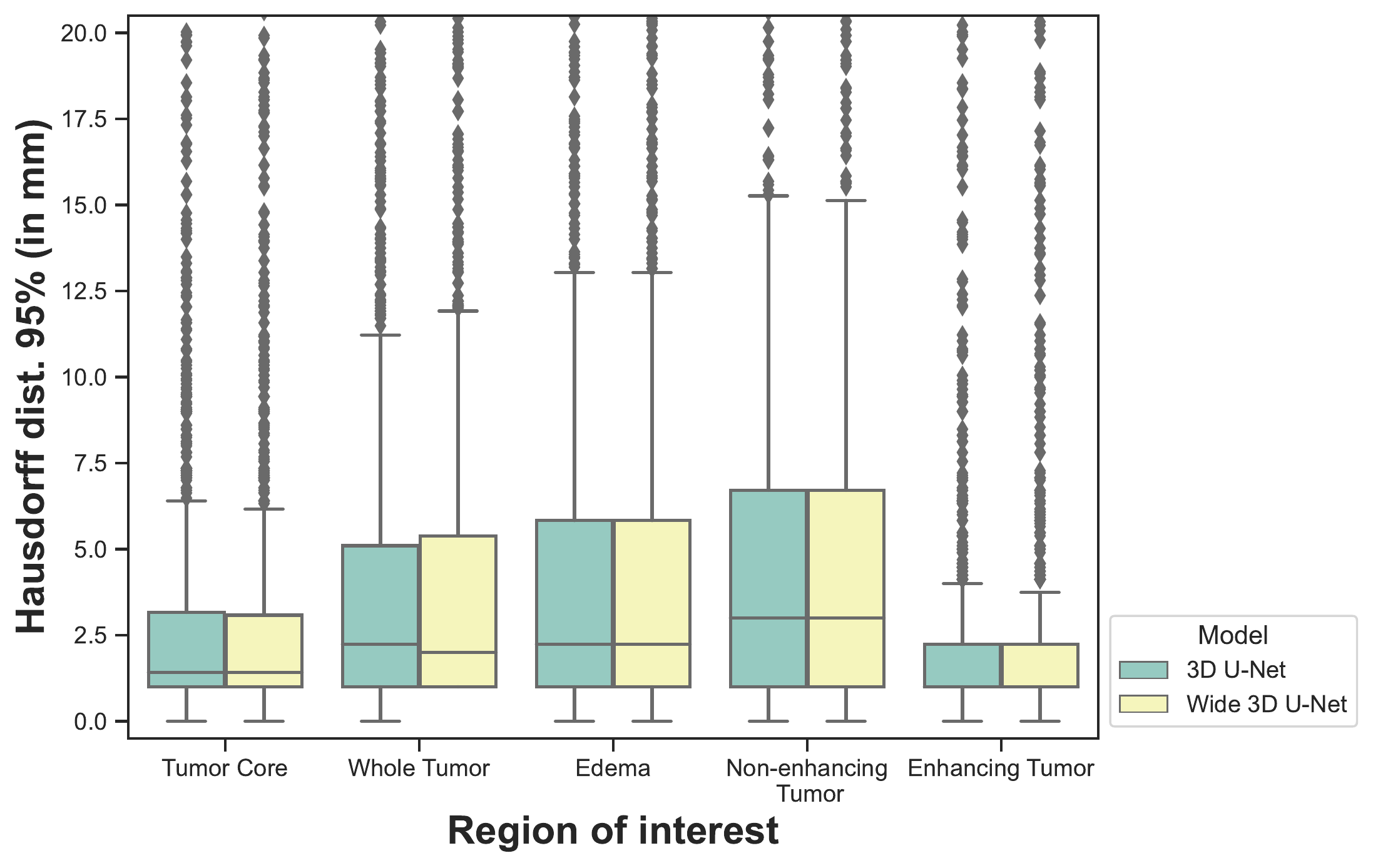}
    \end{subfigure}
    \caption{
    \textbf{Evaluating the capacity of the 3D U-Net architecture on the BraTS 2021 training dataset ($1251$ cases).}
    Wide 3D U-Net has the same architecture as 3D U-Net with twice as many features in every internal layers, i.e. it has approximately $4$ times more trainable parameters.
    There is no difference of segmentation performance between the two models which suggests that the 3D U-Net is already large enough for training on this dataset.
    Each box plot results from 5-fold cross validation on the $1251$ cases.
    Box limits are the first quartiles and third quartiles. The central ticks are the median values. The whiskers extend the boxes to show the rest of the distribution, except for points that are determined to be outliers.
    Outliers are data points outside the range median $\pm 1.5\times$ interquartile range.
    }
\end{figure}
\FloatBarrier

\newpage
\section{Proofs of \chapref{chap:dro}}

\subsection{Summary of the Notations used in the Proofs}\label{s:notations}

For the ease of reading the proofs we first summarize our notations.

\subsubsection{Probability Theory Notations}
\begin{itemize}
    \item $\Delta_n = \left\{\left(p_i\right)_{i=1}^n \in [0,1]^n, \,\,
        \sum_i p_i = 1\right\}$
    \item Let $\vq=(q_i) \in \Delta_n$, and $f$ a function, we denote $\E_{\vq}[f(\rx)]:=\sum_{i=1}^n q_i f(\rx_i)$.
    \item Let $\vq \in \Delta_n$, and $f$ a function, we denote $\V_{\vq}[f(\rx)]:=\sum_{i=1}^n q_i\norm{f(\rx_i) - \E_q[f(\rx)]}^2$.
    \item $\vp_{\textrm{train}}$ is the uniform training data distribution, i.e. $\vp_{\textrm{train}}=\left(\frac{1}{n}\right)_{i=1}^n \in \Delta_n$
\end{itemize}

\subsubsection{Machine Learning Notations}
\begin{itemize}
    \item n is the number of training examples
    \item d is the dimension of the output
    \item $\gd$ is the dimension of the input
    \item m is the number of nodes in each layer
    \item Training data: $\{(\rx_i, \ry_i)\}_{i=1}^n$, where for all $i\in \{1,\ldots,n\}$, $\rx_i \in \sR^{\gd}$ and $\ry_i \in \sR^d$
    \item $h: \rx \mapsto \ry$ is the predictor (deep neural network)
    \item $\vtheta$ is the set of parameters of the predictor
    \item For all $i$, $h_i: \vtheta \mapsto h(\rx_i;\vtheta)$ is the output of the network for example $i$ as a function of $\vtheta$
    \item $\mathcal{L}$ is the objective function
    \item $\mathcal{L}_i: \vv \mapsto \mathcal{L}(\vv, \ry_i)$ is the objective function for example $i$.
    \item We denote by $\vL$ the vector-valued function $\vL: (\vv_i)_{i=1}^n \mapsto (\cL_i(\vv_i))_{i=1}^n$
    \item $b \in \{1, \ldots, n\}$ is the batch size
    \item $\eta > 0$ is the learning rate
    \item ERM is short for Empirical Risk Minimization
\end{itemize}

\subsubsection{Distributionally Robust Optimisation Notations}

\begin{itemize}
    \item Forall $\vtheta$, $R(\vL(h(\vtheta))) 
        = \max_{\vq \in \Delta_n} \E_{\vq} \left[\cL\left(h(\rx; \vtheta), \ry\right)\right]
        - \frac{1}{\de} D_{\phi}(\vq \Vert \vp_{\textrm{train}})$ is the \textbf{Distributionally Robust Loss} evaluated at $\vtheta$, where $\de>0$ is the parameter that adjusts the distributionally robustness.
        For short, we also used the terms \textbf{distributionally robust loss} or just \textbf{robust loss} for $R(\vL(h(\vtheta)))$
    \item DRO is short for Distributionally Robust Optimisation
\end{itemize}

\subsubsection{Miscellaneous}
% \begin{itemize}
%     \item 
    By abuse of notation, and similarly to \cite{allen-zhu19a}, we use the Bachmann-Landau notations to hide constants that do not depend on our main hyper-parameters.
    Let $f$ and $g$ be two scalar functions, we note:
    \[
    \left\{
    \begin{array}{ccccc}
        f \leq O(g)      & \iff & \exists c > 0 & \textup{ s.t. } & f \leq c g\\
        f \geq \Omega(g) & \iff & \exists c > 0 & \textup{ s.t. } & f \geq c g\\
        f = \Theta(g)    & \iff & \exists c_1 > 0 \textup{ and } \exists c_2 > c_1 & \textup{ s.t. } & c_1 g \leq f \leq c_2 g\\
    \end{array}
    \right.
    \]
% \end{itemize}
\subsection{Evaluation of the Influence of $\de$ on the Segmentation Performance for BraTS}\label{s:more_results_brats}

\begin{table}[H]
	\centering
	\caption{\textbf{Detailed evaluation on the BraTS 2019 online validation set (125 cases).}
	All the models in this table were trained using the default \textbf{SGD with Nesterov momentum} of nnU-Net~\cite{isensee2021nnu}.
	Dice scores were computed using the BraTS online plateform for evaluation \protect\url{https://ipp.cbica.upenn.edu/}.
	ERM: Empirical Risk Minimization,
	DRO: Distributionally Robust Optimization,
	IS: Importance Sampling is used,
	IQR: Interquartile range.
    The best values are in bold.
	}
	\begin{tabularx}{\textwidth}{c *{9}{Y}}
		\toprule
        \textbf{Optimization}
        & \multicolumn{3}{c}{Enhancing Tumor} 
        & \multicolumn{3}{c}{Whole Tumor}
        & \multicolumn{3}{c}{Tumor Core}\\
        \cmidrule(lr){2-4} \cmidrule(lr){5-7} \cmidrule(lr){8-10}
		\textbf{problem} & Mean & Median & IQR & Mean & Median & IQR & Mean & Median & IQR\\ 
	\midrule
		ERM 
		 & 73.0 & 87.1 & 15.6
		 & 90.7 & 92.6 & \bf5.4
		 & 83.9 & 90.5 & 14.3\\
	\cmidrule(lr){1-10}
        DRO $\de=10$
         & 74.6 & 86.8 & 14.1
         & \bf90.8 & \bf93.0 & 5.9
         & 83.4 & 90.7 & 14.5\\
    \cmidrule(lr){1-10}
        DRO $\de=10$ IS
         & \bf75.3 & 86.0 & \bf13.3
         & 90.0 & 91.9 & 7.0
         & 82.8 & 89.1 & 14.3\\
	\cmidrule(lr){1-10}
        DRO $\de=100$ 
         & 73.4 & 86.7 & 14.3
         & 90.6 & 92.6 & 6.2
         & \bf84.5 & \bf90.9 & 13.7\\
	\cmidrule(lr){1-10}
        DRO $\de=100$ IS
         & 74.5 & \bf87.3 & 13.8
		 & 90.6 & 92.6 & 5.9
		 & 84.1 & 90.0 & \bf12.5\\
	\cmidrule(lr){1-10}
        DRO $\de=1000$
         & 74.5 & 84.2 & 33.0
         & 89.5 & 91.8 & 5.9
         & 71.1 & 87.2 & 41.1\\
    \cmidrule(lr){1-10}
        DRO $\de=1000$ IS
         & 72.2 & 85.7 & 15.0
         & 90.3 & 92.2 & 6.3
         & 81.1 & 89.4 & 15.1\\
	\bottomrule
	\end{tabularx}
	\label{tab:models_all_results}
\end{table}
\subsection{Importance Sampling Approximation in \Algref{alg:1}}\label{appendix:importance_sampling}

In this section, we give additional details about the approximation made in the computation of the importance weights (step 9 of \Algref{alg:1}).

Let $\vtheta$ be the parameters of the neural network $h$, $\vL=\left(L_i\right)_{i=1}^n$ be the stale per-example loss vector, and let $i$ be an index in the current batch $I$.

We start from the definition of the importance weight $w_i$ for example $i$ and use the formula for the hardness weighted sampling probabilities of Example~\ref{ex:softmax}.

\begin{equation}
    \begin{aligned}
        w_i &= \frac{p_i^{new}}{p_i^{old}}\\
            &= \frac{\exp\left(\de L_i^{new}\right)}{\exp\left(\de L_i^{new}\right) + \sum_{j \neq i} \exp\left(\de L_j^{old}\right)}
            \times
            \frac{\sum_{j=1}^n \exp\left(\de L_j^{old}\right)}{\exp\left(\de L_i^{old}\right)}\\
            & \approx \exp\left(\de (L_i^{new} - L_{i}^{old})\right)
    \end{aligned}
\end{equation}

where we have assumed that the two sums of exponentials are approximately equal.
% \section{Proofs}
\subsection{Proof of Example \ref{ex:softmax}: Formula of the Sampling Probabilities for the KL Divergence}\label{s:proof_softmax}

We give here a simple proof of the formula of the sampling probabilities for the KL divergence as $\phi$-divergence (i.e. $\phi: z \mapsto z \log(z) - z + 1$)
\[
\forall \vtheta, \quad 
\bar{p}(\vL(h(\vtheta))) 
= \softmax\left(\de \vL(h(\vtheta))\right)
\]

\paragraph*{Proof:}
For any $\vtheta$, the distributionally robust loss for the KL divergence at $\vtheta$ is given by
\[
\begin{aligned}
    R \circ \vL \circ ~h (\vtheta)
    &= \max_{\vq \in \Delta_n}
    \left(
    \sum_{i=1}^n q_i \cL_i \circ ~h_i (\vtheta)
    - \frac{1}{\de}\sum_{i=1}^n q_i\log\left(n q_i\right)
    \right)\\
    &= \max_{\vq \in \Delta_n}
    \sum_{i=1}^n \left(q_i \cL_i \circ ~h_i (\vtheta)
    - \frac{1}{\de} q_i\log\left(n q_i\right)
    \right)
\end{aligned}
\]
where we have used that $\frac{1}{p_{\textrm{train},i}} = n$ inside the $\log$ function.
To simplify the notations, let us denote 
$\vv=(v_i)_{i=1}^n
:=\left(\cL_i \circ ~h_i (\vtheta)\right)_{i=1}^n$,
and $\bar{\vp} = (\bar{p}_i)_{i=1}^n := \bar{\textbf{p}}(\vL(h(\vtheta)))$.
Thus $\bar{\textbf{p}}(\vL(\vh(\vtheta)))$ is, by definition, solution of the optimization problem
\begin{equation}
    \label{eq:optim_softmax_1}
    \argmax_{\vq \in \Delta_n}
    \sum_{i=1}^n \left(q_i v_i
    - \frac{1}{\de} q_i\log\left(n q_i\right)
    \right)
\end{equation}
First, let us remark that the function $q \mapsto \sum_{i=1}^n q_i\log\left(n q_i\right)$ is strictly convex on the non empty closed convex set $\Delta_n$ as a sum of strictly convex functions.
This implies that the optimization \eqref{eq:optim_softmax_1} has a unique solution and as a result $\bar{p}(\vL(h(\vtheta)))$ is well defined.

We now reformulate the optimization problem \eqref{eq:optim_softmax_1} as a convex smooth constrained optimization problem by writing the condition $\vq \in \Delta_n$ as constraints.
\begin{equation}
    \label{eq:optim_softmax_2}
    \begin{aligned}
        \argmax_{\vq \in \sR^n_+}&
    \sum_{i=1}^n \left(q_i v_i
    - \frac{1}{\de} q_i\log\left(n q_i\right)
    \right)\\
    \text{s.t.}& \sum_{i=1}^n q_i =1 
    \end{aligned}
\end{equation}

There exists a Lagrange multiplier $\lambda \in \sR$, such that the solution $\bar{p}$ of \eqref{eq:optim_softmax_2} is characterized by
\begin{equation}
    \label{eq:optim_softmax_3}
    \left\{
    \begin{aligned}
        \forall i \in \{1, \ldots, n\},\quad & v_i
    - \frac{1}{\de} \left(
        \log\left(n \bar{p}_i\right) + 1
    \right) + \lambda = 0\\
    & \sum_{i=1}^n \bar{p}_i =1 
    \end{aligned}
    \right.
\end{equation}
Which we can rewrite as
\begin{equation}
    \label{eq:optim_softmax_4}
    \left\{
    \begin{aligned}
        \forall i \in \{1, \ldots, n\},\quad & 
        \bar{p}_i = \frac{1}{n} \exp\left(
        \de \left(v_i + \lambda\right) - 1
        \right)\\
    & \frac{1}{n} \sum_{i=1}^n \exp\left(
        \de \left(v_i + \lambda\right) - 1
        \right) = 1 
    \end{aligned}
    \right.
\end{equation}

The last equality gives
\[
    \exp\left(\de \lambda - 1\right) 
        = \frac{n}{\sum_{i=1}^n \exp\left(\de v_i\right)}
\]
And by replacing in the formula of the $\bar{p}_i$
\[
\begin{aligned}
    \forall i \in \{1, \ldots, n\},\quad
        \bar{p}_i =&
        \frac{1}{n} 
        \exp\left(\de v_i\right)
        \exp\left(
        \de \lambda - 1
        \right)\\
        =&\frac{\exp\left(\de v_i\right)}{{\sum_{j=1}^n \exp\left(\de v_j\right)}}
\end{aligned}
\]
Which corresponds to
$
\bar{\vp} = \softmax\left(\de \vv\right)
\,\,\blacksquare
$

\subsection{Proof of Lemma \ref{lemma:R_property}: Regularity Properties of R}\label{s:regularity_R}

For the ease of reading, let us first recall that given a $\phi$-divergence $D_{\phi}$ that satisfies Definition~\ref{def:phi_divergence}, we have defined in \eqref{eq:dro}
\begin{equation}
    \begin{aligned}
        R:\,\,&\sR^n \rightarrow \sR\\
             & \vv \,\mapsto \max_{\vq \in \Delta_n} \sum_i q_i v_i
        - \frac{1}{\de} D_{\phi}(\vq \Vert \vp_{\textrm{train}})
    \end{aligned}
\end{equation}
And in \eqref{eq:G_def}
\begin{equation}
    \begin{aligned}
        G:\,\,&\sR^n \rightarrow \sR\\
             & \vp \,\mapsto \frac{1}{\de} D_{\phi}(\vp \Vert \vp_{\textrm{train}}) + \delta_{\Delta_n}(\vp)
    \end{aligned}
\end{equation}
where $\delta_{\Delta_n}$ is the characteristic function of the closed convex set $\Delta_n$, i.e.
\begin{equation}
    \forall \vp \in \sR^n,\,\, \delta_{\Delta_n}(\vp)=\left\{
\begin{array}{cl}
    0 & \text{if } \vp \in \Delta_n \\
    +\infty & \text{otherwise}
\end{array}
\right.
\end{equation}

We now prove Lemma \ref{lemma:R_property} on the regularity of $R$.

\begin{lemma}[Regularity of R -- Restated from Lemma \ref{lemma:R_property}]
\label{lemma:R_property_re}
    Let $\phi$ that satisfies Definition~\ref{def:phi_divergence}, $G$ and $R$ satisfy
    % , and $G: p \mapsto \frac{1}{\de} D_{\phi}(p\Vert \ptrain) + \delta_{\Delta_n}(p)$
    \begin{equation}
        \label{eq:strong_convexity_G_re}
        G \text{ is} \left(\frac{n\rho}{\de}\right)~\text{-strongly convex}
    \end{equation}
    \begin{equation}
        \label{eq:link_R_and_G_re}
        R(\vL(h(\vtheta)))
        =
        \max_{\vq \in \sR^n} 
            \left(
            \langle \vL(h(\vtheta)), \vq\rangle - G(\vq)
            \right)
        = G^*\left(\vL(h(\vtheta))\right)
    \end{equation}
    % And
    \begin{equation}
        \label{eq:gradient_Lip_R_re}
        R \text{ is } \left(\frac{\de}{n\rho}\right)~\text{-gradient Lipschitz continuous.}
    \end{equation}
\end{lemma}

\paragraph*{Proof:}
$\phi$ is $\rho$-strongly convex on $[0,n]$ so
\begin{equation}
	\label{eq:strong-conv}
	\forall x,y \in [0, n]^2, \forall \lambda \in [0,1],
	\phi\left(\lambda x + (1 - \lambda)y \right) \leq
	\lambda \phi(x) + (1-\lambda)\phi(y) - \frac{\rho \lambda(1-\lambda)}{2} |y-x|^2
\end{equation}
Let $\vp=\left(p_i\right)_{i=1}^n$, $\vq=\left(q_i\right)_{i=1}^n \in \Delta_n$, and $\lambda \in [0,1]$,
using \eqref{eq:strong-conv} and the convexity of $\delta_{\Delta_n}$, we obtain:
\begin{equation}
	\begin{aligned}
	G\left(\lambda \vp + (1 - \lambda)\vq \right) &
	= \frac{1}{\de n} \sum_{i=1}^n \phi\left(n\lambda p_i + n(1 - \lambda)q_i \right) + \delta_{\Delta_n}\left(\lambda \vp + (1 - \lambda)\vq \right)\\
	&\leq \lambda G(\vp) + (1 - \lambda)G(\vq) - \frac{1}{\de n}\sum_{i=1}^n\frac{\rho \lambda(1-\lambda)}{2} |np_i-nq_i|^2\\
	&\leq \lambda G(\vp) + (1 - \lambda)G(\vq) - \frac{n \rho}{\de}\frac{\lambda(1 - \lambda)}{2} \norm{\vp - \vq}^2
	\end{aligned}
\end{equation}

This proves that $G$ is $\frac{n\rho}{\de}$-strongly convex.

Since $G$ is convex, $R=G^*$ is also convex, and $R^*=\left(G^*\right)^*=G$~\cite{hiriart2013convex}.
We obtain \eqref{eq:link_R_and_G_re} using Definition \ref{def:fenchel_conjugate}.

We now show that $R$ is Frechet differentiable on $\sR^n$.
Let $\vv \in \sR^n$.

$G$ is strongly-convex, so in particular $G$ is strictly convex.
This implies that the following optimization problem has a unique solution that we denote $\hat{\vp}(\vv)$.
\begin{equation}
    \label{eq:max_pb_R_grad_lip}
    \argmax_{\vq \in \sR^n} 
            \left(
            \langle \vv, \vq\rangle - G(\vq)
            \right)
\end{equation}

In addition
\[
\begin{aligned}
    \hat{\vp} \in \Delta_n \text{ solution of \eqref{eq:max_pb_R_grad_lip} }
    & \Longleftrightarrow 0 \in \vv - \partial G(\hat{\vp})\\
    & \Longleftrightarrow \vv \in \partial G(\hat{\vp})\\
    & \Longleftrightarrow \hat{\vp} \in \partial G^*(\vv) \\
    % \quad \text{(using \cite[Proposition 6.1.2]{hiriart2013convex})}\\
    & \Longleftrightarrow \hat{\vp} \in \partial R(\vv)
    % \quad \text{(using \eqref{eq:link_R_and_G_re})}
\end{aligned}
\]
where we have used \cite[Proposition 6.1.2 p.39]{hiriart2013convex} for the third equivalence, and \eqref{eq:link_R_and_G_re} for the last equivalence.

As a result, $\partial R(\vv)=\{\hat{\vp}(\vv)\}$.
this implies that $R$ admit a gradient at $\vv$, and
\begin{equation}
    \label{eq:R_differentiable}
    \nabla_{\vv} R(\vv) = \hat{\vp}(\vv)
\end{equation}

Since this holds for any $\vv \in \sR^n$, we deduce that $R$ is Frechet differentiable on $\sR^n$. $\blacksquare$

We are now ready to show that $R$ is $\frac{\de}{n\rho}$-gradient Lipchitz continuous by using the following lemma \cite[Theorem 6.1.2 p.280]{hiriart2013convex}.

\begin{lemma}%[\cite[Theorem 6.1.2 p.280]{hiriart2013convex}]
    A necessary and sufficient condition for a convex function $f\,: \sR^n \rightarrow \sR$ to be $c$-strongly convex on a convex set $C$ is that for all $x_1,x_2 \in C$
    \[
    \langle
    s_2 - s_1, x_2 - x_1
    \rangle
    \geq c \norm{x_2 - x_1}^2 \quad \text{for all } s_i \in \partial f(x_i), i=1,2.
    \]
\end{lemma}
Using this lemma for $f=G$, $c=\frac{n\rho}{\de}$, and $C=\Delta_n$, we obtain:

For all $\vp_1, \vp_2 \in \Delta_n$, for all $\vv_1 \in \partial G(\vp_1)$, $\vv_2 \in \partial G(\vp_2)$,
\[
\langle
\vv_2 - \vv_1, \vp_2 - \vp_1
\rangle
\geq \frac{n\rho}{\de} \norm{\vp_2 - \vp_1}^2
\]
In addition, for $i \in \{1,\,2\}$, $\vv_i \in \partial G(\vp_i) \Longleftrightarrow \vp_i \in \partial R(\vv_i)= \{\nabla_{\vv} R(\vv_i)\}$.

And using Cauchy Schwarz inequality
\[
\norm{\vv_2 - \vv_1} \norm{\vp_2 - \vp_1}
\geq
\langle
\vv_2 - \vv_1, \vp_2 - \vp_1
\rangle
\]
We conclude that
\[
\frac{n\rho}{\de} \norm{\nabla_{\vv} R(\vv_2) - \nabla_{\vv} R(\vv_1)} \leq \norm{\vv_2 - \vv_1}
\]
Which implies that $R$ is $\frac{\de}{n\rho}$-gradient Lipchitz continuous. $\blacksquare$
\subsection{Proof of Lemma \ref{lemma:robust_loss_sg}: Formula of the Distributionally Robust Loss Gradient}\label{s:robust_loss_sg}
We prove Lemma \ref{lemma:robust_loss_sg} that we restate here for the ease of reading.
\begin{lemma}[Stochastic Gradient of the DRO Loss -- Restated from Lemma \ref{lemma:robust_loss_sg}]
\label{lemma:robust_loss_sg_re}
For all $\vtheta$, we have
% $\bar{p}(\cL(h(\vtheta)))$ satisfies:
    \begin{equation}
        \label{eq:p_re}
        \bar{p}(\vL(h(\vtheta))) = \nabla_{\vv} R(\vL(h(\vtheta)))
    \end{equation}
    \begin{equation}
        \label{eq:sg_re}
        \nabla_\vtheta (R \circ \vL \circ ~h)(\vtheta) 
        = \E_{\bar{\vp}(\cL(\vh(\vtheta)))}\left[\nabla_\vtheta \cL\left(h(\rx; \vtheta), y\right)\right]
    \end{equation}
\end{lemma}
where $\nabla_\vv R$ is the gradient of $R$ with respect to its input.

\paragraph*{Proof:}
For a given $\vtheta$,
equality \eqref{eq:p_re} is a special case of \eqref{eq:R_differentiable} for $\vv = \cL(\vh(\vtheta))$.

Then using the chain rule and \eqref{eq:p_re},
\[
\begin{aligned}
    \nabla_{\vtheta} (R \circ \vL \circ ~h)(\vtheta)
    &= \sum_{i=1}^n \frac{\partial R}{\partial v_i}(\vL \circ~h(\vtheta))) \nabla_{\vtheta} (\cL_i \circ ~h_i)(\vtheta)\\
    &= \sum_{i=1}^n \bar{p}_i(\vL(h(\vtheta))) \nabla_{\vtheta} (\cL_i \circ ~h_i)(\vtheta)\\
    &= \E_{\bar{\vp}(\cL(\vh(\vtheta)))}\left[\nabla_\vtheta \cL\left(h(\rx; \vtheta), y\right)\right]
\end{aligned}
\]
Which concludes the proof. $\blacksquare$
\subsection{Proof of Theorem \ref{th:hard_example_mining}: Distributionally Robust Optimization as Principled Hard Example Mining}\label{s:hard_example_mining}

In this section, we demonstrate that the proposed hardness weighted sampling can be interpreted as a principled hard example mining method.

Let $D_{\phi}$ an $\phi$-divergence satisfying Definition~\ref{def:phi_divergence},
and $\vv=\left(v_i\right)_{i=1}^n \in \sR^n$.
% a pseudo loss vector.
$\vv$ will play the role of a generic loss vector.

$\phi$ is strongly convex, and $\Delta_n$ is closed and convex, so the following optimization problem has one and only one solution
\begin{equation}
    \max_{\vp=\left(p_i\right)_{i=1}^n \in \Delta_n} \langle \vv, \vp \rangle - \frac{1}{\de n}\sum_{i=1}^n \phi(n p_i)
\end{equation}
Making the constraints associated with $p \in \Delta_n$ explicit, this can be rewritten as
\begin{equation}
\begin{aligned}
    \max_{\vp=\left(p_i\right)_{i=1}^n \in \sR^n} &\langle \vv, \vp \rangle - \frac{1}{\de n}\sum_{i=1}^n \phi(n p_i)\\
    \text{ s.t. } &\forall i \in \{1, \ldots, n\},\,\, p_i \geq 0\\
    \text{ s.t. } &\sum_{i=1}^n p_i = 1
\end{aligned}
\end{equation}
There exists KKT multipliers $\lambda \in \sR$ and $\forall i,\, \mu_i \geq 0$ such that the solution $\bar{\vp}=\left(\bar{p}_i\right)_{i=1}^n$ satisfies
\begin{equation}
\label{eq:hard_proof}
    \left\{
    \begin{aligned}
        \forall i \in \{1, \ldots, n\}, \quad& v_i - \frac{1}{\de}\phi'(n\bar{p}_i) + \lambda - \mu_i = 0\\
        \forall i \in \{1, \ldots, n\},\quad& \mu_i p_i = 0\\
        \forall i \in \{1, \ldots, n\},\quad& p_i \geq 0\\
        & \sum_{i=1}^n \bar{p}_i = 1
    \end{aligned}
    \right.
\end{equation}
Since $\phi$ is continuously differentiable and strongly convex, we have $\left(\phi'\right)^{-1} = \left(\phi^*\right)'$, where $\phi^*$ is the Fenchel conjugate of $\phi$~\cite[Proposition 6.1.2]{hiriart2013convex}.
As a result, \eqref{eq:hard_proof} can be rewritten as
\begin{equation}
\label{eq:hard_proof_1}
    \left\{
    \begin{aligned}
        \forall i \in \{1, \ldots, n\}, \quad& \bar{p}_i = \frac{1}{n}\left(\phi^*\right)'\left(\de(v_i + \lambda - \mu_i)\right)\\
        \forall i \in \{1, \ldots, n\},\quad& \mu_i p_i = 0\\
        \forall i \in \{1, \ldots, n\},\quad & p_i \geq 0\\
        & \frac{1}{n}\sum_{i=1}^n \left(\phi^*\right)'\left(\de(v_i + \lambda - \mu_i)\right) = 1
    \end{aligned}
    \right.
\end{equation}
We now show that the KKT multipliers are uniquely defined.
\paragraph{\textbf{The $\mu_i$'s are uniquely defined by $\vv$ and $\lambda$}:}\ \\
Since $\forall i \in \{1, \ldots, n\},\,\, \mu_i p_i = 0,\,\, p_i \geq 0$ and $\mu_i \geq 0$,
for all $\forall i \in \{1, \ldots, n\}$, either $p_i=0$ or $\mu_i=0$.
In the case $p_i=0$ and using \eqref{eq:hard_proof_1} it comes $\left(\phi^*\right)'\left(\de(v_i + \lambda - \mu_i)\right)=0$.

According to Definition~\ref{def:phi_divergence}, $\phi$ is strongly convex and continuously differentiable, so $\phi'$ and $(\phi^*)'=(\phi')^{-1}$ are continuous and strictly increasing functions.
As a result, it exists a unique $\mu_i$ (dependent to $\vv$ and $\lambda$) such that:
\[
\left(\phi^*\right)'\left(\de(v_i + \lambda - \mu_i)\right)=0
\]
And \eqref{eq:hard_proof_1} can be rewritten as
\begin{equation}
\label{eq:hard_proof_2}
    \left\{
    \begin{aligned}
        \forall i \in \{1, \ldots, n\}, \quad& \bar{p}_i = \relu\left( \frac{1}{n}\left(\phi^*\right)'\left(\de(v_i + \lambda)\right)
        \right)
        =\frac{1}{n}\relu\left(\left(\phi^*\right)'\left(\de(v_i + \lambda)\right)
        \right)\\
        & \frac{1}{n}\sum_{i=1}^n \relu\left(\left(\phi^*\right)'\left(\de(v_i + \lambda)\right)\right) = 1\\
    \end{aligned}
    \right.
\end{equation}

\paragraph{\textbf{The KKT multiplier $\lambda$ is uniquely defined by $\vv$ and a continuous function of $\vv$}:}\ \\
Let $\lambda \in \sR$ that satisfies \eqref{eq:hard_proof_2}.
We have $\frac{1}{n}\sum_{i=1}^n \relu\left(\left(\phi^*\right)'\left(\de(v_i + \lambda)\right)\right) = 1$.
So there exists at least one index $i_0$ such that
\[
\relu\left(\left(\phi^*\right)'\left(\de(v_{i_0} + \lambda)\right)\right)=\left(\phi^*\right)'\left(\de(v_{i_0} + \lambda)\right) \geq 1
\]
Since $(\phi^*)^{-1}$ is continuous and striclty increasing, 
$\lambda'\mapsto \relu\left(\left(\phi^*\right)'\left(\de(v_{i_0} + \lambda')\right)\right)$ is continuous and strictly increasing on a neighborhood of $\lambda$.
In addition $\relu$ is continuous and increasing,
so for all $i \in \{1,\ldots,n\}$, 
$\lambda'\mapsto \relu\left(\left(\phi^*\right)'\left(\de(v_{i} + \lambda')\right)\right)$ is a continuous and increasing function.

As a result, $\lambda'\mapsto \frac{1}{n}\sum_{i=1}^n \relu\left(\left(\phi^*\right)'\left(\de(v_i + \lambda')\right)\right)$ is a continuous function that is increasing on $\sR$, and strictly increasing on a neighborhood of $\lambda$.
This implies that $\lambda$ is uniquely defined by $\vv$, and that $\vv \mapsto \lambda(\vv)$ is continuous.

\subsubsection{Link between Hard Weighted Sampling and Hard Example Mining}

For any pseudo loss vector $\vv=\left(v_i\right)_{i=1}^n \in \sR^n$, there exists a unique KKT multiplier $\lambda$ and a unique $\bar{\vp}$ that satisfies \eqref{eq:hard_proof_2}, so we can define the mapping:
\begin{equation}
    \begin{aligned}
        \bar{\vp}:\,\,\sR^n &\rightarrow \Delta_n\\
                    \vv & \mapsto \bar{\vp}(\vv;\lambda(\vv))\\
    \end{aligned}
\end{equation}
where for all $\vv$, $\lambda(\vv)$ is the unique $\lambda \in \sR$ satisfying \eqref{eq:hard_proof_2}.

We will now demonstrate that each $\bar{\vp}_{i_0}(\vv)$ for $i_0 \in \{1,\ldots, n\}$ is an increasing function of $v_i$ and a decreasing function of the $v_i$ for $i\neq i_0$.
Without loss of generality we assume $i_0=1$.

Let $\vv=\left(v_i\right)_{i=1}^n \in \sR^n$, and $\epsilon >0$.
Let us define $\vv'=\left(v_i'\right)_{i=1}^n \in \sR^n$, such that $v_1'=v_1+\epsilon$ and $\forall i \in \{2,\ldots, n\},\,\, v_i'=v_i$.
Similarly as in the proof of the uniqueness of $\lambda$ above, we can show that there exists $\eta > 0$ such that the function
\[
F:\lambda'\mapsto \frac{1}{n}\sum_{i=1}^n \relu\left(\left(\phi^*\right)'\left(\de(v_i + \lambda')\right)\right)
\]
is continuous and strictly increasing on $[\lambda(\vv)-\eta, \lambda(\vv)+\eta]$, and $F(\lambda(\vv))=1$.

$v \mapsto \lambda(\vv)$ is continuous, so for $\epsilon$ small enough $\lambda(\vv') \in [\lambda(v)-\eta, \lambda(\vv)+\eta]$.

Let us now prove by contradiction that $\lambda(\vv') \leq \lambda(\vv)$. Therefore,
let us assume that $\lambda(\vv') > \lambda(\vv)$. Then, as $\relu \circ \left(\phi^*\right)'$ is an increasing function and $F$ is strictly increasing on $[\lambda(\vv)-\eta, \lambda(\vv)+\eta]$, and $\epsilon > 0$ we obtain
\[
\begin{aligned}
    1 &= \frac{1}{n}\sum_{i=1}^n \relu\left(\left(\phi^*\right)'\left(\de(v_i' + \lambda(\vv'))\right)\right)\\
    & \geq \frac{1}{n}\sum_{i=1}^n \relu\left(\left(\phi^*\right)'\left(\de(v_i + \lambda(\vv'))\right)\right)\\
    & \geq F(\lambda(\vv'))\\
    & > F(\lambda(\vv))\\
    & >1
\end{aligned}
\]
which is a contradiction.
As a result
\begin{equation}
    \label{eq:hard_proof_3}
    \lambda(\vv') \leq \lambda(\vv)
\end{equation}

Using equations \eqref{eq:hard_proof_2} and \eqref{eq:hard_proof_3}, and the fact that $\relu \circ \left(\phi^*\right)'$ is an increasing function, we obtain for all $i \in \{2, \ldots, n\}$
\begin{equation}
\begin{aligned}
    \bar{p}_i(\vv') 
 &= \frac{1}{n}\relu\left(\left(\phi^*\right)'\left(\de(v_i' + \lambda(\vv'))\right)\right)\\
 &= \frac{1}{n}\relu\left(\left(\phi^*\right)'\left(\de(v_i + \lambda(\vv'))\right)\right)\\
 & \leq \frac{1}{n}\relu\left(\left(\phi^*\right)'\left(\de(v_i + \lambda(\vv))\right)\right)\\
 & \leq \bar{p}_i(\vv)
\end{aligned}
\end{equation}

In addition
\[
\sum_{i=1}^n \bar{p}_i(\vv') = 1 = \sum_{i=1}^n \bar{\vp}_i(\vv)
\]
So necessarily
\begin{equation}
    \bar{\vp}_1(\vv') \geq \bar{\vp}_1(\vv)
\end{equation}
This holds for any $i_0$ and any $\vv$, which concludes the proof. $\blacksquare$

\subsection{Proof of Equivalence between \eqref{eq:expvar} and \eqref{eq:dro2}: Link between DRO and Percentile Loss}\label{s:proof_dro_and_percentile}

In the DRO optimization problem of equation \eqref{eq:dro2}, the optimal $\vq$ for any $\vtheta$ has the closed-form formula as shown in Appendix~\ref{s:proof_softmax}
\begin{equation*}
    \forall \vtheta,\quad
    \vq^*\left(\vtheta\right) = \softmax\left(
        \left(\de \cL\left(h(\vx_i; \vtheta), \vy_i\right)\right)_{i=1}^n
    \right)
\end{equation*}
By injecting this in equation \eqref{eq:dro2}, we obtain
\begin{align*}
    \min_{\vtheta}\,& \max_{\vq \in \Delta_n}
        \left(
        \sum_{i=1}^n q_i \cL\left(h(\vx_i; \vtheta), \vy_i\right)
        - \frac{1}{\de} D_{KL}\left(\vq\, \biggr\Vert\, \frac{1}{n}\mathbf{1}\right)
        \right)\\
    % \iff
    = \min_{\vtheta}\,&
        \left(
        \sum_{i=1}^n q^*_i(\vtheta) \cL\left(h(\vx_i; \vtheta), \vy_i\right)
        - \frac{1}{\de} 
            \sum_{i=1}^n q^*_i(\vtheta)
                \log\left(\frac{
                    \exp\left(\de \cL\left(h(\vx_i; \vtheta), \vy_i\right)\right)}{
                    \frac{1}{n}\sum_{j=1}^n\exp\left(\de \cL\left(h(\vx_j; \vtheta), \vy_j\right)\right)}\right)
        \right)\\
    = \min_{\vtheta}\,&
        \left(
        \sum_{i=1}^n q^*_i(\vtheta) \cL\left(h(\vx_i; \vtheta), \vy_i\right)
        - \sum_{i=1}^n q^*_i(\vtheta)
                \frac{1}{\de} \log\left(\exp\left(\de \cL\left(h(\vx_i; \vtheta), \vy_i\right)\right)\right)
        \right.\\
    & + \frac{1}{\de}
        \left.
            \left(\sum_{i=1}^n q^*_i(\vtheta)\right) \times
            \log\left(
            \frac{1}{n}\sum_{j=1}^n\exp\left(\de \cL\left(h(\vx_j; \vtheta), \vy_j\right)\right)
            \right)
        \right)\\
    %
    % \min_{\vtheta}\,& \frac{1}{\de} \log\left(
    %         \sum_{j=1}^n\exp\left(\de \cL\left(f(\vx_j; \vtheta), \vy_j\right)\right)
    %         \right)
    %         - \frac{1}{\de} \log\left(n\right)
\end{align*}
Since the first two terms cancel each other and $\sum_{i=1}^n q^*_i(\vtheta)=1$, we obtain
\begin{align*}
    \min_{\vtheta}\,& \max_{\vq \in \Delta_n}
        \left(
        \sum_{i=1}^n q_i \cL\left(h(\vx_i; \vtheta), \vy_i\right)
        - \frac{1}{\de} D_{KL}\left(\vq\, \biggr\Vert\, \frac{1}{n}\mathbf{1}\right)
        \right)\\
    = \min_{\vtheta}\,& \frac{1}{\de} \log\left(
            \sum_{j=1}^n\exp\left(\de \cL\left(h(\vx_j; \vtheta), \vy_j\right)\right)
            \right)
            - \frac{1}{\de} \log\left(n\right)\\
    = \min_{\vtheta}\,& \frac{1}{\de} \log\left(
            \sum_{j=1}^n\exp\left(\de \cL\left(h(\vx_j; \vtheta), \vy_j\right)\right)
            \right)
\end{align*}
which is equivalent to the optimization problem \eqref{eq:expvar} because the term $\frac{1}{\de} \log\left(n\right)$ above and the term $\frac{1}{\de} \log\left(\alpha n\right)$ in \eqref{eq:expvar} are independent of $\vtheta$ $\blacksquare$
\subsection{Proof of Theorem~\ref{th:convergence_dro}: convergence of SGD with Hardness Weighted Sampling for Over-parameterized Deep Neural Networks with ReLU}\label{s:convergence_detailed}

In this section, we provide the proof of Theorem~\ref{th:convergence_dro}.
This generalizes the convergence of SGD for empirical risk minimization in~\cite[Theorem 2]{allen-zhu19a} to the convergence of SGD and our proposed hardness weighted sampler for distributionally robust optimization.

We start by describing in details the assumptions made for our convergence result in Section~\ref{s:assumptions}.

In Section~\ref{s:convergence_theorem2}, we restate Theorem~\ref{th:convergence_dro} using the assumptions and notations previously introduced in Section~\ref{s:notations}.

In Section~\ref{s:proof_convergence}, we give the proof of the convergence theorem. 
We focus on providing theoretical tools that could be used to generalize any convergence result for ERM using SGD to DRO using SGD with hardness weighted sampling as described in \Algref{alg:1}.

\subsubsection{Assumptions}\label{s:assumptions}
Our analysis is based on the results developed in~\cite{allen-zhu19a} which is a simplified version of~\cite{allen2018convergence}.
Improving on those theoretical results would automatically improve our results as well.

In the following we state our assumptions on the neural network $h$, and the per-example loss function $\cL$.
\begin{assumption}[Deep Neural Network]
\label{as:2}
In this section, we use the following notations and assumptions similar to \cite{allen-zhu19a}:
    \begin{itemize}
        \item h is a fully connected neural network with $L+2$ layers, $\relu$ as activation functions, and $m$ nodes in each hidden layer
        \item For all $i \in \{1, \ldots, n\}$, we denote $h_i: \vtheta \mapsto h_i(\rx_i;\vtheta)$ the $d$-dimensional output scores of $h$ applied to example $\rx_i$ of dimension $\gd$.
        \item For all $i \in \{1, \ldots, n\}$, we denote $\cL_i: h \mapsto \cL\left(h, \ry_i\right)$ where $\ry_i$ is the ground truth associated to example $i$.
        \item $\vtheta=\left(\vtheta_l\right)_{l=0}^{L+1}$ is the set of parameters of the neural network h, where $\vtheta_l$ is the set of weights for layer $l$ with $\vtheta_0 \in \sR^{\gd \times m}$, $\vtheta_{L+1} \in \sR^{m \times d}$, and $\vtheta_l \in \sR^{m \times m}$ for any other $l$.
        \item (Data separation) It exists $\delta > 0$ such that for all $i,j \in \{1, \ldots, n\}$, if $i \neq j, \norm{x_i - x_j} \geq \delta$.
        \item We assume $m \geq \Omega(d \times \textup{poly}(n,L,\delta^{-1}))$ for some sufficiently large polynomial $\textup{poly}$, and $\delta \geq O\left(\frac{1}{L}\right)$. We refer the reader to \cite{allen-zhu19a} for details about the polynomial $\textup{poly}$.
        \item The parameters $\vtheta=\left(\vtheta_l\right)_{l=0}^{L+1}$ are initialized at random such that:
        \begin{itemize}
            \item $\left[\vtheta_0\right]_{i,j} \sim \mathcal{N}\left(0, \frac{2}{m}\right)$ for every $(i,j) \in \{1, \ldots, m\}\times\{1, \ldots, \gd\}$
            \item $\left[\vtheta_l\right]_{i,j} \sim \mathcal{N}\left(0, \frac{2}{m}\right)$ for every $(i,j) \in \{1, \ldots, m\}^2$ and $l \in \{1, \ldots, L\}$
            \item $\left[\vtheta_{L+1}\right]_{i,j} \sim \mathcal{N}\left(0, \frac{1}{d}\right)$ for every $(i,j) \in \{1, \ldots, d\}\times\{1, \ldots, m\}$
        \end{itemize}
    \end{itemize}
\end{assumption}

\begin{assumption}[Regularity of $\cL$]
    \label{as:3}
    For all i, $\cL_i$ is a $C(\nabla \cL)$-gradient Lipschitz continuous, $C(\cL)$-Lipschitz continuous, and bounded (potentially non-convex) function.
\end{assumption}

\subsubsection{Convergence theorem (restated)}\label{s:convergence_theorem2}

In this section, we restate the convergence Theorem~\ref{th:convergence_dro} for SGD with hardness weighted sampling and stale per-example loss vector.

As an intermediate step, we will first 
generalize the convergence of SGD in \cite[Theorem 2]{allen-zhu19a} to the minimization of the distributionally robust loss using SGD and an \emph{exact} hardness weighted sampling~\eqref{eq:p}, i.e. with an exact per-example loss vector.
\begin{theorem}[Convergence with exact per-example loss vector]
    \label{th:conv_sgd_exact_loss_history}
    Let batch size $1 \leq b \leq n$, and $\epsilon > 0$.
    Suppose there exists constants $C_1,\, C_2,\, C_3 > 0$ such that
    the number of hidden units satisfies 
    $m \geq C_1 (d \epsilon^{-1} \times \textup{poly}(n,L,\delta^{-1}))$,
    $\delta \geq \left(\frac{C_2}{L}\right)$,
    and the learning rate be 
    $\eta_{exact} = C_3\left(
        \min \left(1,\,
        \frac{\alpha n^2 \rho}{\de C(\cL)^2+ 2 n \rho C(\nabla \cL)}
        \right)
        \times \frac{b\delta d}{\textup{poly}(n,L)m\log^2(m)}
        \right)$.
    There exists constants $C_4,\,C_5 >0$ such that
    with probability at least $1 - \exp\left(-C_4(\log^2(m))\right)$ over the randomness of the initialization and the mini-batches, SGD with hardness weighted sampling and exact per-example loss vector guarantees
    $\norm{\nabla_{\vtheta} (R\circ \vL \circ h)(\vtheta)} \leq \epsilon$ after 
    $T=C_5\left(\frac{L n^3}{\eta_{exact} \delta \epsilon^2}\right)$ iterations.
\end{theorem}
The proof can be found in Appendix~\ref{s:proof_exact_loss_history}.

$\alpha = \min_{\vtheta} \min_{i} \bar{p}_i(\vL(\vtheta))$ is a lower bound on the sampling probabilities.
For the Kullback-Leibler $\phi$-divergence, and for any $\phi$-divergence satisfying Definition~\ref{def:phi_divergence} with a robustness parameter $\de$ small enough, we have $\alpha > 0$.
We refer the reader to \cite[Theorem 2]{allen-zhu19a} for the values of the constants $C_1,\, C_2,\, C_3,\, C_4,\, C_5$ and the definitions of the polynomials.

Compared to \cite[Theorem 2]{allen-zhu19a} only the learning rate differs.
The $\min(1, \,.\,)$ operation in the formula for $\eta_{exact}$ allows us to guarantee that $\eta_{exact} \leq \eta '$ where $\eta '$ is the learning rate of \cite[Theorem 2]{allen-zhu19a}.

It is worth noting that for the KL $\phi$-divergence, $\rho=\frac{1}{n}$.
In addition, in the limit $\de \rightarrow 0$, which corresponds to ERM, we have $\alpha \rightarrow \frac{1}{n}$.
As a result, we recover exactly Theorem 2 of \cite{allen-zhu19a} as extended in their Appendix A for any smooth loss function $\cL$ that satisfies assumption $\ref{as:3}$ with $C(\nabla\cL)=1$.

% When the amount of distributionally robustness increases the sampling differs more and more from the uniform sampling and becomes more sensitive to changes of the loss distribution. One way to mitigate this issue is to reduce the learning rate.
%
% The conditions of Theorem~\ref{th:conv_sgd_exact_loss_history} are consistent with this observation since when $\de$ increases, $\alpha$ and $\eta_{exact}$ decreases.

% In practice, we have access only to a stale per-example loss vector.
%
We now restate the convergence of SGD with hardness weighted sampling and a stale per-example loss vector as in \Algref{alg:1}.
\begin{theorem}[Convergence with a stale per-example loss vector]
\label{th:conv_sgd_stale_loss_history}
Let batch size $1 \leq b \leq n$, and $\epsilon > 0$.
Under the conditions of Theorem \ref{th:conv_sgd_exact_loss_history}, 
the same notations,
and with the learning rate
$\eta_{stale} = C_6
\min \left(1,\,
\frac{\alpha \rho d^{3/2} \delta b \log\left(\frac{1}{1-\alpha}\right)}{\de C(\cL) A(\nabla \cL)L m^{3/2} n^{3/2} \log^2(m)}
\right)
\times \eta_{exact}$ for a constant $C_6 > 0$.
With probability at least $1 - \exp\left(-C_4(\log^2(m))\right)$ over the randomness of the initialization and the mini-batches, SGD with hardness weighted sampling and stale per-example loss vector guarantees
$\norm{\nabla_{\vtheta} (R\circ \vL \circ h)(\vtheta)} \leq \epsilon$ after 
$T=C_5\left(\frac{L n^3}{\eta_{stale} \delta \epsilon^2}\right)$ iterations.
\end{theorem}
The proof can be found in Appendix~\ref{s:proof_stale_loss_history}.

$C(\cL) > 0$ is a constant such that $\cL$ is $C(\cL)$-Lipschitz continuous,
and $A(\nabla \cL) > 0$ is a constant that bounds the gradient of $\cL$ with respect to its input.
$C(\cL)$ and $A(\nabla \cL)$ are guaranteed to exist under assumptions \ref{as:2}.

Compared to Theorem \ref{th:conv_sgd_exact_loss_history} only the learning rate differs.
Similarly to Theorem \ref{th:conv_sgd_exact_loss_history}, when $\de$ tends to zero we recover Theorem 2 of \cite{allen-zhu19a}.

It is worth noting that when $\de$ increases, $\frac{\alpha \rho d^{3/2} \delta b \log\left(\frac{1}{1-\alpha}\right)}{\de C(\cL) A(\nabla \cL) L m^{3/2} n^{3/2} \log^2(m)}$ decreases.
This implies that $\eta_{stale}$ decreases faster than $\eta_{exact}$ when $\de$ increases.
This was to be expected since the error that is made by using the stale per-example loss vector instead of the exact loss increases when $\de$ increases.

\subsubsection{Proofs of convergence}\label{s:proof_convergence}

In this section, we prove the results of Therem \ref{th:conv_sgd_exact_loss_history} and \ref{th:conv_sgd_stale_loss_history}.

For the ease of reading the proof, we remind here the chain rules for the distributionally robust loss that we are going to use intensively in the following proofs.

\paragraph{\textbf{Chain rule for the derivative of $R \circ \vL$ with respect to the network outputs $h$:}}\ \\
\begin{equation}
    \label{eq:reminder_chain_rule_h}
    \begin{aligned}
        \nabla_h (R \circ \vL)(h(\vtheta)) 
        &= \left(\nabla_{h_i} (R \circ \vL)(h(\vtheta))\right)_{i=1}^n\\
        \forall i \in \{1,\ldots n\},\quad \nabla_{h_i} (R \circ \vL)(h(\vtheta))
        &=\sum_{j=1}^n \frac{\partial R}{\partial v_j}(\vL(h(\vtheta))) \nabla_{h_i}\cL_j(h_j(\vtheta)) \\
        &= \bar{p}_i(\vL(h(\vtheta)))\nabla_{h_i}\cL_i(h_i(\vtheta))
    \end{aligned}
\end{equation}

\paragraph{\textbf{Chain rule for the derivative of $R \circ \vL \circ h$ with respect to the network parameters $\vtheta$:}}\ \\
\begin{equation}
\label{eq:reminder_chain_rule_theta}
    \begin{aligned}
        \nabla_{\vtheta} (R \circ \vL \circ h)(\vtheta) 
        &= \sum_{i=1}^n \nabla_\theta h_i(\vtheta) \nabla_{h_i}(R \circ \vL)(h(\vtheta))\\
        &= \sum_{i=1}^n \bar{p}_i(\vL(h(\vtheta))) \nabla_\theta h_i(\vtheta) \nabla_{h_i}\cL_i(h_i(\vtheta))\\
        &= \sum_{i=1}^n \bar{p}_i(\vL(h(\vtheta))
        \nabla_{\vtheta}(\cL_i \circ h_i)(\vtheta))\\
    \end{aligned}
\end{equation}

where for all $i \in \{1,\ldots n\}$,  $\nabla_\theta h_i(\vtheta)$ is the transpose of the Jacobian matrix of $h_i$ as a function of $\vtheta$.

%%%%%%%%%%%%%%%%%%%%%%%%%%%%%%%%%%%%%%%%%%%%%%%%%%%%%%
\subsubsection{Proof that R o L is one-sided gradient Lipchitz}
This property that $R\circ \vL$ is one-sided gradient Lipschitz is a key element for the proof of the semi-smoothness theorem for the distributionally robust loss Theorem \ref{th:semi-smoothness}.

Under Definition~\ref{def:phi_divergence} for the $\phi$-divergence, we have shown that $R$ is $\frac{\de}{n\rho}$-gradient Lipchitz continuous (Lemma~\ref{lemma:R_property}).
And under assumption \ref{as:3}, for all $i$, $\cL_i$ is $C(\cL)$-Lipschitz continuous and $C(\nabla \cL)$-gradient Lipschitz continuous.

Let $\vz=(z_i)_{i=1}^n, \vz'=(z_i')_{i=1}^n \in \sR^{dn}$.

We want to show that $R\circ \vL$ is one-sided gradient Lipschitz, i.e. we want to prove the existence of a constant $C>0$, independent to $z$ and $z'$, such that:
\[
\langle \nabla_{\vz} (R\circ \vL)(\vz) - \nabla_z (R\circ \vL)(\vz'), \vz - \vz'\rangle 
\leq C \norm{\vz - \vz'}^2
\]
We have
\begin{equation}
	\label{eq:one-sided1}
	\begin{aligned}
	    \langle \nabla_{\vz} (R\circ \vL)(\vz) - \nabla_{\vz} (&R\circ \vL)(\vz'), \vz - \vz'\rangle \\
	    & = \sum_{i=1}^n \langle \nabla_{z_i} (R\circ \vL)(\vz) - \nabla_{z_i} (R\circ \vL)(\vz'), z_i - z_i'\rangle\\
	    & = \sum_{i=1}^n \langle \bar{p}_i(\vL(\vz))\nabla_{z_i} \cL_i(z_i) - \bar{p}_i(\vL(\vz'))\nabla_{z_i}\cL_i(z_i'), z_i - z_i'\rangle\\
	    & = \sum_{i=1}^n \bar{p}_i(\vL(\vz))\langle \nabla_{z_i} \cL_i(z_i) - \nabla_{z_i}\cL_i(z_i'), z_i - z_i'\rangle\\
	    & \quad + \sum_{i=1}^n \left(\bar{p}_i(\vL(\vz)) - \bar{p}_i(\vL(\vz'))\right)\langle \nabla_{z_i}\cL_i(z_i'), z_i - z_i'\rangle\\
	\end{aligned}
\end{equation}

Where for all $i \in \{1,\ldots,n\}$ we have used the chain rule 
\[
\nabla_{z_i} (R\circ \vL)(\vz) = \sum_{j=1}^n \frac{\partial R}{\partial v_j}(\cL(\vz)) \nabla_{z_i}\cL_{j}(z_j) = \bar{p}_i(\vL(\vz))\nabla_{z_i} \cL_i(z_i)
\]

Let
\[
A = \left|\sum_{i=1}^n \bar{p}_i(\vL(\vz))\langle \nabla_{z_i} \cL_i(z_i) - \nabla_{z_i}\cL_i(z_i'), z_i - z_i'\rangle\right|
\]

For all $i$, $\cL_i$ is $C(\nabla \cL)$-gradient Lipchitz continuous, so using Cauchy-Schwarz inequality
\begin{equation}
	\label{eq:one-sided2}
	    A
	    \leq \sum_{i=1}^n C(\nabla \cL) \norm{z_i - z_i'}^2
	    = C(\nabla \cL) \norm{\vz - \vz'}^2
\end{equation}

Let 
\[
B = \left|\sum_{i=1}^n \left(\bar{p}_i(\vL(\vz)) - \bar{p}_i(\vL(\vz'))\right)\langle \nabla_{z_i}\cL_i(z_i'), z_i - z_i'\rangle\right|
\]
Using the triangular inequality:
\begin{equation}
	\label{eq:one-sided3}
	\begin{aligned}
		B
		& \leq \left|\sum_{i=1}^n \left(\bar{p}_i(\vL(\vz)) - \bar{p}_i(\vL(\vz'))\right)(\cL_i(z_i) - \cL_i(z_i')\right|\\
		& \quad + \left|\sum_{i=1}^n \left(\bar{p}_i(\vL(\vz)) - \bar{p}_i(\vL(\vz'))\right)(\cL_i(z_i') + \langle \nabla_{z_i}\cL_i(z_i'), z_i - z_i'\rangle - \cL_i(z_i)\right|\\
		& \leq \left\langle \nabla_{\vL} R(\vL(\vz)) - \nabla_{\vL} R(\vL(\vz')), \cL(\vz) - \cL(\vz')\right\rangle\\
		& \quad + 2 \sum_{i=1}^n\left|\cL_i(z_i') + \langle \nabla_{z_i}\cL_i(z_i'), z_i - z_i'\rangle - \cL_i(z_i)\right|\\
		& \leq \frac{\de}{n \rho}\norm{\vL(\vz) - \vL(\vz')}^2 
		+ 2\frac{C(\nabla \cL)}{2}\norm{\vz - \vz'}^2\\
		& \leq \left(\frac{\de C(\cL)^2}{n \rho}
		+ C(\nabla \cL)\right)\norm{\vz - \vz'}^2
	\end{aligned}
\end{equation}

Combining equations \eqref{eq:one-sided1}, \eqref{eq:one-sided2} and \eqref{eq:one-sided3} we finally obtain
\begin{equation}
	\langle \nabla_{\vz} (R\circ \vL)(\vz) - \nabla_{\vz} (R\circ \vL)(\vz'), \vz - \vz'\rangle
	\leq \left(\frac{\de C(\cL)^2}{n \rho}
	+ 2C(\nabla \cL)\right)\norm{\vz - \vz'}^2
\end{equation}

From there, we can obtain the following inequality that will be used for the proof of the semi-smoothness property in Theorem \ref{th:semi-smoothness}
\begin{equation}
    \label{eq:lipchitz}
	\begin{aligned}
	    &R(\vL(\vz')) - R(\vL(\vz)) - \langle \nabla_{\vz} (R\circ \vL)(\vz), \vz' - \vz \rangle\\
	    &\quad = \int_{t=0}^{1} \langle\nabla_{\vz} (R\circ \vL)\left(\vz + t(\vz' - \vz)\right) - \nabla_{\vz} (R\circ \vL)(\vz), \vz' - \vz\rangle dt\\
	    &\quad \leq \frac{1}{2}\left(\frac{\de C(\cL)^2}{n \rho}
	    + 2C(\nabla \cL)\right)\norm{\vz - \vz'}^2
	\end{aligned}
\end{equation}

%%%%%%%%%%%%%%%%%%%%%%%%%%%%%%%%%%%%%%%%%%%%%%%%%%%%%%%%%%%%
\subsubsection{Semi-smoothness property of the distributionally robust loss}

We prove the following lemma which is a generalization of Theorem 4 in \cite{allen-zhu19a} for the distributionally robust loss.

\begin{theorem}[Semi-smoothness of the distributionally robust loss]\ \\
    \label{th:semi-smoothness}
    Let $\omega \in \left[\Omega\left(\frac{d^{3/2}}{m^{3/2}L^{3/2}\log^{3/2}(m)}\right),
    O\left(\frac{1}{L^{4.5}\log^{3}(m)}\right)\right]$,
    and the $\vtheta^{(0)}$ being initialized randomly as described in assumption \ref{as:2}.
    With probability as least $1 - \exp{(-\Omega(m\omega^{3/2}L))}$ over the initialization, we have for all $\vtheta, \vtheta' \in \left(\sR^{m \times m}\right)^L$ with $\norm{\vtheta - \vtheta^{(0)}}_2 \leq \omega$,
    and $\norm{\vtheta - \vtheta'}_2 \leq \omega$
    \begin{equation}
        \begin{aligned}
            R(\vL(h(\vtheta ')) & 
            \leq
            R(\vL(h(\vtheta)) + 
            \langle 
            \nabla_{\vtheta} (R \circ \vL \circ h)(\vtheta), \vtheta' - \vtheta
            \rangle \\
            & + \norm{\nabla_h (R \circ \vL)(h(\vtheta))}_{2,1} O\left(\frac{L^2\omega^{1/3}\sqrt{m\log(m)}}{\sqrt{d}}\right)\norm{\vtheta' - \vtheta}_{2, \infty}\\
            & + O\left(\left(\frac{\de C(\cL)^2}{n \rho}
	        + 2C(\nabla \cL)\right)\frac{n L^2 m}{d}\right) \norm{\vtheta' - \vtheta}_{2, \infty}^2\\
	    %
	       % \leq
        %     R(\vL(h(\vtheta)) + 
        %     \langle 
        %     \nabla_{\vtheta} (R \circ \vL \circ h)(\vtheta), \vtheta' - \vtheta
        %     \rangle \\
        %     & + \sqrt{n}\norm{\nabla_h (R \circ \vL)(h(\vtheta))}_{2,2} O\left(\frac{L^2\omega^{1/3}\sqrt{m\log(m)}}{\sqrt{d}}\right)\norm{\vtheta' - \vtheta}_{2, \infty}\\
        %     & + O\left(\left(\frac{\de C(\cL)^2}{n \rho}
	       % + 2C(\nabla \cL)\right)\frac{n L^2 m}{d}\right) \norm{\vtheta' - \vtheta}_{2, \infty}^2\\
        \end{aligned}
    \end{equation}
\end{theorem}

where for all layer $l \in \{1,\ldots, L\}$, $\vtheta_l$ is the vector of parameters for layer $l$, and
\[
\begin{aligned}
    \norm{\vtheta' - \vtheta}_{2, \infty} 
    &= \max_{l} \norm{\vtheta_l' - \vtheta_l}_2\\
    \norm{\vtheta' - \vtheta}_{2, \infty}^2 
    &= \left(\max_{l} \norm{\vtheta_l' - \vtheta_l}_2^2\right)^2= \max_{l} \norm{\vtheta_l' - \vtheta_l}_2^2\\
    \norm{\nabla_h (R \circ \vL)(h(\vtheta))}_{2,1} 
    &=
        \sum_{i=1}^n \norm{\nabla_{h_i} (R \circ \vL)(h(\vtheta))}_2\\
    &= \sum_{i=1}^n \norm{\bar{p}_i(\vL(h(\vtheta)))\nabla_{h_i}\cL_i(h_i(\vtheta))}_2 \quad \text{ (chain rule \eqref{eq:reminder_chain_rule_h}) }\\
    %
    % \norm{\nabla_h (R \circ \vL)(h(\vtheta))}_{2,2} 
    % &= \left(
    %     \sum_{i=1}^n \norm{\nabla_{h_i} (R \circ \vL)(h(\vtheta))}_2^2
    %     \right)^{1/2}\\
    % %
    % &= \left(
    %     \sum_{i=1}^n \norm{\bar{p}_i(\vL(h(\vtheta)))\nabla_{h_i}\cL_i(h_i(\vtheta))}_2^2
    %     \right)^{1/2} \quad \text{ (chain rule \eqref{eq:reminder_chain_rule_h}) }\\
\end{aligned}
\]

To compare this semi-smoothness result to the one in \cite[Theorem 4]{allen-zhu19a}, let us first remark that
\[
\begin{aligned}
    \norm{\nabla_h (R \circ \vL)(h(\vtheta))}_{2,1}
    & \leq \sqrt{n} \norm{\nabla_h (R \circ \vL)(h(\vtheta))}_{2,2}\\
\end{aligned}
\]

As a result, our result is analogous to \cite[Theorem 4]{allen-zhu19a}, up to an additional multiplicative factor $\left(\frac{\de C(\cL)^2}{n \rho}+ 2C(\nabla \cL)\right)$ in the last term of the right-hand side.
It is worth noting that there is also implicitly an additional multiplicative factor $C(\nabla \cL)$ in Theorem 3 of \cite{allen-zhu19a} since \cite{allen-zhu19a} make the assumption that $C(\nabla \cL)=1$ \cite[Appendix A]{allen-zhu19a}.

Let $\vtheta, \vtheta' \in \left(\sR^{m \times m}\right)^L$ verifying the conditions of Theorem \ref{th:semi-smoothness}.

Let
% \[
$
A = R(\vL(h(\vtheta ')) - R(\vL(h(\vtheta)) 
    - \langle \nabla_{\vtheta} (R \circ \vL \circ h)(\vtheta), \vtheta' - \vtheta\rangle
$
% \]
, the quantity we want to bound.

Using \eqref{eq:lipchitz} for $\vz=h(\vtheta)$ and $\vz'=h(\vtheta')$, we obtain
\begin{equation}
\begin{aligned}
    % R(\vL(h(\vtheta ')) - R(\vL(h(\vtheta)) 
    % - \langle \nabla_{\vtheta} (R \circ \vL \circ h)(\vtheta), \vtheta' - \vtheta\rangle
    A
    & \leq \frac{1}{2}\left(\frac{\de C(\cL)^2}{n \rho}
	    + 2C(\nabla \cL)\right)\norm{h(\vtheta') - h(\vtheta)}_2^2\\
	& \quad + \langle \nabla_h (R\circ \vL)(h(\vtheta)), h(\vtheta') - h(\vtheta) \rangle\\
	& \quad - \langle \nabla_{\vtheta} (R \circ \vL \circ h)(\vtheta), \vtheta' - \vtheta\rangle\\
\end{aligned}
\end{equation}

Then using the chain rule \eqref{eq:reminder_chain_rule_theta}
\begin{equation}
\label{eq:11_3}
\begin{aligned}
    % R(\vL(h(\vtheta ')) - R(\vL(h(\vtheta)) 
    % - \langle \nabla_{\vtheta} (R \circ \vL \circ h)(\vtheta), \vtheta' - \vtheta\rangle
    A
    & \leq \frac{1}{2}\left(\frac{\de C(\cL)^2}{n \rho}
	    + 2C(\nabla \cL)\right)\norm{h(\vtheta') - h(\vtheta)}_2^2\\
	& \quad + \sum_{i=1}^n\langle \nabla_{h_i} (R\circ \vL)(h(\vtheta)), 
	h_i(\vtheta') - h_i(\vtheta) - \left(\nabla_{\vtheta}h_i(\vtheta)\right)^T(\vtheta'-\vtheta)\rangle\\
\end{aligned}
\end{equation}

For all $i \in \{1, \ldots, n\}$, let us denote
% $\breve{\mathup{loss}}_i = \nabla_{h_i} (R\circ \vL)(h(\vtheta))$
$\breve{loss}_i := \nabla_{h_i} (R\circ \vL)(h(\vtheta))$
to match the notations used in \cite{allen-zhu19a} for the derivative of the loss with respect to the output of the network for example i of the training set.

With this notation, we obtain exactly equation (11.3) in \cite{allen-zhu19a} up to the multiplicative factor $\left(\frac{\de C(\cL)^2}{n \rho} + 2C(\nabla \cL)\right)$ for the distributionally robust loss.

From there the proof of Theorem 4 in \cite{allen-zhu19a} being independent to the formula for $\breve{loss}_i$, we can conclude the proof of our Theorem \ref{th:semi-smoothness} as in \cite[Appendix A]{allen-zhu19a}.

%%%%%%%%%%%%%%%%%%%%%%%%%%%%%%%%%%%%%%%%%%%%%%%%%%%
\subsubsection{Gradient bounds for the distributionally robust loss}

We prove the following lemma which is a generalization of Theorem 3 in \cite{allen-zhu19a} for the distributionally robust loss.

\begin{theorem}[Gradient Bounds for the Distributionally Robust Loss]\ \\
\label{th:gradient_bound}
    Let $\omega \in O\left(\frac{\delta^{3/2}}{n^{9/2}L^6\log^3(m)}\right)$,  and $\vtheta^{(0)}$ being initialized randomly as described in assumption~\ref{as:2}.
    With probability as least $1 - \exp{(-\Omega(m\omega^{3/2}L))}$ over the initialization, we have for all $\vtheta \in \left(\sR^{m \times m}\right)^L$ with $\norm{\vtheta - \vtheta^{(0)}}_2 \leq \omega$
    \begin{equation}
        \begin{aligned}
            &\forall i \in \{1,\ldots,n\},\,\, \forall l \in \{1, \ldots, L\},\,\, \forall \hat{\vL} \in \sR^n\\
            %
            % &\norm{\bar{p}_i(\vL(h(\vtheta))) \nabla_{\vtheta_l}(\cL_i \circ h_i)(\vtheta)}_2^2 
            % \leq O\left(\frac{m}{d}\norm{\bar{p}_i(\vL(h(\vtheta))) \nabla_{h_i}\cL_i (h_i(\vtheta))}_2^2\right)\\
            &\norm{\bar{p}_i(\hat{\vL}) \nabla_{\vtheta_l}(\cL_i \circ h_i)(\vtheta)}_2^2 
            \leq O\left(\frac{m}{d}\norm{\bar{p}_i(\hat{\vL}) \nabla_{h_i}\cL_i (h_i(\vtheta))}_2^2\right)\\
            & \forall l \in \{1, \ldots, L\},\,\, \forall \hat{\vL} \in \sR^n\\
            %
            % &\norm{\sum_{i=1}^n\bar{p}_i(\vL(h(\vtheta))) \nabla_{\vtheta_l}(\cL_i \circ h_i)(\vtheta)}_2^2 
            % \leq O\left(\frac{m n}{d}\sum_{i=1}^n\norm{\bar{p}_i(\vL(h(\vtheta))) \nabla_{h_i}\cL_i (h_i(\vtheta))}_2^2\right)\\
            &\norm{\sum_{i=1}^n\bar{p}_i(\hat{\vL}) \nabla_{\vtheta_l}(\cL_i \circ h_i)(\vtheta)}_2^2 
            \leq O\left(\frac{m n}{d}\sum_{i=1}^n\norm{\bar{p}_i(\hat{\vL}) \nabla_{h_i}\cL_i (h_i(\vtheta))}_2^2\right)\\
            %
            % & \norm{\sum_{i=1}^n\bar{p}_i(\vL(h(\vtheta))) \nabla_{\vtheta_L}(\cL_i \circ h_i)(\vtheta)}_2^2
            % \geq \Omega\left(\frac{m \delta}{d n^2}\sum_{i=1}^n\norm{\bar{p}_i(\vL(h(\vtheta))) \nabla_{h_i}\cL_i (h_i(\vtheta))}_2^2\right)\\
            & \norm{\sum_{i=1}^n\bar{p}_i(\hat{\vL}) \nabla_{\vtheta_L}(\cL_i \circ h_i)(\vtheta)}_2^2
            \geq \Omega\left(\frac{m \delta}{d n^2}\sum_{i=1}^n\norm{\bar{p}_i(\hat{\vL}) \nabla_{h_i}\cL_i (h_i(\vtheta))}_2^2\right)\\
        \end{aligned}
    \end{equation}
\end{theorem}

It is worth noting that the loss vector $\hat{\vL}$ used for computing the robust probabilities $\bar{\vp}(\hat{\vL})=\left(\bar{p}_i(\hat{\vL})\right)_{i=1}^n$ does not have to be equal to $\vL(h(\vtheta))$.

We will use this for the proof of the Robust SGD with stale per-example loss vector.

The adaptation of the proof of Theorem 3 in \cite{allen-zhu19a} is straightforward.

Let $\vtheta \in \left(\sR^{m \times m}\right)^L$ satisfying the conditions of Theorem \ref{th:gradient_bound}, and $\hat{\vL} \in \sR^n$.

Let us denote $\vv := \left(\bar{p}_i(\hat{\vL}) \nabla_{h_i}\cL_i(h_i(\vtheta))\right)_{i=1}^n$, applying the proof of Theorem 3 in \cite{allen-zhu19a} to our $\vv$ gives:
\[
\begin{aligned}
    &\forall i \in \{1,\ldots,n\},\,\, \forall l \in \{1, \ldots, L\},\\
            &\norm{\bar{p}_i(\hat{\vL}) \nabla_{\vtheta_l}(\cL_i \circ h_i)(\vtheta)}_2^2 
            \leq O\left(\frac{m}{d}\norm{\bar{p}_i(\hat{\vL}) \nabla_{h_i}\cL_i (h_i(\vtheta))}_2^2\right)\\
            & \forall l \in \{1, \ldots, L\},\,\, \forall \hat{\vL} \in \sR^n\\
            &\norm{\sum_{i=1}^n\bar{p}_i(\hat{\vL}) \nabla_{\vtheta_l}(\cL_i \circ h_i)(\vtheta)}_2^2 
            \leq O\left(\frac{m n}{d}\sum_{i=1}^n\norm{\bar{p}_i(\hat{\vL}) \nabla_{h_i}\cL_i (h_i(\vtheta))}_2^2\right)\\
            & \norm{\sum_{i=1}^n\bar{p}_i(\hat{\vL}) \nabla_{\vtheta_L}(\cL_i \circ h_i)(\vtheta)}_2^2
            \geq \Omega\left(\frac{m \delta}{d n}\max_{i}\left(\norm{\bar{p}_i(\hat{\vL}) \nabla_{h_i}\cL_i (h_i(\vtheta))}_2^2\right)\right)\\
\end{aligned}
\]

In addition
\[
\max_{i}\left(\norm{\bar{p}_i(\hat{\vL}) \nabla_{h_i}\cL_i (h_i(\vtheta))}_2^2\right)
\geq \frac{1}{n} \sum_{i=1}^n \norm{\bar{p}_i(\hat{\vL}) \nabla_{h_i}\cL_i (h_i(\vtheta))}_2^2
\]
This allows us to conclude the proof of our Theorem \ref{th:gradient_bound}. $\blacksquare$

%%%%%%%%%%%%%%%%%%%%%%%%%%%%%%%%%%%%%%%%%%%%%%%%%%%%%%%%%%%%%%%%%
\subsubsection{Convergence of SGD with Hardness Weighted Sampling and exact per-example loss vector}\label{s:proof_exact_loss_history}

We can now prove Theorem \ref{th:conv_sgd_exact_loss_history}.
% \begin{theorem}[Convergence of Robust SGD with exact per-example loss vector -- Restated from Theorem \ref{th:conv_sgd_exact_loss_history}]
%     \label{th:conv_sgd_exact_loss_history2}
%     Suppose batch size $1 \leq b \leq n$, number of hidden units 
%     $m \geq \Omega(d \epsilon^{-1} \times \textup{poly}(n,L,\delta^{-1}))$, 
%     % $\de \leq O(\frac{n\rho C(\nabla\cL)}{C(\cL)^2})$, 
%     and $\delta \geq O\left(\frac{1}{L}\right)$.
%     Let $\epsilon > 0$, and the learning rate be 
%     % $\eta = \Theta\left(\min\left(1, \frac{n}{\alpha C(\nabla \cL)}\right)\frac{b\delta d}{\textup{poly}(n,L)m\log^2(m)}\right)$,
%     % $\eta = \Theta\left(\frac{n}{\alpha C(\nabla \cL)} \times \frac{b\delta d}{\textup{poly}(n,L)m\log^2(m)}\right)$,
%     $\eta_{exact} =\Theta\left(
%         \frac{\alpha n^2 \rho}{\de C(\cL)^2+ 2 n \rho C(\nabla \cL)} 
%         \times \frac{b\delta d}{\textup{poly}(n,L)m\log^2(m)}
%         \right)$,
%     with probability at least $1 - \exp\left(-\Omega(\log^2(m))\right)$ over the randomness of the initialization and the mini-batches, Robust SGD with exact loss vector finds 
%     $\norm{\nabla_{\vtheta} (R\circ f \circ h)(\vtheta)} \leq \epsilon$ after 
%     $T=O\left(\frac{L n^3}{\eta \delta \epsilon^2}\right)$ iterations.
% \end{theorem}

Similarly to the proof of the convergence of SGD for the mean loss (Theorem 2 in \cite{allen-zhu19a}), the convergence of SGD for the distributionally robust loss will mainly rely on the semi-smoothness property (Theorem \ref{th:semi-smoothness}) and the gradient bound (Theorem \ref{th:gradient_bound}) that we have proved previously for the distributionally robust loss.

Let $\vtheta \in \left(\sR^{m \times m}\right)^L$ satisfying the conditions of Theorem \ref{th:conv_sgd_exact_loss_history},
and $\hat{\vL}$ be the exact per-example loss vector at $\vtheta$, i.e.
\begin{equation}
    \label{eq:exact_loss_history}
    \hat{\vL}=\left(\cL_i(h_i(\vtheta))\right)_{i=1}^n
\end{equation}
For the batch size $b \in \{1, \ldots, n\}$, let $S=\{i_j\}_{j=1}^b$ a batch of indices drawn from $\bar{\vp}(\hat{\vL})$ without replacement, i.e.
\begin{equation}
    \forall j \in \{1, \ldots b\}, \,\, i_j \overset{\text{i.i.d.}}{\sim} \bar{\vp}(\hat{\vL})
\end{equation}

Let $\vtheta' \in \left(\sR^{m \times m}\right)^L$ be the values of the parameters after a stochastic gradient descent step at $\vtheta$ for the batch $S$, i.e.
\begin{equation}
    \label{eq:next_theta}
    \vtheta' = \vtheta - \eta \frac{1}{b} \sum_{i \in S} \nabla_{\vtheta}(\cL_i \circ h_i)(\vtheta) 
\end{equation}
where $\eta >0$ is the learning rate.

Assuming that $\vtheta$ and $\vtheta'$ satisfies the conditions of Theorem \ref{th:semi-smoothness}, we obtain
\begin{equation}
\label{eq:cv_sgd_exact_1}
    \begin{aligned}
            R(\vL(h(\vtheta ')) \leq &
            R(\vL(h(\vtheta)) 
            - \eta \langle \nabla_{\vtheta} (R \circ \vL \circ h)(\vtheta),\frac{1}{b} \sum_{i \in S} \nabla_{\vtheta}(\cL_i \circ h_i)(\vtheta)\rangle \\
            & + \eta \sqrt{n}\norm{\nabla_h (R \circ \vL)(h(\vtheta))}_{2,2} O\left(\frac{L^2\omega^{1/3}\sqrt{m\log(m)}}{\sqrt{d}}\right)
            \norm{
            \frac{1}{b} \sum_{i \in S} \nabla_{\vtheta}(\cL_i \circ h_i)(\vtheta)
            }_{2, \infty}\\
            & + \eta^2 O\left(\left(\frac{\de C(\cL)^2}{n \rho}
	    + 2C(\nabla \cL)\right)\frac{n L^2 m}{d}\right) 
	    \norm{
	    \frac{1}{b} \sum_{i \in S} \nabla_{\vtheta}(\cL_i \circ h_i)(\vtheta)
	    }_{2, \infty}^2
        \end{aligned}
\end{equation}
where we refer to \eqref{eq:reminder_chain_rule_theta} for the form of $\nabla_{\vtheta} (R \circ \vL \circ h)(\vtheta)$ and to \eqref{eq:reminder_chain_rule_h} for the form of $\nabla_h (R \circ \vL)(h(\vtheta))$.

In addition, we make the assumption that for the set of values of $\vtheta$ considered the hardness weighted sampling probabilities admit an upper-bound
\begin{equation}
    \label{eq:alpha}
    \alpha = \min_{\vtheta} \min_{i} \bar{p}_i(\vL(\vtheta)) > 0
\end{equation}
Which is always satisfied under assumption \ref{as:3} for Kullback-Leibler $\phi$-divergence, and for any $\phi$-divergence satisfying Definition~\ref{def:phi_divergence} with a robustness parameter $\de$ small enough.

Let $\E_S$ be the expectation with respect to $S$.
Applying $\E_S$ to \eqref{eq:cv_sgd_exact_1}, we obtain
\begin{equation}
\label{eq:cv_sgd_exact_3}
    \begin{aligned}
        \E_S&\left[R(\vL(h(\vtheta '))\right] \\
        \leq &
        R(\vL(h(\vtheta)) 
        - \eta \norm{\nabla_{\vtheta} (R \circ \vL \circ h)(\vtheta)}_{2,2}^2 \\
        & + \eta
        \norm{
        \nabla_h (R \circ \vL)(h(\vtheta))
        }_{2,2}
        O\left(\frac{n L^2\omega^{1/3}\sqrt{m\log(m)}}{\sqrt{d}}\right)
        \sqrt{
        \sum_{i=1}^n \max_{l} \norm{\bar{p}_i(\hat{\vL})\nabla_{\vtheta_l}(\cL_i \circ h_i)(\vtheta)}^2
        }\\
        & + \eta^2 O\left(\left(\frac{\de C(\cL)^2}{n \rho}
	    + 2C(\nabla \cL)\right)\frac{n L^2 m}{d}\right) 
	    \frac{1}{\alpha}\sum_{i=1}^n \max_{l} \norm{\bar{p}_i(\hat{\vL})\nabla_{\vtheta_l}(\cL_i \circ h_i)(\vtheta)}^2
    \end{aligned}
\end{equation}
where we have used the following results:
\begin{itemize}
    \item For any integer $k \geq 1$, and all $\left(\va_i\right)_{i=1}^n \in \left(\sR^k\right)^n$, we have (see the proof in \ref{p:tech_lemma_1})
    \begin{equation}
        \label{eq:expectation}
        \begin{aligned}
            \E_S\left[\frac{1}{b}\sum_{i \in S} \va_i\right]
            &= \E_{\bar{p}(\hat{\vL})}\left[\va_{i}\right]
        \end{aligned}
    \end{equation}
    \item Using \eqref{eq:expectation} for $\left(\va_i\right)_{i=1}^n=\left(\nabla_{\vtheta}(\cL_i \circ h_i)(\vtheta)\right)_{i=1}^n$, and the chain rule \eqref{eq:reminder_chain_rule_theta}
    \begin{equation}
        \E_S\left[\frac{1}{b}\sum_{i \in S} \nabla_{\vtheta}(\cL_i \circ h_i)(\vtheta)\right]
            = \sum_{i=1}^n\bar{p}_i(\hat{\vL})\nabla_{\vtheta}(\cL_i \circ h_i)(\vtheta)
            = \nabla_{\vtheta} (R \circ \vL \circ h)(\vtheta)
    \end{equation}
    \item Using the triangular inequality
    \begin{equation}
    \label{eq:tr_ineq}
    \begin{aligned}
        \norm{
        \frac{1}{b} \sum_{i \in S} \nabla_{\vtheta}(\cL_i \circ h_i)(\vtheta)
        }_{2, \infty}
        & \leq \frac{1}{b} \sum_{i \in S} \norm{\nabla_{\vtheta}(\cL_i \circ h_i)(\vtheta)}_{2, \infty}
    \end{aligned}
    \end{equation}
    And using \eqref{eq:expectation} for $\left(a_i\right)_{i=1}^n=\left(\norm{\nabla_{\vtheta}(\cL_i \circ h_i)(\vtheta)}_{2,\infty}\right)_{i=1}^n$,
    \begin{equation}
    \begin{aligned}
        \E_S\left[
        \norm{
        \frac{1}{b} \sum_{i \in S} \nabla_{\vtheta}(\cL_i \circ h_i)(\vtheta)
        }_{2, \infty}
        \right]
        & \leq \sum_{i=1}^n\bar{p}_i(\hat{\vL}) \norm{\nabla_{\vtheta}(\cL_i \circ h_i)(\vtheta)}_{2, \infty}\\
        & \leq \sum_{i=1}^n 
        \max_{l}\norm{\nabla_{\vtheta_l}(\bar{p}_i(\hat{\vL})\cL_i \circ h_i)(\vtheta)}_{2}\\
        & \leq \sqrt{n}\sqrt{
        \sum_{i=1}^n\max_{l}\norm{\nabla_{\vtheta_l}(\bar{p}_i(\hat{\vL})\cL_i \circ h_i)(\vtheta)}_{2}^2
        }
    \end{aligned}
    \end{equation}
    where we have used Cauchy-Schwarz inequality for the last inequality.
    \item Using \eqref{eq:tr_ineq} and the convexity of the function $x \mapsto x^2$
    \begin{equation}
    \begin{aligned}
        \norm{
        \frac{1}{b} \sum_{i \in S} \nabla_{\vtheta}(\cL_i \circ h_i)(\vtheta)
        }_{2, \infty}^2
        & \leq \frac{1}{b} \sum_{i \in S} \norm{\nabla_{\vtheta}(\cL_i \circ h_i)(\vtheta)}_{2, \infty}^2
    \end{aligned}
    \end{equation}
    And using \eqref{eq:expectation} for $\left(a_i\right)_{i=1}^n=\left(\norm{\nabla_{\vtheta}(\cL_i \circ h_i)(\vtheta)}_{2,\infty}^2\right)_{i=1}^n$,
    \begin{equation}
    \label{eq:bad_ineq}
    \begin{aligned}
        \E_S\left[
        \norm{
        \frac{1}{b} \sum_{i \in S} \nabla_{\vtheta}(\cL_i \circ h_i)(\vtheta)
        }_{2, \infty}^2
        \right]
        & \leq \sum_{i=1}^n\bar{p}_i(\hat{\vL}) \norm{\nabla_{\vtheta}(\cL_i \circ h_i)(\vtheta)}_{2, \infty}^2\\
        & \leq \sum_{i=1}^n
        \frac{1}{\bar{p}_i(\hat{\vL})}
        \max_{l}\norm{\nabla_{\vtheta_l}(\bar{p}_i(\hat{\vL})\cL_i \circ h_i)(\vtheta)}_{2}^2\\
        & \leq \frac{1}{\alpha}
        \sum_{i=1}^n\max_{l}\norm{\nabla_{\vtheta_l}(\bar{p}_i(\hat{\vL})\cL_i \circ h_i)(\vtheta)}_{2}^2
    \end{aligned}
    \end{equation}
\end{itemize}

\paragraph{\textbf{Important Remark}:}\label{rk:important}
It is worth noting in \eqref{eq:bad_ineq} the apparition of $\alpha$ defined in \eqref{eq:alpha}.
If we were using a uniform sampling as for ERM (i.e. for DRO in the limit $\de \rightarrow 0$), we would have $\alpha = \frac{1}{n}$.
So although our inequality \eqref{eq:bad_ineq} may seem crude, it is consistent with equation (13.2) in \cite{allen-zhu19a} and the corresponding inequality in the case of ERM.

The rest of the proof of convergence will consist in proving that $\eta \norm{\nabla_{\vtheta} (R \circ \vL \circ h)(\vtheta)}_{2,2}^2$ dominates the two last terms in \eqref{eq:cv_sgd_exact_1}.
As a result, we can already state that either the robustness parameter $\de$, or the learning rate $\eta$ will have to be small enough to control $\alpha$. 
%
% This is consistent with what we observed in our experiments.

Indeed, combining \eqref{eq:cv_sgd_exact_1} with the chain rule \eqref{eq:reminder_chain_rule_theta}, and the gradient bound Theorem \ref{th:gradient_bound} where we use our $\hat{\vL}$ defined in \eqref{eq:exact_loss_history}
\begin{equation}
\label{eq:cv_sgd_exact_2}
    \begin{aligned}
        \E_S\left[R(\vL(h(\vtheta '))\right]  
        & \leq
        R(\vL(h(\vtheta)) 
        - \Omega\left(\frac{\eta m \delta}{d n^2}\right)
        \sum_{i=1}^n 
            \norm{
            \bar{p}_i(\hat{\vL})\nabla_{h_i}\cL_i(h_i(\vtheta))
            }_{2}^2 \\
        & + \eta
        O\left(\frac{n L^2\omega^{1/3}\sqrt{m\log(m)}}{\sqrt{d}}\right)
        O\left(\sqrt{\frac{m}{d}}\right)
        \sum_{i=1}^n 
            \norm{
            \bar{p}_i(\hat{\vL})\nabla_{h_i}\cL_i(h_i(\vtheta))
            }_{2}^2 \\
        & + \eta^2 O\left(\left(\frac{\de C(\cL)^2}{n \rho}
	    + 2C(\nabla \cL)\right)\frac{n L^2 m}{d}\right) 
	     O\left(\frac{m}{d\alpha}\right)
	    \sum_{i=1}^n 
            \norm{
            \bar{p}_i(\hat{\vL})\nabla_{h_i}\cL_i(h_i(\vtheta))
            }_{2}^2 \\
        & \leq
        R(\vL(h(\vtheta)) 
        - \Omega\left(\frac{\eta m \delta}{d n^2}\right)
        \sum_{i=1}^n 
            \norm{
            \bar{p}_i(\hat{\vL})\nabla_{h_i}\cL_i(h_i(\vtheta))
            }_{2}^2 \\
        & +
        O\left(
        \frac{\eta n L^2 m \omega^{1/3}\sqrt{\log(m)}}{d}
        + K \frac{\eta^2 (n / \alpha) L^2 m^2}{d^2}
        \right)
        \sum_{i=1}^n 
            \norm{
            \bar{p}_i(\hat{\vL})\nabla_{h_i}\cL_i(h_i(\vtheta))
            }_{2}^2 \\
    \end{aligned}
\end{equation}
where we have used
\begin{equation}
    K := \frac{\de C(\cL)^2}{n \rho} + 2C(\nabla \cL)
\end{equation}
There are only two differences with equation (13.2) in \cite{allen-zhu19a}:
\begin{itemize}
    \item in the last fraction we have $n / \alpha$ instead of $n^2$ (see remark \ref{rk:important} for more details), and an additional multiplicative term $K$.
    So in total, this term differs by a multiplicative factor $\frac{\alpha n}{K}$ from the analogous term in the proof of \cite{allen-zhu19a}.
    \item we have $\sum_{i=1}^n 
            \norm{
            \bar{p}_i(\hat{\vL})\nabla_{h_i}\cL_i(h_i(\vtheta))
            }_{2}^2$ instead of $F(\mathbf{W}^{(t)})$. 
            In fact they are analogous since in  equation (13.2) in \cite{allen-zhu19a}, $F(\mathbf{W}^{(t)})$ is the squared norm of the mean loss for the $L^2$ loss.
            We don't make such a strong assumption on the choice of $\cL$ (see assumption \ref{as:3}).
            It is worth noting that the same analogy is used in \cite[Appendix A]{allen-zhu19a} where they extend their result to the mean loss with other objective function than the $L^2$ loss.
\end{itemize}

% Using the condition on $\de$ in our Theorem \ref{th:conv_sgd_exact_loss_history}:
% \begin{equation}
%     \de \leq O(\frac{n\rho C(\nabla\cL)}{C(\cL)^2})
% \end{equation}

% As a result
% \begin{equation}
%     K = O\left(C(\nabla \cL)\right)
% \end{equation}
% 
Our choice of learning rate in Theorem \ref{th:conv_sgd_stale_loss_history} can be rewritten as
% \begin{equation}
%     \eta = \Theta\left(\frac{n}{\alpha C(\nabla \cL)} \times \frac{b\delta d}{\textup{poly}(n,L)m\log^2(m)}\right)
% \end{equation}
\begin{equation}
    \begin{aligned}
         \eta_{exact} 
         &= \Theta\left(
        \frac{\alpha n^2 \rho}{\de C(\cL)^2+ 2 n \rho C(\nabla \cL)} 
        \times \frac{b\delta d}{\textup{poly}(n,L)m\log^2(m)}
        \right)\\
        &= \Theta\left(
        \frac{\alpha n}{K} 
        \times \frac{b\delta d}{\textup{poly}(n,L)m\log^2(m)}
        \right)\\
        & \leq \frac{\alpha n}{K} \times \eta'\\
        % &\leq \eta'
    \end{aligned}
\end{equation}
And we also have
\begin{equation}
    \eta_{exact} \leq \eta'
\end{equation}
%
% Which corresponds to $\eta = \frac{n}{\alpha C(\nabla \cL)} \times \eta'$, 
where $\eta'$ is the learning rate chosen in the proof of Theorem 2 in \cite{allen-zhu19a}.
We refer the reader to \cite{allen-zhu19a} for the details of the constant in "$\Theta$" and the exact form of the polynomial $\textup{poly}(n,L)$.

As a result, for $\eta=\eta_{exact}$, the term $\Omega\left(\frac{\eta m \delta}{d n^2}\right)$ dominates the other term of the right-hand side of inequality \eqref{eq:cv_sgd_exact_2}
% the end of the proof of the convergence is the same 
as in the proof of Theorem 2 in \cite{allen-zhu19a}.

This implies that the conditions of Theorem \ref{th:gradient_bound} are satisfied for all  $\vtheta^{(t)}$, and that we have for all iteration $t > 0$
\begin{equation}
    \label{eq:cv_sgd_exact_4}
    \E_{S_t}\left[R(\vL(h(\vtheta^{(t+1)}))\right]  
        \leq
        R(\vL(h(\vtheta^{(t)})) 
        - \Omega\left(\frac{\eta m \delta}{d n^2}\right)
        \sum_{i=1}^n 
            \norm{
            \bar{p}_i(\hat{\vL})\nabla_{h_i}\cL_i(h_i(\vtheta^{(t)}))
            }_{2}^2
\end{equation}

And using a result in Appendix A of \cite{allen-zhu19a}, since under assumption \ref{as:3} the distributionally robust loss is non-convex and bounded, we obtain for all $\epsilon'>0$
\begin{equation}
    \norm{\nabla_h (R\circ \vL)(h(\vtheta^{(T)}))}_{2,2} \leq \epsilon'
    \quad \textup{if} \quad
    T = O\left(\frac{d n^2}{\eta \delta m \epsilon'^2}\right)
\end{equation}
where according to \eqref{eq:reminder_chain_rule_h}
\begin{equation}
    \norm{\nabla_h (R\circ \vL)(h(\vtheta^{(T)}))}_{2,2}
    =
    \sum_{i=1}^n 
            \norm{
            \bar{p}_i(\hat{\vL})\nabla_{h_i}\cL_i(h_i(\vtheta^{(t)}))
            }_{2}^2
\end{equation}
However, we are interested in a bound on $\norm{\nabla_{\vtheta} (R\circ \vL \circ h)(\vtheta^{(T)})}_{2,2}$, rather than a bound on $\norm{\nabla_h (R\circ \vL)(h(\vtheta^{(T)}))}_{2,2}$.
Using the gradient bound of Theorem \ref{th:gradient_bound} and the chain rules \eqref{eq:reminder_chain_rule_theta} and \eqref{eq:reminder_chain_rule_h}
\begin{equation}
    \norm{\nabla_{\vtheta} (R\circ \vL \circ h)(\vtheta^{(T)})}_{2,2}
    \leq c_1 \sqrt{\frac{L m n}{d}} \norm{\nabla_h (R\circ \vL)(h(\vtheta^{(T)}))}_{2,2}
\end{equation}
where $c_1 > 0$ is the constant hidden in $O\left(\sqrt{\frac{L m n}{d}}\right)$.

So with $\epsilon' = \frac{1}{c_1}\sqrt{\frac{d}{L m n}}\epsilon$, we finally obtain
\begin{equation}
    \begin{aligned}
        \norm{\nabla_{\vtheta} (R\circ \vL \circ h)(\vtheta^{(T)})}_{2,2}
        &\leq c_1 \sqrt{\frac{L m n}{d}} 
        \norm{\nabla_h (R\circ \vL)(h(\vtheta^{(T)}))}_{2,2}\\
        &\leq c_1 \sqrt{\frac{L m n}{d}} \epsilon'\\
        & \leq \epsilon
    \end{aligned}
\end{equation}
If
\begin{equation}
    T = O\left(\frac{d n^2}{\eta \delta m \epsilon'^2}\right)
    = O\left(\frac{d n^2}{\eta \delta m}\frac{L m n}{d \epsilon^2}\right)
    = O\left(\frac{L n^3}{\eta \delta \epsilon^2}\right)
\end{equation}
which concludes the proof. $\blacksquare$

%%%%%%%%%%
% \newpage

\subsubsection{Proof of technical lemma 1}\label{p:tech_lemma_1}\ \\
For any integer $k \geq 1$, and all $\left(\va_i\right)_{i=1}^n \in \left(\sR^k\right)^n$, we have 
\begin{equation}
    \begin{aligned}
        \E_S\left[\frac{1}{b}\sum_{i \in S} \va_i\right] 
        &= \sum_{1 \leq i_1,\ldots, i_b \leq n} 
        \left[ 
        \left(\prod_{k=1}^n \bar{p}_{i_k}(\hat{\vL})\right)
        \frac{1}{b}\sum_{j=1}^b \va_{i_j}
        \right]\\
        &= \frac{1}{b} \sum_{1 \leq i_1,\ldots, i_b \leq n} 
        \left[ 
        \sum_{j=1}^b \bar{p}_{i_j}(\hat{\vL})\, \va_{i_j}
        \left(\prod_{\substack{k=1\\ k\neq j}}^n \bar{p}_{i_k}(\hat{\vL})\right)
        \right]\\
        &= \frac{1}{b}\sum_{j=1}^b 
        \left[
        \sum_{1 \leq i_1,\ldots, i_b \leq n}\bar{p}_{i_j}(\hat{\vL})\, \va_{i_j}
        \left(\prod_{\substack{k=1\\ k\neq j}}^n \bar{p}_{i_k}(\hat{\vL})\right)
        \right]\\
        &= \frac{1}{b}\sum_{j=1}^b 
        \left[
        \left(
        \sum_{i_j=1}^n\bar{p}_{i_j}(\hat{\vL})\, \va_{i_j}
        \right)
        \prod_{\substack{k=1\\ k\neq j}}^n
        \left(
        \sum_{i_k=1}^n\bar{p}_{i_k}(\hat{\vL})
         \right)
        \right]\\
        &= \frac{1}{b}\sum_{j=1}^b
        \left(
        \sum_{i=1}^n\bar{p}_{i}(\hat{\vL})\, \va_{i}
        \right)\\
        &= \sum_{i=1}^n\bar{p}_{i}(\hat{\vL})\, \va_{i}\\
        &= \E_{\bar{\vp}(\hat{\vL})}\left[\va_{i}\right]
    \end{aligned}
\end{equation}

%%%%%%%%%%%%%%%%%%%%%%%%%%%%%%%%%%%%%%%%%%%%%%%%%%%%%%%%%%%%%%%%%
\subsubsection{Convergence of SGD with Hardness Weighted Sampling and stale per-example loss vector}\label{s:proof_stale_loss_history}

The proof of the convergence of \Algref{alg:1} under the conditions of Theorem \ref{th:conv_sgd_stale_loss_history} follows the same structure as the proof of the convergence of Robust SGD with exact per-example loss vector \ref{s:proof_exact_loss_history}.
We will reuse the intermediate results of \ref{s:proof_exact_loss_history} when possible and focus on the differences between the two proofs due to the inexactness of the per-example loss vector.

Let $t$ be the iteration number, and let $\vtheta^{(t)} \in \left(\sR^{m \times m}\right)^L$ be the parameters of the deep neural network at iteration $t$.
We define the stale per-example loss vector at iteration $t$ as
\begin{equation}
    \label{eq:stale_loss_history}
    \hat{\vL}=\left(\cL_i(h_i(\vtheta^{(t_i(t))}))\right)_{i=1}^n
\end{equation}
where for all $i$, $t_i(t) < t$ corresponds to the latest iteration before $t$ at which the per-example loss value for example $i$ has been updated. 
Or equivalently, it corresponds to the last iteration before $t$ when example $i$ was drawn to be part of a mini-batch.

We also define the exact per-example loss vector that is unknown in \Algref{alg:1}, as
\begin{equation}
    \label{eq:ecat_loss_history2}
    \breve{\vL}=\left(\cL_i(h_i(\vtheta^{(t)}))\right)_{i=1}^n
\end{equation}
Similarly to \eqref{eq:next_theta} we define
\begin{equation}
    \label{eq:next_theta2}
    \vtheta^{(t+1)} = \vtheta^{(t)} - \eta \frac{1}{b} \sum_{i \in S} \nabla_{\vtheta}(\cL_i \circ h_i)(\vtheta^{(t)}) 
\end{equation}
and using Theorem \ref{th:semi-smoothness}, similarly to \eqref{eq:cv_sgd_exact_1}, we obtain 
\begin{equation}
\label{eq:cv_sgd_stale_1}
    \begin{aligned}
            R(\vL(h(\vtheta^{(t+1)})) \leq &
            R(\vL(h(\vtheta^{(t)})) 
            - \eta \langle \nabla_{\vtheta} (R \circ \vL \circ h)(\vtheta^{(t)}),\frac{1}{b} \sum_{i \in S} \nabla_{\vtheta}(\cL_i \circ h_i)(\vtheta^{(t)})\rangle \\
            %
            % & + \eta \sqrt{n}\norm{\nabla_h (R \circ \vL)(h(\vtheta^{(t)}))}_{2,2} O\left(\frac{L^2\omega^{1/3}\sqrt{m\log(m)}}{\sqrt{d}}\right)
            & + \eta \norm{\nabla_h (R \circ \vL)(h(\vtheta^{(t)}))}_{1,2} O\left(\frac{L^2\omega^{1/3}\sqrt{m\log(m)}}{\sqrt{d}}\right)
            \norm{
            \frac{1}{b} \sum_{i \in S} \nabla_{\vtheta}(\cL_i \circ h_i)(\vtheta^{(t)})
            }_{2, \infty}\\
            & + \eta^2 O\left(\left(\frac{\de C(\cL)^2}{n \rho}
	    + 2C(\nabla \cL)\right)\frac{n L^2 m}{d}\right) 
	    \norm{
	    \frac{1}{b} \sum_{i \in S} \nabla_{\vtheta}(\cL_i \circ h_i)(\vtheta^{(t)})
	    }_{2, \infty}^2
        \end{aligned}
\end{equation}
We can still define $\alpha$ as in \eqref{eq:alpha}
\begin{equation}
    \label{eq:alpha2}
    \alpha = \min_{\vtheta} \min_{i} \bar{p}_i(\vL(\vtheta)) > 0
\end{equation}
where we are guaranteed that $\alpha > 0$ under assumptions \ref{as:2}.

Since Theorem \ref{th:gradient_bound} is independent to the choice of $\hat{\vL}$, taking the expectation with respect to $S$, similarly to \eqref{eq:cv_sgd_exact_2}, we obtain

\begin{equation}
\label{eq:cv_sgd_stale_2}
    \begin{aligned}
        \E_S\left[R(\vL(h(\vtheta^{(t+1)}))\right]  
        & \leq
        R(\vL(h(\vtheta^{(t)})) 
        - \eta 
        \langle \nabla_{\vtheta} (R \circ \vL \circ h)(\vtheta^{(t)}),
        \sum_{i=1}^n 
            \bar{p}_i(\hat{\vL})\nabla_{\vtheta}(\cL_i \circ h_i)(\vtheta^{(t)}))
        \rangle \\
        & + \eta
         \norm{
        \nabla_h (R \circ \vL)(h(\vtheta^{(t)}))
        }_{1,2}
        O\left(\frac{L^2\omega^{1/3}\sqrt{nm\log(m)}}{\sqrt{d}}\right)
        % O\left(\sqrt{\frac{m}{d}}\right)
        \sqrt{
        \sum_{i=1}^n 
            \norm{
            \bar{p}_i(\hat{\vL})\nabla_{h_i}\cL_i(h_i(\vtheta^{(t)}))
            }_{2}^2
        }\\
        & + \eta^2 O\left(\left(\frac{\de C(\cL)^2}{n \rho}
	    + 2C(\nabla \cL)\right)\frac{n L^2 m}{d}\right) 
	     O\left(\frac{m}{d\alpha}\right)
	    \sum_{i=1}^n 
            \norm{
            \bar{p}_i(\hat{\vL})\nabla_{h_i}\cL_i(h_i(\vtheta^{(t)}))
            }_{2}^2 \\
    \end{aligned}
\end{equation}
where the differences with respect to \eqref{eq:cv_sgd_exact_2} comes from the fact that $\hat{\vL}$ is not the exact per-example loss vector here, i.e. $\hat{\vL} \neq \breve{\vL}$,
which leads to
\begin{equation}
    \begin{aligned}
        \nabla_{\vtheta} (R \circ \vL \circ h)(\vtheta^{(t)}) 
        &=
        \sum_{i=1}^n 
            \hat{p}_i(\breve{\vL})\nabla_{\vtheta}(\cL_i \circ h_i)(\vtheta^{(t)}))\\
        & \neq
        \sum_{i=1}^n 
            \bar{p}_i(\hat{\vL})\nabla_{\vtheta}(\cL_i \circ h_i)(\vtheta^{(t)}))
    \end{aligned}
\end{equation}
and
\begin{equation}
    \begin{aligned}
        \norm{
            \nabla_{h} (R \circ \vL)(h(\vtheta^{(t)}))
        }_{1,2}
        &=
        \sum_{i=1}^n
            \norm{
                \hat{p}_i(\breve{\vL})
                \nabla_{h_i}\cL_i(h_i(\vtheta^{(t)})))
            }_2\\
        & \neq
        \sum_{i=1}^n
            \norm{
                \hat{p}_i(\hat{\vL})
                \nabla_{h_i}\cL_i(h_i(\vtheta^{(t)})))
            }_2\\
    \end{aligned}
\end{equation}
Let 
\begin{equation}
    K' = C(\cL) A(\nabla \cL)
    \,O\left(
    \frac{\de L m^{3/2} \log^2(m)}{\alpha n^{1/2} \rho d^{3/2} b \log\left(\frac{1}{1-\alpha}\right)}
    \right)
\end{equation}
Where $C(\cL) > 0$ is a constant such that $\cL$ is $C(\cL)$-Lipschitz continuous,
and $A(\nabla \cL) > 0$ is a constant that bound the gradient of $\cL$ with respect to its input.
$C(\cL)$ and $A(\nabla \cL)$ are guaranteed to exist under assumptions \ref{as:2}.

We can prove that, with probability at least 
$1 - \exp\left(-\Omega\left(\log^2(m)\right)\right)$,
\begin{itemize}
    \item according to lemma~\ref{p:tech_lemma_2}
    \begin{equation}
    \norm{\hat{p}(\hat{\vL}) - \hat{p}(\breve{\vL})}_2
    = 
    \sqrt{
    \sum_{i=1}^n \left(\hat{p}_i(\hat{\vL}) - \hat{p}_i(\breve{\vL})\right)^2
    }
    \leq 
    \eta \alpha K'
    \end{equation}
    \item according to lemma~\ref{p:tech_lemma_3}
    \begin{equation}
        \begin{aligned}
            \left|
            \langle
            \nabla_{\vtheta} (R \circ \vL \circ h)(\vtheta^{(t)})
            - \sum_{i=1}^n 
            \bar{p}_i(\hat{\vL})\nabla_{\vtheta}(\cL_i \circ h_i)(\vtheta^{(t)})),
            \sum_{i=1}^n 
            \bar{p}_i(\hat{\vL})\nabla_{\vtheta}(\cL_i \circ h_i)(\vtheta^{(t)}))
            \rangle
            \right|\\
            \leq 
            \eta \frac{m}{d} K'
            \sum_{i=1}^n
            \norm{
            \bar{p}_i(\hat{\vL})\nabla_{\vtheta}(\cL_i \circ h_i)(\vtheta^{(t)}))}_2^2
        \end{aligned}
    \end{equation}
    \item according to lemma~\ref{p:tech_lemma_4}
    \begin{equation}
        \norm{
            \nabla_{h} (R \circ \vL)(h(\vtheta^{(t)}))
        }_{1,2}
        \leq
        \left(\sqrt{n} + \eta K'\right)
        \sqrt{
            \sum_{i=1}^n
            \norm{
            \bar{p}_i(\hat{\vL})\nabla_{\vtheta}(\cL_i \circ h_i)(\vtheta^{(t)}))}_2^2
        }
    \end{equation}
\end{itemize}
Combining those three inequalities with \eqref{eq:cv_sgd_stale_2} we obtain
\begin{equation}
    \begin{aligned}
        \E_S\left[R(\vL(h(\vtheta^{(t+1)}))\right]& - R(\vL(h(\vtheta^{(t)}))
        \leq
        \\
        % & 
        % \leq
        \eta& \left[
            - \Omega\left(\frac{m \delta}{d n^2}\right)
            + O\left(
            \frac{n L^2 m \omega^{1/3}\sqrt{\log(m)}}{d}
            \right)
            \right]
            \sum_{i=1}^n 
            \norm{
            \bar{p}_i(\hat{\vL})\nabla_{h_i}\cL_i(h_i(\vtheta^{(t)}))
            }_{2}^2
            \\
        % &
        \eta^2 &
            O\left(
            K \frac{(n / \alpha) L^2 m^2}{d^2}
            + \left(1 + \frac{m}{d}\right) K'
            \right)
        \sum_{i=1}^n 
            \norm{
            \bar{p}_i(\hat{\vL})\nabla_{h_i}\cL_i(h_i(\vtheta^{(t)}))
            }_{2}^2
            \\
    \end{aligned}
\end{equation}
One can see that compared to \eqref{eq:cv_sgd_exact_2}, there is only the additional term $\left(1 + \frac{m}{d}\right) K'$.

Using our choice of $\eta$,
\begin{equation}
    \eta = \eta_{stale} \leq O \left(
                        \frac{\delta}{n^2 K'}\eta_{exact}
                    \right)
\end{equation}
where $\eta_{exact}$ is the learning rate of Theorem \ref{th:conv_sgd_exact_loss_history},
we have
\begin{equation}
    \Omega\left(\frac{\eta m \delta}{d n^2}\right) 
    \geq 
    O\left(\eta ^2 \left(1 + \frac{m}{d}\right) K'\right)
\end{equation}
As a result, $\eta^2 \left(1 + \frac{m}{d}\right) K'$ is dominated by the term $\Omega\left(\frac{\eta m \delta}{d n^2}\right)$ 

In addition, since $\eta_{stale} \leq \eta_{exact}$, $\Omega\left(\frac{\eta m \delta}{d n^2}\right)$ still dominates also the ther terms as in the proof of Theorem \ref{th:conv_sgd_exact_loss_history}.

As a consequence, we obtain as in \eqref{eq:cv_sgd_exact_4} that for any iteration $t > 0$
\begin{equation}
    \label{eq:cv_sgd_stale_4}
    \E_{S_t}\left[R(\vL(h(\vtheta^{(t+1)}))\right]
        \leq
        R(\vL(h(\vtheta^{(t)})) 
        - \Omega\left(\frac{\eta m \delta}{d n^2}\right)
        \sum_{i=1}^n 
            \norm{
            \bar{p}_i(\hat{\vL})\nabla_{h_i}\cL_i(h_i(\vtheta^{(t)}))
            }_{2}^2
\end{equation}
This concludes the proof using the same arguments as in the end of the proof of Theorem \ref{th:conv_sgd_exact_loss_history} starting from \eqref{eq:cv_sgd_exact_4}. $\blacksquare$

%%%%%%%%%%%%%%
\subsubsection{Proof of technical lemma 2}\label{p:tech_lemma_2}\ \\
Using Lemma~\ref{lemma:robust_loss_sg} and Lemma~\ref{lemma:R_property} we obtain
\begin{equation}
    \begin{aligned}
        \norm{\hat{\vp}(\hat{\vL}) - \hat{\vp}(\breve{\vL})}_2
        &= 
        % \sqrt{
        % \sum_{i=1}^n \left(\hat{p}_i(\hat{\vL}) - \hat{p}_i(\breve{\vL})\right)^2
        % }\\
        \norm{\nabla_v R(\hat{\vL}) - \nabla_v R(\breve{\vL})}_2\\
        & \leq \frac{\de}{n \rho} \norm{\hat{\vL} - \breve{\vL}}_2
    \end{aligned}
\end{equation}
Using assumptions~\ref{as:3} and \cite[Claim 11.2]{allen-zhu19a}
\begin{equation}
    \begin{aligned}
        \norm{\hat{\vp}(\hat{\vL}) - \hat{\vp}(\breve{\vL})}_2
        &\leq 
        \frac{\de}{n \rho} 
        \sqrt{
        \sum_{i=1}^n \left(\cL_i\circ~h_i(\vtheta^{(t)}) - \cL_i\circ~h_i(\vtheta^{(t_i(t))})\right)^2
        }\\
        & \leq 
        \frac{\de}{n \rho}
        C(\cL)C(h)
        \sqrt{
        \sum_{i=1}^n \norm{\vtheta^{(t)} - \vtheta^{(t_i(t))}}_{2,2}^2
        }\\
        & \leq 
        C(\cL)
        O\left(\frac{\de L m^{1/2}}{n \rho d^{1/2}}\right)
        \sqrt{
        \sum_{i=1}^n \norm{\vtheta^{(t)} - \vtheta^{(t_i(t))}}_{2,2}^2
        }\\
    \end{aligned}
\end{equation}
Where $C(\cL)$ is the constant of Lipschitz continuity of the per-example loss $\cL$ (see assumptions~\ref{as:3}) and $C(h)$ is the constant of Lipschitz continuity of the deep neural network $h$ with respect to its parameters $\vtheta$.

By developing the recurrence formula of $\vtheta^{(t)}$~\eqref{eq:next_theta2}, we obtain
\begin{equation*}
    \begin{aligned}
        \norm{\hat{\vp}(\hat{\vL}) - \hat{\vp}(\breve{\vL})}_2
        & \leq 
        C(\cL)
        O\left(\frac{\de L m^{1/2}}{n \rho d^{1/2}}\right)
        \sqrt{
        \sum_{i=1}^n \norm{
        \vtheta^{(t_i(t))} 
        - \left(\sum_{\tau = t_i(t)}^{t-1} \frac{\eta}{b} \sum_{j \in S_{\tau}} \nabla_{\vtheta} (\cL_j \circ~h_j)(\vtheta^{(\tau)})
        \right)
        - \vtheta^{(t_i(t))}
        }_{2,2}^2
        }\\
         & \leq 
         \eta
        C(\cL)
        O\left(\frac{\de L m^{1/2}}{n \rho d^{1/2}}\right)
        \sqrt{
        \sum_{i=1}^n \norm{
        \sum_{\tau = t_i(t)}^{t-1} \frac{1}{b} \sum_{j \in S_{\tau}} \nabla_{\vtheta} (\cL_j \circ~h_j)(\vtheta^{(\tau)})
        }_{2,2}^2
        }\\
    \end{aligned}
\end{equation*}
Let $A(\nabla \cL)$ a bound on the gradient of the per-example loss function.
Using Theorem~\ref{th:gradient_bound} and the chain rule
\begin{equation}
    \forall j,\, \forall \tau \quad 
    \norm{
    \nabla_{\vtheta} (\cL_j \circ~h_j)(\vtheta^{(\tau)})
    }_{2,2}
    \leq 
    A(\nabla \cL) O \left( \frac{m}{d}\right)
\end{equation}
And using the triangular inequality
\begin{equation}
    \begin{aligned}
        \norm{
        \sum_{\tau = t_i(t)}^{t-1} \frac{1}{b} \sum_{j \in S_{\tau}} \nabla_{\vtheta} (\cL_j \circ~h_j)(\vtheta^{(\tau)})
        }_{2,2}
        &\leq
        \sum_{\tau = t_i(t)}^{t-1} \frac{1}{b} \sum_{j \in S_{\tau}} \norm{\nabla_{\vtheta} (\cL_j \circ~h_j)(\vtheta^{(\tau)})}_{2,2}\\
        & \leq
        \sum_{\tau = t_i(t)}^{t-1} A(\nabla \cL) O \left( \frac{m}{d}\right)\\
        & \leq 
        A(\nabla \cL) O \left( \frac{m}{d}\right) (t - t_i(t))
    \end{aligned}
\end{equation}
As a result, we obtain
\begin{equation}
    \begin{aligned}
        \norm{\hat{\vp}(\hat{\vL}) - \hat{\vp}(\breve{\vL})}_2
         & \leq 
         \eta
        C(\cL)
        A(\nabla \cL)
        O\left(\frac{\de L m^{3/2}}{n \rho d^{3/2}}\right)
        \sqrt{
        \sum_{i=1}^n (t - t_i(t))^2
        }\\
    \end{aligned}
\end{equation}
For all $i$ and for any $\tau$ the probability that the sample $i$ is not in batch $S_{\tau}$ is lesser than 
$\left(1 - \alpha\right)^b$.

Therefore, for any $k \geq 1$ and for any $t$,
\begin{equation}
    P\left(t - t_i(t) \geq k\right) \leq \left(1 - \alpha\right)^{kb}
\end{equation}
For 
$k\geq \frac{1}{b}\Omega\left(\frac{\log^2(m)}{\log\left(\frac{1}{1-\alpha}\right)}\right)$,
we have
$\left(1 - \alpha\right)^{kb} \leq \exp\left(-\Omega\left(\log^2(m)\right)\right)$,
and thus with probability at least 
$1 - \exp\left(-\Omega\left(\log^2(m)\right)\right)$,
\begin{equation}
    \forall t,\quad t - t_i(t) \leq O\left(\frac{\log^2(m)}{b\log\left(\frac{1}{1-\alpha}\right)}\right)
\end{equation}
As a result, we finally obtain that with probability at least 
$1 - \exp\left(-\Omega\left(\log^2(m)\right)\right)$,
\begin{equation}
    \begin{aligned}
        \norm{\hat{\vp}(\hat{\vL}) - \hat{\vp}(\breve{\vL})}_2
         & \leq 
         \eta
        C(\cL)
        A(\nabla \cL)
        O\left(\frac{\de L m^{3/2}}{n \rho d^{3/2}}\right)
        \sqrt{n}
        O\left(\frac{\log^2(m)}{b\log\left(\frac{1}{1-\alpha}\right)}\right)
        \\
        & \leq
        \eta \alpha 
        O\left(
        \frac{\de L m^{3/2} \log^2(m)}{\alpha n^{1/2} \rho d^{3/2} b \log\left(\frac{1}{1-\alpha}\right)}
        \right)\\
        & \leq
        \eta \alpha K'
    \end{aligned}
\end{equation}

%%%%%%%%%%%%%%
\subsubsection{Proof of technical lemma 3}\label{p:tech_lemma_3}\ \\
Let us first denote
\begin{equation}
    \begin{aligned}
        A &= \left|
            \langle
            \nabla_{\vtheta} (R \circ \vL \circ h)(\vtheta^{(t)})
            - \sum_{i=1}^n 
            \bar{p}_i(\hat{\vL})\nabla_{\vtheta}(\cL_i \circ h_i)(\vtheta^{(t)})),
            \sum_{i=1}^n 
            \bar{p}_i(\hat{\vL})\nabla_{\vtheta}(\cL_i \circ h_i)(\vtheta^{(t)}))
            \rangle
            \right|\\
        &=
        \left|
            \langle
            \sum_{i=1}^n
            \left(
                \bar{p}_i(\breve{\vL})
                - \bar{p}_i(\hat{\vL})
            \right)
            \nabla_{\vtheta}(\cL_i \circ h_i)(\vtheta^{(t)})),
            \sum_{i=1}^n 
            \bar{p}_i(\hat{\vL})\nabla_{\vtheta}(\cL_i \circ h_i)(\vtheta^{(t)}))
            \rangle
            \right|\\
    \end{aligned}
\end{equation}
Using Cauchy-Schwarz inequality
\begin{equation}
    \begin{aligned}
        A 
        &=
        \left|
        \sum_{i=1}^n
        \left(
            \bar{p}_i(\breve{\vL})
            - \bar{p}_i(\hat{\vL})
        \right)
        \langle
            \nabla_{\vtheta}(\cL_i \circ h_i)(\vtheta^{(t)})),
            \sum_{j=1}^n 
            \bar{p}_j(\hat{\vL})\nabla_{\vtheta}(\cL_j \circ h_j)(\vtheta^{(t)}))
        \rangle
        \right|\\
        &\leq 
        \norm{\hat{\vp}(\hat{\vL}) - \hat{\vp}(\breve{\vL})}_2
        \sqrt{
            \sum_{i=1}^n\left(
                \langle
                \nabla_{\vtheta}(\cL_i \circ h_i)(\vtheta^{(t)})),
                \sum_{j=1}^n 
                \bar{p}_j(\hat{\vL})\nabla_{\vtheta}(\cL_j \circ h_j)(\vtheta^{(t)}))
                \rangle
            \right)^2
        }
    \end{aligned}
\end{equation}
Let 
\begin{equation}
    B = 
    % \sqrt{
    %         \sum_{i=1}^n\left(
                \langle
                \nabla_{\vtheta}(\cL_i \circ h_i)(\vtheta^{(t)})),
                \sum_{j=1}^n 
                \bar{p}_j(\hat{\vL})\nabla_{\vtheta}(\cL_j \circ h_j)(\vtheta^{(t)}))
                \rangle
        %     \right)^2
        % }
\end{equation}
Using again Cauchy-Schwarz inequality
\begin{equation}
    B
    % \langle
    % \nabla_{\vtheta}(\cL_i \circ h_i)(\vtheta^{(t)})),
    %         \sum_{j=1}^n 
    %         \bar{p}_j(\hat{\vL})\nabla_{\vtheta}(\cL_j \circ h_j)(\vtheta^{(t)}))
    % \rangle
    \leq 
    \norm{\nabla_{\vtheta}(\cL_i \circ h_i)(\vtheta^{(t)}))}_{2,2}
    \norm{\sum_{j=1}^n 
            \bar{p}_j(\hat{\vL})\nabla_{\vtheta}(\cL_j \circ h_j)(\vtheta^{(t)}))}_{2,2}
\end{equation}
As a result, $A$ becomes
\begin{equation}
    \begin{aligned}
        A
        & \leq
        \norm{\hat{\vp}(\hat{\vL}) - \hat{\vp}(\breve{\vL})}_2
        \norm{\sum_{j=1}^n 
            \bar{p}_j(\hat{\vL})\nabla_{\vtheta}(\cL_j \circ h_j)(\vtheta^{(t)}))
        }_{2,2}
        \sqrt{
            \sum_{i=1}^n
                \norm{\nabla_{\vtheta}(\cL_i \circ h_i)(\vtheta^{(t)}))}_{2,2}^2
        }\\
        & \leq
        \norm{\hat{\vp}(\hat{\vL}) - \hat{\vp}(\breve{\vL})}_2
        \norm{\sum_{j=1}^n 
            \bar{p}_j(\hat{\vL})\nabla_{\vtheta}(\cL_j \circ h_j)(\vtheta^{(t)}))
        }_{2,2}
        \sqrt{
            \sum_{i=1}^n
                \frac{1}{\alpha^2}\norm{\bar{p}_j(\hat{\vL})\nabla_{\vtheta}(\cL_i \circ h_i)(\vtheta^{(t)}))}_{2,2}^2
        }\\
        & \leq
        \frac{1}{\alpha}
        \norm{\hat{\vp}(\hat{\vL}) - \hat{\vp}(\breve{\vL})}_2
        \norm{\sum_{j=1}^n 
            \bar{p}_j(\hat{\vL})\nabla_{\vtheta}(\cL_j \circ h_j)(\vtheta^{(t)}))
        }_{2,2}^2
    \end{aligned}
\end{equation}
Using the triangular inequality, Theorem~\ref{th:gradient_bound}, and Lemma~\ref{p:tech_lemma_2}, we finally obtain
\begin{equation}
    \begin{aligned}
        A
        & \leq
        \frac{m}{\alpha d}
        \norm{\hat{\vp}(\hat{\vL}) - \hat{\vp}(\breve{\vL})}_2
        \sum_{j=1}^n 
            \norm{\bar{p}_j(\hat{\vL})\nabla_{h_j}\cL_j (h_j(\vtheta^{(t)}))
        }_{2,2}^2\\
        &\leq
        \eta 
        \frac{m}{d}
        K'
        \sum_{j=1}^n 
            \norm{\bar{p}_j(\hat{\vL})\nabla_{h_j}\cL_j (h_j(\vtheta^{(t)}))
        }_{2,2}^2
    \end{aligned}
\end{equation}

%%%%%%%%%%%%%%
\subsubsection{Proof of technical lemma 4}\label{p:tech_lemma_4}\ \\
We have
\begin{equation}
    \begin{aligned}
        \norm{
            \nabla_{h} (R \circ \vL)(h(\vtheta^{(t)}))
        }_{1,2}
        &=
        \sum_{j=1}^n
            \bar{p}_j(\breve{\vL})
            \norm{\nabla_{h_j}\cL_j (h_j(\vtheta^{(t)}))
        }_{2,2}\\
        &=
        \sum_{j=1}^n
            \bar{p}_j(\hat{\vL})
            \norm{\nabla_{h_j}\cL_j (h_j(\vtheta^{(t)}))
        }_{2,2}\\
        &
        + \sum_{j=1}^n
            \left(\frac{\bar{p}_j(\breve{\vL}) - \bar{p}_j(\hat{\vL})}{\bar{p}_j(\hat{\vL})}\right)
            \bar{p}_j(\hat{\vL})
            \norm{\nabla_{h_j}\cL_j (h_j(\vtheta^{(t)}))
        }_{2,2}\\
    \end{aligned}
\end{equation}
Using Cauchy-Schwarz inequality
\begin{equation}
    \begin{aligned}
        \norm{
            \nabla_{h} (R \circ \vL)(h(\vtheta^{(t)}))
        }_{1,2}
        &\leq
        \left(
        \sqrt{n} +
        \sqrt{
        \sum_{j=1}^n
            \left(\frac{\bar{p}_j(\breve{\vL}) - \bar{p}_j(\hat{\vL})}{\bar{p}_j(\hat{\vL})}\right)^2
            }
        \right)
        \sqrt{
        \sum_{j=1}^n
            \norm{\bar{p}_j(\hat{\vL})\nabla_{h_j}\cL_j (h_j(\vtheta^{(t)}))
            }_{2,2}^2
        }\\
    \end{aligned}
\end{equation}
Using Lemma~\ref{p:tech_lemma_2}
\begin{equation}
    \begin{aligned}
        \sum_{j=1}^n
            \left(\frac{\bar{p}_j(\breve{\vL}) - \bar{p}_j(\hat{\vL})}{\bar{p}_j(\hat{\vL})}\right)^2
        & \leq 
        \frac{1}{\alpha}
        \norm{\hat{\vp}(\hat{\vL}) - \hat{\vp}(\breve{\vL})}_2\\
        & \leq
        \eta K'
    \end{aligned}
\end{equation}
Therefore, we finally obtain
\begin{equation}
    \begin{aligned}
        \norm{
            \nabla_{h} (R \circ \vL)(h(\vtheta^{(t)}))
        }_{1,2}
        &\leq
        \left(
        \sqrt{n} + \eta K'
        \right)
        \sqrt{
        \sum_{j=1}^n
            \norm{\bar{p}_j(\hat{\vL})\nabla_{h_j}\cL_j (h_j(\vtheta^{(t)}))
            }_{2,2}^2
        }\\
    \end{aligned}
\end{equation}

% \newpage
% \input{appendices/incompressible/main}
\newpage
\section{Additional Results for \chapref{chap:twai}}
\subsection{Class-wise Quantitative Evaluation.}\label{appendix:class_results}

\begin{figure}[!htb]
    \centering
    \includegraphics[width=\linewidth,trim=0cm 0cm 0cm 3.8cm,clip]{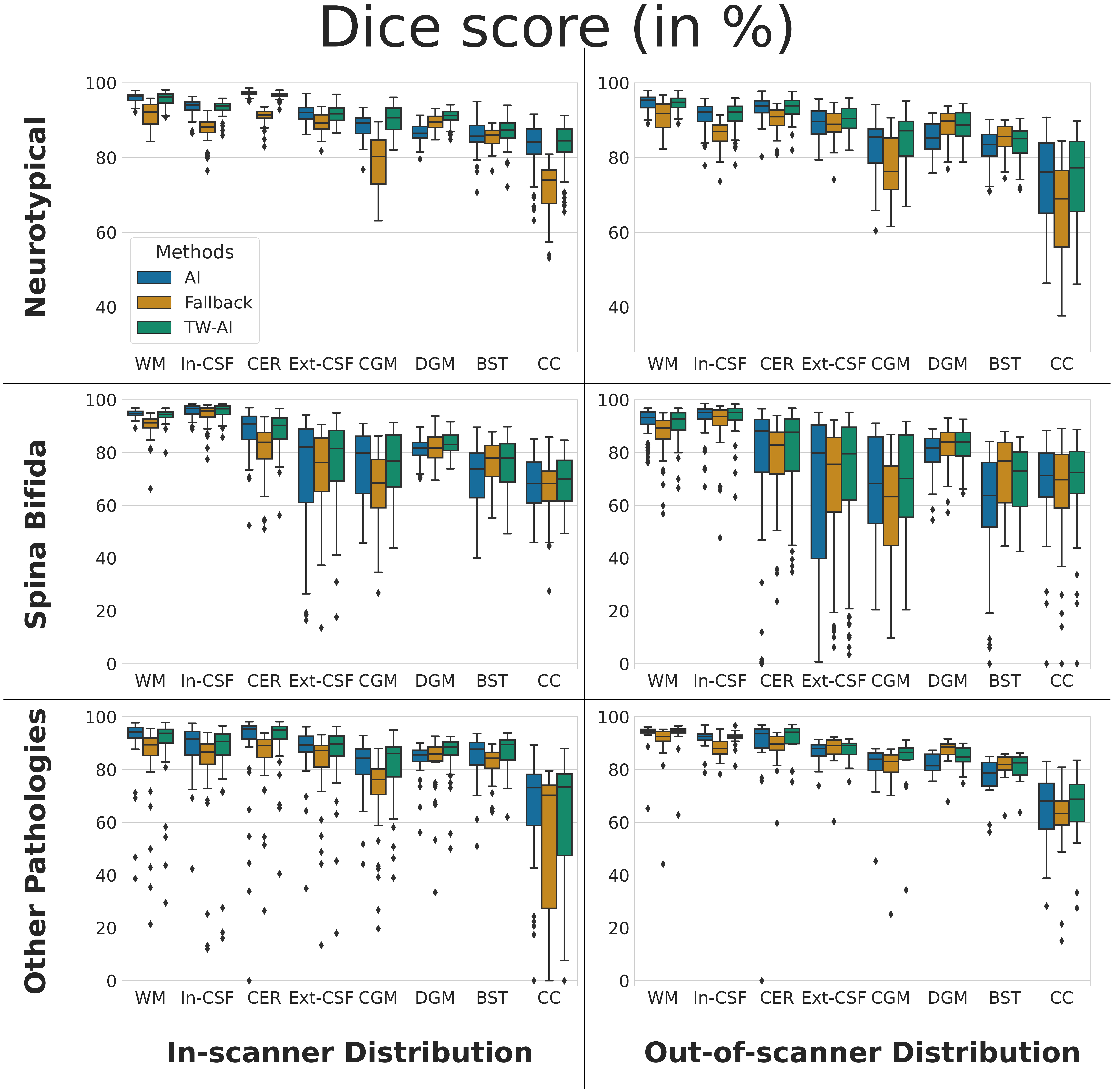}
    \caption{
    \textbf{Dice score (in \%) comparison of our AI, fallback, and trustworthy AI segmentation algorithms for fetal brain 3D MRI segmentation.}
    Dice scores are reported for all 3D MRI for $7$ tissue types:
    white matter (\textbf{WM}),
    intra-axial cerebrospinal fluid (\textbf{in-CSF}),
    cerebellum (\textbf{Cer}),
    extra-axial cerebrospinal fluid (\textbf{Ext-CSF}),
    cortical gray matter (\textbf{CGM}),
    deep gray matter (\textbf{DGM}),
    brainstem (\textbf{BST}),
    and corpus callosum (\textbf{CC}).
    Box limits are the first quartiles and third quartiles. The central ticks are the median values. The whiskers extend the boxes to show the rest of the distribution, except for points that are determined to be outliers.
    Outliers are data points outside the range median $\pm 1.5\times$ interquartile range.
    Fig.~\ref{fig:twai-dice_GA} was obtained from the same data after averaging the scores across regions of interest for each 3D MRI.
    }
    \label{fig:twai-dice_roi}
\end{figure}

\begin{figure}[!htb]
    \centering
    \includegraphics[width=\linewidth,trim=0cm 0cm 0cm 3.8cm,clip]{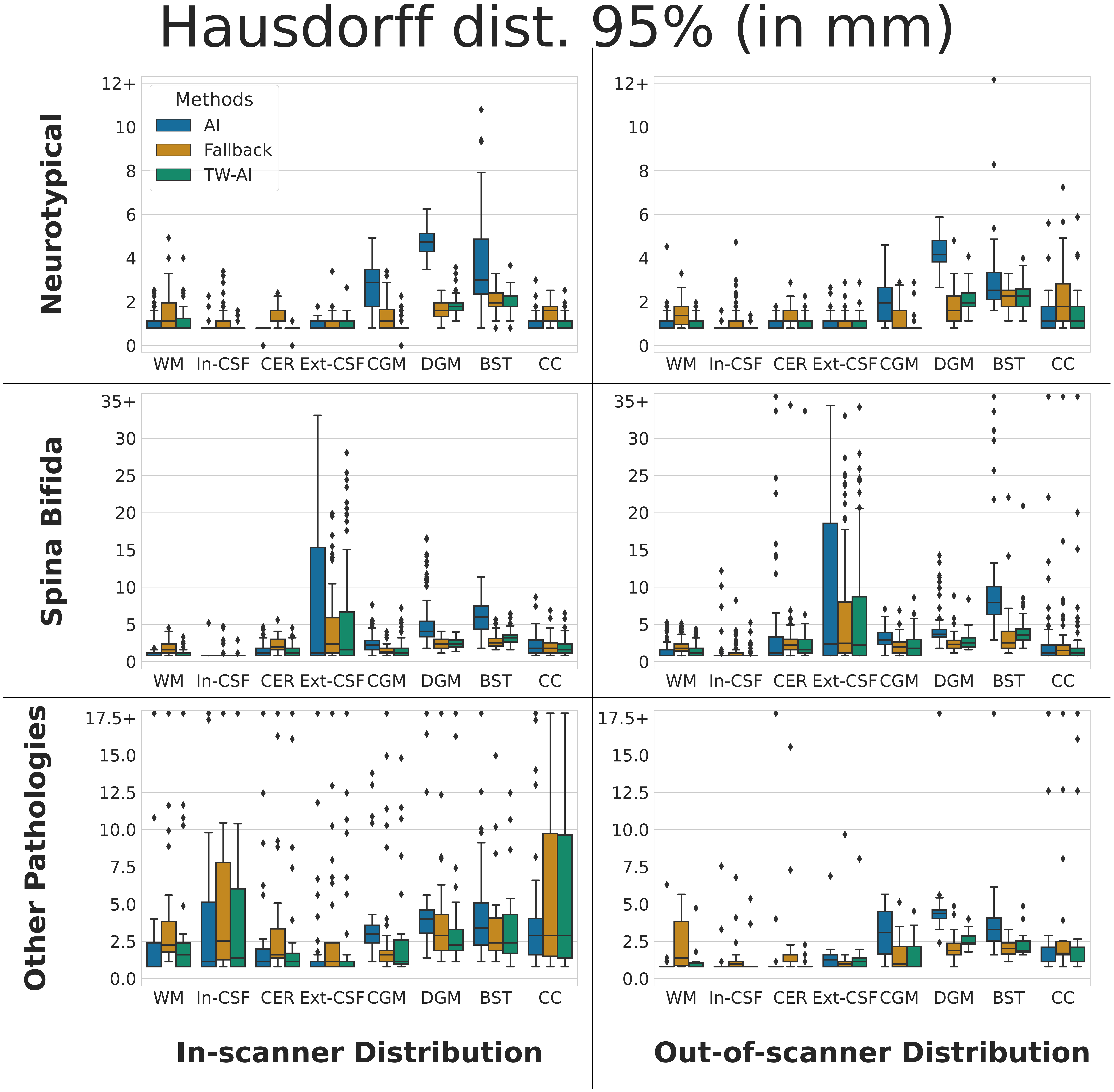}
    \caption{
    \textbf{Hausdorff distance (in mm) comparison of our AI, fallback, and trustworthy AI segmentation algorithms for fetal brain 3D MRI segmentation.}
    The organization and legend of this figure is the same as in Fig.~\ref{fig:twai-dice}, except that here the Hausdorff distance at $95\%$ percentile is reported in place of the Dice score.
    To improve the visualization we have clipped the distances to a maximum value. The clipped outliers are still visible on the top of each boxplot.
    Fig.~\ref{fig:twai-hausdorff} was obtained from the same data after averaging the scores across regions of interest for each 3D MRI.
    Box limits are the first quartiles and third quartiles. The central ticks are the median values. The whiskers extend the boxes to show the rest of the distribution, except for points that are determined to be outliers.
    Outliers are data points outside the range median $\pm 1.5\times$ interquartile range.
    }
    \label{fig:twai-hausdorff_roi}
\end{figure}

\begin{figure}[!htb]
    \centering
    \includegraphics[width=\linewidth,trim=0cm 0cm 0cm 3cm,clip]{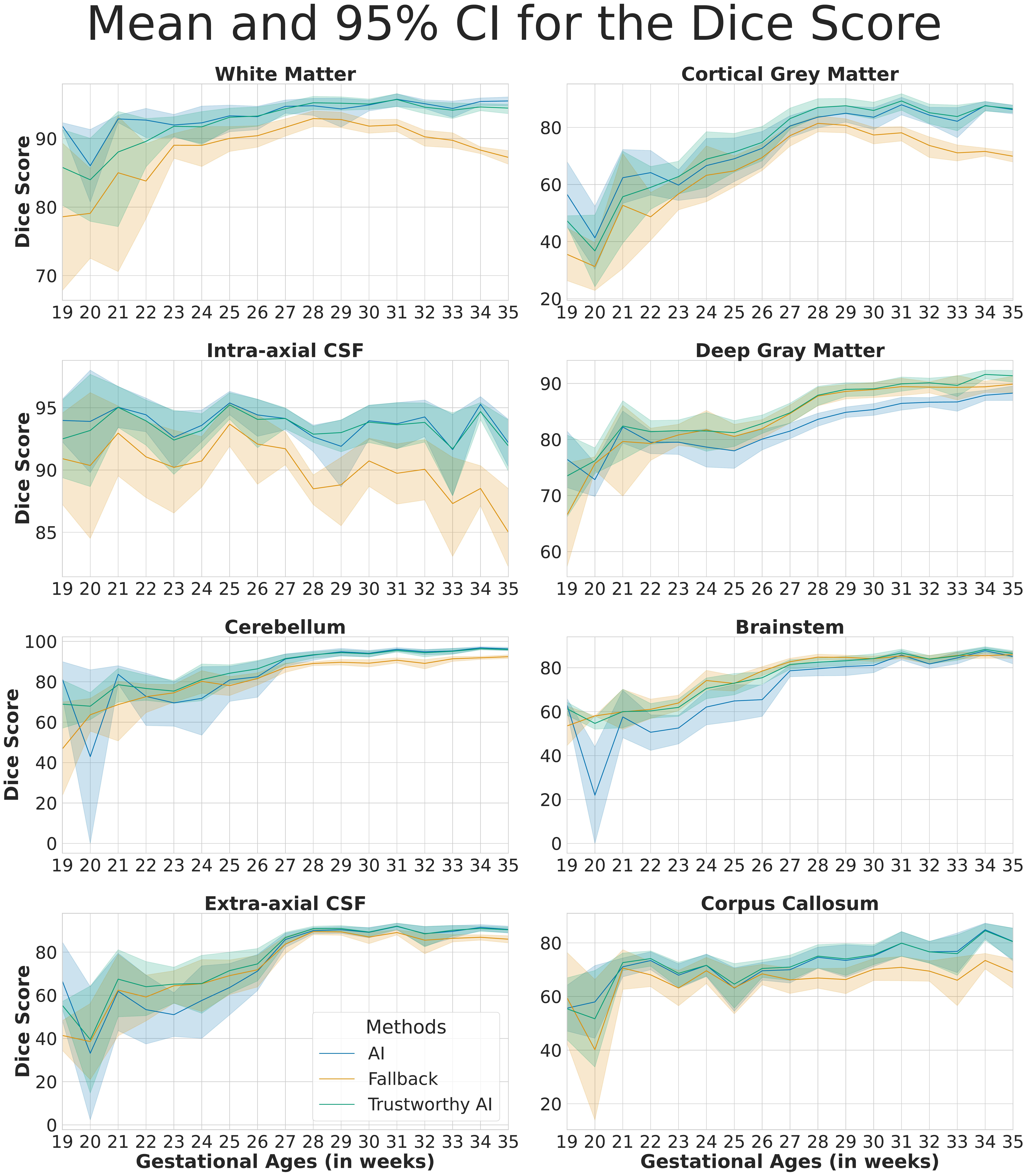}
    \caption{
    \textbf{Mean Dice score (in \%) and $95\%$ confidence interval as a function of the gestational age.}
    Here, we have used all the neurotypical and the spina bifida cases of the testing dataset.
    The trustworthy AI (in green) algorithm achieves similar or higher Dice scores than the best of the backbone AI (in blue) and the fallback (in orange) for all tissue type and all gestational age.
    Fig.~\ref{fig:twai-dice_GA} was obtained from the same data after averaging the scores across regions of interest for each 3D MRI.
    Box limits are the first quartiles and third quartiles. The central ticks are the median values. The whiskers extend the boxes to show the rest of the distribution, except for points that are determined to be outliers.
    Outliers are data points outside the range median $\pm 1.5\times$ interquartile range.
    }
    \label{fig:twai-dice_GA_roi}
\end{figure}

\begin{figure}[!htb]
    \centering
    \includegraphics[width=\linewidth,trim=0cm 0cm 0cm 3cm,clip]{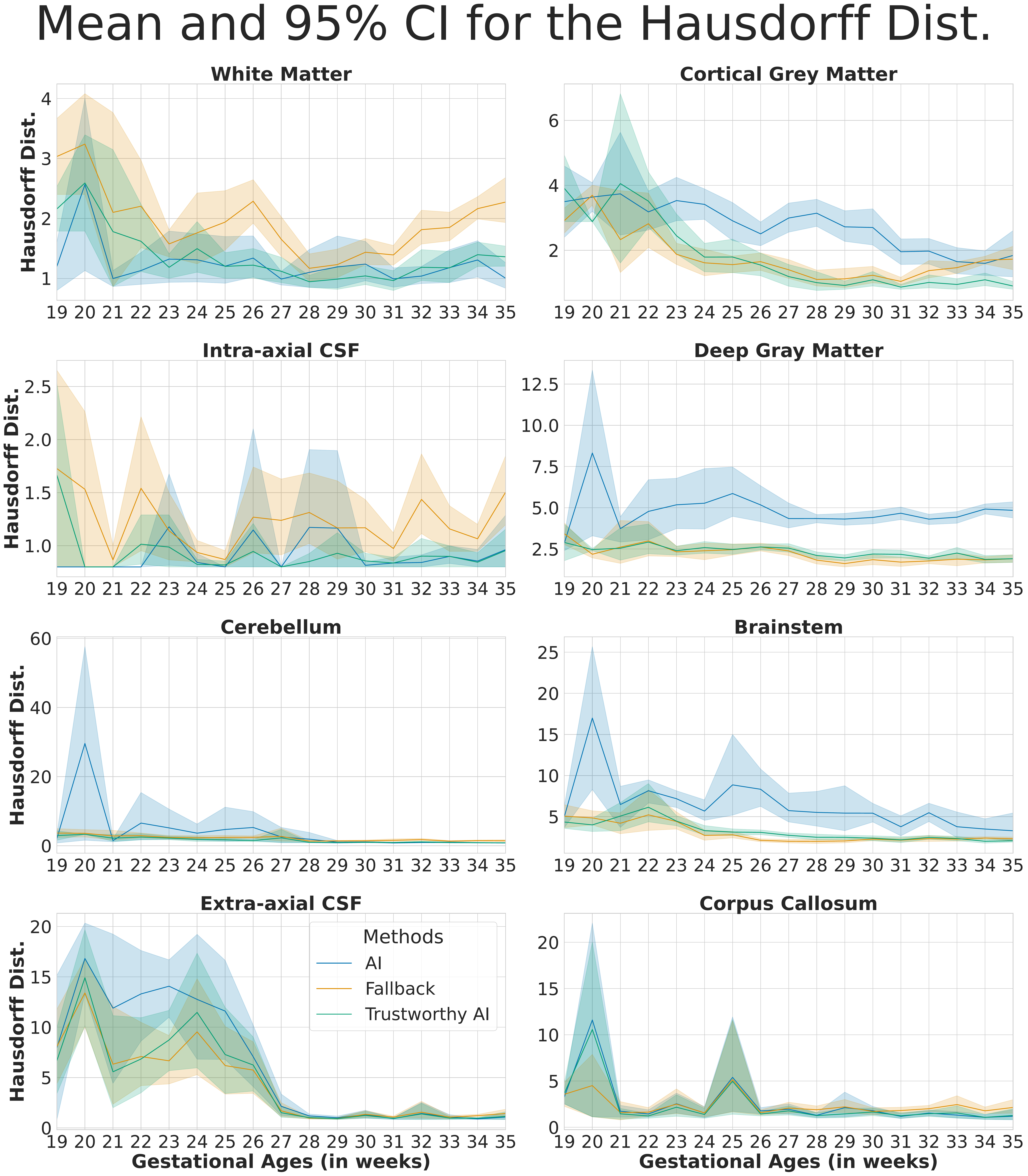}
    \caption{
    \textbf{Mean Hausdorff distance (in mm) and $95\%$ confidence interval as a function of the gestational age.}
    Here, we have used all the neurotypical and the spina bifida cases of the testing dataset.
    The trustworthy IA (in green) algorithm achieves similar or lower Hausdorff distance than the best of the backbone AI (in blue) and the fallback (in orange) for all tissue type and all gestational age.
    Fig.~\ref{fig:twai-hausdorff_GA} was obtained from the same data after averaging the scores across regions of interest for each 3D MRI.
    }
    \label{fig:twai-hausdorff_GA_roi}
\end{figure}

\begin{figure}
    \centering
    \includegraphics[width=\linewidth,trim=0cm 0cm 0cm 3cm,clip]{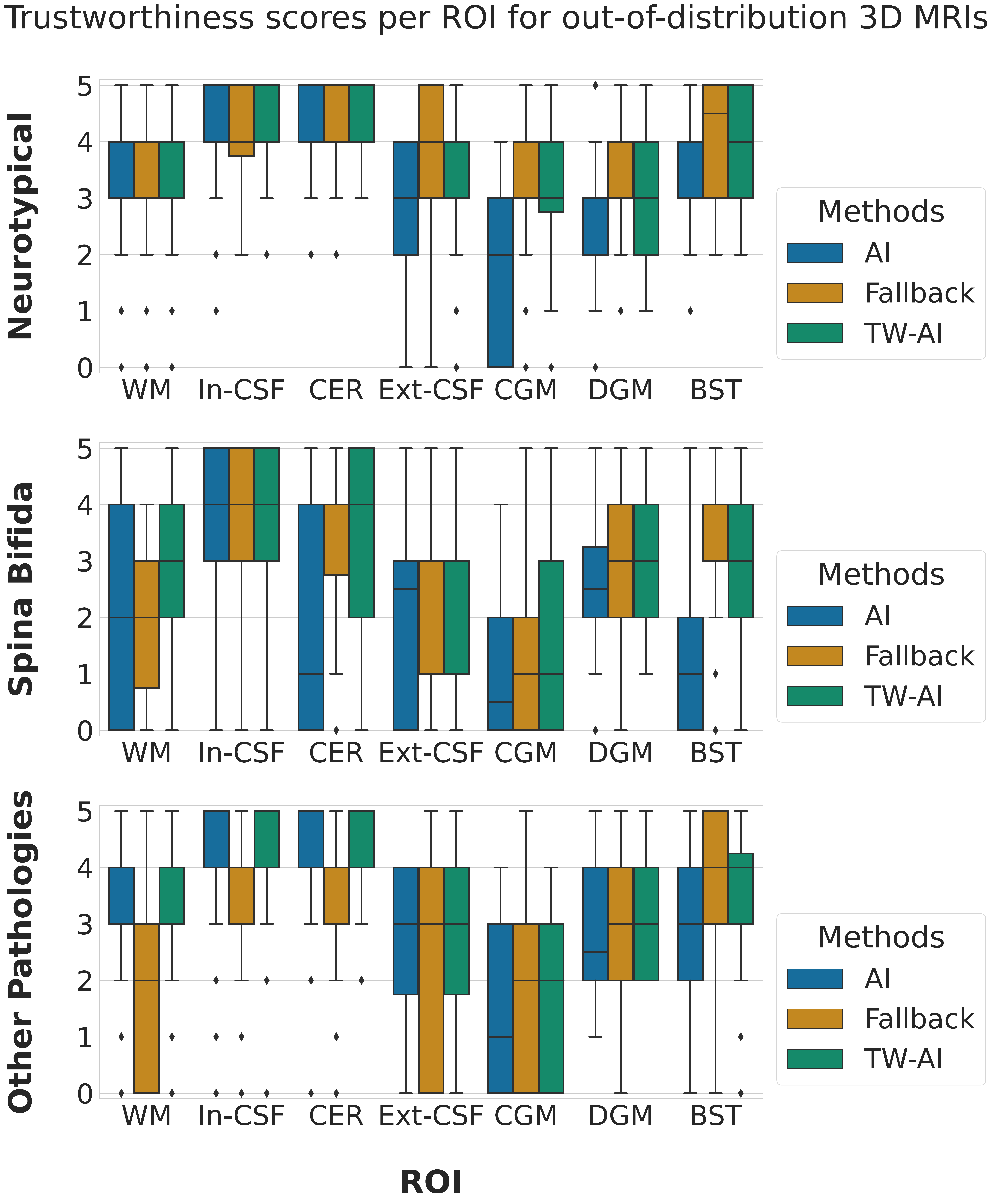}
    \caption{
    \textbf{Experts scores for out-of-distribution 3D MRIs.}
    % The scores were evaluated by a panel of \numraters{} experts
    % for the three segmentation algorithms for each case and for all the region of interests.
    % 
    A panel of \numraters{} experts performed \numscoring{} independent scoring for each ROI.
    The scores of different experts for a given region of interest and for a given 3D MRI were aggregated using averaging.
    Here, we have used $50$ out-of-scanner distribution 3D MRIs from the FeTA dataset ($20$ neurotypical, $20$ spina bifida, and $10$ other pathologies).
    Box limits are the first quartiles and third quartiles. The central ticks are the median values. The whiskers extend the boxes to show the rest of the distribution, except for points that are determined to be outliers.
    Outliers are data points outside the range median $\pm 1.5\times$ interquartile range.
    Fig.~\ref{fig:twai-scores} was obtained from the same data after averaging the scores across regions of interest for each 3D MRI.
    }
    \label{fig:twai-scores_roi}
\end{figure}

\FloatBarrier
\subsection{Tuning of the Registration Parameters}\label{appendix:registration_tuning}

\begin{table*}[h]
	\caption{\textbf{Tuning of the registration parameters.}
	We report the population average of the mean-class Dice score (DSC) in percentages.
    We also report the average number of volumes that need to be registered for each configuration. This number is approximately proportional to the computational time for the segmentation computation.
    The 3D MRIs used were the fold 0 of the training dataset.
    The row highlighted in \textcolor{blue}{\textbf{blue}} (resp. \textcolor{orange}{\textbf{orange}}) correspond to the value of $\Delta$GA selected for the neurotypical cases (resp. the spina bifida cases).
    A higher value of $\Delta$GA leads to using more volumes in the fallback, registration-based segmentation method.
    Hence, this difference of $\Delta$GA reflects the use of two neurotypical atlases~\cite{gholipour2017normative,wu2021age} while only one spina bifida atlas~\cite{fidon2021atlas} is available.
    }
	\begin{tabularx}{\linewidth}{*{6}{Y}}
	    \toprule
		\textbf{Atlas fusion} & \textbf{Atlas selection} & 
		$\Delta$GA &
		DSC \mbox{Control} & DSC \mbox{Spina bifida}
		& Average \mbox{\#volumes}\\
		\midrule
		Mean & Condition & 0 & 83.9 & 70.0 & 1.6\\
		Mean & Condition & 1 & 84.7 & 72.4 & 4.8\\
		Mean & Condition & 2 & 84.8 & 73.1 & 7.7\\
		Mean & Condition & 3 & 84.9 & 73.3 & 10.4\\
		Mean & Condition & 4 & 85.1 & 73.3 & 13.0\\
		Mean & All       & 0 & 82.0 & 66.4 & 3.1\\
		Mean & All       & 1 & 84.6 & 67.8 & 9.4\\
		Mean & All       & 2 & 85.0 & 66.9 & 15.3\\
		Mean & All       & 3 & 85.1 & 66.8 & 20.8\\
		Mean & All       & 4 & 85.2 & 66.8 & 26.0\\
		GIF  & Condition & 0 & 84.0 & 72.0 & 1.6\\
		\textcolor{blue}{\textbf{GIF}}  & \textcolor{blue}{\textbf{Condition}} & \textcolor{blue}{\textbf{1}} & 84.8 & 76.1 & 4.8\\
		GIF  & Condition & 2 & 85.0 & 76.9 & 7.7\\
		\textcolor{orange}{\textbf{GIF}}  & \textcolor{orange}{\textbf{Condition}} & \textcolor{orange}{\textbf{3}} & 85.2 & 77.6 & 10.4\\
		GIF  & Condition & 4 & 85.4 & 77.7 & 13.0\\
		GIF  & All       & 0 & 84.1 & 72.0 & 3.1\\
		GIF  & All       & 1 & 84.9 & 74.6 & 9.4\\
		GIF  & All       & 2 & 85.2 & 75.1 & 15.3\\
		GIF  & All       & 3 & 85.3 & 75.4 & 20.8\\
		GIF  & All       & 4 & 85.5 & 75.3 & 26.0\\
		\bottomrule
	\end{tabularx}
\end{table*}

\section{Proofs of \chapref{chap:twai}}
\subsection{Proof of the Formula \eqref{eq:anatomical_BPA} for the Anatomical BPA.}\label{appendix:proof_anatomical_BPA}

In this section we give the proof for the formula \eqref{eq:anatomical_BPA}:
For all voxel $\textbf{x}$ and all $\mathbf{C}' \subset \mathbf{C}$,
\begin{equation}
    \label{eq:anatomical_BPA2}
    \begin{aligned}
        m^{\anatomy}_{I, \textbf{x}}(\mathbf{C}\setminus \mathbf{C}') 
        &=\left(\bigoplus_{c \in \mathbf{C}} m^{(c)}_{\textbf{x}}\right)(\mathbf{C}\setminus \mathbf{C}')\\
        &=
        \prod_{c\in \mathbf{C}}
        \left(
        \delta_{c}(\mathbf{C}')m^{(c)}_{\textbf{x}}(\mathbf{C} \setminus \{c\}) 
        + (1 - \delta_{c}(\mathbf{C}'))m^{(c)}_{\textbf{x}}(\mathbf{C})
        \right)
    \end{aligned}
\end{equation}

To simplify the notations and without loss of generality, in this proof we assume that 
$\mathbf{C}=\left\{1,\ldots,K\right\}$ with $K$ the number of classes.
This simply amounts to renaming the classes by the numbers from $1$ to $K$.

Equation \eqref{eq:anatomical_BPA2}, that we want to prove, can then be rewritten as
\begin{equation}
    m^{\anatomy}_{I, \textbf{x}}(\mathbf{C}\setminus \mathbf{C}') =
    \prod_{c=1}^K
    \left(
        \delta_{c}(\mathbf{C}')m^{(c)}_{\textbf{x}}(\mathbf{C} \setminus \{c\}) 
        + (1 - \delta_{c}(\mathbf{C}'))m^{(c)}_{\textbf{x}}(\mathbf{C})
    \right)
\end{equation}

Let us first give the reader an intuition of the formula that we will prove by computing the Dempster's rule of combination for the first two BPAs $m_{\textbf{x}}^{(1)}$ and $m_{\textbf{x}}^{(2)}$.
To simplify the calculations, we will write the Dempster's rule of combination for complements of sets like in \eqref{eq:anatomical_BPA2}.

Let $\mathbf{C}' \subsetneq \mathbf{C}$, using the definition of Dempster's rule of combination \eqref{eq:ds}
and using the relation $\forall \mathbf{G},\mathbf{H} \subset \mathbf{C},\,\, (\mathbf{C}\setminus \mathbf{G})\cap (\mathbf{C}\setminus \mathbf{H}) = \mathbf{C}\setminus (\mathbf{G}\cup \mathbf{H})$
\begin{equation}
    \label{eq:complement_drc}
    m_{\textbf{x}}^{(1)} \oplus m_{\textbf{x}}^{(2)}(\mathbf{C}\setminus \mathbf{C}') =
    \frac{\sum_{\mathbf{G},\mathbf{H}\subset \mathbf{C}|\mathbf{G} \cup \mathbf{H} = \mathbf{C}'} m_{\textbf{x}}^{(1)}(\mathbf{C}\setminus \mathbf{G})m_{\textbf{x}}^{(2)}(\mathbf{C}\setminus \mathbf{H})}{1 - \sum_{\mathbf{G},\mathbf{H}\subset \mathbf{C}|\mathbf{G} \cup \mathbf{H} = \mathbf{C}} m_{\textbf{x}}^{(1)}(\mathbf{C}\setminus \mathbf{G})m_{\textbf{x}}^{(2)}(\mathbf{C}\setminus \mathbf{H})}
\end{equation}
Using the definition of $m_{\textbf{x}}^{(1)}$ and $m_{\textbf{x}}^{(2)}$ in \eqref{eq:anatomical_bpa},
$m_{\textbf{x}}^{(1)}(\mathbf{C}\setminus \mathbf{G})=0$ if $\mathbf{G}\not\in \left\{\emptyset, \{1\}\right\}$
and 
$m_{\textbf{x}}^{(2)}(\mathbf{C}\setminus \mathbf{H})=0$ if $\mathbf{H}\not\in \left\{\emptyset, \{2\}\right\}$.

This implies that the sum in the denominator is equal to zeros and that there are only four possible values of $\mathbf{C}'$ such that the numerator is non zeros, i.e. 
$\mathbf{C}'\in \{\emptyset, \{1\}, \{2\}, \{1, 2\}\}$. This gives
\begin{equation}
    \begin{aligned}
        \left(m_{\textbf{x}}^{(1)} \oplus m_{\textbf{x}}^{(2)}\right)(\mathbf{C}) &=
        m_{\textbf{x}}^{(1)}(\mathbf{C}) m_{\textbf{x}}^{(2)}(\mathbf{C})
        \\
        \left(m_{\textbf{x}}^{(1)} \oplus m_{\textbf{x}}^{(2)}\right)(\mathbf{C}\setminus \{1\}) &=
        m_{\textbf{x}}^{(1)}(\mathbf{C}\setminus \{1\}) m_{\textbf{x}}^{(2)}(\mathbf{C})
        \\
        \left(m_{\textbf{x}}^{(1)} \oplus m_{\textbf{x}}^{(2)}\right)(\mathbf{C}\setminus \{1\}) &=
        m_{\textbf{x}}^{(1)}(\mathbf{C}) m_{\textbf{x}}^{(2)}(\mathbf{C}\setminus \{2\})
        \\
        \left(m_{\textbf{x}}^{(1)} \oplus m_{\textbf{x}}^{(2)}\right)(\mathbf{C}\setminus \{1,2\}) &=
        m_{\textbf{x}}^{(1)}(\mathbf{C}\setminus \{1\}) m_{\textbf{x}}^{(2)}(\mathbf{C}\setminus \{2\})
    \end{aligned}
\end{equation}
and $\left(m_{\textbf{x}}^{(1)} \oplus m_{\textbf{x}}^{(2)}\right)(\mathbf{C}\setminus \mathbf{C}')=0$
for all other values of $\mathbf{C}'$.

A general formula is given by, $\mathbf{C}' \subset \mathbf{C}$,
\begin{equation}\tag{$\textup{H}_2$}
    \label{eq:h2}
    \begin{aligned}
        &\left(m_{\textbf{x}}^{(1)} \oplus m_{\textbf{x}}^{(2)}\right)(\mathbf{C}\setminus \mathbf{C}')=
        \\
        &\quad\prod_{c=3}^K\left(1 - \delta_{c}(\mathbf{C}')\right)
        \prod_{c=1}^2
        \left(
        \delta_{c}(\mathbf{C}')m^{(c)}_{\textbf{x}}(\mathbf{C} \setminus \{c\}) 
        + (1 - \delta_{c}(\mathbf{C}'))m^{(c)}_{\textbf{x}}(\mathbf{C})
        \right)
    \end{aligned}
\end{equation}
For clarity we remind that for all $c \in \mathbf{C}$, $\delta_c$ is the Dirac measure associated with $c$ defined as
\begin{equation}
    \forall \mathbf{C}' \subset \mathbf{C}, \quad \delta_c(\mathbf{C}') =
    \left\{
    \begin{array}{cc}
        1 & \textup{if}\,\, c \in \mathbf{C}'\\
        0 & \textup{if}\,\, c \not \in \mathbf{C}'
    \end{array}
    \right.
\end{equation}

The idea of the proof is to generalize formula \eqref{eq:h2} to all the combinations of the first $k$ anatomical BPAs until reaching $k=K$.

For all $k \in \{2,\ldots, K\}$, we defined \eqref{eq:hk} as
\begin{equation}\tag{$\textup{H}_k$}
    \label{eq:hk}
    \begin{aligned}
        &\left(\bigoplus_{c =1}^k m^{(c)}_{\textbf{x}}\right)(\mathbf{C}\setminus \mathbf{C}') =
        \\
        &\quad\prod_{c=k+1}^K\left(1 - \delta_{c}(\mathbf{C}')\right)
        \prod_{c=1}^k
        \left(
        \delta_{c}(\mathbf{C}')m^{(c)}_{\textbf{x}}(\mathbf{C} \setminus \{c\}) 
        + (1 - \delta_{c}(\mathbf{C}'))m^{(c)}_{\textbf{x}}(\mathbf{C})
        \right)
    \end{aligned}
\end{equation}
When $k=K$, the set of indices for the first product is empty and the product is equal to $1$ by convention.
Therefore, $\textup{H}_K$ is exactly the same as relation \eqref{eq:anatomical_BPA2} that we want to prove.
We will prove this equality by induction on the variable $k$ for $k$ from $2$ to $K$.

We have already proven \eqref{eq:h2}.
It remains to demonstrate that, for all $k$ from $2$ to $K-1$, $\textup{H}_k$ holds true implies that $\textup{H}_{k+1}$ also holds true.

Let $k \in \{2, \ldots, K-1\}$, let us assume that $\textup{H}_k$ is true.

Let $\mathbf{C}' \subsetneq \mathbf{C}$, using the same formula as in \eqref{eq:complement_drc}
\begin{equation}
    \label{eq:induction1}
    \begin{aligned}
    &\left(\bigoplus_{c =1}^k m^{(c)}_{\textbf{x}}\right) \oplus m_{\textbf{x}}^{(k+1)}(\mathbf{C}\setminus \mathbf{C}') =\\
    &\quad\quad\frac{\sum_{\mathbf{G},\mathbf{H}\subset \mathbf{C}|\mathbf{G} \cup \mathbf{H} = \mathbf{C}'} \left(\bigoplus_{c =1}^k m^{(c)}_{\textbf{x}}\right)(\mathbf{C}\setminus \mathbf{G})m_{\textbf{x}}^{(k+1)}(\mathbf{C}\setminus \mathbf{H})}{1 - \sum_{\mathbf{G},\mathbf{H}\subset \mathbf{C}|\mathbf{G} \cup \mathbf{H} = \mathbf{C}} \left(\bigoplus_{c =1}^k m^{(c)}_{\textbf{x}}\right)(\mathbf{C}\setminus \mathbf{G})m_{\textbf{x}}^{(k+1)}(\mathbf{C}\setminus \mathbf{H})}
    \end{aligned}
\end{equation}

Let us denote
\begin{equation}
    N = \sum_{\mathbf{G},\mathbf{H}\subset \mathbf{C}|\mathbf{G} \cup \mathbf{H} = \mathbf{C}} \left(\bigoplus_{c =1}^k m^{(c)}_{\textbf{x}}\right)(\mathbf{C}\setminus \mathbf{G})m_{\textbf{x}}^{(k+1)}(\mathbf{C}\setminus \mathbf{H})
\end{equation}
Let us first demonstrate that $N=0$.
Using the definition of $m_{\textbf{x}}^{(k+1)}$ in \eqref{eq:anatomical_bpa},
$m_{\textbf{x}}^{(k+1)}(\mathbf{C}\setminus \mathbf{G})=0$ if $\mathbf{G}\not\in \left\{\emptyset, \{k+1\}\right\}$.
Therefore, we need only to study the cases
$G\in \{\mathbf{C}, \mathbf{C}\setminus \{k+1\}\}$.

For $G = \mathbf{C}$, 
$\left(\bigoplus_{c =1}^k m^{(c)}_{\textbf{x}}\right)(\mathbf{C}\setminus \mathbf{G})
= \left(\bigoplus_{c =1}^k m^{(c)}_{\textbf{x}}\right)(\emptyset)
= 0
$ like for every basic probability assignment (BPA).

For $G = \mathbf{C}\setminus \{k+1\}$,
according to \eqref{eq:hk}, that we have assumed true,
\begin{equation}
\label{eq:N0}
    \begin{aligned}
        \left(\bigoplus_{c =1}^k m^{(c)}_{\textbf{x}}\right)(\mathbf{C}\setminus \mathbf{G}) 
        &= \left(\bigoplus_{c =1}^k m^{(c)}_{\textbf{x}}\right)(\{k+1\})
        \\
        &=\prod_{c=k+1}^K\left(1 - \delta_{c}(\mathbf{C}\setminus \{k+1\})\right)
        \prod_{c=1}^k m^{(c)}_{\textbf{x}}(\mathbf{C} \setminus \{c\})
    \end{aligned}
\end{equation}
We have to consider two cases, $k+1<K$ and $k+1=K$.

If $k+1<K$,
the second term in the first product of \eqref{eq:N0} is equal to $0$
and therfore
$\left(\bigoplus_{c =1}^k m^{(c)}_{\textbf{x}}\right)(\{k+1\})m^{(k+1)}_{\textbf{x}}(\mathbf{C}\setminus \{k+1\})=0$

If $k+1=K$,
we have
\begin{equation}
    \left(\bigoplus_{c =1}^k m^{(c)}_{\textbf{x}}\right)(\{k+1\})m^{(k+1)}_{\textbf{x}}(\mathbf{C}\setminus \{k+1\})
    = \prod_{c=1}^K m^{(c)}_{\textbf{x}}(\mathbf{C} \setminus \{c\})
\end{equation}
Voxel $\textbf{x}$ belongs to at least one of the class binary masks.
% because the mask partition the space of the image. 
Let us denote $c_0$ the binary mask to which is voxel $\textbf{x}$ belongs to.
Using the definition of $m_{\textbf{x}}^{(c_0)}$ in \eqref{eq:anatomical_bpa}, we have 
$m_{\textbf{x}}^{(c_0)}(\mathbf{C}\setminus \{c_0\})=0$.
Therefore, the product above is equal to $0$.
This allows us to conclude, in every case, that $N=0$.

Therefore, \eqref{eq:induction1} becomes
\begin{equation}
    \label{eq:induction2}
    \begin{aligned}
    &\left(\bigoplus_{c =1}^k m^{(c)}_{\textbf{x}}\right) \oplus m_{\textbf{x}}^{(k+1)}(\mathbf{C}\setminus \mathbf{C}') =\\
    &\quad\quad\sum_{\mathbf{G},\mathbf{H}\subset \mathbf{C}|\mathbf{G} \cup \mathbf{H} = \mathbf{C}'} \left(\bigoplus_{c =1}^k m^{(c)}_{\textbf{x}}\right)(\mathbf{C}\setminus \mathbf{G})m_{\textbf{x}}^{(k+1)}(\mathbf{C}\setminus \mathbf{H})
    \end{aligned}
\end{equation}
Similarly as before, due to the definition of $m_{\textbf{x}}^{(k+1)}$, we only need to study the cases of the sets $\textbf{G}$ that are solutions of
$\textbf{G}\cap \emptyset = \textbf{C}'$ or 
$\textbf{G}\cap \{k+1\} = \textbf{C}'$.
The first equality has the unique solution $\textbf{G}=\textbf{C}'$
and the second equality has either no solution, if $k+1\not\in \textbf{C}'$,
or two solutions $G\in\{\textbf{C}'\setminus \{k+1\}, \textbf{C}'\}$ if $k+1\in \textbf{C}'$.
Using the Dirac measure, we can treat all the cases at once and \eqref{eq:induction2} becomes
\begin{equation}
    \label{eq:induction3}
    \begin{aligned}
    &\left(\bigoplus_{c =1}^k m^{(c)}_{\textbf{x}}\right) \oplus m_{\textbf{x}}^{(k+1)}(\mathbf{C}\setminus \mathbf{C}') =
    \left(\bigoplus_{c =1}^k m^{(c)}_{\textbf{x}}\right)(\mathbf{C}\setminus \mathbf{C}')m_{\textbf{x}}^{(k+1)}(\mathbf{C})
    \\
    &\quad\quad\quad+ \delta_{k+1}(\textbf{C}') \left(\bigoplus_{c =1}^k m^{(c)}_{\textbf{x}}\right)(\mathbf{C}\setminus \mathbf{C}')m_{\textbf{x}}^{(k+1)}(\mathbf{C}\setminus \{k+1\})
    \\
    &\quad\quad\quad+ \delta_{k+1}(\textbf{C}') \left(\bigoplus_{c =1}^k m^{(c)}_{\textbf{x}}\right)(\mathbf{C}\setminus (\mathbf{C}'\setminus \{k+1\}))m_{\textbf{x}}^{(k+1)}(\mathbf{C}\setminus \{k+1\})
    \end{aligned}
\end{equation}
Using \eqref{eq:hk}, we can rewrite the second term of \eqref{eq:induction3} as
\begin{equation}
\begin{aligned}
    &\delta_{k+1}(\textbf{C}') \left(\bigoplus_{c =1}^k m^{(c)}_{\textbf{x}}\right)(\mathbf{C}\setminus \mathbf{C}') =
    \\
    &\quad
    \delta_{k+1}(\textbf{C}')
    \prod_{c=k+1}^K\left(1 - \delta_{c}(\mathbf{C}')\right)
        \prod_{c=1}^k
        \left(
        \delta_{c}(\mathbf{C}')m^{(c)}_{\textbf{x}}(\mathbf{C} \setminus \{c\}) 
        + (1 - \delta_{c}(\mathbf{C}'))m^{(c)}_{\textbf{x}}(\mathbf{C})
        \right)
\end{aligned}
\end{equation}
The product of the first two terms of the product on the right-hand side is
$\delta_{k+1}(\textbf{C}')(1-\delta_{k+1}(\textbf{C}'))=0$, independently to the value of $\textbf{C}'$.
Therefore, the second term of \eqref{eq:induction3} is equal to zeros.
Using \eqref{eq:hk}, and by remarking that 
\begin{equation}
    \begin{aligned}
        \delta_{k+1}(\mathbf{C}'\setminus \{k+1\})
        &=0\\
        \forall c \in \textbf{C}\setminus \{K+1\},\quad \delta_{k+1}(\mathbf{C}'\setminus \{k+1\})
        &=\delta_{k+1}(\mathbf{C}')
    \end{aligned}
\end{equation}
we obtain
\begin{equation}
\begin{aligned}
    &\left(\bigoplus_{c =1}^k m^{(c)}_{\textbf{x}}\right)(\mathbf{C}\setminus (\mathbf{C}'\setminus \{k+1\})) =
    \\
    &\quad
    \prod_{c=k+2}^K\left(1 - \delta_{c}(\mathbf{C}')\right)
        \prod_{c=1}^k
        \left(
        \delta_{c}(\mathbf{C}')m^{(c)}_{\textbf{x}}(\mathbf{C} \setminus \{c\}) 
        + (1 - \delta_{c}(\mathbf{C}'))m^{(c)}_{\textbf{x}}(\mathbf{C})
        \right)
\end{aligned}
\end{equation}

Using this equality and \eqref{eq:hk}, we can rewrite \eqref{eq:induction3} as
\begin{equation}
    \label{eq:induction4}
    \begin{aligned}
    &\left(\bigoplus_{c =1}^k m^{(c)}_{\textbf{x}}\right) \oplus m_{\textbf{x}}^{(k+1)}(\mathbf{C}\setminus \mathbf{C}') = 
    \\
    &
    m_{\textbf{x}}^{(k+1)}(\mathbf{C})
    (1 -\delta_{k+1}(\textbf{C}'))
    \prod_{c=k+2}^K\left(1 - \delta_{c}(\mathbf{C}')\right)
        \prod_{c=1}^k
        \left(
        \delta_{c}(\mathbf{C}')m^{(c)}_{\textbf{x}}(\mathbf{C} \setminus \{c\}) 
        + (1 - \delta_{c}(\mathbf{C}'))m^{(c)}_{\textbf{x}}(\mathbf{C})
        \right)
    \\
    &+ 
    \delta_{k+1}(\textbf{C}')
    m_{\textbf{x}}^{(k+1)}(\mathbf{C}\setminus \{k+1\})
    \prod_{c=k+2}^K\left(1 - \delta_{c}(\mathbf{C}')\right)
        \prod_{c=1}^k
        \left(
        \delta_{c}(\mathbf{C}')m^{(c)}_{\textbf{x}}(\mathbf{C} \setminus \{c\}) 
        + (1 - \delta_{c}(\mathbf{C}'))m^{(c)}_{\textbf{x}}(\mathbf{C})
        \right)
    \end{aligned}
\end{equation}
By grouping the two terms we eventually obtain that $\textup{H}_{k+1}$ holds true, i.e.
\begin{equation}
    \begin{aligned}
    &\left(\bigoplus_{c =1}^{k+1} m^{(c)}_{\textbf{x}}\right)(\mathbf{C}\setminus \mathbf{C}') = \\
    &\quad
    \prod_{c=k+2}^K\left(1 - \delta_{c}(\mathbf{C}')\right)
        \prod_{c=1}^{k+1}
        \left(
        \delta_{c}(\mathbf{C}')m^{(c)}_{\textbf{x}}(\mathbf{C} \setminus \{c\}) 
        + (1 - \delta_{c}(\mathbf{C}'))m^{(c)}_{\textbf{x}}(\mathbf{C})
        \right)
    \end{aligned}
\end{equation}

We have proved that $\textup{H}_{2}$ is true and we have proved that for all $k$ from $2$ to $K-1$, $\textup{H}_k$ holds true implies that $\textup{H}_{k+1}$ also holds true.
Therefore, using the induction principle, we conclude that $\textup{H}_{K}$ is true.$\quad \blacksquare$

\newpage
\subsection{Proof of Equality \eqref{eq:DRC_proba_anatomical_BPA}.}\label{appendix:proof_DRC_proba_anatomical_BPA}

In this section, we give a proof of \eqref{eq:DRC_proba_anatomical_BPA}.
It states that for all $c \in \mathbf{C}$,
\begin{equation}
    \label{eq:DRC_proba_anatomical_BPA2}
    \left(
    p_{I,\textbf{x}} \oplus m^{\anatomy}_{I,\textbf{x}}
    \right)(c) =
    \frac{p_{I,\textbf{x}}(c)m^{(c)}_{\textbf{x}}(\mathbf{C})}{\sum_{c'\in \mathbf{C}}p_{I,\textbf{x}}(c')m^{(c')}_{\textbf{x}}(\mathbf{C})}
\end{equation}

We start the proof from the Dempster's rule of combination for a probability and a BPA \eqref{eq:drc_proba}.
\begin{equation}
    \left(
        p_{I,\textbf{x}} \oplus m^{\anatomy}_{I,\textbf{x}}
    \right)(c) =
    \frac{p_{I,\textbf{x}}(c)\sum_{\textbf{F}\subset \mathbf{C}|c \in \textbf{F}} m^{\anatomy}_{I,\textbf{x}}(\textbf{F})}{1 - \sum_{c' \in \mathbf{C}} \sum_{\textbf{F} \subset \left(\mathbf{C}\setminus \{c'\}\right)} p_{I,\textbf{x}}(c')m^{\anatomy}_{I,\textbf{x}}(\textbf{F})}
\end{equation}
We now rewrite this equation using complement sets in the numerator to be able to use the formula \eqref{eq:anatomical_BPA} for the anatomical BPAs.
\begin{equation}
\label{eq:drc_proof1}
    \left(
        p_{I,\textbf{x}} \oplus m^{\anatomy}_{I,\textbf{x}}
    \right)(c) =
    \frac{p_{I,\textbf{x}}(c)\sum_{\textbf{G}\subset (\mathbf{C}\setminus \{c\})} m^{\anatomy}_{I,\textbf{x}}(\textbf{C}\setminus\textbf{G})}{1 - \sum_{c' \in \mathbf{C}} \sum_{\textbf{F} \subset \left(\mathbf{C}\setminus \{c'\}\right)} p_{I,\textbf{x}}(c')m^{\anatomy}_{I,\textbf{x}}(\textbf{F})}
\end{equation}
Let us first simplify the numerator.
Using \eqref{eq:anatomical_BPA} we obtain, for all $\textbf{G}\subset (\mathbf{C}\setminus \{c\}$,
\begin{equation}
\begin{aligned}
    &
    % \sum_{\textbf{G}\subset (\mathbf{C}\setminus \{c\})} 
    m^{\anatomy}_{I,\textbf{x}}(\textbf{C}\setminus\textbf{G})
    % \\
    % & 
    % \quad
    =
    % \sum_{\textbf{G}\subset (\mathbf{C}\setminus \{c\})}
    \prod_{c'\in \mathbf{C}}
    \left(
        \delta_{c'}(\mathbf{C}')m^{(c)}_{\textbf{x}}(\mathbf{C} \setminus \{c\}) 
        + (1 - \delta_{c'}(\mathbf{C}'))m^{(c')}_{\textbf{x}}(\mathbf{C})
    \right)
    \\
    & 
    \quad
    =
    m^{(c)}_{\textbf{x}}(\mathbf{C})
    \left(
    % \sum_{\textbf{G}\subset (\mathbf{C}\setminus \{c\})}
    \prod_{c'\in (\mathbf{C}\setminus \{c\})}
    \left(
        \delta_{c'}(\mathbf{C}')m^{(c')}_{\textbf{x}}(\mathbf{C} \setminus \{c'\}) 
        + (1 - \delta_{c'}(\mathbf{C}'))m^{(c')}_{\textbf{x}}(\mathbf{C})
    \right)
    \right)
\end{aligned}
\end{equation}
Therefore the term $m^{(c)}_{\textbf{x}}(\mathbf{C})$ can be factorized outside of the sum in the numerator of the right-hand side of \eqref{eq:drc_proof1}.
Let us denote the sum of the numerator, after factorization, as
\begin{equation}
    A_c = \sum_{\textbf{G}\subset (\mathbf{C}\setminus \{c\})}
    \prod_{c'\in (\mathbf{C}\setminus \{c\})}
    \left(
        \delta_{c'}(\mathbf{C}')m^{(c')}_{\textbf{x}}(\mathbf{C} \setminus \{c'\}) 
        + (1 - \delta_{c'}(\mathbf{C}'))m^{(c')}_{\textbf{x}}(\mathbf{C})
    \right)
\end{equation}
One can remark that the terms of the product are all independent of $c$.
In addition, the $c'$th terms of the product is either $m^{(c')}_{\textbf{x}}(\mathbf{C} \setminus \{c'\})$ or $m^{(c')}_{\textbf{x}}(\mathbf{C})$ depending on $\textbf{G}$ and when summing over all
$\textbf{G}\subset (\mathbf{C}\setminus \{c\})$ we obtain all the possible products.
Therefore, the sum can be factorized as
\begin{equation}
    A_c =
    \prod_{c'\in (\mathbf{C}\setminus \{c\})}
    \left(
        m^{(c')}_{\textbf{x}}(\mathbf{C} \setminus \{c'\}) 
        + m^{(c')}_{\textbf{x}}(\mathbf{C})
    \right)
\end{equation}
Using the definition of the anatomical BPAs, we have, for all $c \in \textbf{C}$,
$m^{(c')}_{\textbf{x}}(\mathbf{C} \setminus \{c'\}) + m^{(c')}_{\textbf{x}}(\mathbf{C})=1$.
As a result, we obtain $A_c = 1$.

This proves, that 
\begin{equation}
    \left(
        p_{I,\textbf{x}} \oplus m^{\anatomy}_{I,\textbf{x}}
    \right)(c) \propto p_{I,\textbf{x}}(c) m^{(c)}_{\textbf{x}}(\mathbf{C})
\end{equation}
And since
$\sum_{c \in \textbf{C}}\left(p_{I,\textbf{x}} \oplus m^{\anatomy}_{I,\textbf{x}}\right)(c) = 1$,
we can conclude without additional calculations that 
\begin{equation}
    \left(
    p_{I,\textbf{x}} \oplus m^{\anatomy}_{I,\textbf{x}}
    \right)(c) =
    \frac{p_{I,\textbf{x}}(c)m^{(c)}_{\textbf{x}}(\mathbf{C})}{\sum_{c'\in \mathbf{C}}p_{I,\textbf{x}}(c')m^{(c')}_{\textbf{x}}(\mathbf{C})}
\end{equation}
$\blacksquare$

\end{document}